\documentclass[a4paper,11pt]{article}
\pdfoutput=1 
             
\usepackage{jheppub} 
                     
\usepackage{mathrsfs}
\usepackage[T1]{fontenc} 
\usepackage{amsmath}
\usepackage{natbib}
\usepackage{graphicx}
\usepackage{subfigure}
\usepackage{mathrsfs}
\usepackage{multirow}
\usepackage{array}
\usepackage{dcolumn}
\usepackage{bm}
\usepackage[section]{placeins}

\title{\boldmath OCTOPUS: A Versatile, User-Friendly, and Extensible Public Code for General-Relativistic Ray-Tracing in Spherically Symmetric and Static Spacetimes}

\author[a,b,1]{Shiyang Hu,\note{Corresponding author.}}
\author[a]{Shijie Tan,}
\author[a,b]{Dan Li,}
\author[c]{Lina Zhang,}
\author[d]{Chen Deng,}
\author[e]{and Wenfu Cao}

\affiliation[a]{School of Mathematics and Physics, University of South China, \\ Hengyang, 421001, People's Republic of China}
\affiliation[b]{Hunan Provincial Key Laboratory of Mathematical Modeling and Scientific Computing, University of South China, \\ Hengyang, 421001, People's Republic of China}
\affiliation[c]{College of Science, Hunan Institute of Technology, \\ Hengyang, 421002, People's Republic of China}
\affiliation[d]{School of Astronomy and Space Science, Nanjing University, \\ Nanjing, 210023, People's Republic of China}
\affiliation[e]{School of Physics and Technology, University of Jinan, \\ Jinan, 250022, People's Republic of China}

\emailAdd{husy$\_$arcturus@163.com}
\emailAdd{sjtan2009@163.com}
\emailAdd{danli@usc.edu.cn}
\emailAdd{linazhang@hnit.edu.cn}
\emailAdd{dengchen@smail.nju.edu.cn}
\emailAdd{wfcao0622@163.com}

\abstract{This paper presents OCTOPUS, a relativistic ray-tracing algorithm developed within a Fortran-based, OpenMP-accelerated framework, designed for asymptotically flat, spherically symmetric curved spacetimes. The code efficiently and accurately computes key relativistic features---including the black hole event horizon, photon rings, critical curves, and innermost stable circular orbits---and simulates black hole shadows, redshift factor distributions, accretion disk images, toroidal images, as well as gravitational lensing, light curves, and gravitational radiation from hot-spots. OCTOPUS provides an automated, modular solution for qualitative studies of black hole observables and multi-messenger correlations between electromagnetic and gravitational signals in curved spacetime. Its implementation requires only the metric potential and its first-, second-, and third-order radial derivatives as input, ensuring low user barriers while remaining highly extensible and adaptable. Using a Schwarzschild black hole surrounded by a Dehnen-type dark matter halo, we thoroughly validate the algorithm's precision, efficiency, and functionality, and investigate how dark matter halo parameters affect observational signatures. Our results demonstrate that increasing the scale and density of the dark matter halo strengthens the spacetime's gravitational field, an effect clearly reflected in black hole images and supported by hot-spot light curve signatures. A future version of OCTOPUS, with expanded capabilities for axisymmetric spacetimes, is planned for release.}

\begin{document}
\maketitle
\flushbottom

\section{Introduction}
The groundbreaking images of supermassive black holes in M87$^{*}$ and Sagittarius A$^{*}$, released by the Event Horizon Telescope collaboration \cite{Akiyama et al. (2019),Akiyama et al. (2022)}, have sparked unprecedented enthusiasm for simulations of compact object shadows \cite{Gralla et al. (2019),Narayan et al. (2019),Tian and Zhu (2019),Zeng et al. (2020),Li et al. (2020),Zhang et al. (2020),Hou et al. (2021a),Peng et al. (2021),Bronzwaer and Falcke (2021),Li and He (2021),Vincent et al. (2022),Afrin and Ghosh (2022),Zeng et al. (2022),Hu et al. (2022a),Hou et al. (2022a),Hou et al. (2022b),Vagnozzi et al. (2023),Hu et al. (2023),Heydari-Fard et al. (2023a),Meng et al. (2023),Wang et al. (2023),Yang et al. (2023),Gao et al. (2023),Cao et al. (2023),Heydari-Fard et al. (2023b),Zhang et al. (2023),Li et al. (2024a),Li et al. (2024b),Hu et al. (2024),Zhang et al. (2024a),Zhang et al. (2024b),Chen et al. (2025a),Hu et al. (2025a),Li et al. (2025),Yang et al. (2025),Guo et al. (2025),Liang et al. (2025),Wang et al. (2025a),Chen et al. (2025b),Xiong et al. (2025),He et al. (2025a),He et al. (2025b),Zeng et al. (2025a)}, Polarized images \cite{Chen et al. (2020),Zhang et al. (2021),Hu et al. (2022b),Qin et al. (2022),Chen et al. (2022),Zhang et al. (2024c),Shi et al. (2024),Chen et al. (2024a),Hou et al. (2025),Angelov et al. (2025),Wang et al. (2025b)}, holographic images \cite{He et al. (2025c),Zeng et al. (2025b),Chen et al. (2025c),Chen et al. (2025d),Zeng et al. (2025c)}, hot-spot images \cite{Huang et al. (2024a),Chen et al. (2024b),Wu and Chen (2024),Zhu (2025a),Zhu (2025b),Zhang et al. (2025)}, and lensing phenomena \cite{Gralla and Lupsasca (2021),Gao and Xie (2021),Wu et al. (2021),He et al. (2024),He et al. (2025d),Wang et al. (2025c),Wang et al. (2025d),Huang et al. (2025)}. This achievement has opened new observational windows into curved spacetimes, representing not merely a milestone in probing black hole physics and extreme gravitational environments but, more profoundly, a paradigm of perfect integration between theoretical prediction and observational verification.

To establish a comprehensive framework for matching observational black hole images with theoretical predictions, a complete numerical simulation pipeline is required. This typically involves three interconnected processes: First, simulating the accretion environment around the black hole using general relativistic magnetohydrodynamics (GRMHD) \cite{Porth et al. (2019)} to obtain stabilized dynamical parameters and fluid properties of the accretion flow. Second and third, the parallel procedures of general relativistic radiative transfer (GRRT) \cite{Wu et al. (2006),Younsi et al. (2012),Gammie and Leung (2012),Dexter (2016),Takahashi and Umemura (2017),Pelle et al. (2022),Kawashima et al. (2023),Sharma et al. (2023),Huang et al. (2024b)} and ray-tracing---the former integrates the radiative transfer equation along light paths through the emitting medium to determine the specific intensity based on emission mechanisms, while the latter bridges the source (accretion disk) and observer by simulating photon propagation to resolve image features such as the shadow and light rings. It should be emphasized that both GRMHD and GRRT simulations are exceptionally sophisticated, complex, and computationally demanding processes. Existing public codes in the scientific community present significant accessibility barriers, with resource requirements often exceeding the capacity of individual research groups. To address this challenge, several researchers have developed innovative solutions based on time-averaged GRMHD image fitting, which cleverly reduces disk radiation to analytical functions of source coordinates \cite{Gralla et al. (2020),Chael et al. (2021),Hou et al. (2021b)}. This approach not only preserves qualitatively reliable features of black hole images but also circumvents the computational complexity of full GRMHD/GRRT simulations. In other words, ray-tracing can be decoupled from GRRT and employed as an independent tool for qualitatively simulating the observational characteristics of compact celestial objects. The computation of the specific intensity along light rays depends on the intersection points between the rays and the radiation sources, as well as on the adopted analytical radiation models. Thus, the focus of the following discussion turns to the general-relativistic ray-tracing code.

From a physical perspective, light rays emitted by accreting matter around black holes propagate through spacetime and are eventually captured by distant observers. This fundamental process enables the construction of forward ray-tracing schemes \cite{Zhou et al. (2025)}. However, it is crucial to emphasize that such schemes typically require the black hole spacetime to be integrable. In non-integrable spacetimes, the absence of sufficient constants of motion prevents efficient and accurate bridging between emission sources and observers, rendering forward ray-tracing ineffective. Remarkably, the principle of time-reversal invariance in photon propagation offers an alternative approach: by launching light rays from the observer's position and integrating backward along the time axis, one achieves equivalent tracing results. This foundational principle underpins the widely adopted backward ray-tracing algorithm in contemporary scientific practice.

The backward ray-tracing algorithm has undergone extensive development over the years, with numerous distinctive and innovative advances made by the scientific community. Karas et al. classified light trajectories into distinct categories based on black hole spin and observer inclination, and employed Chebyshev polynomials to fit these trajectories, thereby establishing an efficient ray-tracing algorithm tailored to Kerr spacetime \cite{Karas et al. (1992)}. This approach is particularly useful for astrophysical problems such as modeling photometric light curves of hot-spots. GYOTO \cite{Vincent et al. (2011)}, a public general relativistic ray-tracing code developed by Vincent et al., is applicable to both analytical Kerr spacetime and numerically metric based on the $3+1$ decomposition. It is capable of simulating black hole images, accretion disk spectra, and trajectories of massive particles. The code employs a fourth-order Runge-Kutta method (RK4) for integrating geodesic equations and uses the Hamiltonian constraint to calibrate outputs, thereby improving numerical accuracy. Notably, GYOTO features a graphical user interface (GUI), significantly enhancing user accessibility. Psaltis and Johannsen have conducted a series of studies on ray-tracing advancements \cite{Johannsen and Psaltis (2010a),Johannsen and Psaltis (2010b),Psaltis and Johannsen (2012),Baubock et al. (2012)}, employing their algorithm to investigate the influence of quadrupole moments on the observational signatures of quasi-Kerr black holes and neutron stars. Their method implements an innovative adaptive step-size strategy for geodesic integration, dynamically adjusting the step-size based on the rate of change of the light vector, thereby improving the accuracy of ray-tracing. This adaptive step-size approach has also been incorporated into GRay, a GPU-accelerated curved-spacetime ray-tracing algorithm developed by Chen et al. \cite{Chen et al. (2013),Chen et al. (2018)}. In this method, the geodesic integrator is configured to automatically switch from the RK4 to the forward Euler scheme while reducing the step-size when rays pass near the black hole's polar regions, ensuring stable integration through these numerically challenging areas. Building upon the ray initialization scheme developed by Psaltis and Johannsen, Bambi developed a ray-tracing algorithm capable of simulating black hole images and computing accretion disk emission lines within the Novikov-Thorne theoretical framework \cite{Bambi (2012)}. Developed within the MATLAB GUI environment by Chen et al., KERTAP \cite{Chen et al. (2015)} is specialized for image simulation, redshift, and polarization calculations in Kerr spacetime. Under specific configurations, it achieves a computational speed of $97$ light rays per second. The integrability of Kerr spacetime allows the null geodesic equations to be expressed analytically in terms of elliptic functions. Leveraging this property, Yang and Wang developed an analytic ray-tracing algorithm for Kerr spacetime---YNOGK \cite{Yang and Wang (2013)}---demonstrating its performance across various astrophysical applications, including accretion disk imaging, gravitational lensing, hot-spot light curves, and disk spectra. For further reference on elliptic-function-based analytic ray-tracing methods, see \cite{Gralla and Lupsasca (2021),Luminet (1979),Alejandro et al. (2023),Wang et al. (2024),Guo et al. (2024)}. Building on the ray initialization scheme described in \cite{Younsi et al. (2016)}, Pu et al. developed ODYSSEY \cite{Pu et al. (2016)}, a ray-tracing algorithm for Kerr spacetime capable of mapping spacetime structures, simulating black hole shadows and images, and modeling hot-spot light curves. With GPU acceleration, ODYSSEY achieves high computational efficiency and has been widely adopted in astrophysical research \cite{Kimpson et al. (2019),Lin et al. (2022)}. Working within the Zero Angular Momentum Observer (ZAMO) framework, Cunha et al. developed a ray-tracing algorithm and conducted in-depth investigations into how scalar hair influences black hole images \cite{Cunha et al. (2015),Cunha et al. (2016)}. Similarly, Hu et al., also working within the ZAMO reference frame, introduced a fisheye camera projection method for black hole imaging and developed a corresponding ray-tracing algorithm \cite{Hu et al. (2021)}. This algorithm has been widely applied in black hole image simulations \cite{Hou et al. (2022a),Zhang et al. (2024a),Zhang et al. (2024b),Zhong et al. (2021)} and has been further refined into Coport \cite{Huang et al. (2024b)}, which includes polarization calculations. 

Furthermore, the scientific community has developed numerous relativistic ray-tracing algorithms tailored to different scenarios \cite{Dexter and Agol (2009),Muller and Frauendiener (2012),Muller (2014),James et al. (2015),Ayzenberg and Yunes (2018),Huang et al. (2018),Cadavid et al. (2022),Davelaar and Haiman (2022a),Garnier (2023)}. However, it is worth noting that these sophisticated algorithms often present notable accessibility barriers, especially when adapting to different black hole spacetimes. Moreover, individual implementations typically offer limited functionality. To address these challenges, we aimed to develop a highly modular, efficient, and versatile relativistic ray-tracing algorithm that is easy to modify and extend, enabling users to customize target black hole spacetimes with minimal programming overhead. This forms the primary motivation behind OCTOPUS. On a related note, while the current version of OCTOPUS is designed for asymptotically flat, spherically symmetric spacetimes, a version supporting axisymmetric spacetimes is planned for future release.

Developing relativistic ray-tracing codes for spherically symmetric spacetimes is justified by multiple compelling rationales. First, from a pedagogical perspective, these spacetimes represent the most fundamental scenario encountered by beginners, and creating efficient, user-friendly codes for such systems contributes meaningfully to scientific education and training. Second, astrophysically, the existence of non-spinning or extremely low-spin black holes cannot be ruled out; indeed, observational evidence suggests such objects may exist \cite{Gou et al. (2010),Steiner et al. (2012),Steiner et al. (2014)}, making dedicated studies of these systems scientifically valuable. Meanwhile, current observational techniques require further refinement to unambiguously extract the influence of spin on black hole image features. Additionally, the mathematical simplicity of spherical symmetry ensures low algorithmic complexity, ease of dissemination, and broad applicability across research contexts.

The remainder of this paper is organized as follows. Section 2 provides a detailed description of the algorithm's architecture, underlying theoretical foundations, key mathematical expressions, and functional capabilities. Section 3 presents comprehensive tests of the algorithm using a Schwarzschild spacetime surrounded by a Dehnen-type dark matter halo, evaluating its precision, functionality, and computational efficiency. This section also reveals characteristic observational signatures of the target black hole and examines how dark matter halo parameters influence spacetime properties. Finally, section 4 offers concluding remarks.
\section{Code description}
Our code efficiently and accurately computes key relativistic features, including the black hole event horizon, photon ring, critical impact parameter, and innermost stable circular orbit (ISCO), in asymptotically flat spherically symmetric spacetimes. It performs four primary functions: (1) numerical simulation of black hole images with an optically thin, geometrically thin accretion disk; (2) modeling of gravitational lensing phenomena, such as Einstein rings, for point sources; (3) simulating light curves of hot-spots; and (4) projecting hot-spot trajectories onto the observer's screen. To ensure both user-friendliness and extensibility, the implementation requires only the target black hole metric potential and its first-, second-, and third-order radial derivatives as essential inputs. Let us begin this journey with the spacetime line element.
\subsection{Target spacetime}
In the coordinate system $x^{\alpha}=(t,r,\theta,\varphi)$, the general form of the asymptotically flat spherically symmetric black hole metric line element can be expressed as
\begin{equation}\label{1}
\textrm{d}s^{2} = g_{tt}\textrm{d}t^{2} + g_{rr}\textrm{d}r^{2} + g_{\theta\theta}\textrm{d}\theta^{2} + g_{\varphi\varphi}\textrm{d}\varphi^{2},
\end{equation}
where $g_{\mu\nu}$ represents the covariant metric tensor. In particular, for most cases, $g_{tt}=-1/g_{rr}=-f(r)$, where $f(r)$ denotes the metric potential\footnote{It should be noted that certain black hole models, such as deformed Schwarzschild black holes, do not satisfy the condition $g_{tt}g_{rr}=-1$. Nevertheless, our code can be readily adapted to these cases with minor modifications, while maintaining full functionality \cite{Li et al. (2024b)}.}. On the other hand, the metric components satisfy $g_{\theta\theta}=A(r)r^{2}$ and $g_{\varphi\varphi}=A(r)r^{2}\sin^{2}\theta$, with $A(r) \sim 1$ at spatial infinity to preserve the spherically symmetric asymptotic structure. Indeed, for most asymptotically flat, spherically symmetric black holes \cite{Vagnozzi et al. (2023)}, the angular components simplify to $g_{\theta\theta}=r^{2}$ and $g_{\varphi\varphi}=r^{2}\sin^{2}\theta$. Consequently, our target black hole metric takes the explicit form:
\begin{equation}\label{2}
\textrm{d}s^{2} = -f(r)\textrm{d}t^{2} + \frac{\textrm{d}r^{2}}{f(r)} + r^{2}\textrm{d}\theta^{2} + r^{2}\sin^{2}\theta\textrm{d}\varphi^{2},
\end{equation}
which represents a common class of metrics in both astrophysical applications and modified gravity theories. This formulation ensures that our algorithm maintains computational efficiency while allowing flexible adaptation to different theoretical models, thereby guaranteeing broad applicability.

The event horizon radius $r_{\textrm{eh}}$ satisfies the condition $f(r_{\textrm{eh}})=0$. In our code, we employ the Newton iteration method, 
\begin{equation}\label{3}
r_{\textrm{eh}}^{n} = r_{\textrm{eh}}^{n-1} - \frac{f(r_{\textrm{eh}}^{n-1})}{f^{\prime}(r_{\textrm{eh}}^{n-1})}, 
\end{equation}
to solve this equation efficiently and with high numerical precision. Here, $r_{\textrm{eh}}^{n}$ and $r_{\textrm{eh}}^{n-1}$ represent the intermediate values of the event horizon radius at the $n$-th and $(n-1)$-th steps, respectively, where the prime symbol denotes the first-order derivative with respect to $r$. The iteration converges when the difference between successive values satisfies $|r_{\textrm{eh}}^{n}-r_{\textrm{eh}}^{n-1}| \leq 10^{-15}$, at which point $r_{\textrm{eh}}^{n}$ is identified as the event horizon radius $r_{\textrm{eh}}$. The event horizon constitutes a fundamental parameter in ray-tracing computations, serving dual physical roles: on the one hand, as the ultimate destination for light rays that shape the black hole's shadow, and on the other hand, as the natural inner boundary for accretion disk structure.
\subsection{Critical photon orbit}
In the spacetime described by \eqref{2}, the motion of photons can be governed by the Lagrangian
\begin{eqnarray}\label{4}
\mathscr{L} &=& \frac{1}{2}g_{\mu\nu}\dot{x^{\mu}}\dot{x^{\nu}} \nonumber \\
&=&\frac{1}{2}\left[-f(r)\dot{t}^{2} + \frac{\dot{r}^{2}}{f(r)} + r^{2}\dot{\theta}^{2} + r^{2}\sin^{2}\theta\dot{\varphi}^{2}\right],
\end{eqnarray}
where $\dot{x}^{\mu}$ represents the four-velocity of particles, and its relation to the conjugate momentum $p_{\mu}$ is given by
\begin{eqnarray}
p_{t} &=& g_{tt}\dot{t} = -f(r)\dot{t}, \label{5} \\
p_{r} &=& g_{rr}\dot{r} = \frac{\dot{r}}{f(r)}, \label{6} \\
p_{\theta} &=& g_{\theta\theta}\dot{\theta} = r^{2}\dot{\theta}, \label{7} \\
p_{\varphi} &=& g_{\varphi\varphi}\dot{\varphi} = r^{2}\sin^{2}\theta\dot{\varphi}. \label{8}
\end{eqnarray}

Since the metric does not explicitly depend on the time coordinate $t$ nor the azimuthal coordinate $\varphi$---a manifestation of time-translational and rotational invariance---the quantities $p_{t}$ and $p_{\varphi}$ are conserved in photon motion. These correspond to the photon's specific energy $p_{t}=-\mathscr{E}$ and specific angular momentum $p_{\varphi}=\mathscr{J}$. Furthermore, due to the spherical symmetry of the spacetime, the orbital characteristics can be analyzed in any plane without loss of generality. For this purpose, we restrict our analysis to the equatorial plane ($\theta = \pi/2$) with $\dot{\theta}=\ddot{\theta}=0$.

For photons governed by the Lagrangian constraint $\mathscr{L}=0$, the circular orbit conditions $\dot{r}=\ddot{r}=0$, combined with equation \eqref{4}, lead to the photon effective potential:
\begin{equation}\label{9}
\mathscr{V}_{\textrm{eff}}^{\textrm{N}}=\frac{1}{b}=\sqrt{\frac{-g_{tt}}{g_{\varphi\varphi}}}=\frac{\sqrt{f(r)}}{r}.
\end{equation}
Here, $b$ represents the impact parameter, mathematically defined as the ratio of specific angular momentum to specific energy, $b=\mathscr{J}/\mathscr{E}$. Typically, the photon effective potential exhibits a local maximum with respect to $r$, corresponding to the critical unstable circular photon orbit. This orbit radius satisfies the condition:
\begin{equation}\label{10}
\frac{\partial\mathscr{V}_{\textrm{eff}}^{\textrm{N}}}{\partial r} = 0. 
\end{equation}
Similar to the event horizon radius calculation, our code implements the Newton iteration method to solve this equation, requiring only the metric potential $f(r)$ and its first derivative with respect to $r$ as input. The solution yields two fundamental parameters: the critical photon orbit (photon ring) radius $r_{\textrm{p}}$ and its associated critical impact parameter $b_{\textrm{p}}=r_{\textrm{p}}/\sqrt{f(r_{\textrm{p}})}$. These metric-dependent quantities serve as unique spacetime fingerprints, with $b_{\textrm{p}}$ specifically corresponding to the higher-order bright rings---critical curves \cite{Gralla et al. (2019)}---in black hole images.

It should be emphasized that utilizing EHT shadow observations to constrain spacetime parameters requires matching observational data with $b_{\textrm{p}}$, which necessitates converting $b_{\textrm{p}}$ from geometric units (M) to microarcseconds ($\mu$as). In our code, this conversion is implemented using the relation derived in previous work \cite{Hu et al. (2022a)}:
\begin{equation}\label{11}
\frac{\Theta}{\mu as} = \frac{6.191165 \times 10^{-8}}{\pi}\frac{\Gamma}{D/\textrm{Mpc}}\frac{b_{\textrm{p}}}{\textrm{M}}.
\end{equation}
where $\Gamma$ and $D$ denote the mass ratio between the target black hole and the Sun, and the observation distance in megaparsecs (Mpc), respectively.
\subsection{Accretion disk}
We employ an idealized model of an optically thin, geometrically thin accretion disk, where the accretion mechanism arises from the angular momentum transfer of timelike particles. As these particles undergo quasi-Keplerian circular motion, they gradually radiate angular momentum outward while drifting toward the black hole. Once particles cross the ISCO, they inevitably follow plunging trajectories into the black hole's event horizon. Notably, this entire process occurs strictly within the equatorial plane. Thus, the accretion disk is naturally divided by the ISCO into two distinct regions: the quasi-Keplerian region (outside the ISCO) and the plunging zone (inside the ISCO). Our current objective is to determine the orbital parameters of particles within these regions.

The dynamics of massive particles are likewise governed by the Lagrangian \eqref{4}, subject to the constraint condition $\mathscr{L}=-1/2$, which yields the effective potential for massive particles:
\begin{equation}\label{12}
\mathscr{V}_{\textrm{eff}}^{\textrm{M}}=E=\sqrt{-g_{tt}\left(1+\frac{L^{2}}{g_{\varphi\varphi}}\right)}=\sqrt{f(r)\left(1+\frac{L^{2}}{r^{2}}\right)},
\end{equation}
where $E$ and $L$ denote the specific energy and angular momentum, respectively. In the quasi-Keplerian region, timelike particles execute circular orbits corresponding to minima of the effective potential, characterized by:
\begin{eqnarray}
\frac{\partial\mathscr{V}_{\textrm{eff}}^{\textrm{M}}}{\partial r} = 0, \label{13} \\
\frac{\partial^{2}\mathscr{V}_{\textrm{eff}}^{\textrm{M}}}{\partial r^{2}} > 0. \label{14}
\end{eqnarray}
This leads to the specific angular momentum for circular orbits at radius $r_{\textrm{e}}$:
\begin{equation}\label{15}
L = \frac{r_{\textrm{e}}\sqrt{r_{\textrm{e}}f^{\prime}(r_{\textrm{e}})}}{\sqrt{2f(r_{\textrm{e}})-r_{\textrm{e}}f^{\prime}(r_{\textrm{e}})}}.
\end{equation}
Substituting this analytical expression into equation \eqref{12}, we derive the corresponding specific energy:
\begin{equation}\label{16}
E = \frac{f(r_{\textrm{e}})}{\sqrt{f(r_{\textrm{e}})-\frac{1}{2}r_{\textrm{e}}f^{\prime}(r_{\textrm{e}})}}.
\end{equation}
Furthermore, through the relationship between conjugate momenta and velocities, we obtain the particle's angular velocity:
\begin{equation}\label{17}
\Omega = \frac{\dot{\varphi}}{\dot{t}} = \frac{f^{\prime}(r_{\textrm{e}})}{\sqrt{2r_{\textrm{e}}f^{\prime}(r_{\textrm{e}})}}.
\end{equation}
The complete four-velocity of particles in the quasi-Keplerian region is thus determined as $\dot{x}^{\mu}=(E/f(r_{\textrm{e}}),0,0,L/r_{\textrm{e}}^{2})$.

When particles cross the ISCO, they inevitably follow plunging orbits into the event horizon, ultimately being absorbed by the black hole. During this process, these particles can still emit detectable electromagnetic radiation, similar to those in the quasi-Keplerian region. To simplify the calculation of the four-velocity in this region, we adopt the same convention as in \cite{Hou et al. (2022a),Hu et al. (2024)}: both the specific energy and angular momentum of particles in this zone remain constant, equal to their ISCO values, $E_{\textrm{isco}}$ and $L_{\textrm{isco}}$. Consequently, the $t$- and $\varphi$-components of the four-velocity can be directly determined. Using the Lagrangian constraint, we then solve for the radial velocity:
\begin{equation}\label{18}
\dot{r} = -\sqrt{f(r_{\textrm{e}})\left(\frac{E_{\textrm{isco}}^{2}}{f(r_{\textrm{e}})}-\frac{L_{\textrm{isco}}^{2}}{r_{\textrm{e}}^{2}}-1\right)}.
\end{equation}
where $r_{\textrm{e}}$ denotes the radial coordinate of the emission source, and the negative sign indicates that the source is inspiraling toward the black hole.

Next, we need to determine the critical boundary between the plunging zone and the quasi-Keplerian zone---the ISCO radius. This is defined by the simultaneous conditions:
\begin{eqnarray}
\frac{\partial\mathscr{V}_{\textrm{eff}}^{\textrm{M}}}{\partial r} = 0, \label{19} \\
\frac{\partial^{2}\mathscr{V}_{\textrm{eff}}^{\textrm{M}}}{\partial r^{2}} = 0, \label{20}
\end{eqnarray}
which, when combined with the evolutionary relationship between specific angular momentum and orbital radius in equation \eqref{15}, yields the following equation for calculating the ISCO location $r_{\textrm{isco}}$:
\begin{equation}\label{21}
r_{\textrm{isco}} = \frac{3f(r_{\textrm{isco}})f^{\prime}(r_{\textrm{isco}})}{2f^{\prime}(r_{\textrm{isco}})^{2}-f(r_{\textrm{isco}})f^{\prime\prime}(r_{\textrm{isco}})}.
\end{equation}
In our algorithm, the ISCO radius $r_{\textrm{isco}}$ is computed using Newton iteration method, requiring the metric potential $f(r)$ and its first-, second-, and third-order radial derivatives.

Finally, after establishing the structure and kinematic distribution of the accretion disk, we must specify its radiation profile. Our implementation adopts the GRMHD time-averaged image-fitted radiation model developed by Chael et al. \cite{Chael et al. (2021)}, which achieves both computational efficiency in determining emission intensity at given source positions and qualitatively convincing observational signatures. The model is expressed logarithmically as:
\begin{equation}\label{22}
\log\left[\mathscr{F}(r_{\textrm{e}})\right] = p_{1}\log\left(\frac{r_{\textrm{e}}}{r_{\textrm{eh}}}\right) + p_{2}\left[\log\left(\frac{r_{\textrm{e}}}{r_{\textrm{eh}}}\right)\right]^{2},
\end{equation}
where the frequency-dependent parameters $(p_{1},p_{2})$ take characteristic values of $(0,-3/4)$ for $86$ GHz and $(-2,-1/2)$ for $230$ GHz emission. Notably, OCTOPUS also accommodates arbitrary user-specified emission functions $\mathscr{F}(r_ {\textrm{e}})$ for generalized studies, while defaulting to these empirically verified profiles for standard applications.
\subsection{Light propagation}
In ray-tracing algorithms, achieving both high computational efficiency and numerical precision in determining photon trajectories is essential. This process fundamentally consists of two components: (1) establishing the initial conditions for light rays, and (2) simulating photon geodesic propagation. We begin by introducing the coordinate systems and parameters essential for this computational framework. 

As shown in figure 1, the observer's reference frame is parameterized by a Cartesian coordinate system $xyz$, with the $z$-axis oriented toward $o^{\prime}$. Here, the $\overline{xoy}$ plane represents the observer's screen, which is perpendicular to the line of sight $\overline{oo^{\prime}}$. The black hole is positioned at the origin $o^{\prime}$ of a local coordinate system $x^{\prime}y^{\prime}z^{\prime}$, with its north pole aligned along the $z^{\prime}$-axis. A solid blue disk in the equatorial plane $\overline{x^{\prime}o^{\prime}y^{\prime}}$ models the accretion disk. The viewing angle $\omega$ is defined as the angle between the $z^{\prime}$-axis and $\overline{o^{\prime}o}$, while the azimuthal angle $\varphi_{\textrm{obs}}$ corresponds to the angle between the projection of $\overline{o^{\prime}o}$ onto the equatorial plane and the $x^{\prime}$-axis. The core concept of backward ray-tracing algorithm involves, at each grid point on the observation screen, launching light vectors perpendicular to the viewing plane and directed away from the black hole. These vectors are then numerically integrated backward along the time axis to determine their ultimate propagation fate.
\begin{figure*}%[tbph]
\center{
\includegraphics[width=7cm]{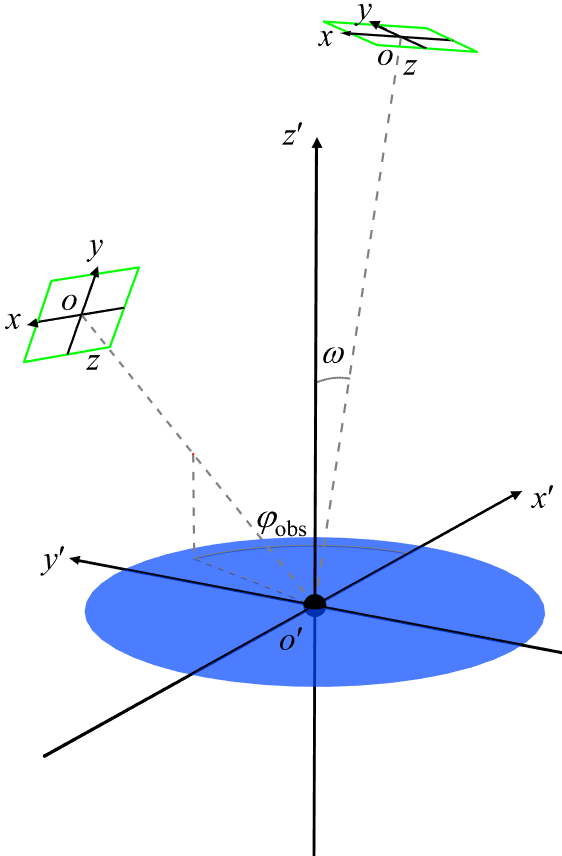}
\caption{Schematic diagram of the ray-tracing coordinate system. The local coordinate systems of the black hole and the observer are denoted by $xyz$ and $x^{\prime}y^{\prime}z^{\prime}$, respectively, with the black hole located at point $o^{\prime}$. The green square represents the $\overline{xoy}$ plane of the $xyz$ system, corresponding to the observation screen. The viewing angle $\omega$ is defined as the angle between the line $\overline{o^{\prime}o}$ and the $z^{\prime}$-axis, while the azimuthal viewing angle $\varphi_{\textrm{obs}}$ is the angle between the projection of $\overline{o^{\prime}o}$ onto the black hole's equatorial plane (blue) and the $x^{\prime}$-axis.}}\label{fig1}
\end{figure*}

We now derive the initial conditions for light rays at each grid point on the observation screen. For sufficiently large observation distances (e.g., $r_{\textrm{obs}}=\overline{oo^{\prime}}=1000$ M), the observer's location can be treated as flat spacetime. By performing coordinate rotation and translation to align the observer's frame with the black hole's local coordinates, the initial light ray position $(x,y,z)$ transforms into $(x^ {\prime},y^{\prime},z^{\prime})$ via:
\begin{eqnarray}
x^{\prime} &=& \mathscr{D}\cos\varphi_{\textrm{obs}} - x\sin\varphi_{\textrm{obs}}, \label{23} \\
y^{\prime} &=& \mathscr{D}\sin\varphi_{\textrm{obs}} + x\cos\varphi_{\textrm{obs}}, \label{24} \\
z^{\prime} &=& \left(r_{\textrm{obs}}-z\right)\cos\omega + y\sin\omega, \label{25}
\end{eqnarray}
with the auxiliary expression:
\begin{equation}\label{26}
\mathscr{D} = \left(r_{\textrm{obs}}-z\right)\sin\omega - y\cos\omega.
\end{equation}
The corresponding initial conditions in spherical coordinates $x^{\alpha}=(t,r,\theta,\varphi)$ are then:
\begin{eqnarray}
t &=& 0, \label{27} \\
r &=& \sqrt{x^{\prime 2}+y^{\prime 2}+z^{\prime 2}}, \label{28} \\
\theta &=& \arccos\left(\frac{z^{\prime}}{r}\right), \label{29} \\
\varphi &=& \textrm{atan}2\left(y^{\prime},x^{\prime}\right). \label{30}
\end{eqnarray}

With the initial positions determined, simulating null geodesics additionally requires specifying the initial velocities (or conjugate momenta). Adopting the forward-time perspective, photons emitted from the accretion disk ultimately arrive at the observation screen $\overline{xoy}$ with trajectories oriented away from the black hole. Given the asymptotically flat spacetime, we can safely assume these light rays approach grid points perpendicular to the screen (i.e., parallel to the $z$-axis). Consequently, in the observer's local coordinates, the photon's spatial velocity is set as $(\dot{x},\dot{y},\dot{z})=(0,0,-1)$ (note that the $z$-axis aligns with $\overline{oo^ {\prime}}$ and points toward the black hole). Differentiating both sides of equations \eqref{23}-\eqref{25} with respect to the affine parameter $\lambda$ and substituting the velocity $(0,0,-1)$ yields:
\begin{eqnarray}
\dot{x^{\prime}} &=& \frac{dx^{\prime}}{d\lambda} = \sin\omega\cos\varphi_{\textrm{obs}}, \label{31} \\
\dot{y^{\prime}} &=& \frac{dy^{\prime}}{d\lambda} = \sin\omega\sin\varphi_{\textrm{obs}}, \label{32} \\
\dot{z^{\prime}} &=& \frac{dz^{\prime}}{d\lambda} = \cos\omega. \label{33}
\end{eqnarray}
Similarly, substituting the above three-velocity into the differential forms of equations \eqref{28}-\eqref{30} yields:
\begin{eqnarray}
\dot{r} &=&  \cos\omega\cos\theta + \sin\omega\cos(\varphi-\varphi_{\textrm{obs}})\sin\theta, \label{34} \\
\dot{\theta} &=& \frac{\sin\omega\cos(\varphi-\varphi_{\textrm{obs}})\cos\theta-\cos\omega\sin\theta}{r}, \label{35} \\
\dot{\varphi} &=& -\frac{\sin\omega\sin(\varphi-\varphi_{\textrm{obs}})}{r\sin\theta}. \label{36}
\end{eqnarray}
From the velocity-conjugate momentum relation $p_{\mu}=g_{\mu\nu}\dot{x}^{\nu}$, the components $p_{r}$, $p_{\theta}$, and $p_{\varphi}$ are directly obtained, where the metric tensor $g_{\mu\nu}$ is evaluated using the results from equations \eqref{28}-\eqref{30}. The $t$-component of the photon's four-velocity remains to be determined, which we derive through the Lagrangian constraint. Substituting $\dot{r}$, $\dot{\theta}$,and $\dot{\varphi}$ into the Lagrangian \eqref{4} yields: 
\begin{equation}\label{37}
\dot{t} = \sqrt{\frac{-g_{rr}\dot{r}^{2}-g_{\theta\theta}\dot{\theta}^{2}-g_{\varphi\varphi}\dot{\varphi}^{2}}{g_{tt}}}.
\end{equation}
This naturally gives the photon's specific energy:
\begin{equation}\label{38}
\mathscr{E} = -p_{t} = -g_{tt}\dot{t}.
\end{equation}
Remarkably, we can normalize the photon's conjugate momentum $p_{\mu}$ using specific energy without altering the photon's trajectory, resulting in:
\begin{eqnarray}
p_{t} &=& -1, \label{39} \\
p_{r} &=& \frac{p_{r}}{\mathscr{E}}, \label{40} \\
p_{\theta} &=& \frac{p_{\theta}}{\mathscr{E}}, \label{41} \\
p_{\varphi} &=& \frac{p_{\varphi}}{\mathscr{E}}. \label{42}
\end{eqnarray}
We are now equipped to determine the complete set of initial conditions $(t,r,\theta,\varphi,-1,p_{r},p_{\theta},p_{\varphi})$ for light rays in the black hole's local coordinate, given any observation parameters $(r_{\textrm{obs}},\omega,\varphi_{\textrm{obs}})$ and observation coordinates $(x,y,0)$.

Through the Legendre transform $\mathscr{H} = p_{\mu}\dot{x}^{\mu} - \mathscr{L}$, we derive the Hamiltonian governing photon dynamics:
\begin{equation}\label{43}
\mathscr{H} = \frac{1}{2}g^{\mu\nu}p_{\mu}p_{\nu},
\end{equation}
where $g^{\mu\nu}$ denotes the contravariant metric tensor, which satisfies $g^{\mu\nu}g_{\mu\nu}=1$ in spherically symmetric spacetimes. This leads to the explicit form:
\begin{equation}\label{44}
\mathscr{H} = \frac{1}{2}\left[-\frac{p_{t}^{2}}{f(r)} + f(r)p_{r}^{2} + \frac{p_{\theta}^{2}}{r^{2}} + \frac{p_{\varphi}^{2}}{r^{2}\sin^{2}\theta}\right].
\end{equation}
Theoretically, Hamilton's canonical equations ($\dot{x}^{\mu}=\partial\mathscr{H}/\partial p_{\mu}$, $\dot{p_{\mu}}=-\partial\mathscr{H}/\partial x_{\mu}$) would permit photon trajectory simulation given initial conditions. However, our backward ray-tracing framework requires additional sign reversals in the equations of motion. The modified photon dynamics are therefore:
\begin{eqnarray}
\dot{t} &=& \frac{p_{t}}{f(r)}, \label{45} \\
\dot{r} &=& -p_{r}f(r), \label{46} \\
\dot{\theta} &=& -\frac{p_{\theta}}{r^{2}}, \label{47} \\
\dot{\varphi} &=& -\frac{p_{\varphi}}{r^{2}\sin^{2}\theta}, \label{48} \\
\dot{p_{t}} &=& 0, \label{49} \\
\dot{p_{r}} &=& \frac{1}{2}\left[\frac{f^{\prime}(r)p_{t}^{2}}{f(r)^{2}} + f^{\prime}(r)p_{r}^{2} - \frac{2p_{\theta}^{2}}{r^{3}} - \frac{2p_{\varphi}^{2}}{r^{3}\sin^{2}\theta}\right], \label{50} \\
\dot{p_{\theta}} &=& -\frac{p_{\varphi}^{2}\cos\theta}{r^{2}\sin^{3}\theta}, \label{51} \\
\dot{p_{\varphi}} &=& 0. \label{52}
\end{eqnarray}
The vanishing derivatives of $p_{t}$ and $p_{\varphi}$ reflect their conserved nature. Crucially, for any black hole spacetime admitting the metric form \eqref{2}, the algorithm only requires the metric potential and its first radial derivative as input---a feature that significantly simplifies applications to alternative black hole spacetimes.

The photon equations of motion \eqref{45}-\eqref{52} require numerical integration, for which our code employs a fifth- and sixth-order Runge-Kutta-Fehlberg integrator (RKF56) with a variable step-size. This scheme efficiently and accurately tracks light vectors over arbitrary simulation times. To further optimize the method, we introduce a position-dependent step-size adaptation, where the initial step-size $h$ fed to the integrator is automatically adjusted based on the photon's radial coordinate as follows:
\begin{equation}\label{53}
h = h_{0}\left(\frac{r}{r_{\textrm{eh}}}\right)^{n}.
\end{equation}
Here, $h_{0} = 0.0002$ specifies the minimum step-size for the integrator input, while $n$ represents the step-size scaling exponent, typically set to $n=1.8$. This adaptive scheme enhances both the convergence rate of the RKF56 integrator and the overall computational efficiency of ray-tracing. Moreover, it is permissible to moderately increase the values of $h_{0}$ and $n$ to further enhance computational efficiency, provided that the numerical precision is maintained.
\subsection{Redshift factor}
The relative motion between light sources and the observer introduces Doppler effects in the observed specific intensity, characterized by the redshift factor $\gamma$. In curved spacetime, this factor takes the general form:
\begin{equation}\label{54}
\gamma = \frac{p_{\mu}u_{\textrm{o}}^{\mu}}{p_{\nu}u_{\textrm{e}}^{\nu}},
\end{equation}
where $u_{\textrm{o}}^{\mu}$ and $u_{\textrm{e}}^{\nu}$ represent the four-velocity of the observer and emission source, respectively, while $p_{\mu}$ denotes the photon's conjugate momentum obtained from ray-tracing method. For a static observer in flat spacetime, the simplification $u_{\textrm{o}}^{\mu}=(1,0,0,0)$ yields:
\begin{equation}\label{55}
\gamma = \frac{p_{t}}{p_{\nu}u_{\textrm{e}}^{\nu}} = \frac{p_{t}}{p_{t}\dot{t_{\textrm{e}}}+p_{r}\dot{r_{\textrm{e}}}+p_{\theta}\dot{\theta_{\textrm{e}}}+p_{\varphi}\dot{\varphi_{\textrm{e}}}},
\end{equation}
reducing the calculation to determining the source's four-velocity components $(\dot{t_{\textrm{e}}},\dot{r_{\textrm{e}}},\dot{\theta_{\textrm{e}}},\dot{\varphi_{\textrm{e}}})$.

As established in the previous section, our accretion disk lies strictly within the equatorial plane and is divided into two distinct regions: the quasi-Keplerian zone and the plunging zone. In both regions, timelike particles exhibit no latitudinal motion, allowing us to set $\dot{\theta_{\textrm{e}}}=0$. 

In the quasi-Keplerian region $(r_{\textrm{isco}} \leq r_{\textrm{e}})$, where the particles' gradual inspiral is negligible, the motion is well-approximated by circular orbits. The corresponding redshift factor is given by:
\begin{equation}\label{56}
\gamma_{\textrm{k}} = \frac{p_{t}}{p_{t}\dot{t_{\textrm{e}}}+p_{\varphi}\dot{\varphi_{\textrm{e}}}} = \frac{p_{t}}{p_{t}\frac{E}{f(r_{\textrm{e}})}+p_{\varphi}\frac{L}{r_{\textrm{e}}^{2}}},
\end{equation}
with $E$ and $L$ are evaluated at $r_{\textrm{e}}$ through equations \eqref{15} and \eqref{16}.

For the plunging zone $(r_{\textrm{eh}} < r_{\textrm{e}} < r_{\textrm{isco}})$, the emission sources maintain their ISCO-acquired specific energy $E_{\textrm{isco}}$ and angular momentum $L_{\textrm{isco}}$ while accelerating inward, with radial velocity governed by equation \eqref{18}. The corresponding redshift factor becomes:
\begin{equation}\label{57}
\gamma_{\textrm{p}} = \frac{p_{t}}{p_{t}\dot{t_{\textrm{e}}}+p_{r}\dot{r_{\textrm{e}}}+p_{\varphi}\dot{\varphi_{\textrm{e}}}} = \frac{p_{t}}{p_{t}\frac{E_{\textrm{isco}}}{f(r_{\textrm{e}})}-p_{r}\sqrt{f(r_{\textrm{e}})\left(\frac{E_{\textrm{isco}}^{2}}{f(r_{\textrm{e}})}-\frac{L_{\textrm{isco}}^{2}}{r_{\textrm{e}}^{2}}-1\right)}+p_{\varphi}\frac{L_{\textrm{isco}}}{r_{\textrm{e}}^{2}}}.
\end{equation}
\subsection{Black hole images}
Simulating black hole images fundamentally involves determining the specific intensity at each pixel on the observation screen, where the intensity originates from accretion disk radiation, is carried by light rays, and is modulated by Doppler effects. For a light ray launched from screen coordinates $(x,y,0)$, its trajectory is computed using equations \eqref{45}--\eqref{52}, with continuous monitoring for intersections with either the accretion disk or the event horizon. The simulation terminates the ray's propagation upon horizon encounter, while disk intersections contribute to the observed specific intensity via the radiative transfer process. The final image is generated as the distribution of these intensity values across the screen.

During geodesic integration, a disk intersection is detected when the light vectors at the $n$-th and $(n+1)$-th steps satisfy
\begin{eqnarray}\label{58}
\cos\theta^{n}\cos\theta^{n+1} \leq 0.
\end{eqnarray}
At this point, the emission intensity and redshift factor are computed at the source radius $r_{\textrm{e}}$ using equations \eqref{22}, \eqref{56}, and \eqref{57}. Theoretically, this crossing event contributes an observed specific intensity at pixel $(x,y,0)$ given by
\begin{equation}\label{59}
I_{\textrm{obs}} = \mathscr{F}\gamma^{3}.
\end{equation}
In practice, a single light ray may cross the disk multiple times, accumulating intensity contributions. Thus, the total observed specific intensity for a given pixel is expressed as:
\begin{equation}\label{60}
I_{\textrm{obs}} = \sum_{i=1}^{N_{\textrm{max}}}\kappa_{i}\mathscr{F}_{i}\gamma_{i}^{3},
\end{equation}
where $\kappa$ is a fudge factor ($\kappa = 1$ for $i=1$, $\kappa = 2/3$ for $i > 1$) \cite{Chael et al. (2021)}, and $N_{\textrm{max}}$ (ranging from $1$ to $4$) denotes the maximum number of permitted disk crossings in the simulation. While $N_{\textrm{max}}=4$ ensures well-defined critical curve structures in the image, smaller values significantly reduce computational cost.

When a light ray reaches the event horizon, it is considered absorbed by the black hole, and the simulation for this ray is terminated. However, in numerical simulations, it is practically impossible to precisely determine when the floating-point radial coordinate $r$ equals $r_{\textrm{eh}}$. To address this, we introduce a small tolerance $\varepsilon=5 \times 10^{-4}$, allowing us to safely conclude that the ray has fallen into the black hole when $r \leq r_{\textrm{eh}}+\varepsilon$.

Once the observation resolution and screen size are specified, our code performs uniform grid discretization, simulating the light vector associated with each grid point to construct the observed specific intensity field $(x,y,I_{\textrm{obs}})$---the definitive black hole image.
\subsection{Gravitational lensing}
Beyond simulating black hole images, our code also models gravitational lensing for a point source in curved spacetime. As illustrated in figure 2, we employ a celestial sphere with a radius of $1500$ M, encompassing both the black hole and observer, to visualize lensing effects, following the methodology outlined in \cite{Cunha et al. (2015),Hu et al. (2021),Huang et al. (2025),Bohn et al. (2015)}. The coordinate systems and parameters are consistent with those defined in figure 1. Photons emitted from the observation screen exhibit three distinct outcomes: (1) those intersecting the white point source at $(x^{\prime},y^{\prime},z^{\prime})$ with emission radius $r_{\textrm{source}}$ form lensed images; (2) rays captured by the event horizon contribute to the black hole shadow; and (3) photons reaching the celestial sphere are color-coded on the screen according to their hitting locations---yellow for $z^{\prime} > 0 \cap y^{\prime} < 0$, green for $z^{\prime} > 0 \cap y^{\prime} > 0$, red for $z^{\prime} < 0 \cap y^{\prime} < 0$, and blue for $z^{\prime} < 0 \cap y^{\prime} > 0$.
\begin{figure*}%[tbph]
\center{
\includegraphics[width=10cm]{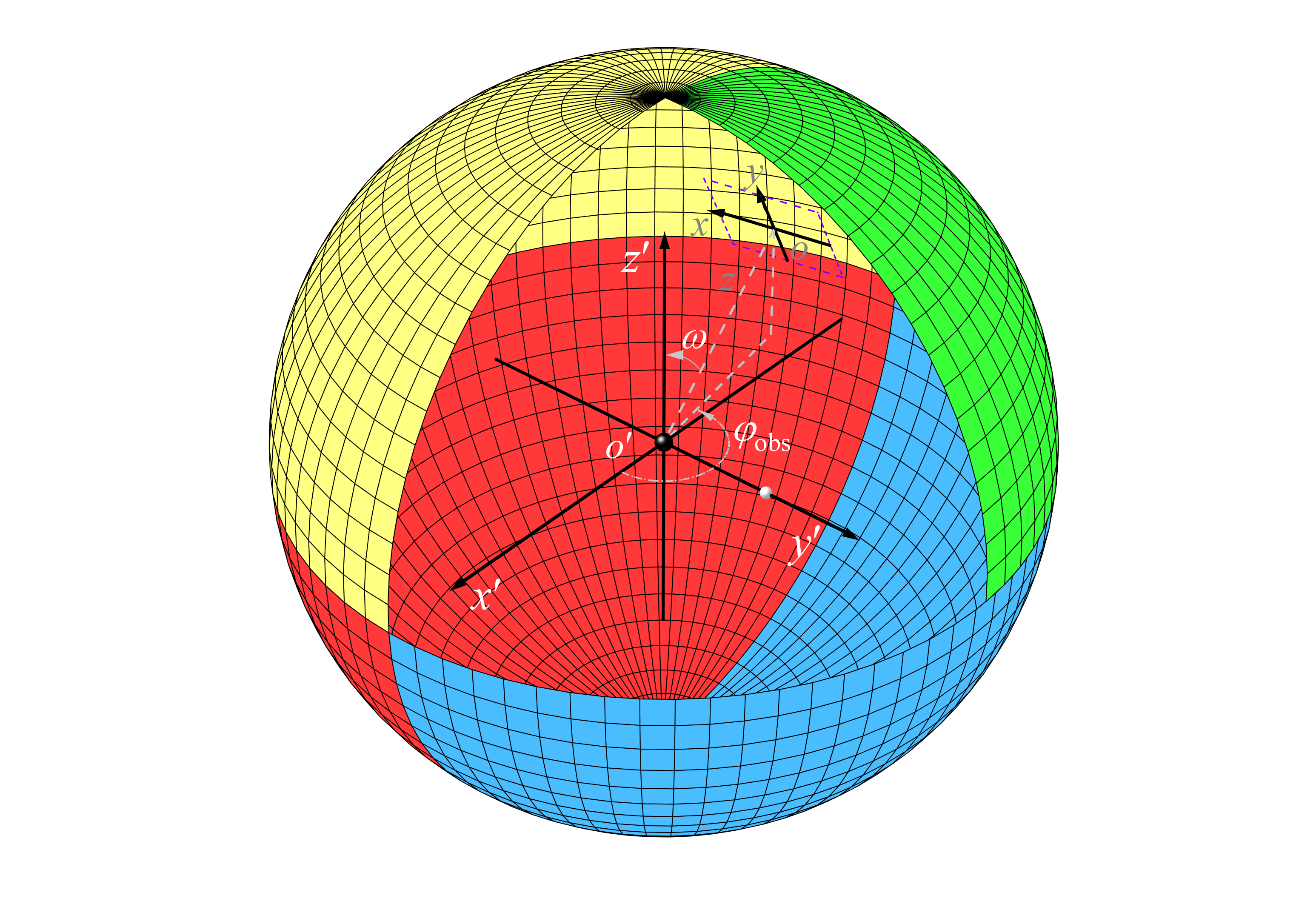}
\caption{Schematic of the celestial sphere and coordinate systems used to simulate gravitational lensing images for static point sources. The local coordinate systems of the black hole and observer, along with all associated parameters, remain consistent with those in figure 1. A stationary point source is represented by a white sphere of non-negligible radius. While shown here positioned along the $y^{\prime}$-axis, it can actually be located anywhere outside the black hole event horizon. The entire system---comprising the black hole and observer---is enclosed by a celestial sphere of radius $1500$ M, with color coding applied to visually enhance the illustration of gravitational lensing effects.}}\label{fig2}
\end{figure*}

Beyond demonstrating lensed images of stationary point sources using idealized geometric light source, our code also supports simplified radiative scenarios where both emission and absorption depend solely on spatial coordinates within the source. Specifically, we model the point source at $(x^{\prime},y^{\prime},z^{\prime})$ as a sphere with radius $r_{\textrm{source}}$, where the emission coefficient $J$ and absorption coefficient $\Upsilon$ are given by:
\begin{eqnarray}
J = j_{0}\textrm{e}^{-\frac{d^{2}}{2\sigma^{2}}}, \label{61} \\
\Upsilon = \alpha_{0}\left(1-\frac{d}{r_{\textrm{source}}}\right)^{2}. \label{62}
\end{eqnarray}
Here, $j_{0}$ and $\alpha_{0}$ quantify the emission and absorption capacities of the medium, respectively (e.g., $\alpha_{0}=0$ corresponds to a perfectly transparent medium). The quantity $d$ denotes the distance from the emission point to the center of the sphere, and $\sigma$ controls the decay rate of emission. For each light ray passing through the sphere, we integrate the radiative transfer equation along its path within the sphere:
\begin{equation}\label{63}
\dot{I} = J - \Upsilon I,
\end{equation}
to compute the specific intensity contribution from that segment. Notably, since the point source remains stationary, the resulting intensity is unaffected by Doppler effects. Furthermore, a single ray may traverse the sphere multiple times, the total observed intensity for a pixel is then the sum of contributions from all such crossings.
\subsection{Trajectories and light curves of Hot-spots}
Timelike particles orbiting near black holes can be treated as radiation sources or hot-spots, geometrically modeled as spheres with non-negligible radii. These particles continuously emit electromagnetic radiation during their orbital motion around the black hole, contributing to observable light curves. Notably, such light curves often carry crucial information about the emitter's dynamics, the black hole's spacetime properties, and the surrounding high-energy environment, making them powerful tools for probing black hole physics \cite{Huang et al. (2024a),Genzel et al. (2003),Li et al. (2014),Baubock et al. (2020),Matsumoto et al. (2020),Ball et al. (2021),Davelaar and Haiman (2022b),Yfantis et al. (2024),Hu et al. (2025b)}. Our code enables the simulation of light curves from hot-spots moving along arbitrary periodic or quasi-periodic orbits. To this end, we first specify the initial conditions of the hot-spot and simulate its trajectory using Hamilton's canonical equations. We then activate the ray-tracing module to determine in real-time whether a light ray intersects the hot-spot's path. When such an intersection occurs, the ray contributes to the light curve count rate at time $T=t_{\textrm{e}}+t$, where $t_{\textrm{e}}$ denotes the coordinate time of the hot-spot at the intersection point, and $t$ represents the light propagation time to that point. It is important to note that the hot-spot's motion is also governed by the Hamiltonian \eqref{44}, but with the constraint $\mathscr{H}=-1/2$. Its equations of motion are derived by removing the negative signs from the photon equations \eqref{45}-\eqref{52}, as the hot-spot evolves forward in time. We now detail the complete procedure.

The required initial conditions for the hot-spot include the starting coordinate time $t_{0}=0$, the release coordinates $(r_{0},\theta_{0},\varphi_{0})$, the specific energy $E=-p_{t0}$, the specific angular momentum $L=p_{\varphi 0}$, the radial momentum $p_{r0}$, and the polar angular momentum $p_{\theta 0}$. Our algorithm supports five distinct methods for specifying these quantities.

\textbf{Method 1} requires the user to provide the release position $(r_{0},\theta_{0},\varphi_{0})$ and the two conserved quantities---specific energy $p_{t0}$ and specific angular momentum $p_{\varphi 0}$. The code then assumes no initial radial motion ($p_{r0}=0$) and solves the Hamiltonian constraint to yield:
\begin{equation}\label{64}
p_{\theta 0} = \sqrt{\frac{-1-g^{tt}p_{t0}^2-g^{\varphi\varphi}p_{\varphi 0}^{2}}{g^{\theta\theta}}}.
\end{equation}
Here, $g^{\mu\nu}$ denotes the contravariant metric tensor. For the metric form in \eqref{2}, this simplifies to $g^{tt}=-1/f(r)$, $g^{\varphi\varphi}=1/(r^{2}\sin^{2}\theta)$, and $g^{\theta\theta}=1/r^{2}$. 

\textbf{Method 2} is similar to the first, but assumes motion confined to the initial orbital plane ($p_{\theta 0}=0$) and analogously derives the initial radial momentum as:
\begin{equation}\label{65}
p_{r0} = \sqrt{\frac{-1-g^{tt}p_{t0}^2-g^{\varphi\varphi}p_{\varphi 0}^{2}}{g^{rr}}},
\end{equation}
where $g^{rr}=f(r)$. 

\textbf{Method 3} initializes the particle by selecting a point on the effective potential curve. Specifically, given a specific angular momentum and initial coordinates, and assuming $p_{r0}=p_{\theta 0}=0$, the temporal conjugate momentum is set as $p_{t}=-\mathscr{V}_{\textrm{eff}}^{\textrm{M}}$.

\textbf{Method 4} computes exact circular orbits. The user need only specify the circular orbit radius $r_{\textrm{e}}$ $(r_{\textrm{e}} \geq r_{\textrm{isco}})$, from which the specific energy and angular momentum are obtained via equations \eqref{15} and \eqref{16}. The remaining initial conditions are fixed as $\theta_{0}=\pi/2$, $\varphi_{0}=0$, $p_{r0} = p_{\theta 0} = 0$.

\textbf{Method 5} grants the user full control: under the sole constraint $\mathscr{H}=-1/2$, all initial conditions are specified directly by the user. This method is suitable for simulating time-like orbits that demand high-precision initial conditions prepared in advance, such as quasi-periodic orbits exhibiting the typical characteristics described in \cite{Levin and Perez-Giz (2008),Liu et al. (2019),Deng (2020a),Deng (2020b),Wang et al. (2022),Lin and Deng (2023),Tu et al. (2023),Huang and Deng (2024)}.

Once the initial conditions for the hot-spot are determined as $(t_{0},r_{0},\theta_{0},\varphi_{0},p_{t0},p_{r0},p_{\theta 0},p_{\varphi 0})$, the code employs a sixth-order Runge-Kutta method (RK6) to simulate its trajectory. This process generates a complete temporal profile of the hot-spot's motion, providing the trajectory data $(t_{\textrm{e}},r_{\textrm{e}},\theta_{\textrm{e}},\varphi_{\textrm{e}})$. Subsequently, light vectors are launched from each pixel on the observation screen, following the methodology outlined in previous sections, with real-time detection of intersections between these rays and the hot-spot's trajectory. Specifically, an intersection is identified when adjacent integration steps place the photon inside and outside the hot-spot's spherical boundary. This event is mathematically expressed as follows:
\begin{eqnarray}
r^{n} &=& \sqrt{\left(x_{p}^{n}-x_{\textrm{e}}\right)^{2}+\left(y_{p}^{n}-y_{\textrm{e}}\right)^{2}+\left(z_{p}^{n}-z_{\textrm{e}}\right)^{2}} \leq r_{\textrm{source}}, \label{66} \\
r^{n+1} &=& \sqrt{\left(x_{p}^{n+1}-x_{\textrm{e}}\right)^{2}+\left(y_{p}^{n+1}-y_{\textrm{e}}\right)^{2}+\left(z_{p}^{n+1}-z_{\textrm{e}}\right)^{2}} \geq r_{\textrm{source}}. \label{67}
\end{eqnarray}
Here, the subscripts ``p'' and ``e'' represent the Cartesian coordinates of the photon and the hot-spot, respectively, while $n$ and $n+1$ denote the photon's vector at the $n$- and $n+1$-th steps. When this condition is satisfied, it indicates that the ray was emitted by the hot-spot at time $t_{\textrm{e}}$ and propagated for a duration $t$ before reaching the observer. Consequently, this ray contributes to the count rate of the light curve at the observed time $T=t_{\textrm{e}}+t$.

Beyond generating light curves, projecting the dynamic hot-spot---including its time-dependent trajectory and radiation---onto the observer's screen provides an exciting opportunity for deeper exploration. This approach enhances our understanding of particle dynamics in curved spacetime and offers potential physical explanations for flares observed around compact sources, such as the Galactic Center. 

To generate such animations, the code first simulates the trajectory of the hot-spot based on user-defined initial conditions. Each point along this trajectory is then treated as a luminous sphere with radius $r_{\textrm{source}}$. As light rays propagate through the sphere, their specific intensity $I$ is computed by integrating the radiative transfer equation \eqref{63}. Unlike the static point source case, the hot-spot's motion introduces Doppler effects, necessitating the modulation of the observed intensity, as described by equation \eqref{59}. Thus, the observed specific intensity becomes:
\begin{equation}\label{68}
I_{\textrm{obs}} = I\gamma^{3},
\end{equation}
where $\gamma$ denotes the redshift factor derived from equation \eqref{54}. The quantity $I_{\textrm{obs}}$ ultimately characterizes the time-evolving specific intensity of each ray at its corresponding pixel on the observation screen.

Moreover, it should be noted that light emitted earlier by the hot-spot may reach the observation screen later due to gravitational lensing. However, this time-delay effect is generally negligible, as the additional path length incurred by light rays orbiting near the black hole is much smaller than the total propagation distance from the screen to the black hole region. Furthermore, such delays can be effectively mitigated during the construction of the time series for dynamic visualizations. As a result, our code does not explicitly account for this temporal offset, and each frame in the animation corresponds directly to the emission coordinate time $t_{\textrm{e}}$ without delay compensation.

Theoretically, a hot-spot and its central compact object form an extreme mass-ratio inspiral (EMRI) system. In addition to electromagnetic emission, such systems are expected to produce gravitational wave signals detectable by space-based laser interferometers such as LISA. To facilitate multi-messenger studies, our algorithm incorporates gravitational wave calculations based on an analytic ``kludge'' waveform model. Specifically, the two polarization states of the gravitational wave are given by \cite{Tu et al. (2023),Huang and Deng (2024),Babak et al. (2007)}:
\begin{eqnarray}
h_{+} &=& -\frac{2\eta}{D_{\textrm{L}}}\frac{G^{2}M^{2}}{c^{4}r_{\textrm{e}}^{2}}\left(1+\cos^{2}\iota\right)\cos\left(2\varphi_{\textrm{e}}+2\zeta\right), \label{69} \\
h_{\times} &=& -\frac{4\eta}{D_{\textrm{L}}}\frac{G^{2}M^{2}}{c^{4}r_{\textrm{e}}^{2}}\cos\iota\sin\left(2\varphi_{\textrm{e}}+2\zeta\right). \label{70}
\end{eqnarray}
Here, $\iota$ and $\zeta$ denote the longitude and inclination of the orbit's periastron, respectively, which are fixed at $\iota=\zeta=\pi/4$. The parameter $\eta=Mm/(M+m)^{2}$ characterizes the EMRI mass ratio, where $m$ and $M$ represent the masses of the hot-spot and central object respectively, and $D_{\textrm{L}}$ is the luminosity distance. Once the hot-spot trajectory is computed, the time-dependent values of $r_{\textrm{e}}$ and $\varphi_{\textrm{e}}$ are substituted into these expressions to obtain the temporal evolution of both gravitational wave polarizations.
\section{Code test in Schwarzschild spacetime endowed with a Dehnen-type dark matter halo}
In this section, we demonstrate the application of OCTOPUS from a user's perspective. It is now widely recognized that dark matter and dark energy are fundamental components of the universe \cite{Bertone et al. (2005),Clifton et al. (2012),Planck (2020),Chakrabarti et al. (2025),Gu et al. (2025)}. In particular, observational evidence indicates the presence of dark matter near the supermassive black hole at the Galactic Center, where it may affect the spacetime geometry. Exploring spacetime characteristics under the influence of dark matter could therefore yield valuable insights into both particle interactions and gravitational theory. Notably, the density profiles of dark matter around black holes exhibit substantial diversity \cite{Xu et al. (2018),Xu et al. (2020)}. For this study, we consider a Schwarzschild black hole surrounded by a Dehnen-type dark matter halo \cite{Gohain et al. (2024),Jha (2024),Al-Badawi and Shaymatov (2024)} as our test case.
\subsection{Spacetime}
In the coordinate system $x^{\alpha}=(t,r,\theta,\varphi)$, the metric of a Schwarzschild black hole embedded in a Dehnen-type dark matter halo retains the form of equation \eqref{2}, with its dimensionless metric potential $f(r)$ given by \cite{Gohain et al. (2024),Jha (2024),Al-Badawi and Shaymatov (2024)}:
\begin{equation}\label{71}
f(r) = 1-\frac{2}{r}-\frac{4\pi\left(r_{\textrm{s}}+2r\right)r_{\textrm{s}}^{3}\rho_{\textrm{s}}}{3\left(r_{\textrm{s}}+r\right)^{2}},
\end{equation}
where $r_{\textrm{s}}$ and $\rho_{\textrm{s}}$ parameterize the scale and density of the dark matter halo, respectively. The spacetime reduces to the standard Schwarzschild case if either parameter vanishes.
\begin{figure*}%[tbph]
\center{
\includegraphics[width=3.5cm]{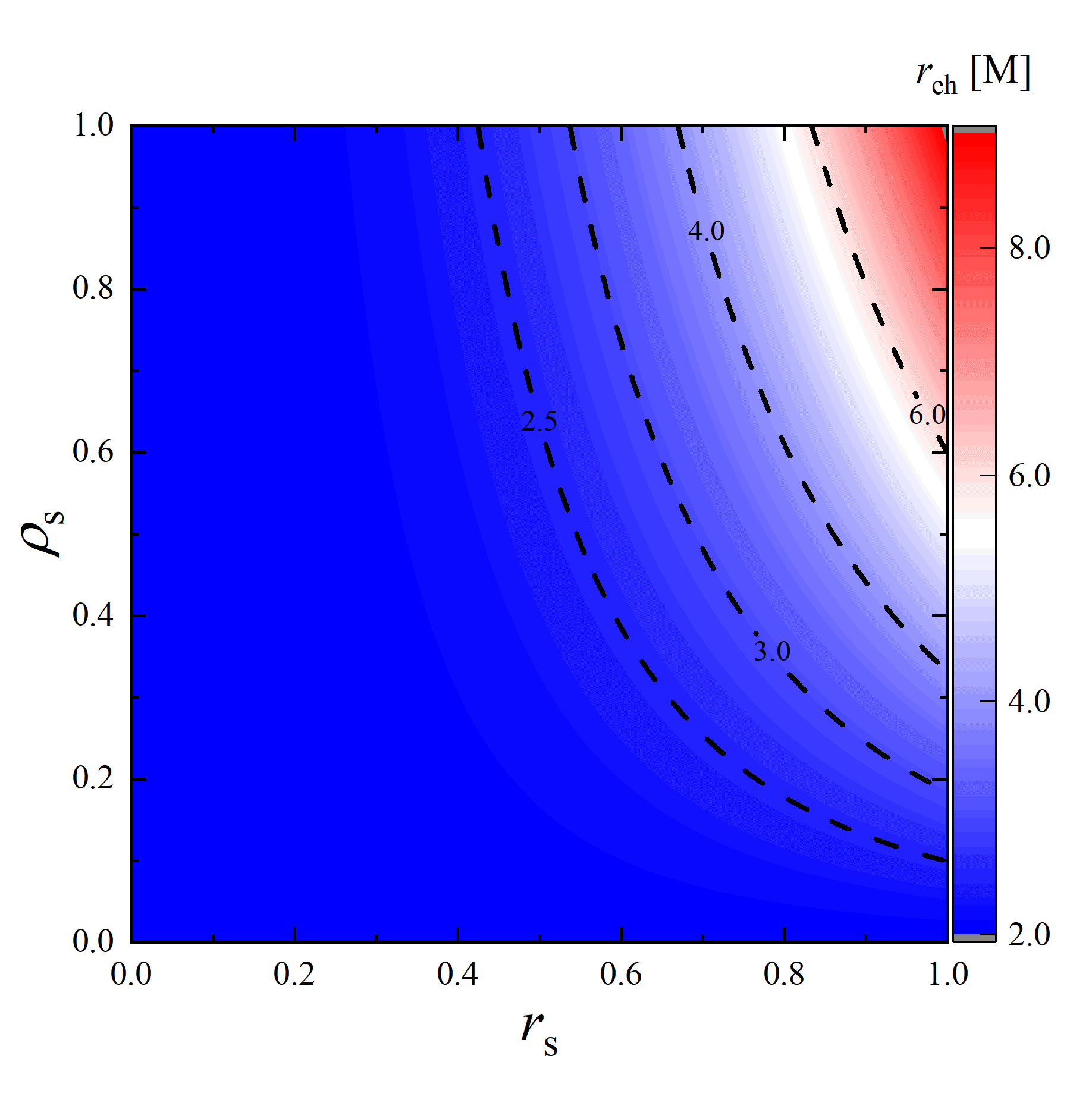}
\includegraphics[width=3.5cm]{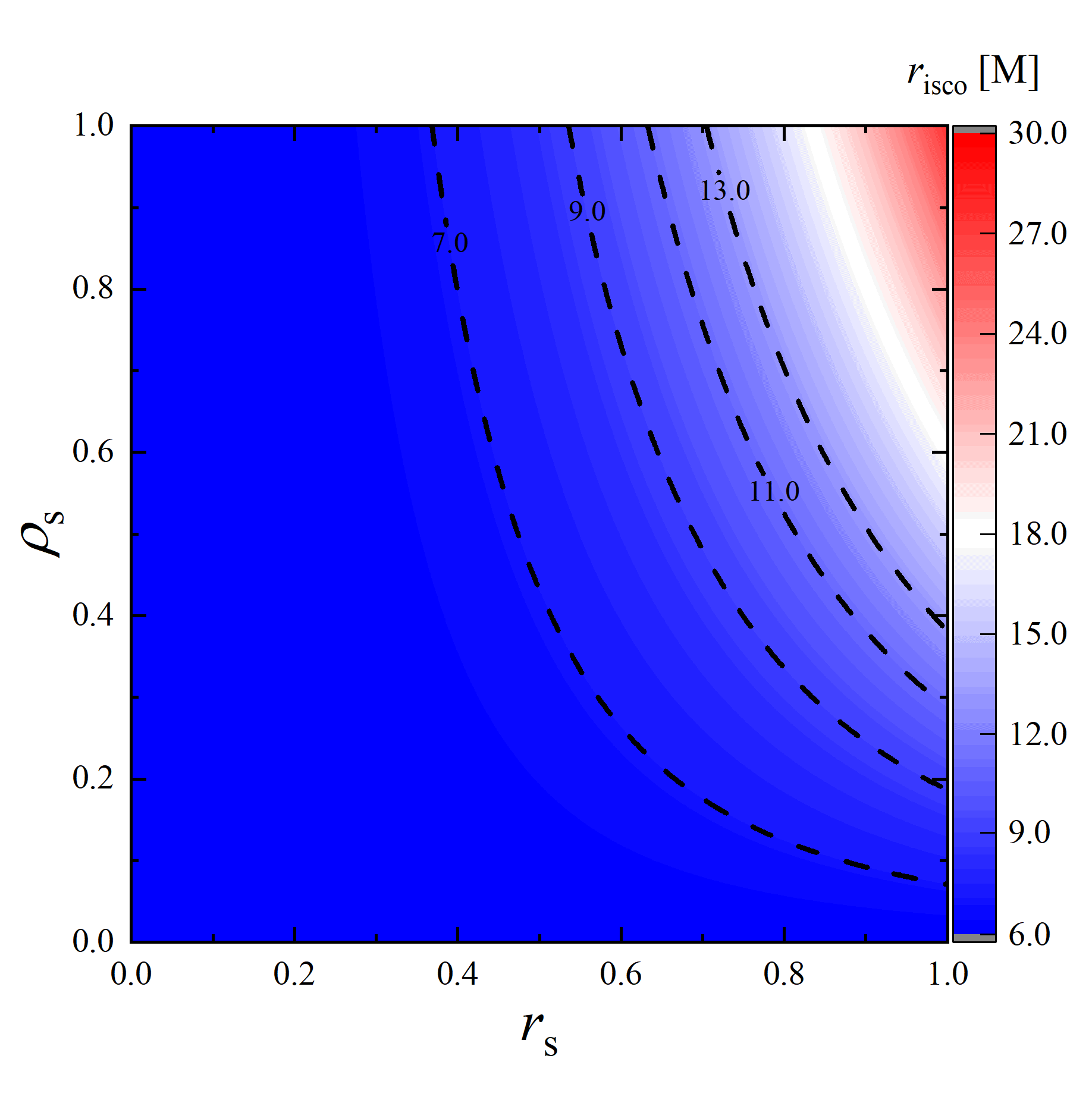}
\includegraphics[width=3.5cm]{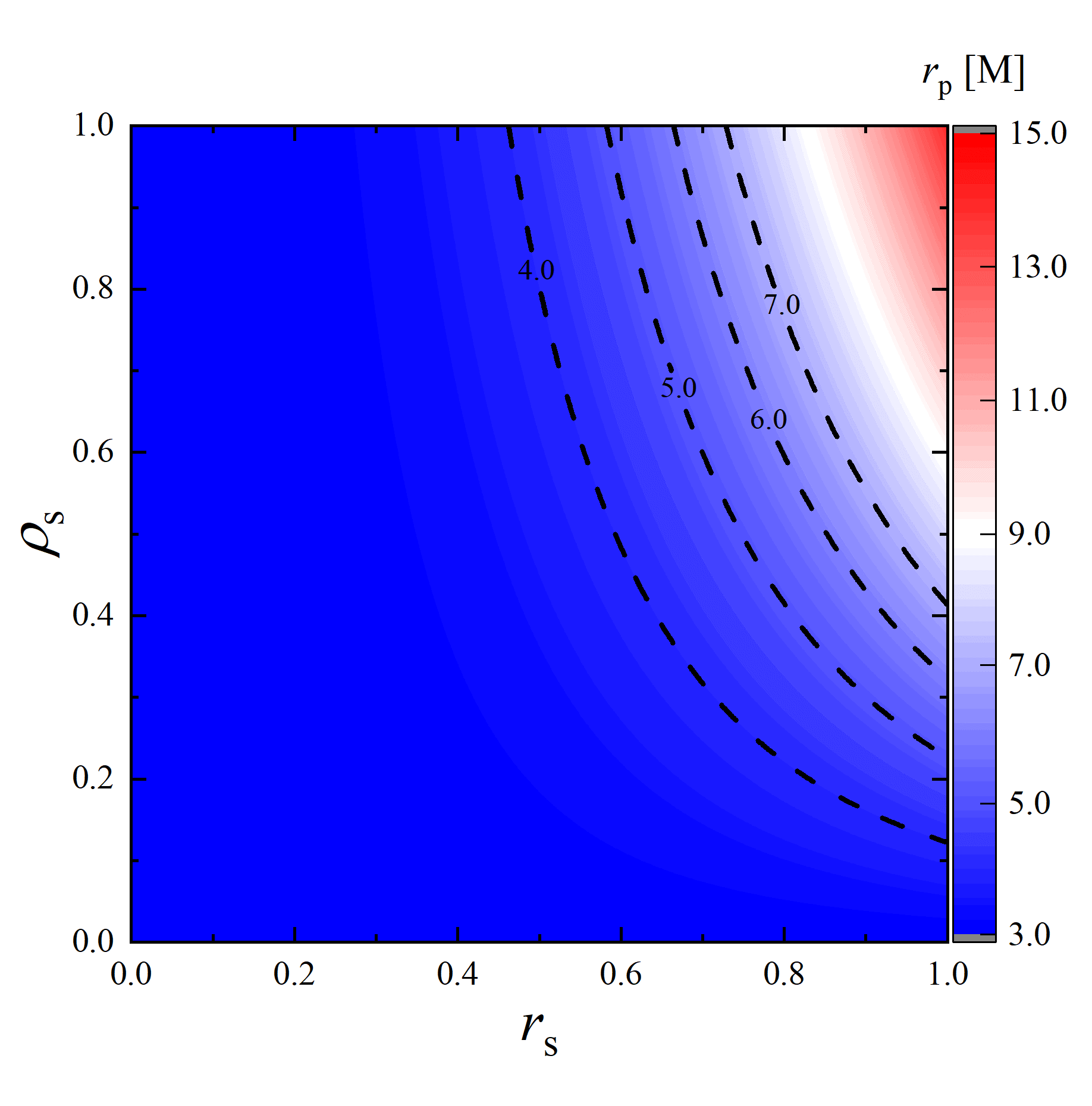}
\includegraphics[width=3.5cm]{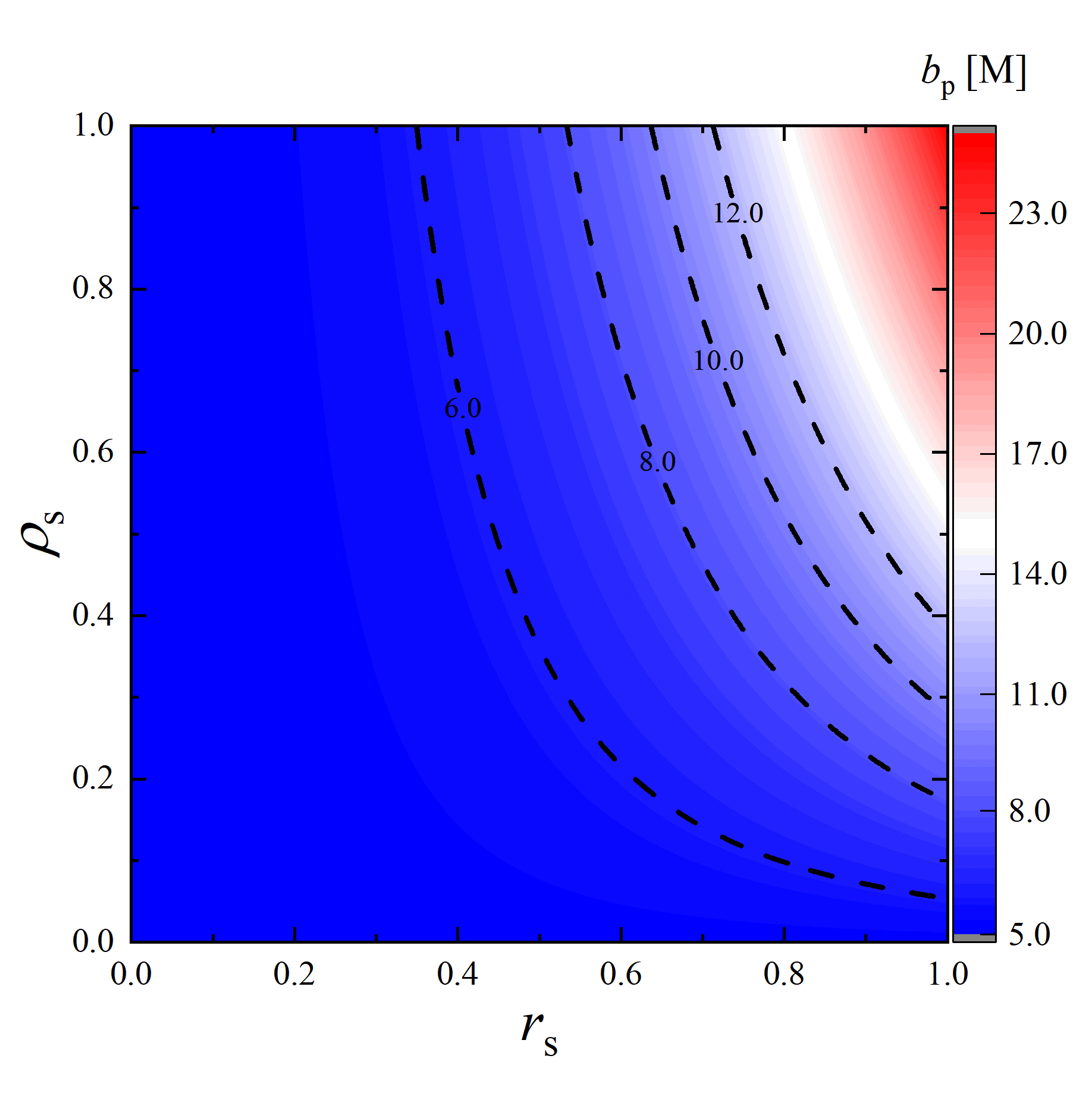}
\caption{From left to right: dependence of the event horizon radius $r_{\textrm{eh}}$, ISCO radius $r_{\textrm{isco}}$, critical photon orbit radius $r_{\textrm{p}}$, and critical impact parameter $b_{\textrm{p}}$ on the dark matter halo parameters. Evidently, increasing $r_{\textrm{s}}$ and $\rho_{\textrm{s}}$ enlarges all of these relativistic parameters, indicating a positive correlation between the dark matter halo and the strength of the gravitational field.}}\label{fig3}
\end{figure*}

To execute our algorithm, one must supply the metric potential $f(r)$ and its first-, second-, and third-order partial derivatives with respect to $r$. Although analytic expressions exist for these derivatives in the chosen spacetime, their explicit forms are omitted here due to their algebraic complexity. Once the model is configured, the algorithm employs a Newton iteration scheme to compute, for given values of $r_{\textrm{s}}$ and $\rho_{\textrm{s}}$, the radii of the event horizon $r_{\textrm{eh}}$, the ISCO $r_{\textrm{isco}}$, the photon ring $r_{\textrm{p}}$, and the critical curve $b_{\textrm{p}}$. The dependence of these relativistic features on the parameters is shown in figure 3. A clear positive correlation emerges between the dark matter halo parameters and all characteristic radii, arising from the additional curvature induced by the halo, which strengthens the gravitational field. Furthermore, the two parameters differ in their influence: $r_{\textrm{s}}$ exerts a stronger effect than $\rho_{\textrm{s}}$, as evidenced by the predominantly vertical contour lines across all four panels. 
\begin{figure*}%[tbph]
\center{
\includegraphics[width=4cm]{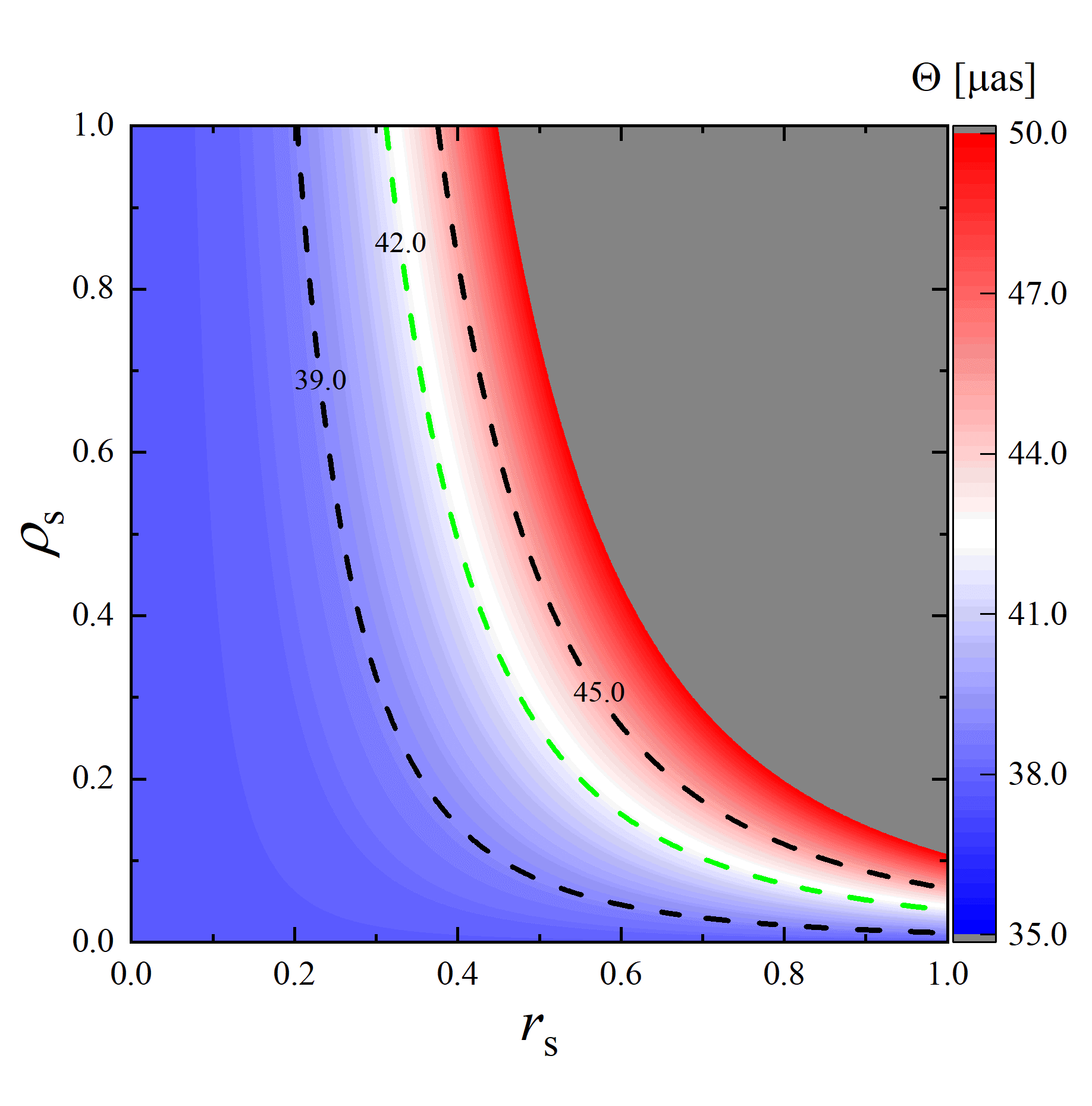}
\includegraphics[width=4cm]{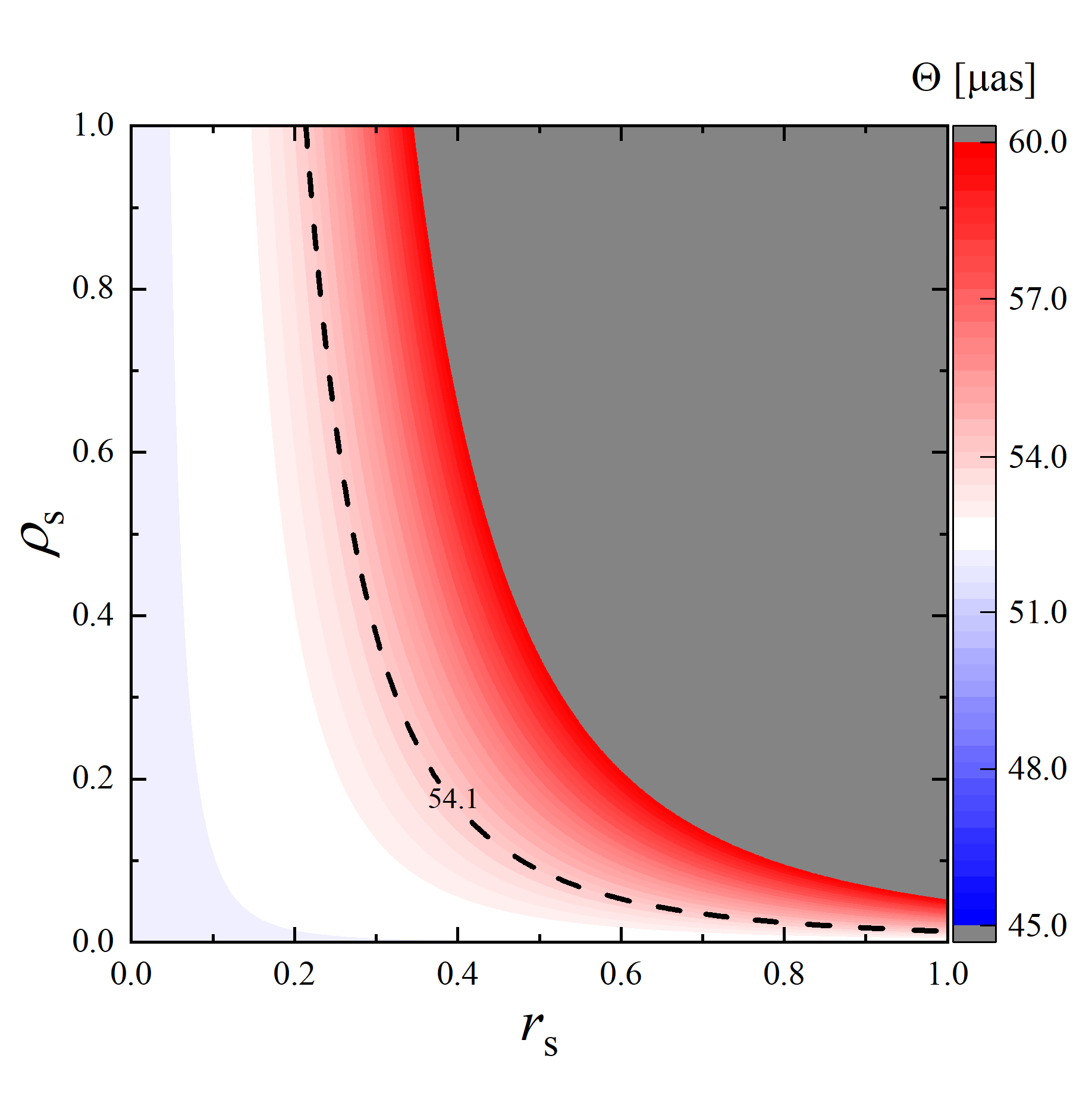}
\caption{Variation of the shadow angular diameter with dark matter halo parameters. The left and right panels correspond to the parameters for M87$^{*}$ and the Galactic Center black hole (Sgr A$^{*}$), respectively. It is observed that the left panel constrains the dark matter halo parameters to a relatively narrow range, while the constraints in the right panel are considerably looser. In summary, however, the dark matter halo model remains consistent with current shadow observations.}}\label{fig4}
\end{figure*}

Employing equation \eqref{11}, the algorithm computes the observable angular diameter of the critical curve for the target black hole, given its mass and distance. Figure 4 illustrates the distribution of the angular diameter $\Theta$ across the parameter space for the target black hole, using the measured parameters of M87$^{*}$ ($M \approx 6.2 \times 10^{9} M_{\odot}$, $D \approx 16.8 \mathrm{Mpc}$) and Sgr A$^{*}$ ($M \approx 4.14 \times 10^{6} M_{\odot}$, $D \approx 8.127 \mathrm{kpc}$). The evolution of $\Theta$ with respect to $r_{\textrm{s}}$ and $\rho_{\textrm{s}}$ is consistent with that of $b_{\textrm{p}}$. The observational data of M87$^{*}$ restrict the dark matter parameters to a crescent-shaped region, whereas the constraints from Sgr A$^{*}$ remain comparatively weak. Taken together, the analysis shows that, within the current angular resolution of the EHT, a Schwarzschild black hole surrounded by a Dehnen-type dark matter halo survives the shadow test and thus ensures its validity in the theory of gravity.
\subsection{Code accuracy}
In the previous section, the determination of the event horizon, photon ring, innermost stable circular orbit, and critical curve of the black hole relied solely on solving algebraic equations, without the need for ray-tracing. By contrast, subsequent applications---such as reconstructing accretion disk structures, tracking the number of disk crossings by photons, computing redshift factor distributions, simulating black hole images, and modeling light curves---require the computation of a large ensemble of light rays. It is therefore imperative to verify the numerical accuracy of the ray-tracing procedure beforehand. To this end, we assess the precision of our algorithm by comparing the critical curve radius $b_{\textrm{p}}$ obtained via ray-tracing with the value derived from the Newton iteration method in equations \eqref{9} and \eqref{10}.
\begin{figure*}%[tbph]
\center{
\includegraphics[width=4cm]{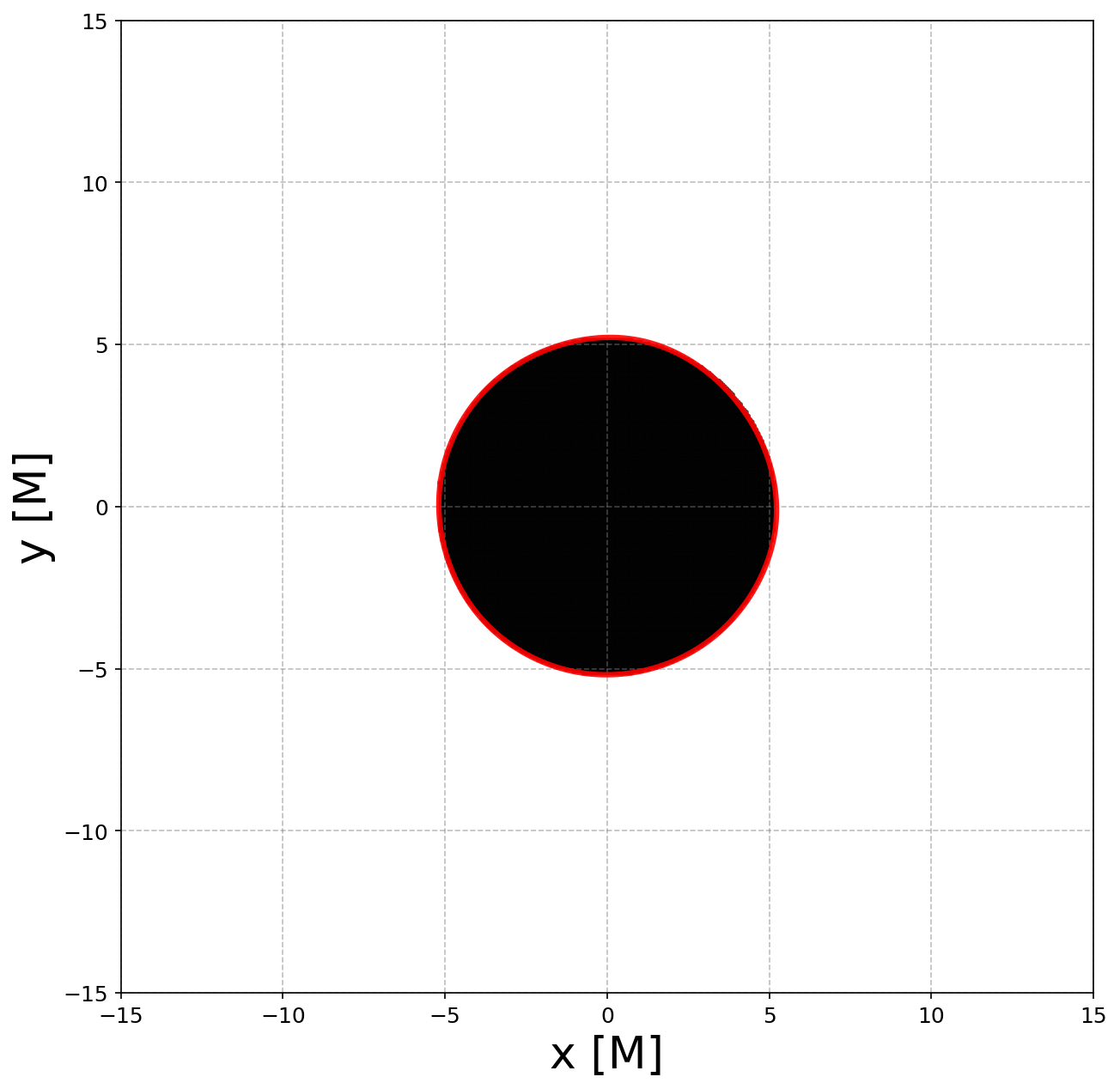}
\caption{The black hole shadow image of a Schwarzschild black hole obtained with OCTOPUS. Black pixels represent rays captured by the black hole. The red closed curve, derived via least-squares fitting, outlines the shadow boundary, with a radius of $5.192 \pm 0.0197$ M, corresponding to a relative error of less than $0.0008$ compared to the theoretical value. It is important to note that this simulation was performed at $500 \times 500$ resolution; higher resolution would further reduce the error.}}\label{fig5}
\end{figure*}

\begin{table}[ht]
\centering
\caption{Theoretical critical impact parameter $b_{\textrm{p}}^{\textrm{A}}$, fitted value $b_{\textrm{p}}^{\textrm{F}}$, and their relative error across different parameter spaces.}
\begin{tabular}{|c|c|c|c|c|}
\hline
\multicolumn{2}{|c|}{Parameters} & Analytical solution $b_{\textrm{p}}^{\textrm{A}}$ & Fitting solution $b_{\textrm{p}}^{\textrm{F}}$ & Relative error $|b_{\textrm{p}}^{\textrm{A}}-b_{\textrm{p}}^{\textrm{F}}|/b_{\textrm{p}}^{\textrm{A}}$\\ \hline
\multirow{4}{*}{$\rho_{\textrm{s}}=0.5$} 
& $r_{\textrm{s}} = 0.1$ & $5.207$ & $5.201 \pm 0.0204$ & 0.0012 \\ 
& $r_{\textrm{s}} = 0.4$ & $5.785$ & $5.779 \pm 0.0195$ & 0.0010 \\ 
& $r_{\textrm{s}} = 0.7$ & $8.234$ & $8.230 \pm 0.0179$ & 0.0005 \\ 
& $r_{\textrm{s}} = 1.0$ & $14.358$ & $14.354 \pm 0.0150$ & 0.0003 \\ \hline
\multirow{4}{*}{$r_{\textrm{s}} = 0.5$} 
& $\rho_{\textrm{s}} = 0.1$ & $5.414$ & $5.411 \pm 0.0211$ & 0.0006 \\ 
& $\rho_{\textrm{s}} = 0.4$ & $6.089$ & $6.085 \pm 0.0189$ & 0.0007 \\ 
& $\rho_{\textrm{s}} = 0.7$ & $6.788$ & $6.783 \pm 0.0215$ & 0.0007 \\ 
& $\rho_{\textrm{s}} = 1.0$ & $7.506$ & $7.499 \pm 0.0193$ & 0.0009 \\ \hline
\end{tabular}
\end{table}

We configure the observation screen with dimensions $x \in [-15,15]$ M and $y \in [-15,15]$ M, with a resolution of $500 \times 500$ pixels. Numerical simulations are first carried out for the Schwarzschild case ($r_{\textrm{s}}=\rho_{\textrm{s}}=0$), with the results shown in figure 5. Black points correspond to light rays captured by the black hole, whose boundary---the critical curve---is extracted via least-squares fitting and indicated by the closed red line. The fitted curve yields a radius of $5.192 \pm 0.0197$ M, differing from the well-known theoretical value $\sqrt{27} \approx 5.196$ M by a relative error of less than $0.0008$.

We further examine the numerical accuracy in the presence of a dark matter halo. Figure 6 displays the black hole shadows and fitted critical curves obtained through ray-tracing under various halo parameters. The fitted curve radii exhibit excellent agreement with results from the Newton iteration method, with relative errors predominantly at the level of $10^{-4}$, as summarized in table 1. Additionally, the critical curve expands with increasing $r_{\textrm{s}}$ or $\rho_{\textrm{s}}$, though the effect of $r_{\textrm{s}}$ is more pronounced than that of the core density $\rho_{\textrm{s}}$. These trends are consistent with the behavior illustrated in panel (d) of figure 3.
\begin{figure*}%[tbph]
\center{
\includegraphics[width=3.5cm]{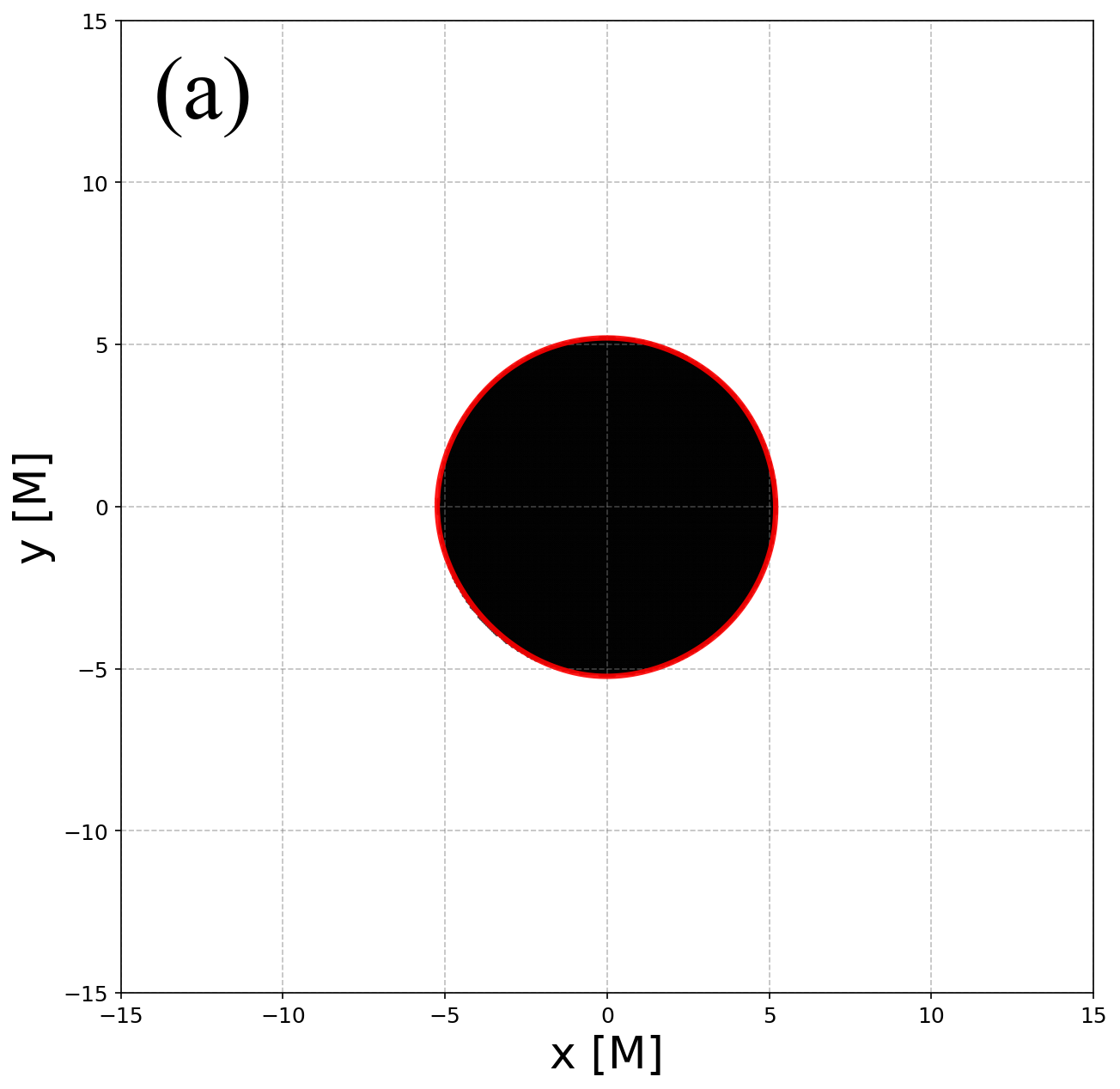}
\includegraphics[width=3.5cm]{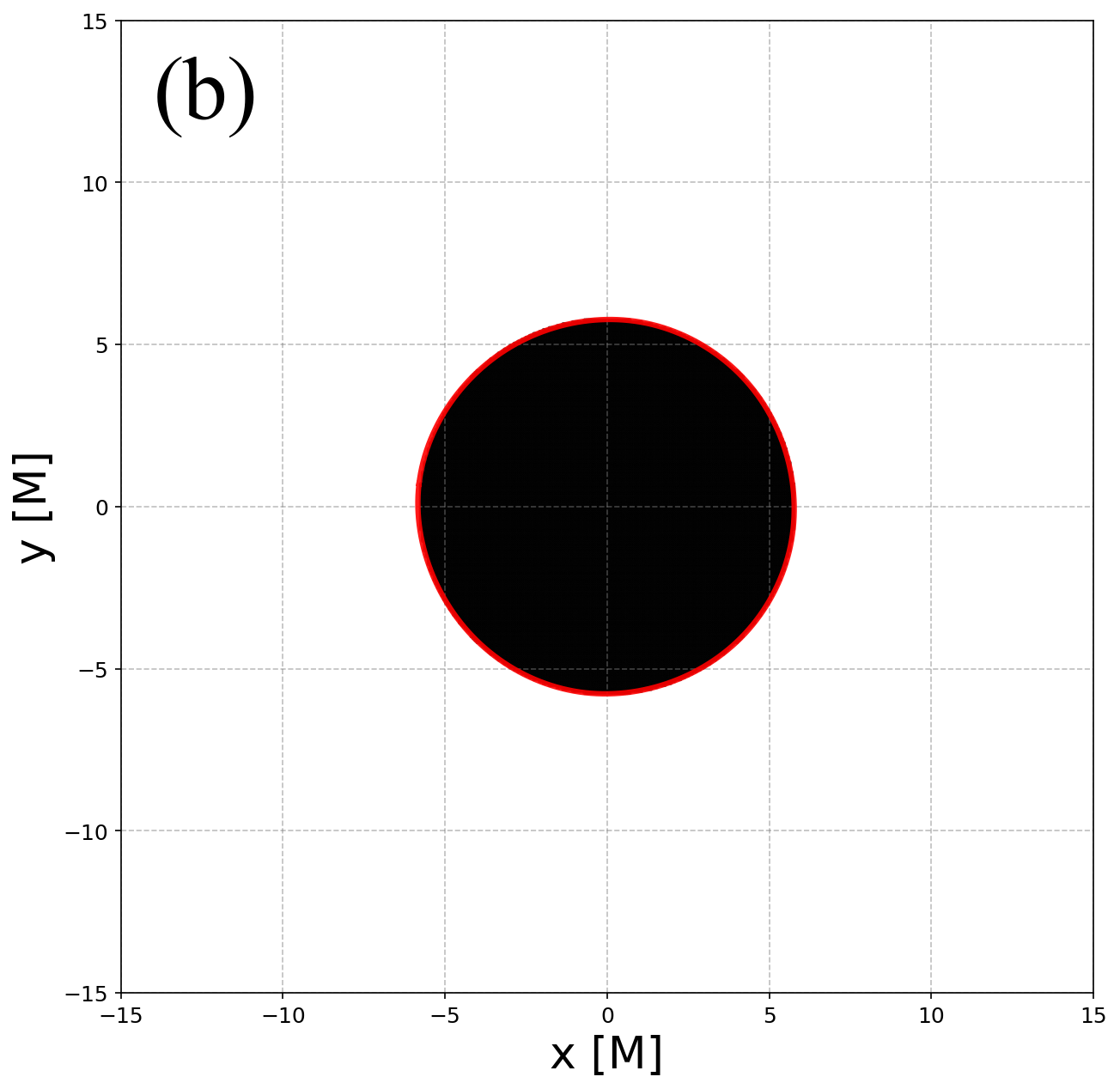}
\includegraphics[width=3.5cm]{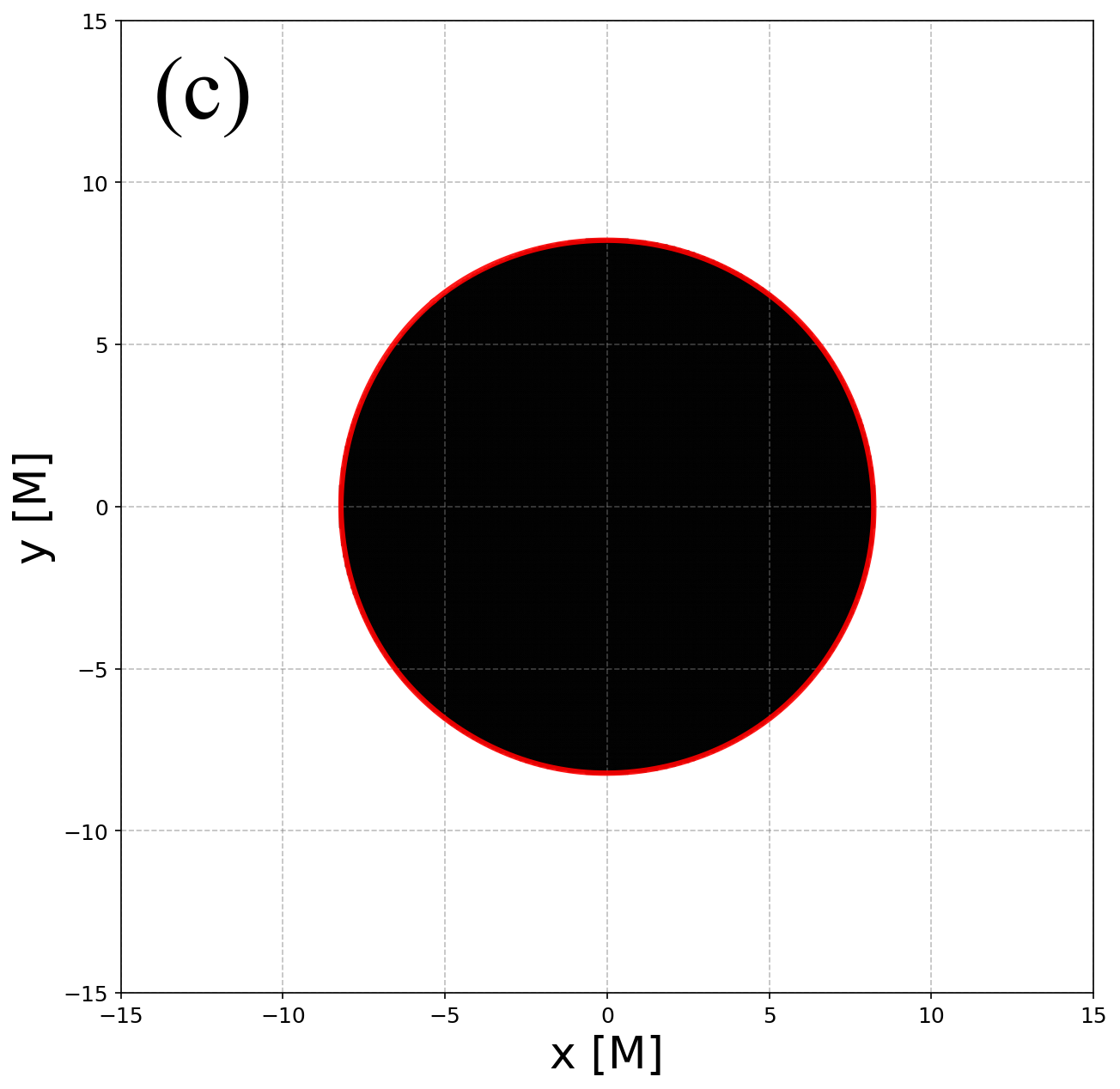}
\includegraphics[width=3.5cm]{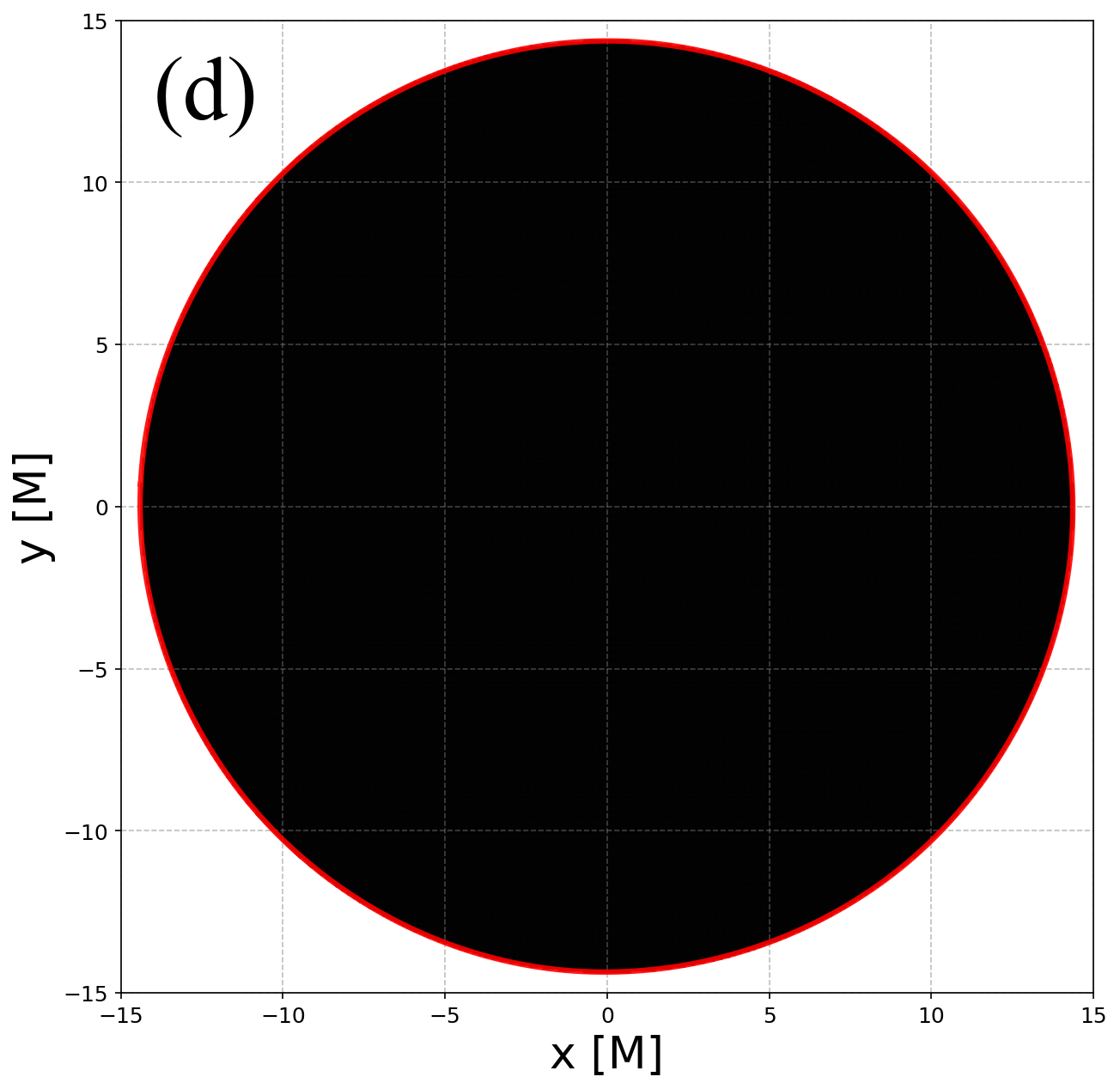}
\includegraphics[width=3.5cm]{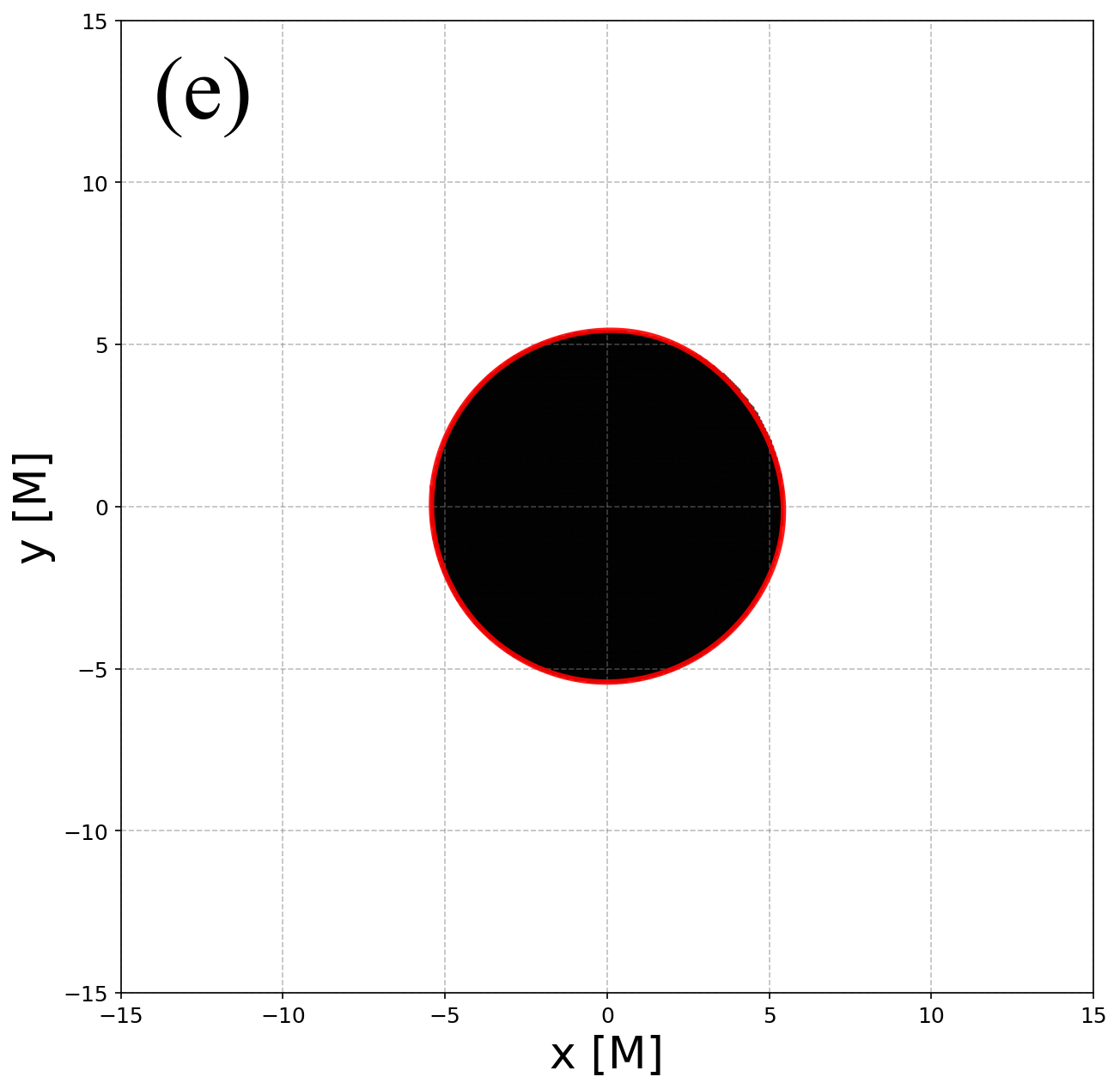}
\includegraphics[width=3.5cm]{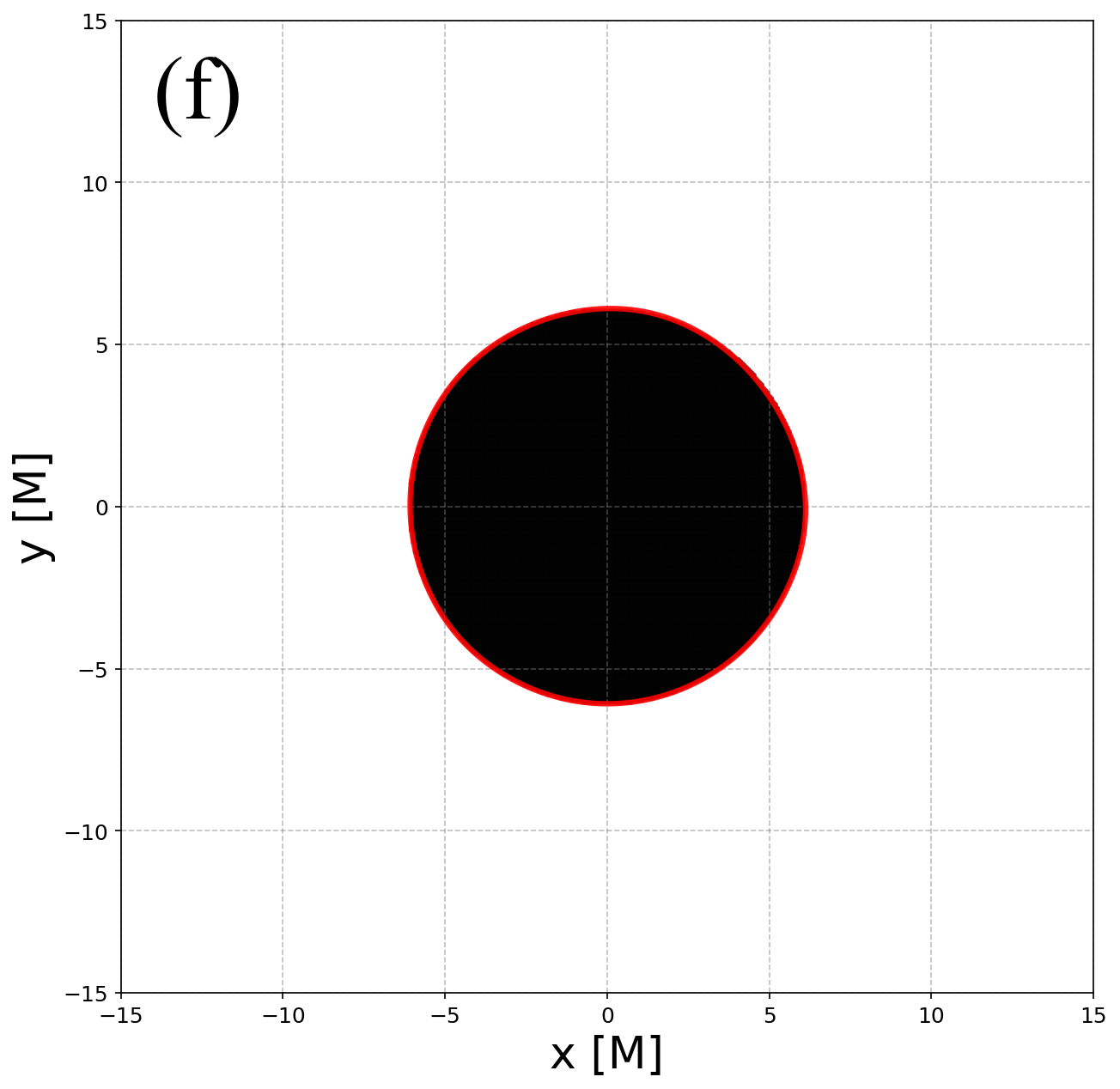}
\includegraphics[width=3.5cm]{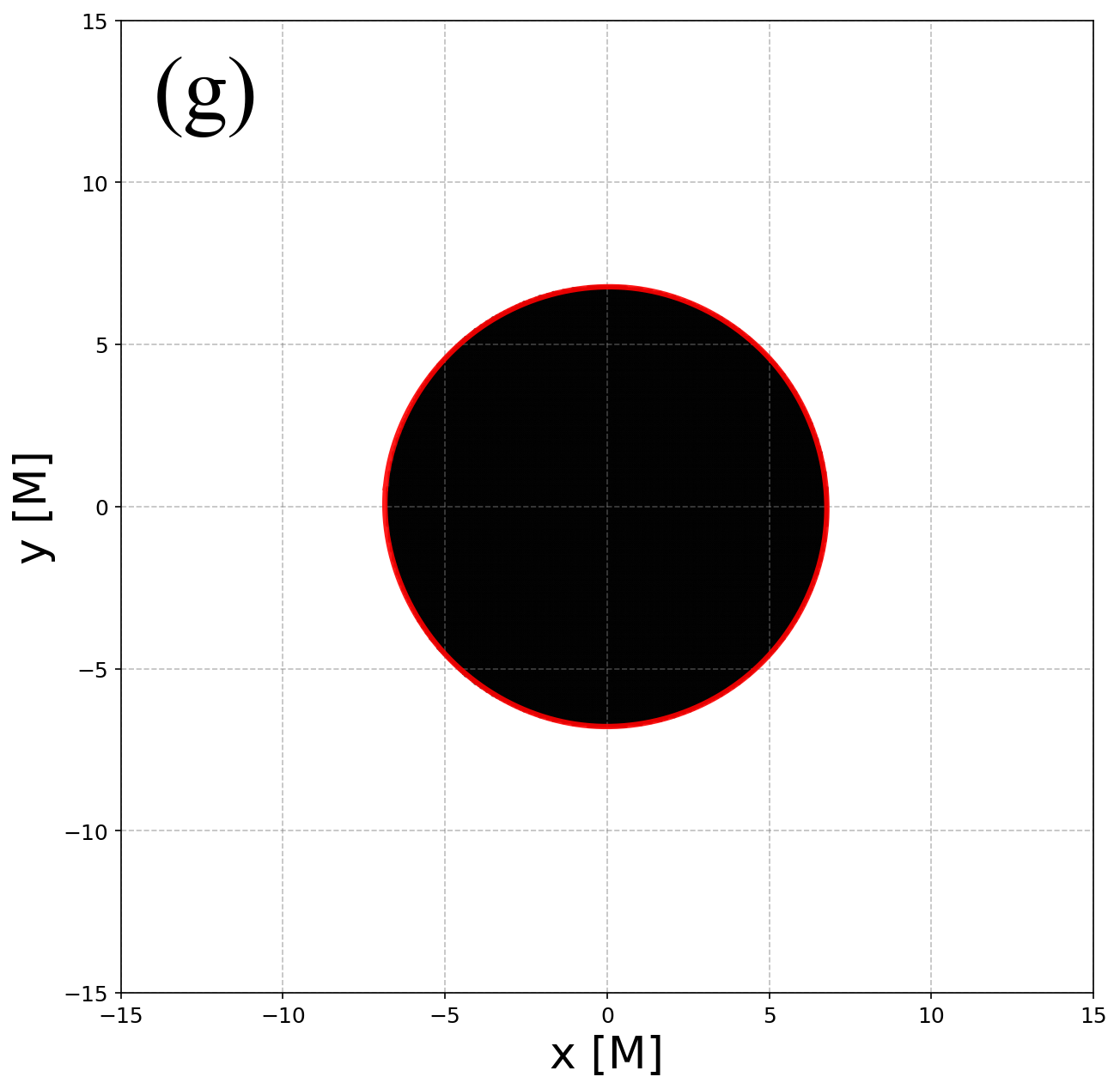}
\includegraphics[width=3.5cm]{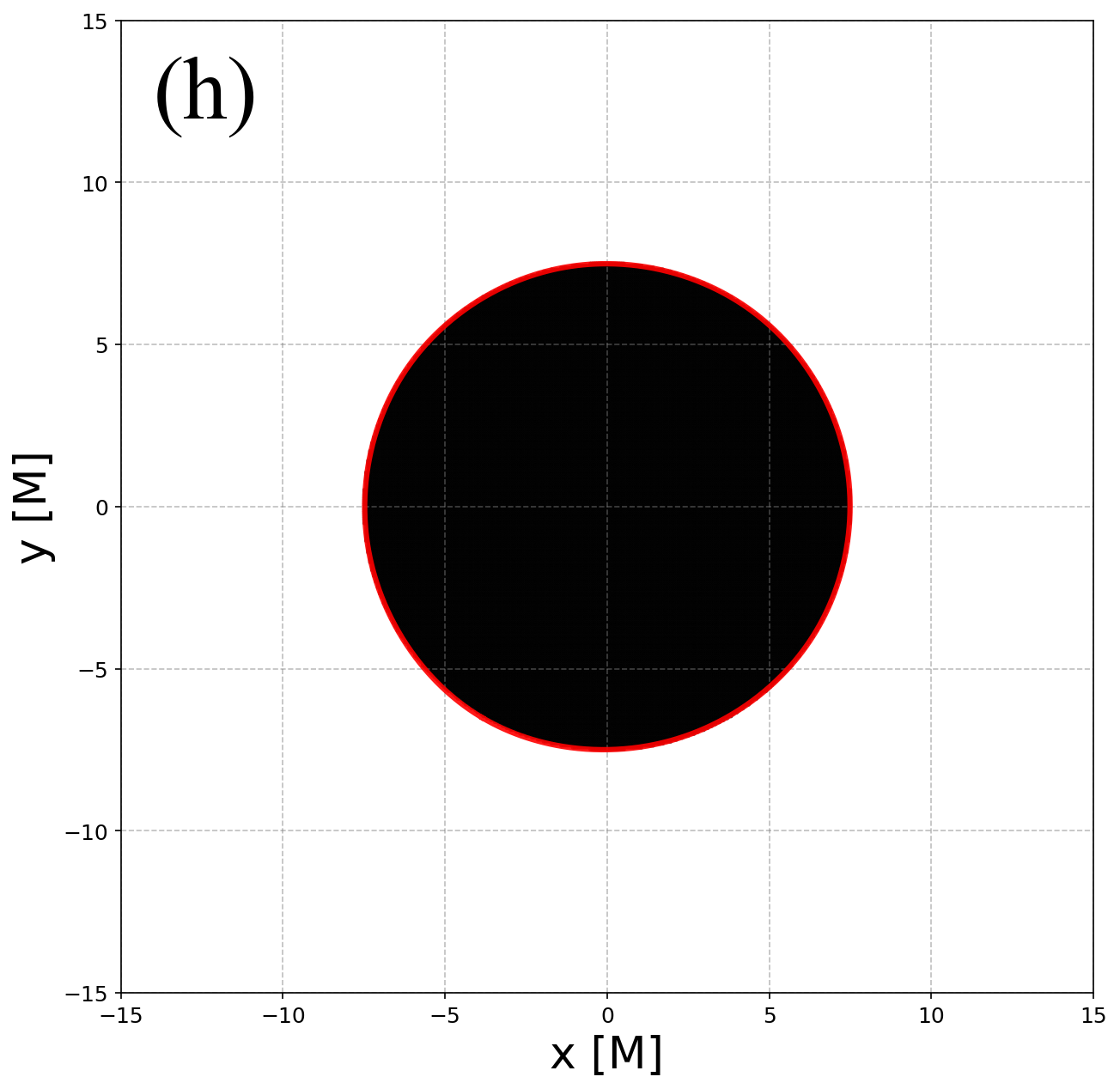}
\caption{Top row: shadow images with a fixed $\rho_{\textrm{s}}=0.5$, corresponding from left to right to $r_{\textrm{s}} = $ $0.1$, $0.4$, $0.7$, $1$. Bottom row: shadow images with a fixed $r_{\textrm{s}} = 0.5$, corresponding from left to right to $\rho_{\textrm{s}} = $ $0.1$, $0.4$, $0.7$, $1$. It can be observed that both parameters increase the shadow size, with $r_{\textrm{s}}$ inducing an approximately exponential expansion, while $\rho_{\textrm{s}}$ exerts a more gradual, linear influence. Here, the resolution is fixed at $500 \times 500$ pixels.}}\label{fig6}
\end{figure*}

It should be emphasized that during photon propagation, the Hamiltonian constraint $\mathscr{H}=0$ must in principle hold at all times. In practice, however, rounding and truncation errors in numerical integration may cause the Hamiltonian $\mathscr{H}_{t}$, evaluated from the canonical variables at time $t$, to deviate from zero. The deviation, defined as $\Delta=\mathscr{H}_{t}-\mathscr{H}=\mathscr{H}_{t}$, therefore provides an additional diagnostic for assessing numerical accuracy.

Figure 7 displays the Hamiltonian error $\Delta$ at the termination of ray propagation for three different inclination angles. For $\omega=0^{\circ}$, the errors for a significant number of rays remain below $10^{-13}$ by the end of integration, approaching machine precision. Near the event horizon, strong light bending requires smaller step-sizes and more integration steps, leading to accumulated rounding errors. Nevertheless, for rays captured by the black hole, the Hamiltonian error remains within $10^{-8}$ to $10^{-6}$. This accuracy is sufficient for reliably determining critical curves, as confirmed in figures 5 and 6.

When the observation inclination is nonzero, as in panels (b) and (c), distinct vertical stripes appear near $x=0$. The Hamiltonian errors of rays in these regions lie between $10^{-12}$ and $10^{-9}$, slightly larger than elsewhere outside the event horizon. This occurs because such rays pass close to or across the black hole's polar regions during propagation, introducing numerical errors in solving equations of motion \eqref{48}, \eqref{50}, and \eqref{51} due to the $\sin\theta$ term in the denominators. Nevertheless, this level of error remains entirely acceptable. Furthermore, the Hamiltonian errors for rays captured by the event horizon remain below $10^{-6}$, consistent with the $\omega=0^{\circ}$ case. Overall, the precision of the integrator employed in our algorithm is fully capable of meeting the requirements for subsequent simulations. 
\begin{figure*}%[tbph]
\center{
\includegraphics[width=4cm]{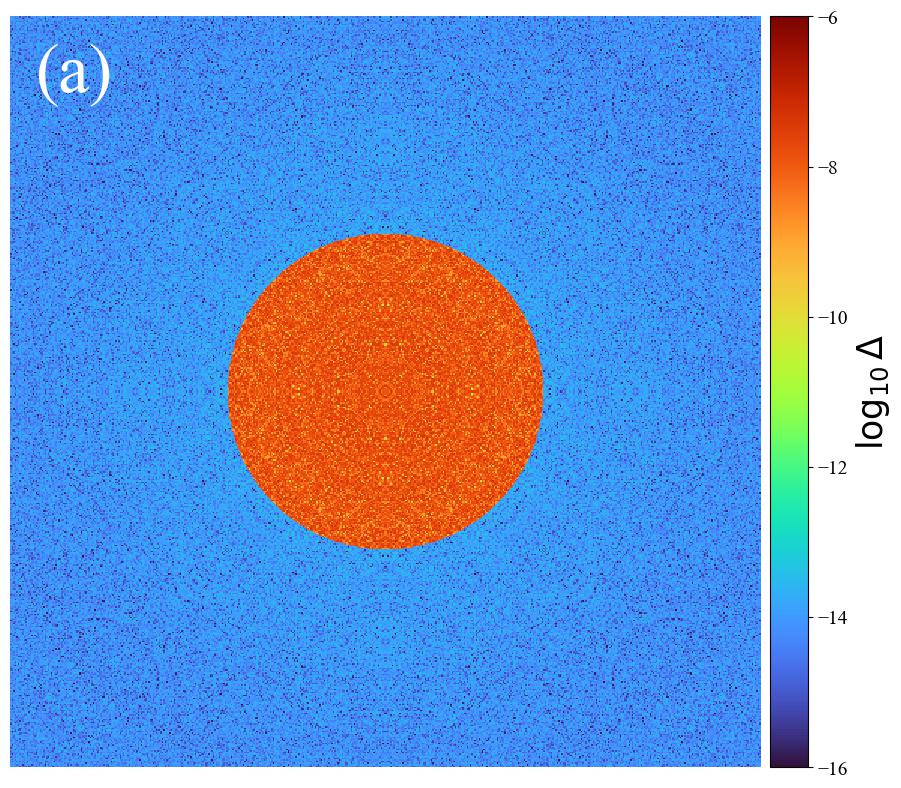}
\includegraphics[width=4cm]{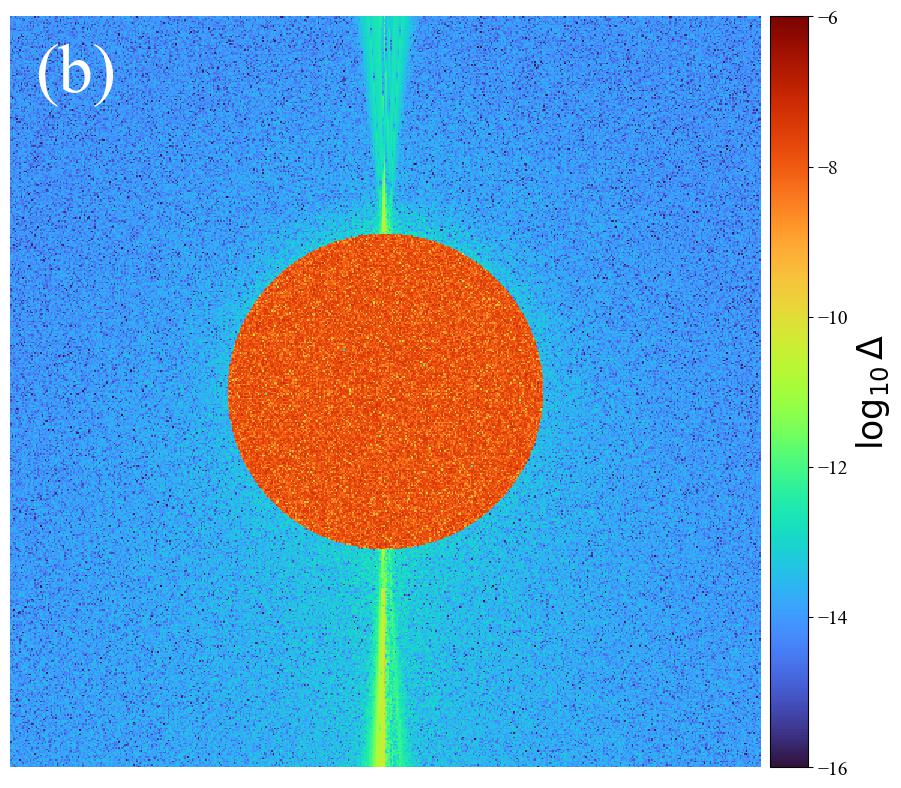}
\includegraphics[width=4cm]{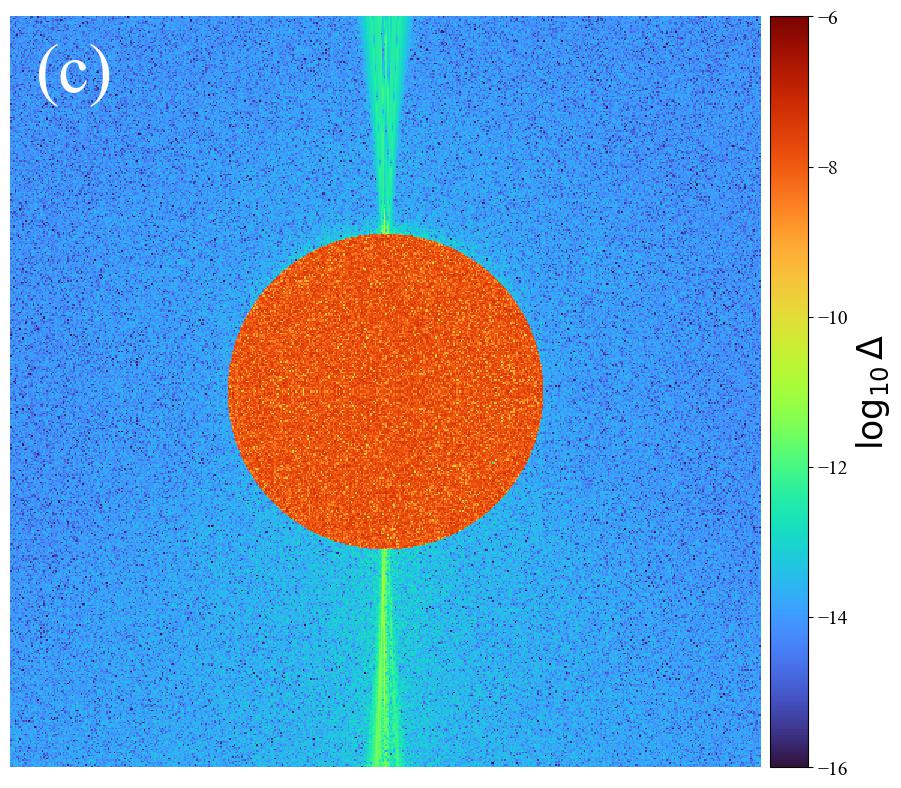}
\caption{Distribution of Hamiltonian errors across the observation screen for different parameter configurations. From left to right, the observation inclinations are $0^{\circ}$, $50^{\circ}$, and $90^{\circ}$, with maximum Hamiltonian errors of $10^{-7.03}$, $10^{-6.93}$, and $10^{-6.85}$, respectively. Here, we fix $r_{\textrm{s}}=\rho_{\textrm{s}}=0.5$ and set the integration parameters $h_{0}=0.0002$ and $n=1.8$.}}\label{fig7}
\end{figure*}
\subsection{Redshift factor}
When the observer is positioned along the black hole's polar axis (i.e., in a face-on view), the rotation of the accretion disk introduces no Doppler effect to the light rays, as the velocity of the emitting particles is orthogonal to the line of sight. However, when the observation inclination satisfies $\omega \neq 0^{\circ}$ or $\omega \neq 180^{\circ}$, Doppler effects induced by the disk dynamics must be considered. Since the momentum of the light ray changes with each disk crossing, the redshift factor $\gamma$ also varies for successive interactions. Therefore, we systematically investigate the distribution of the redshift factor on the observer's screen corresponding to the first, second, and third disk-crossing events, designating them as the first-order, second-order, and third-order redshift factors, respectively.
\begin{figure*}%[tbph]
\center{
\includegraphics[width=3.5cm]{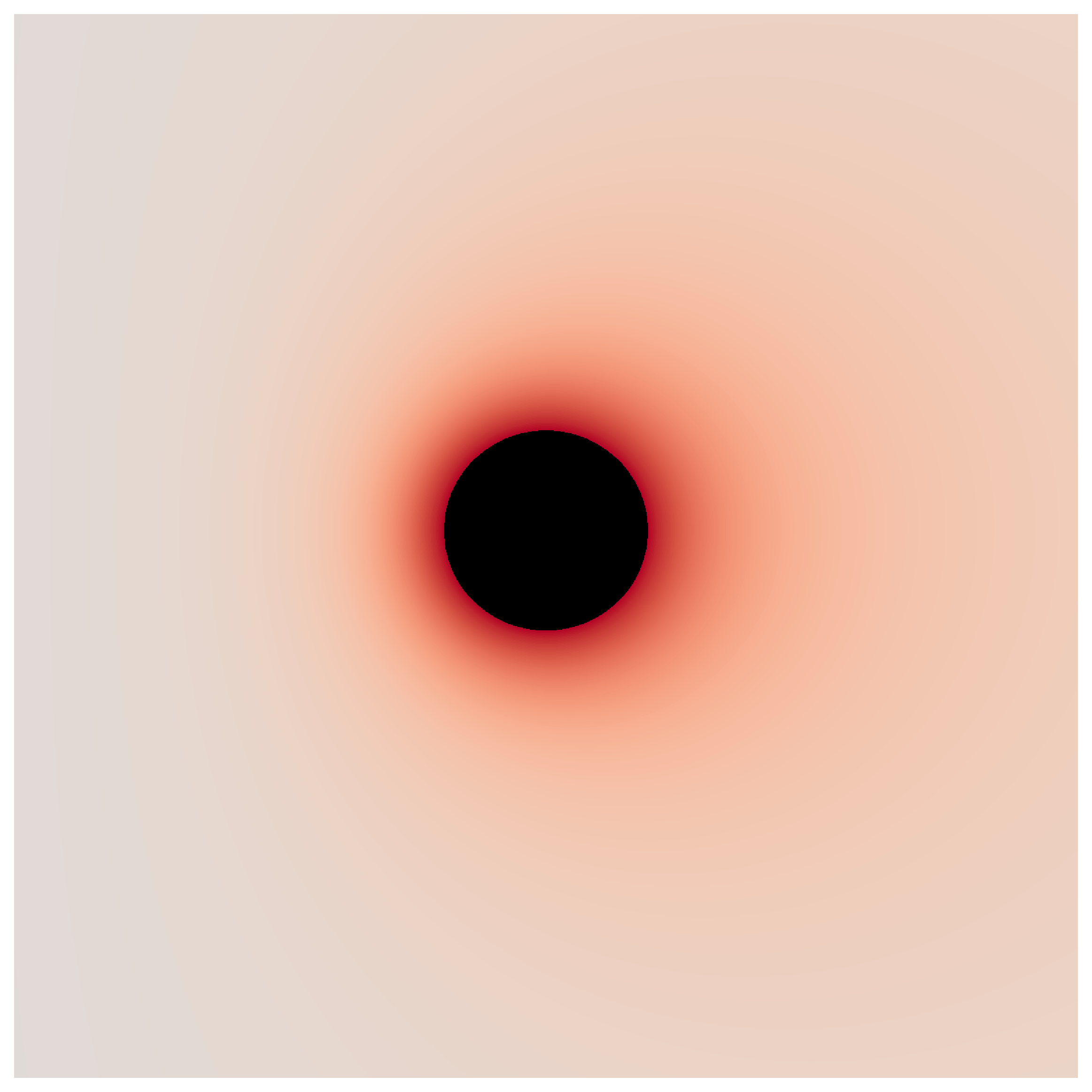}
\includegraphics[width=3.5cm]{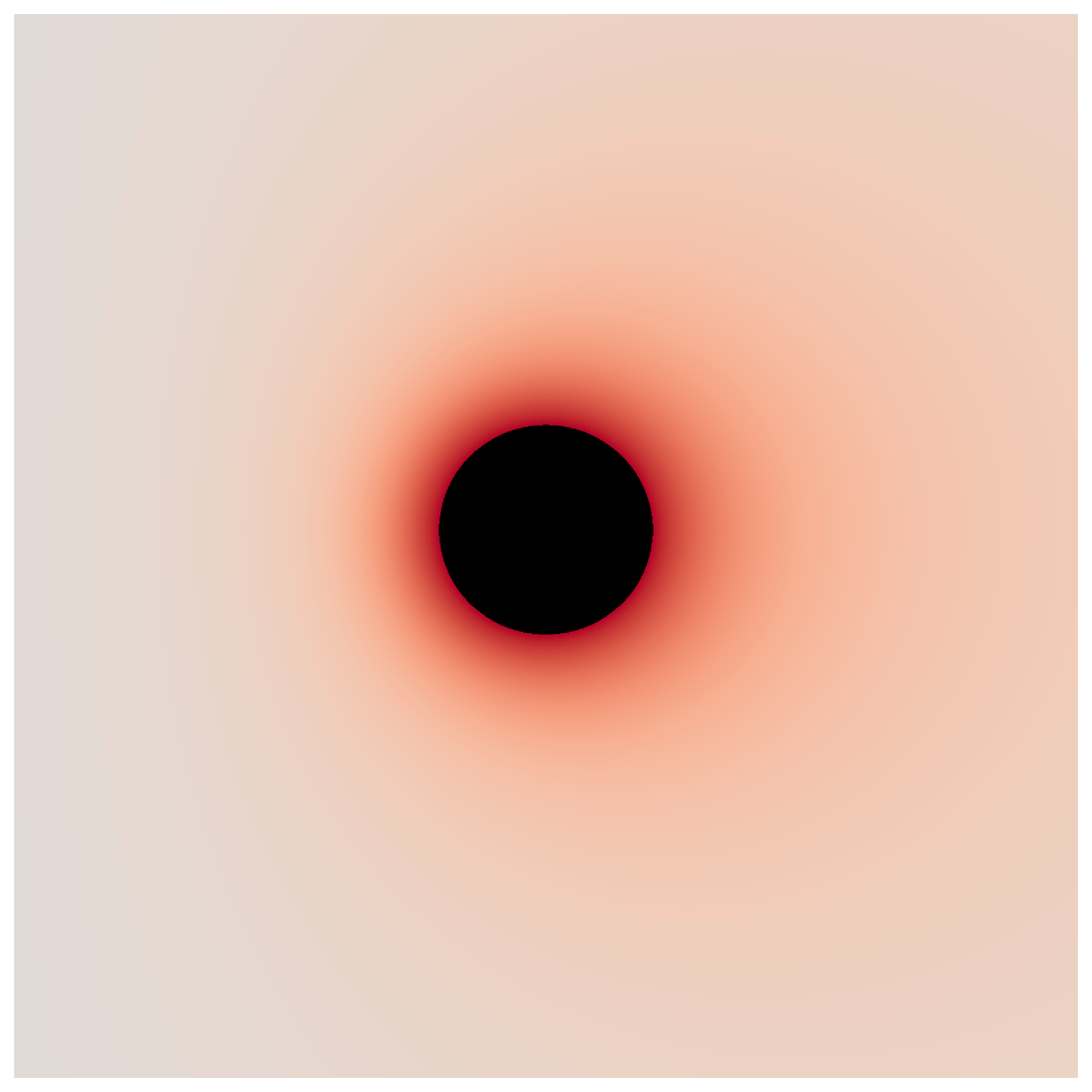}
\includegraphics[width=3.5cm]{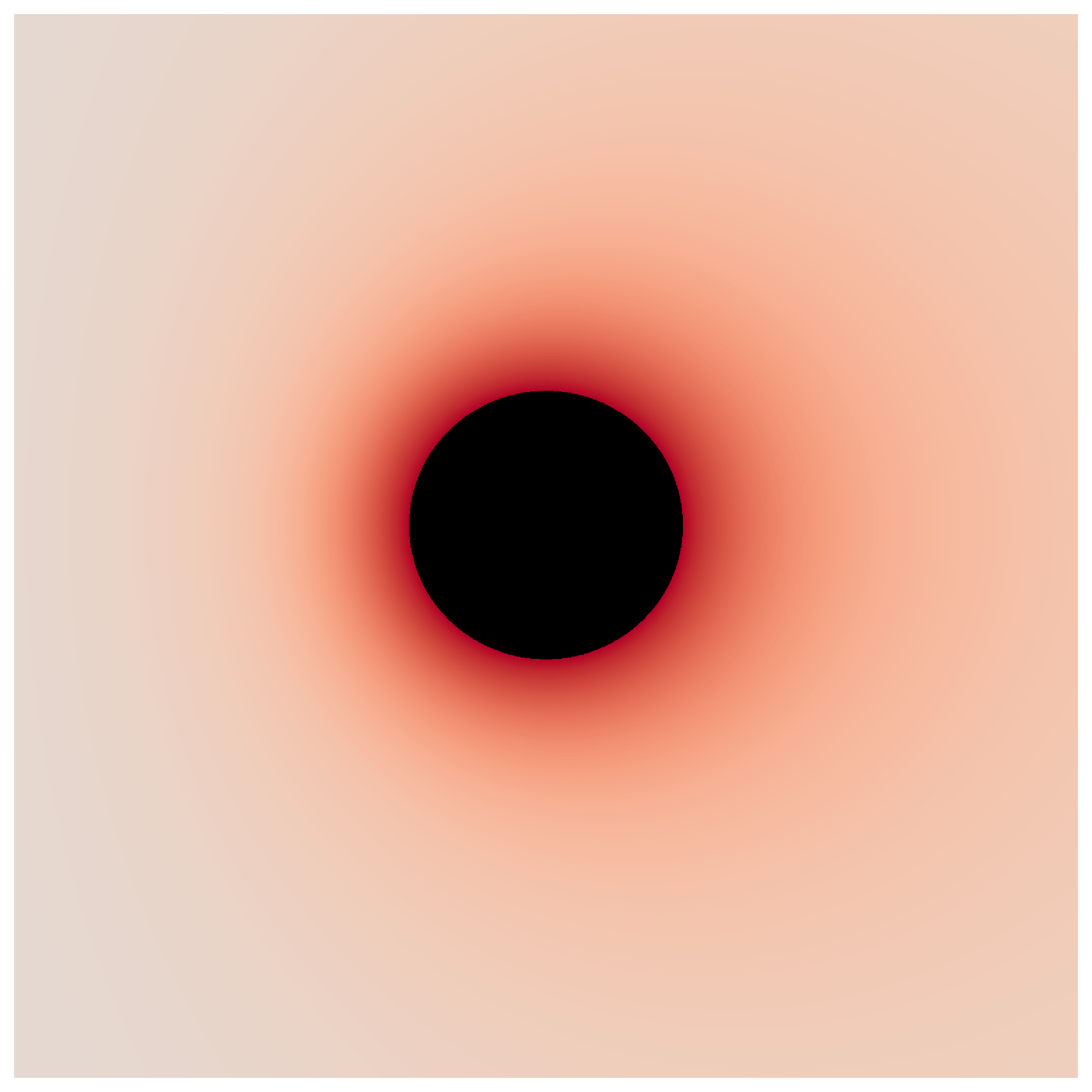}
\includegraphics[width=3.5cm]{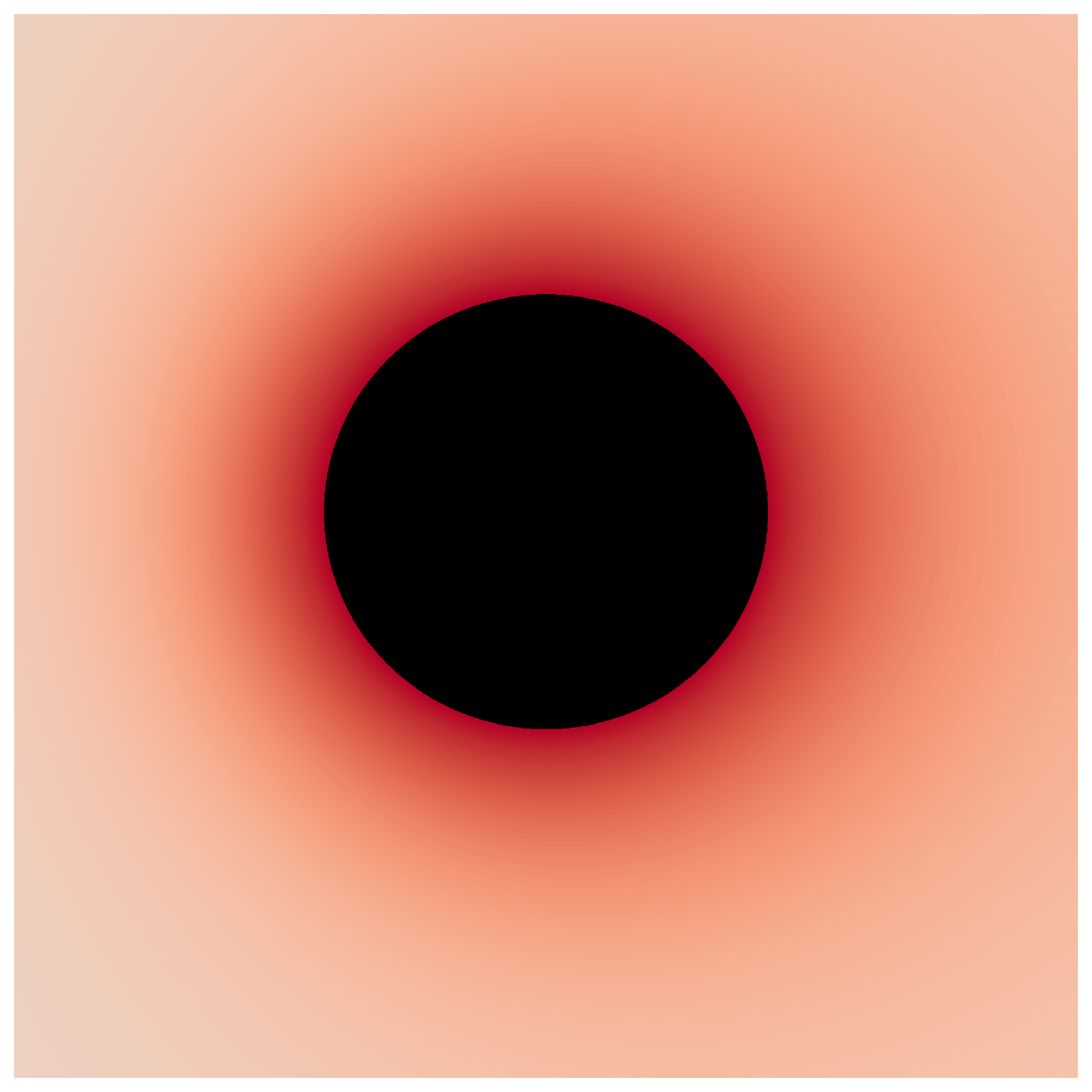}
\includegraphics[width=3.5cm]{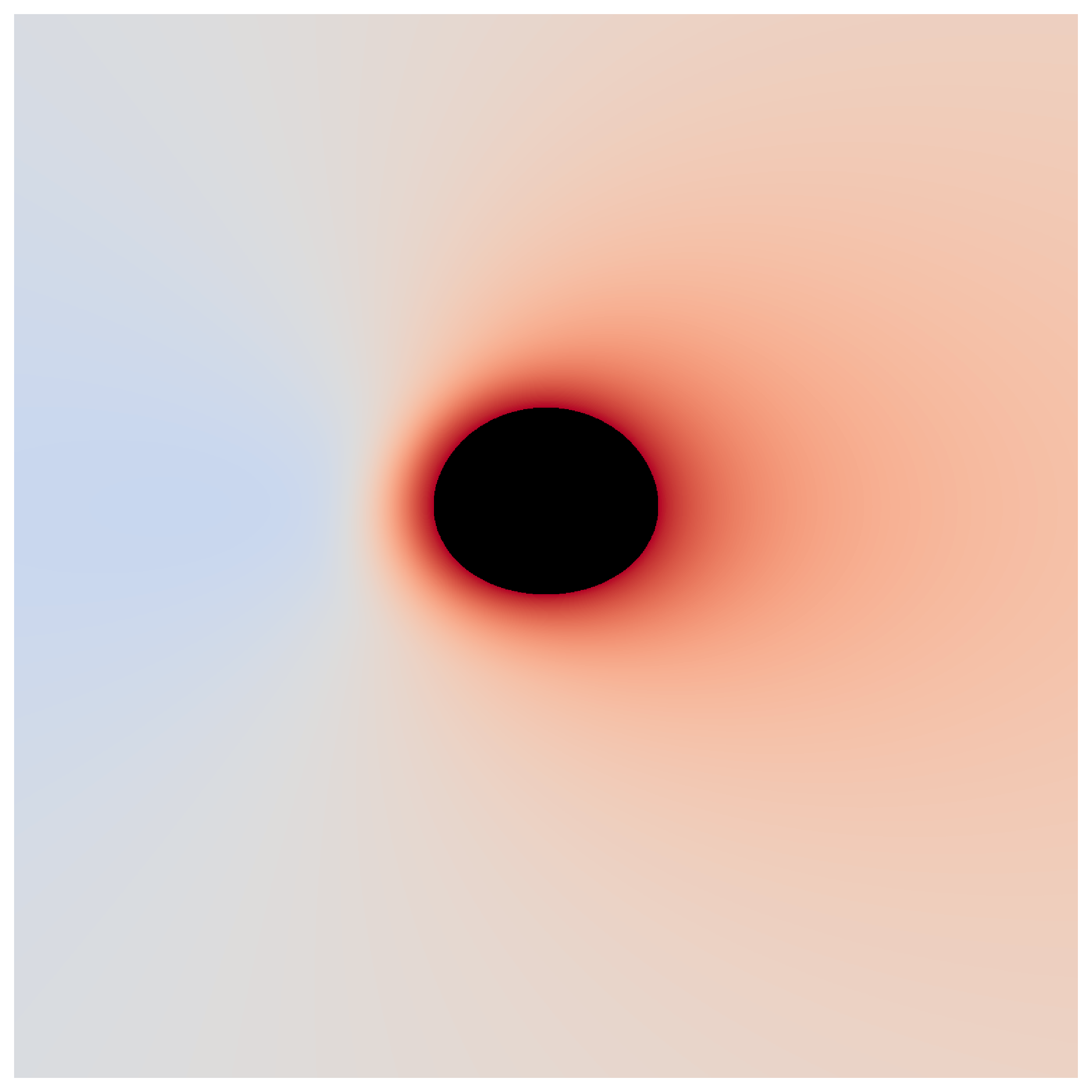}
\includegraphics[width=3.5cm]{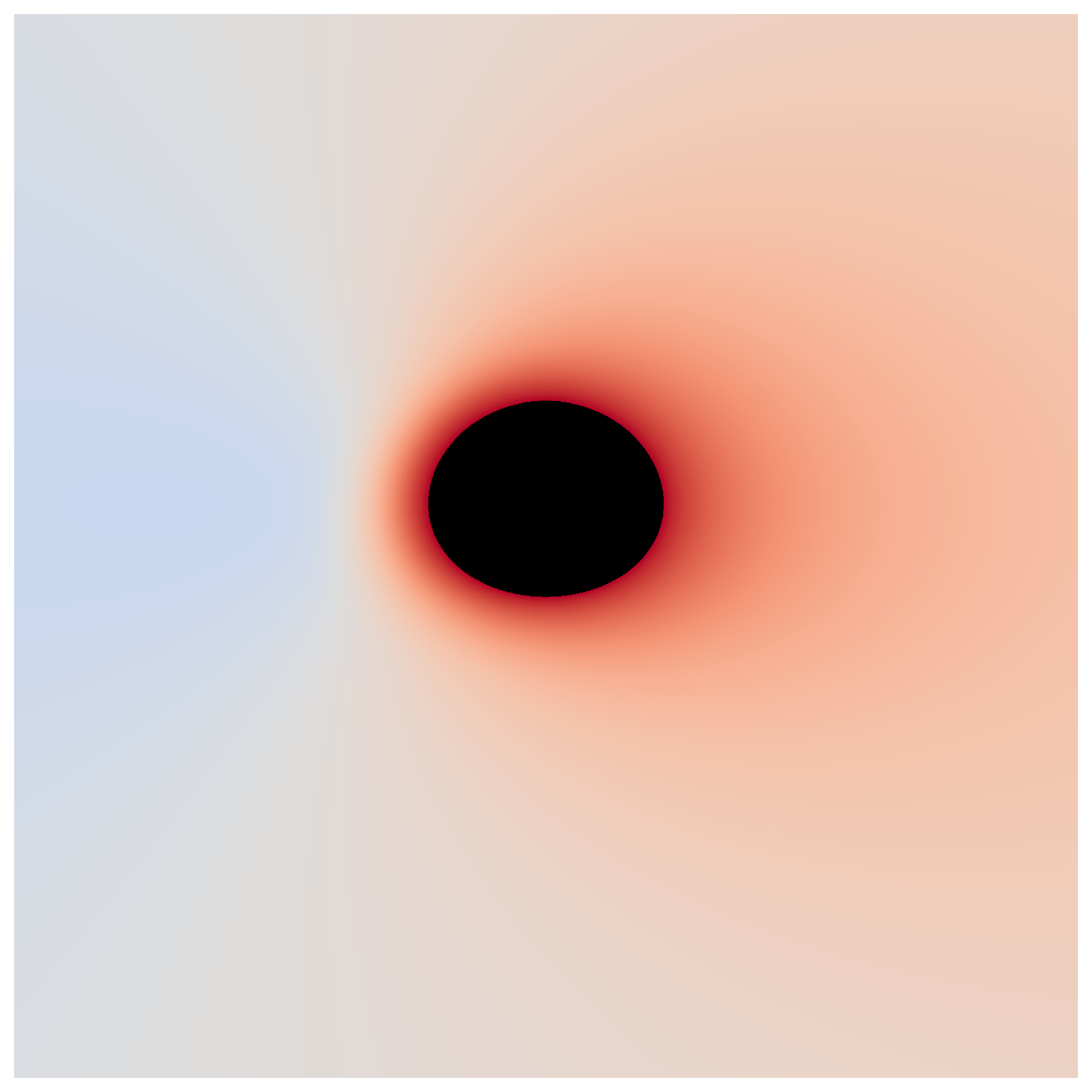}
\includegraphics[width=3.5cm]{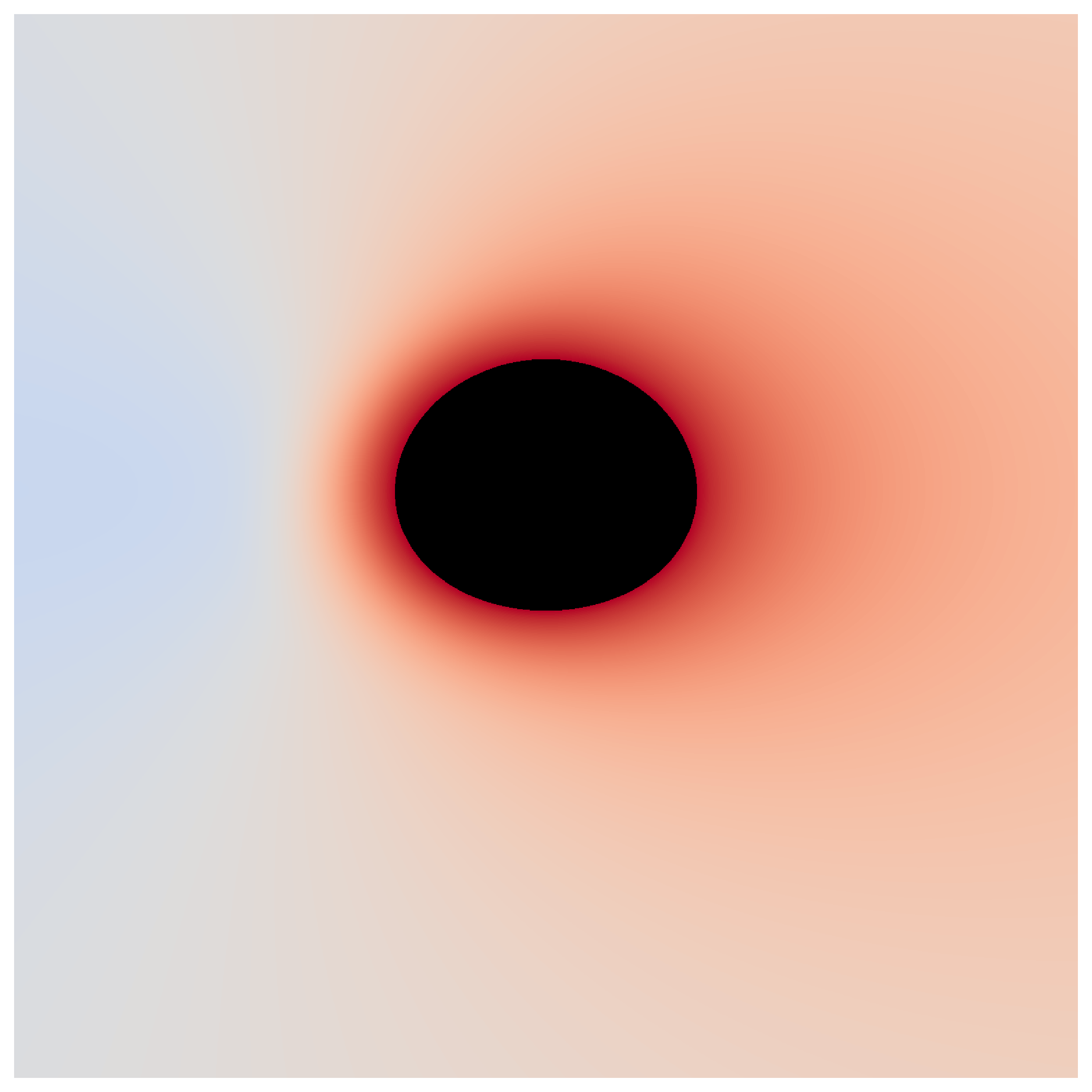}
\includegraphics[width=3.5cm]{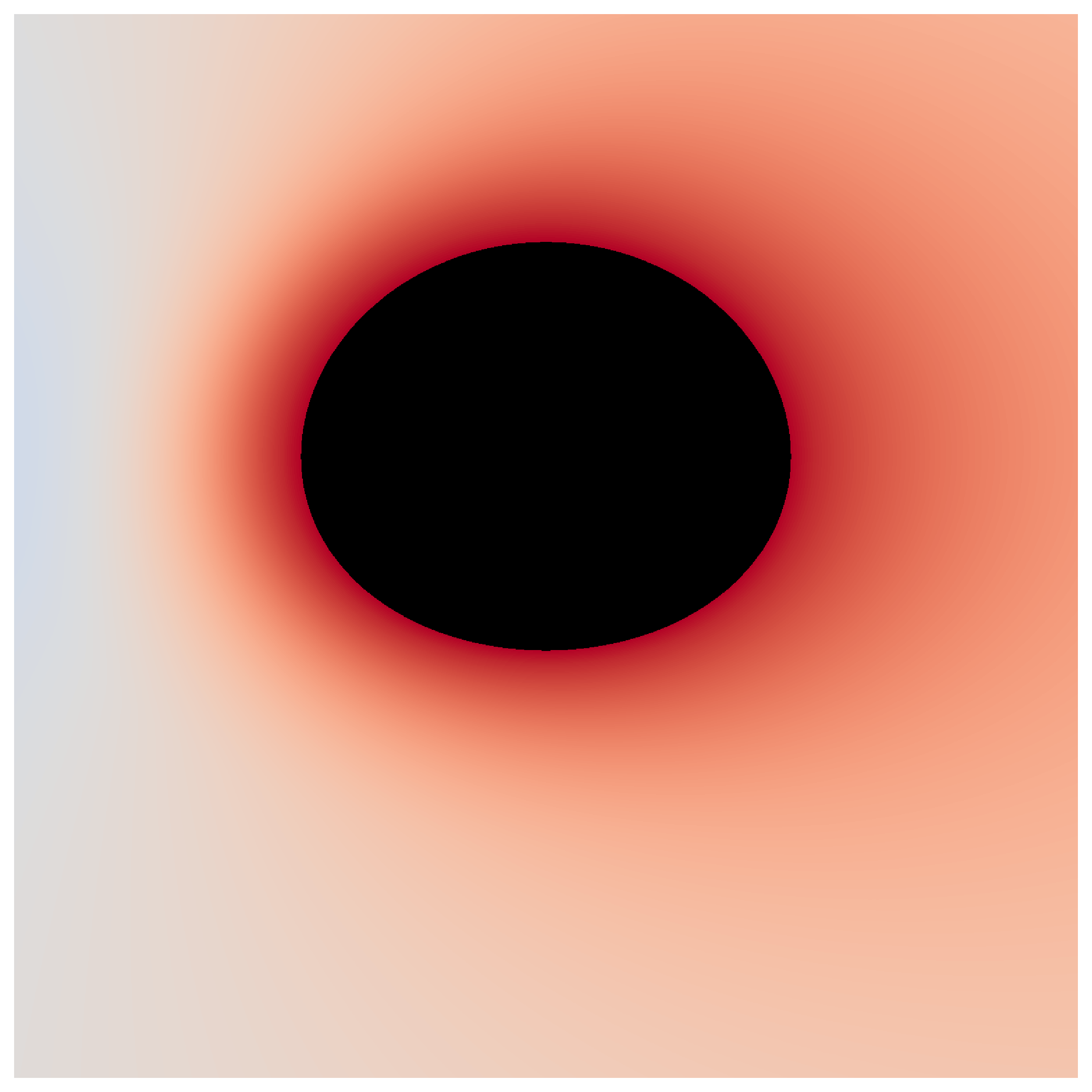}
\includegraphics[width=3.5cm]{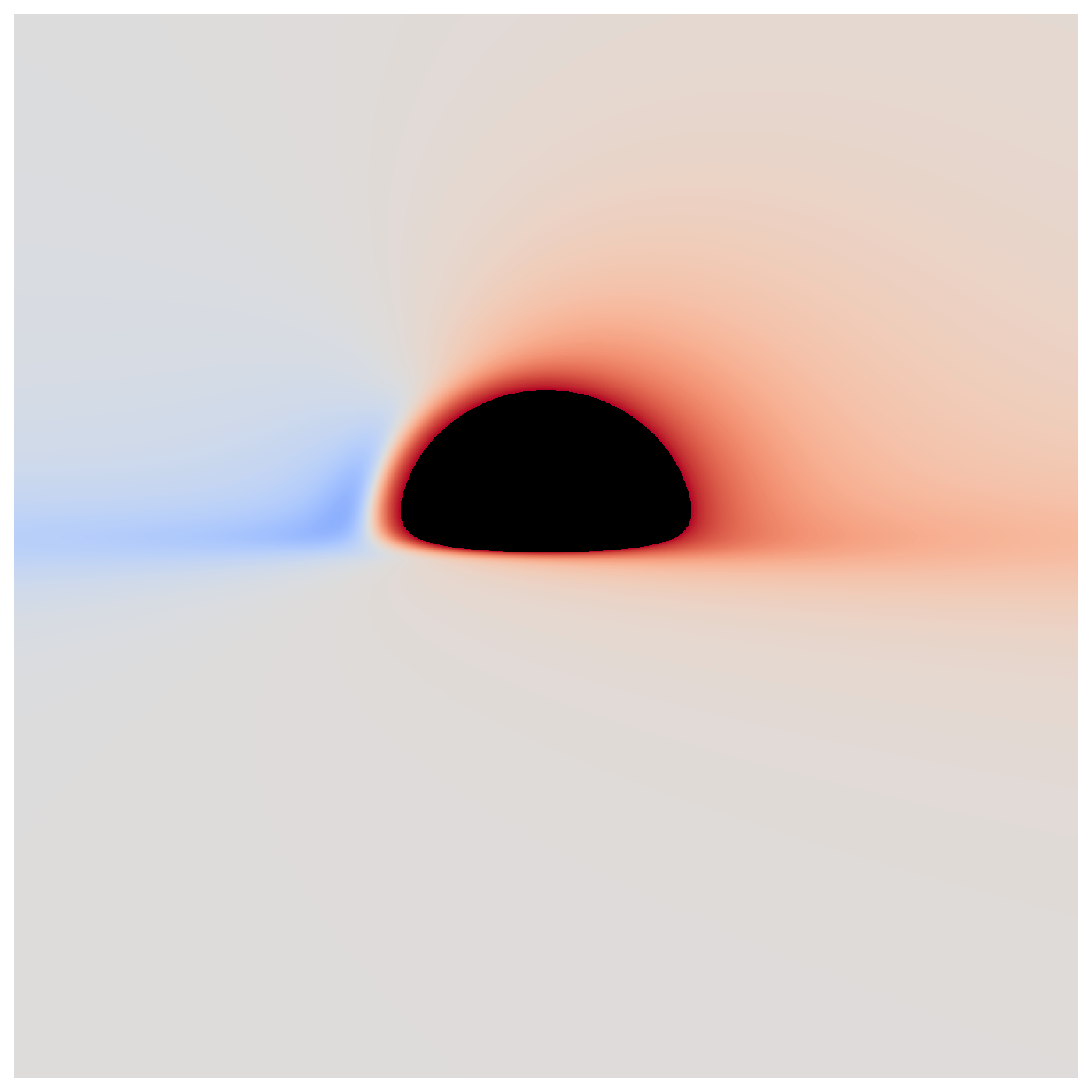}
\includegraphics[width=3.5cm]{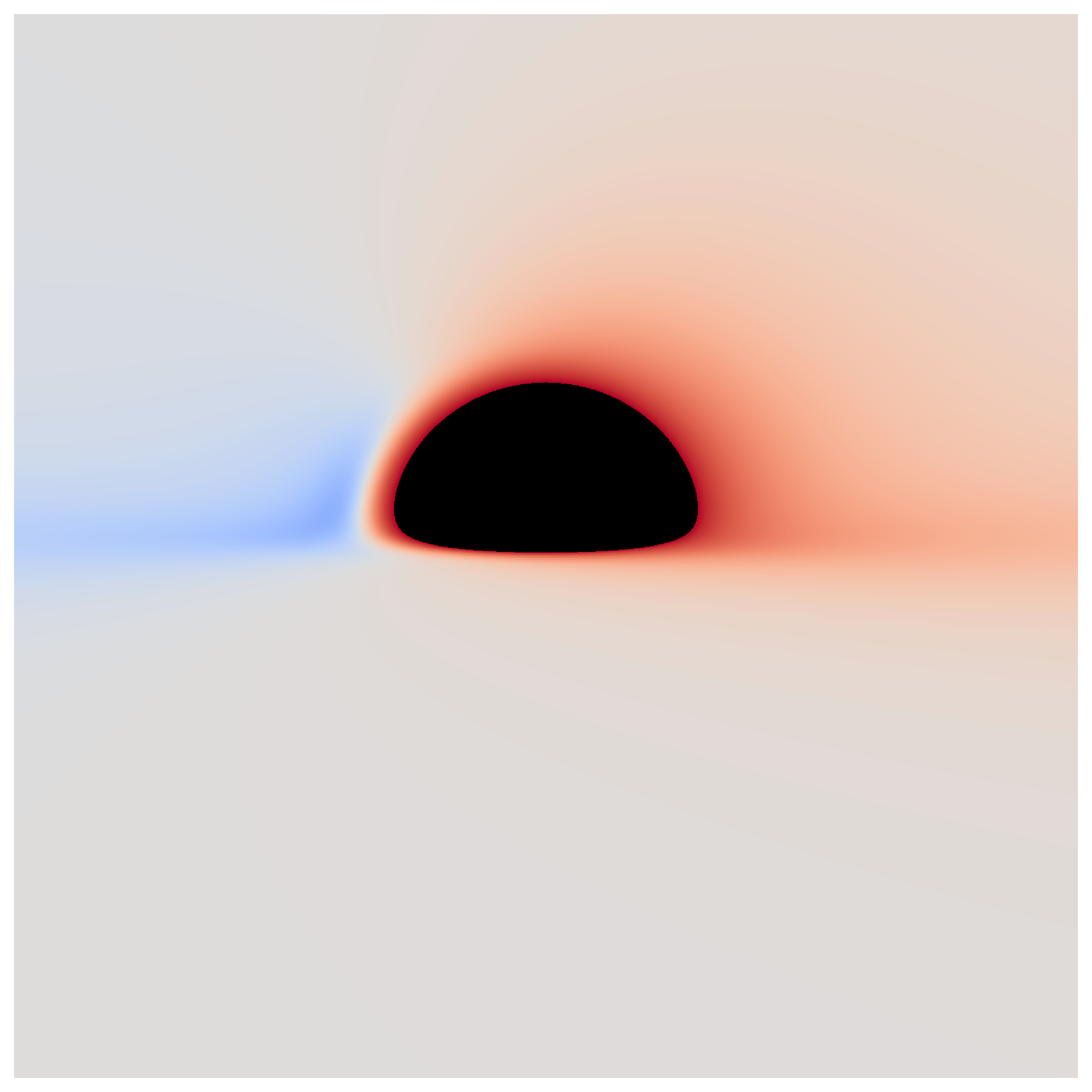}
\includegraphics[width=3.5cm]{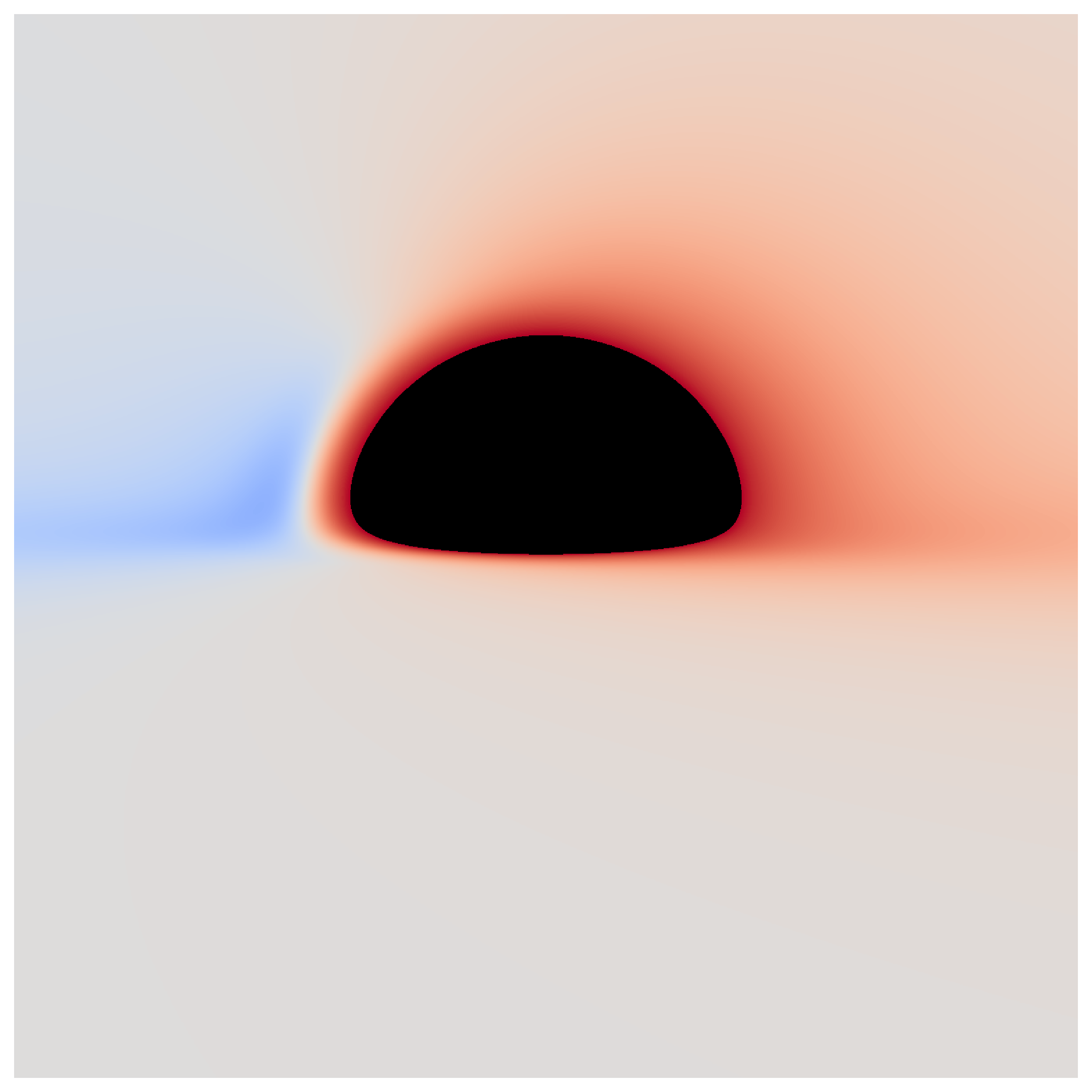}
\includegraphics[width=3.5cm]{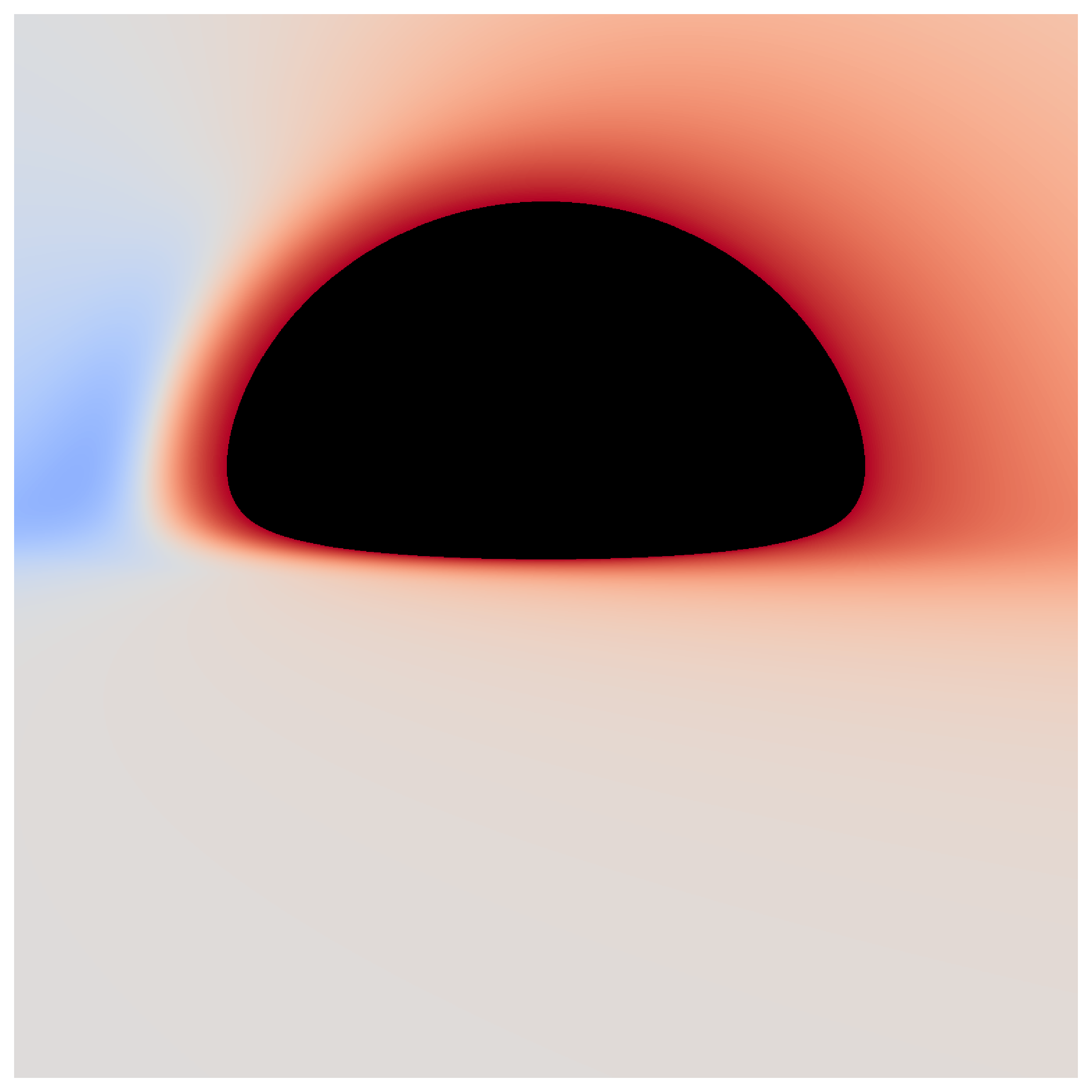}
\caption{Distribution of the first-order redshift factor (direct image contribution from the first disk crossing) across different parameter spaces. From left to right: $r_{\textrm{s}}=$ $0$, $0.3$, $0.6$, $0.9$; from top to bottom: observation inclinations $17^{\circ}$, $50^{\circ}$, and $85^{\circ}$. Here, we fix $\rho_{\textrm{s}} = 0.5$ and set the resolution to $1500 \times 1500$ pixels.}}\label{fig8}
\end{figure*}

We set the field of view to $x \in [-15,15]$ M and $y \in [-15,15]$ M with a resolution of $1500 \times 1500$ pixels, and simulate the distributions of the first-order and second-order redshift factors for various values of the observation inclination $\omega$ and the halo scale parameter $r_{\textrm{s}}$. The results are presented in figures 8 and 9. In these figures, the values of $\gamma \in [0,2]$ are represented by a continuous color map: regions with $\gamma < 1$ (indicating redshift) are shown in red, while regions with $\gamma > 1$ (indicating blueshift) are depicted in blue. For the first-order redshift factor, rays directly captured by the black hole do not produce a measurable redshift factor and contribute to the inner shadow \cite{Chael et al. (2021)}, which is colored black. As $r_{\textrm{s}}$ increases, the inner shadow expands. Additionally, higher inclination angles $\omega$ cause the shadow to stretch horizontally and compress vertically, transforming from a nearly circular shape at low inclinations to a distinct arched morphology at high inclinations. Surrounding the inner shadow is a prominent red ring with extremely low values of $\gamma$. This ring results from material accelerating inward toward the black hole's event horizon, which imparts a strong Doppler redshift to the light rays emitted in these regions. The distribution of the redshift factor beyond the red ring depends on the observation inclination. At small inclination angles, the projected velocity of the accretion disk along the line of sight is insufficient to counteract gravitational redshift, resulting in a predominantly red image. As the inclination increases, the kinematic effects of the disk dynamics become more apparent. The field of view splits into a blueshifted region on the left and a redshifted region on the right, corresponding to the approaching and receding parts of the disk, respectively. Notably, both the redshift and blueshift intensify as $r_{\textrm{s}}$ increases. This enhancement arises from the dark matter halo's contribution to the spacetime's gravitational field, which accelerates the orbital velocity of particles around the black hole. This effect is confirmed by the increase in the accretion disk's angular velocity with $r_{\textrm{s}}$, as demonstrated in figure 10. 
\begin{figure*}%[tbph]
\center{
\includegraphics[width=3.5cm]{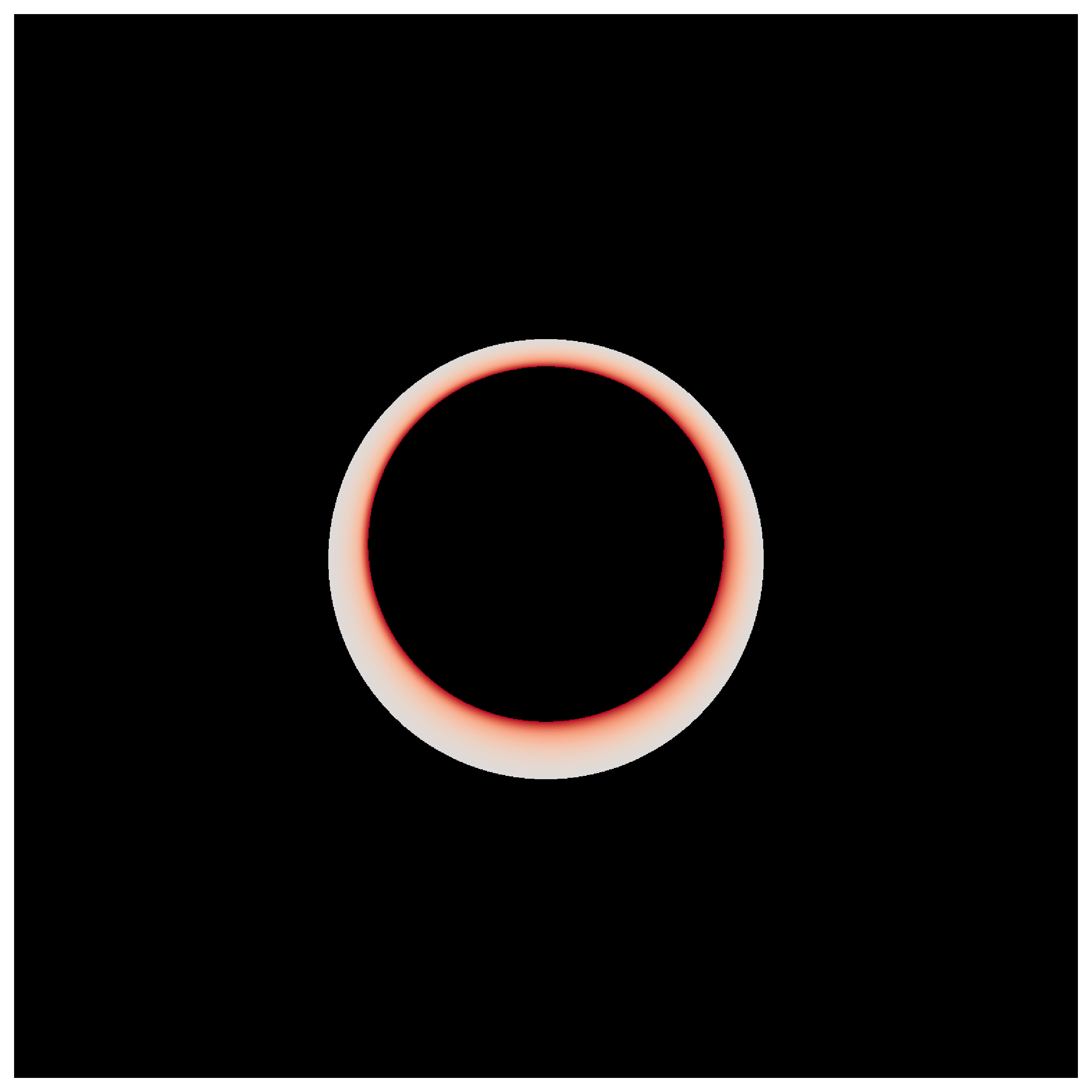}
\includegraphics[width=3.5cm]{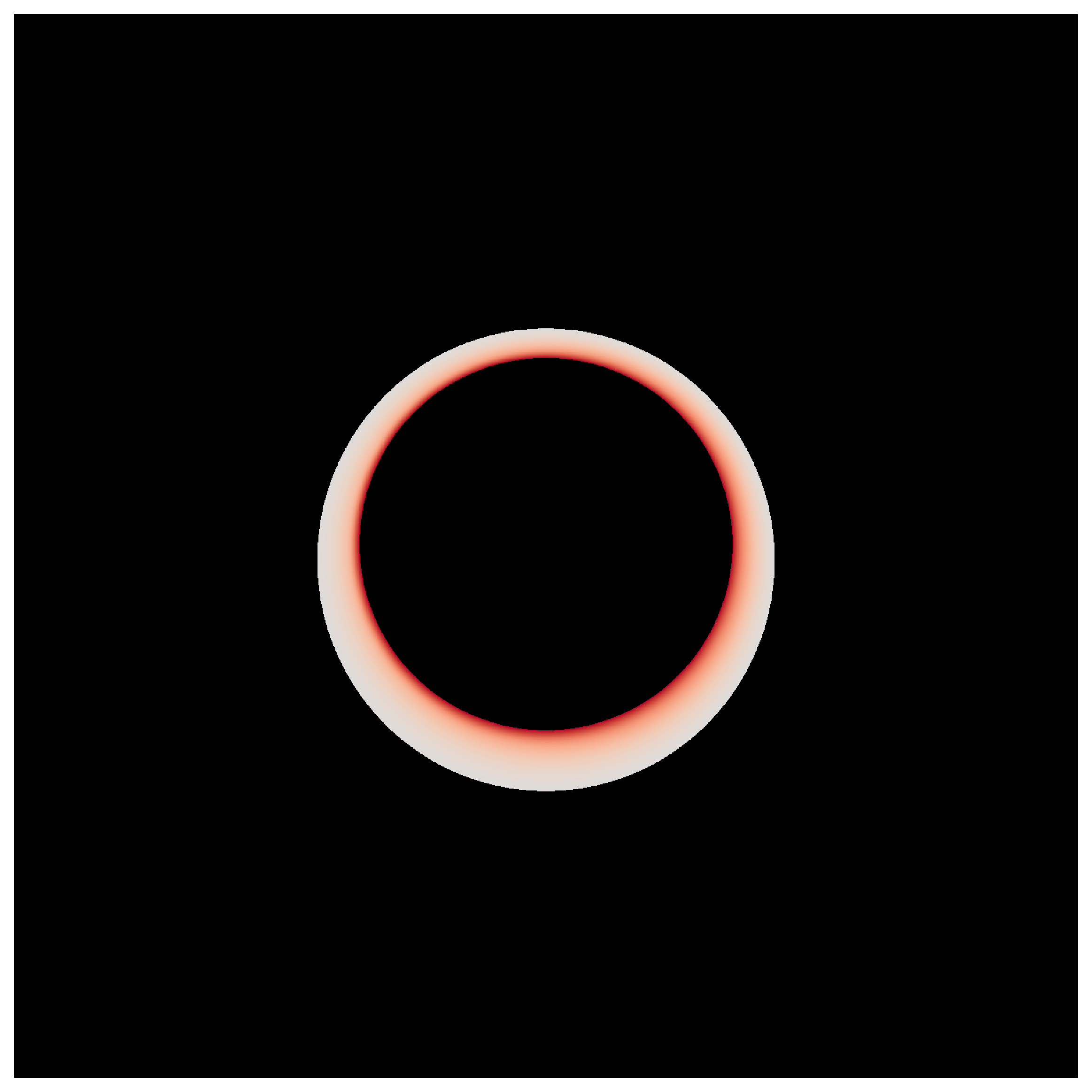}
\includegraphics[width=3.5cm]{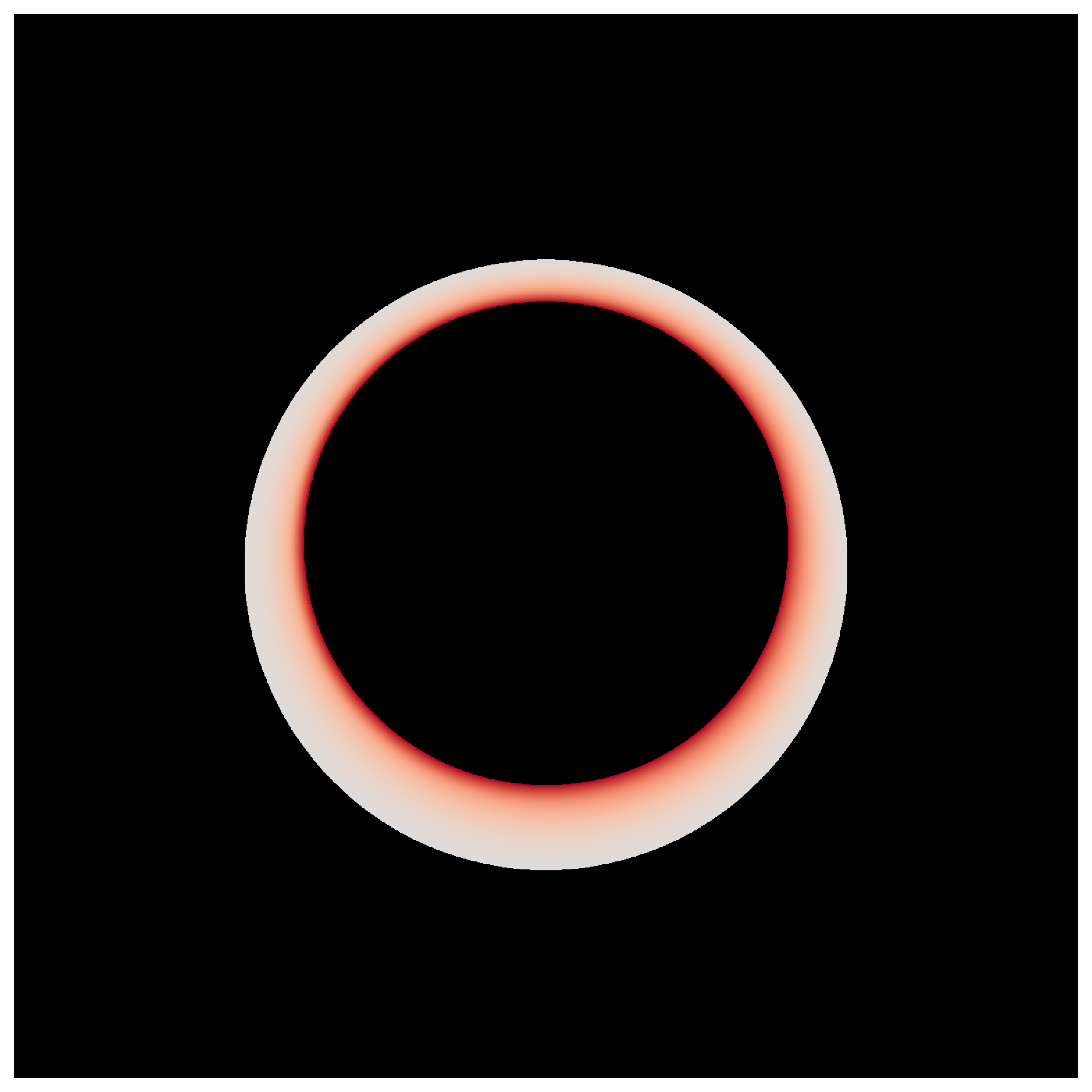}
\includegraphics[width=3.5cm]{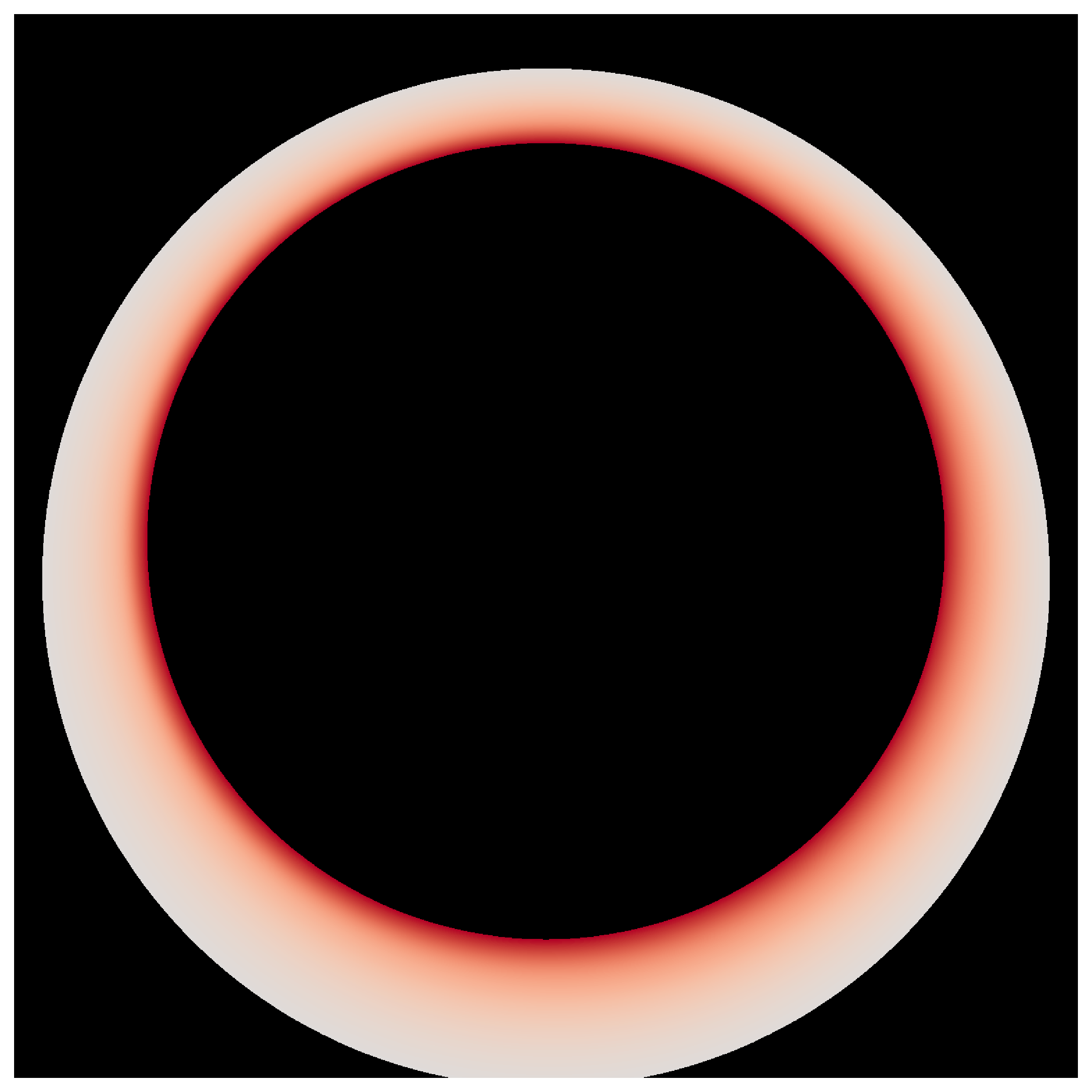}
\includegraphics[width=3.5cm]{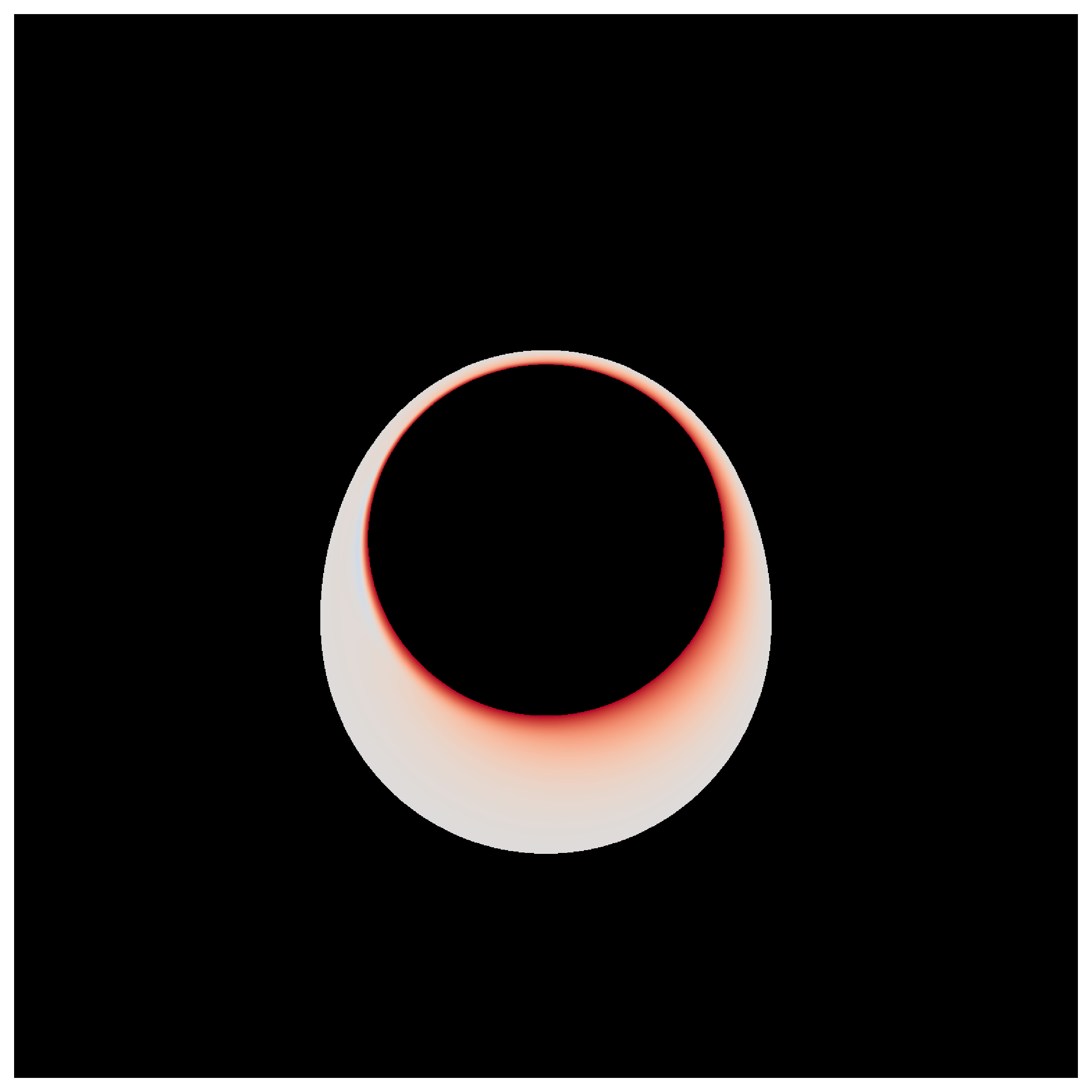}
\includegraphics[width=3.5cm]{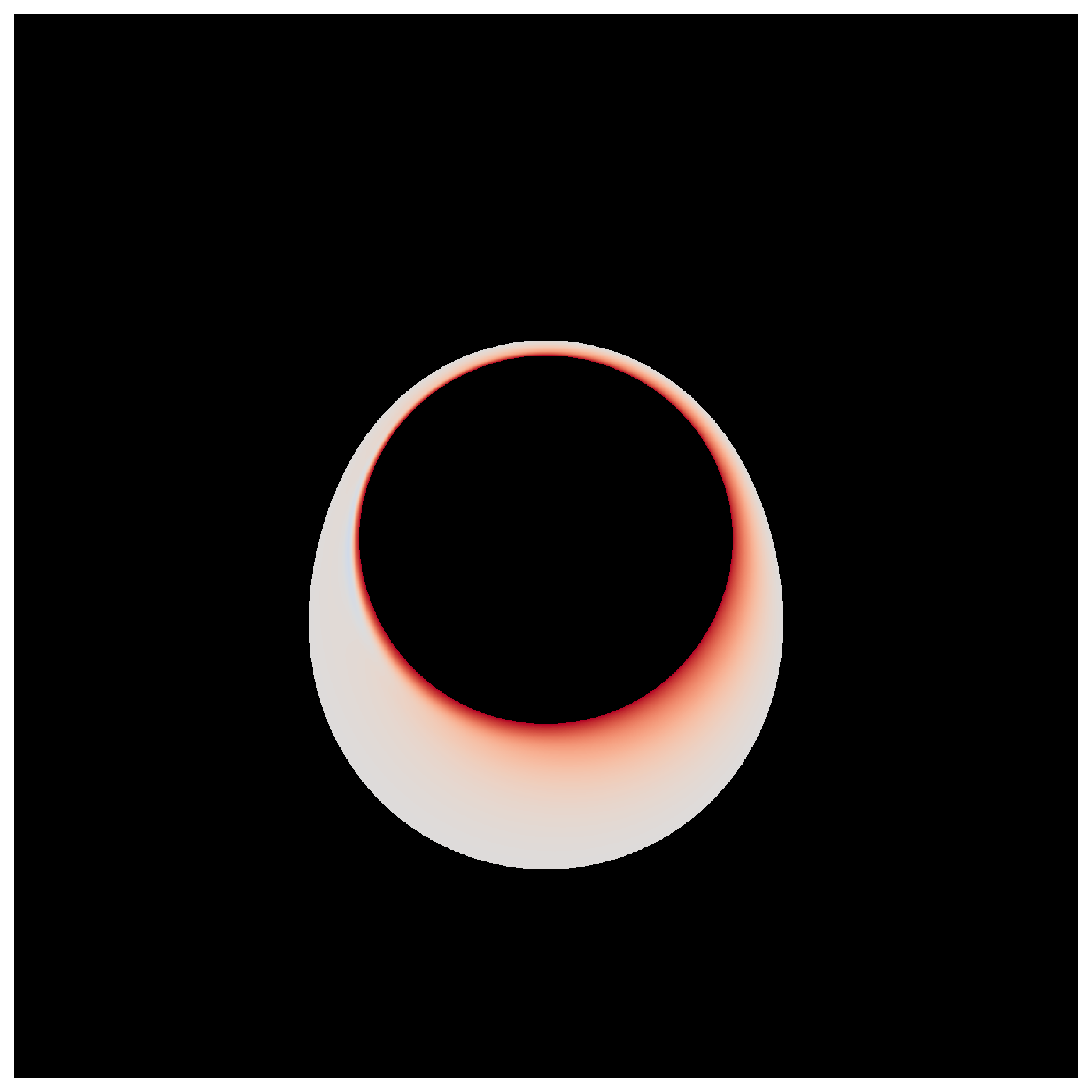}
\includegraphics[width=3.5cm]{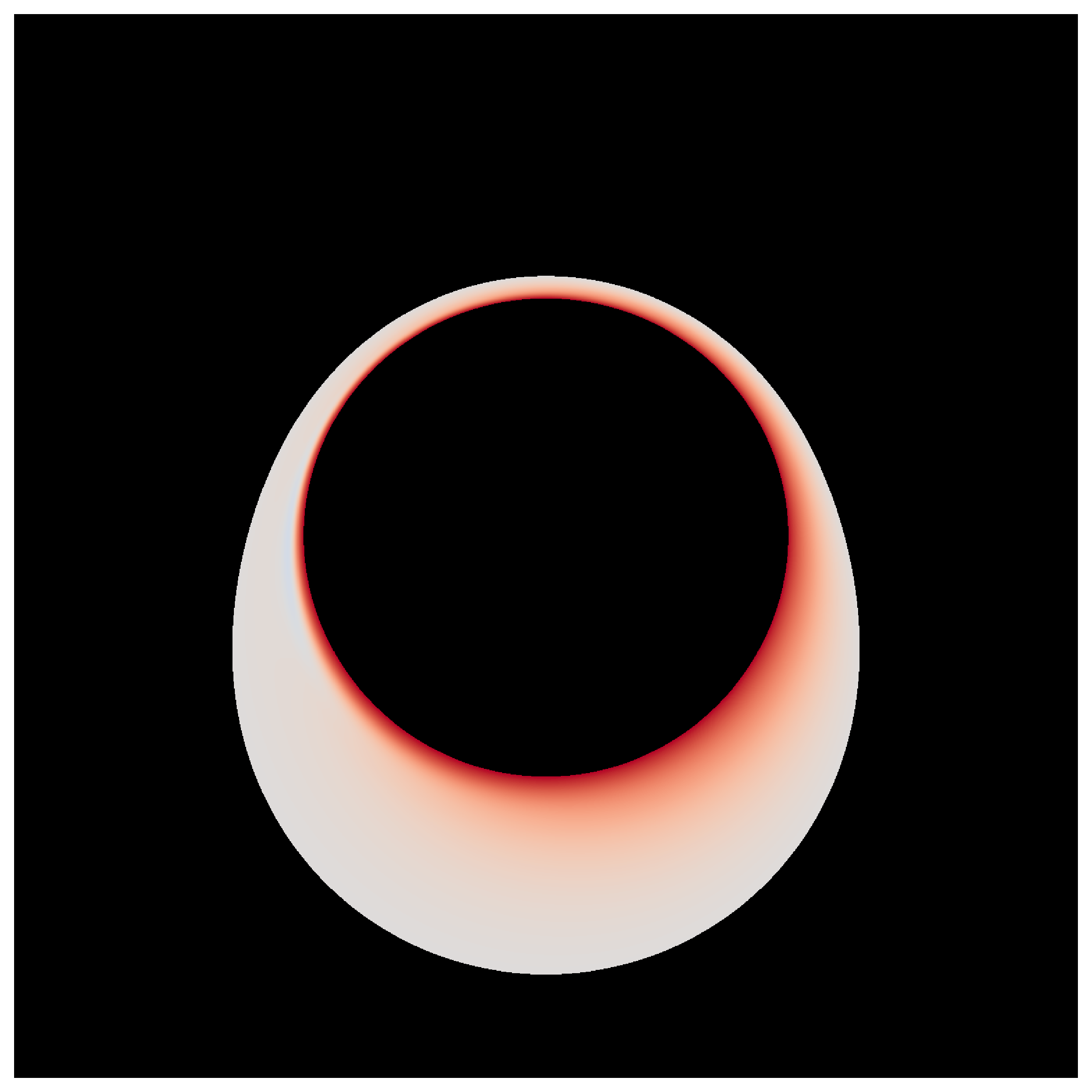}
\includegraphics[width=3.5cm]{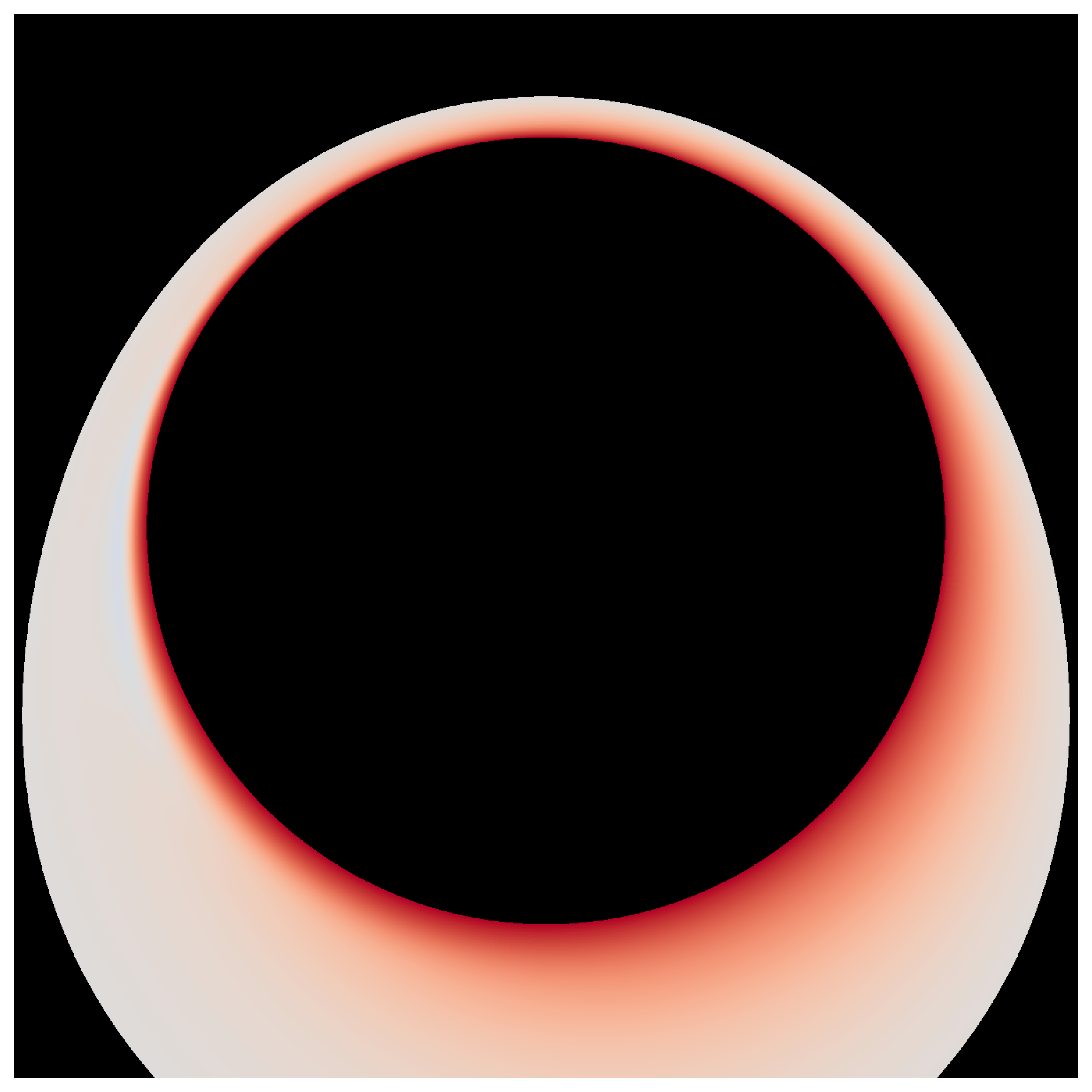}
\includegraphics[width=3.5cm]{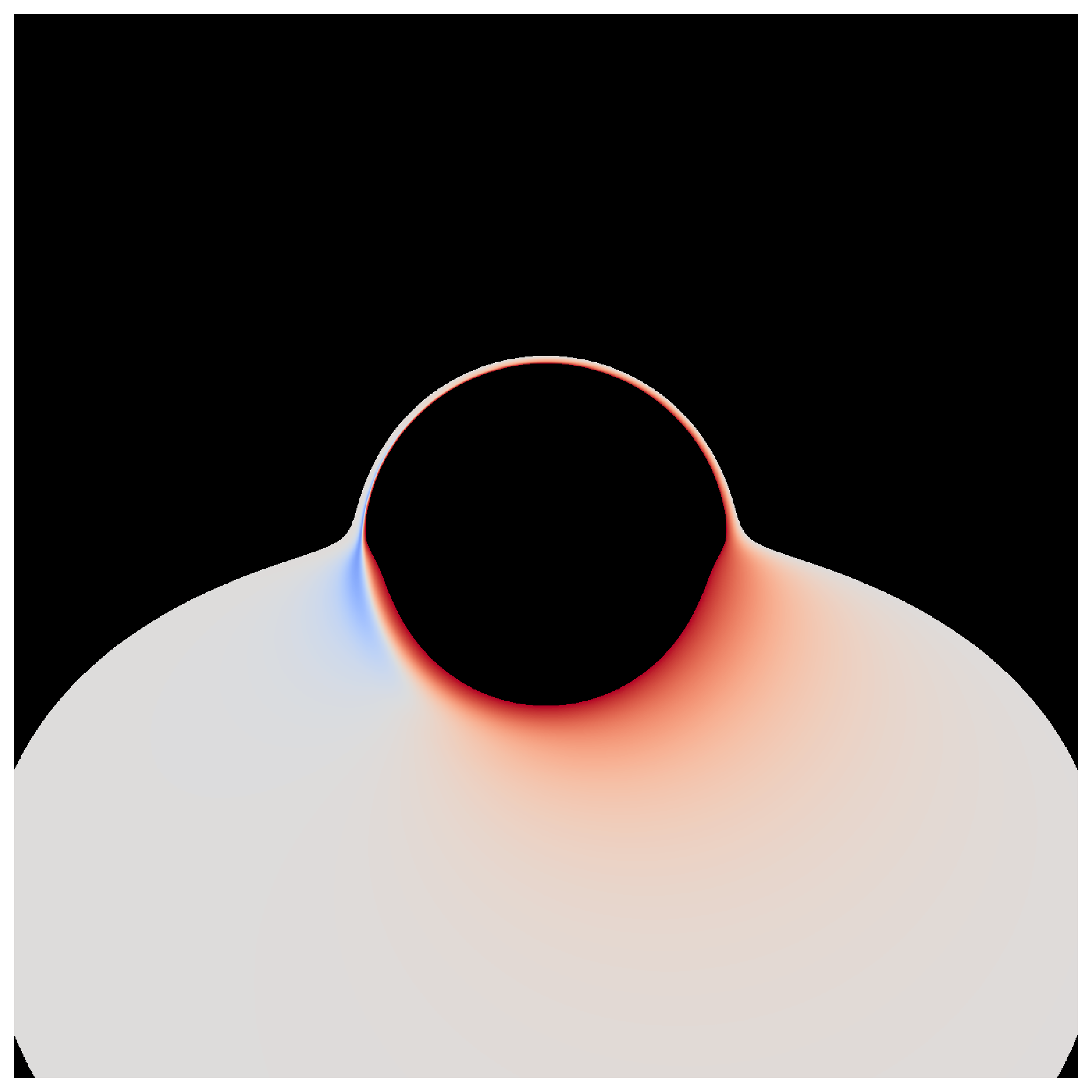}
\includegraphics[width=3.5cm]{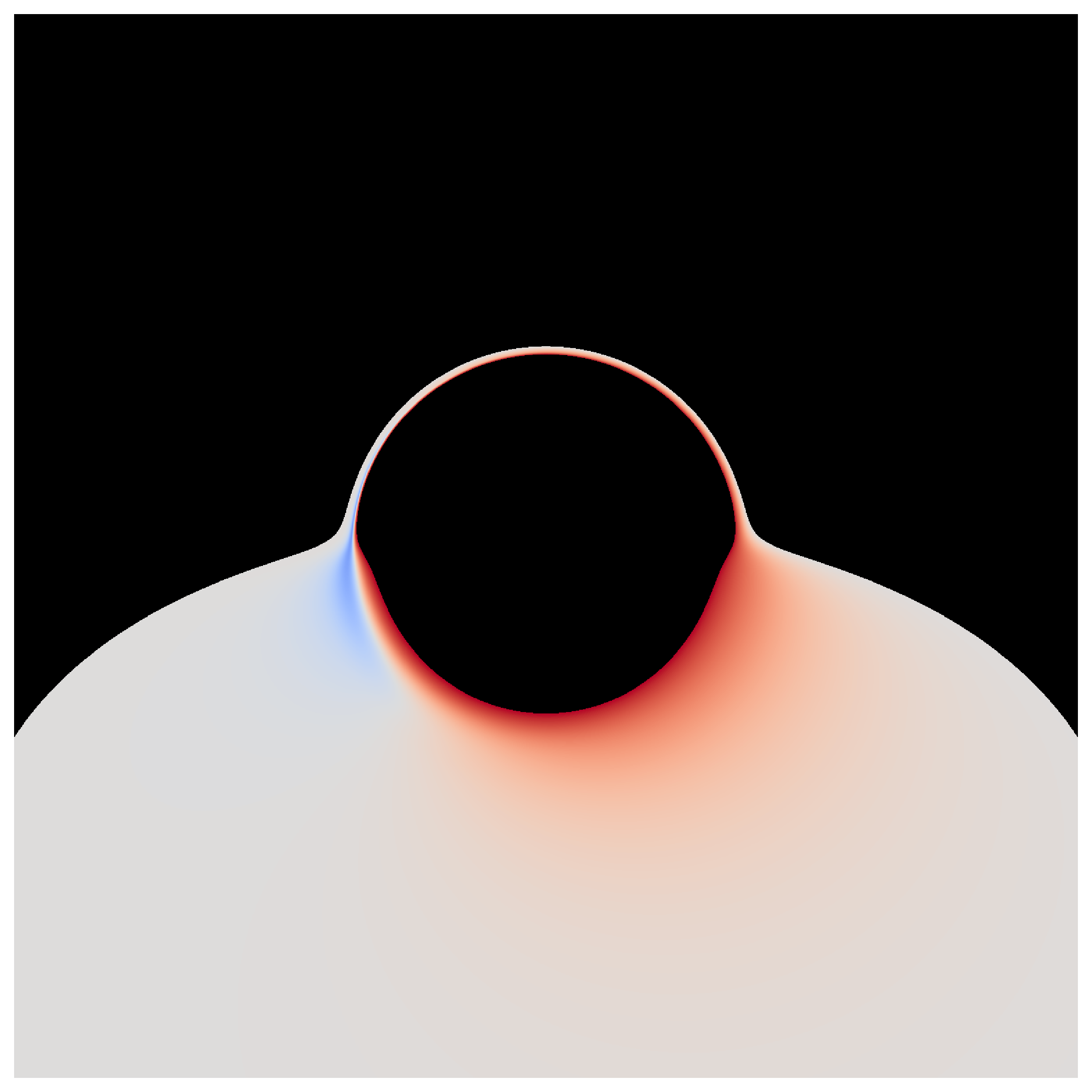}
\includegraphics[width=3.5cm]{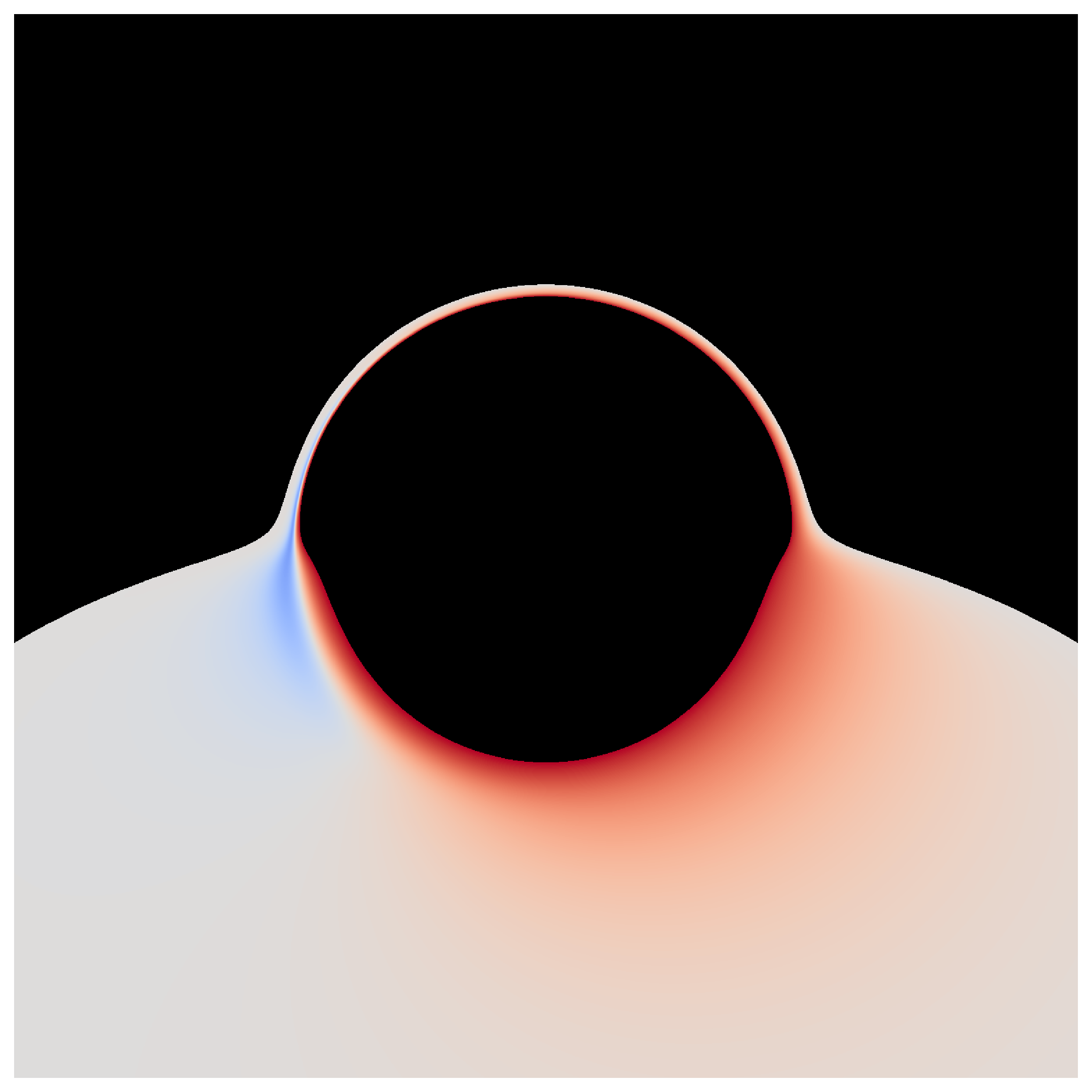}
\includegraphics[width=3.5cm]{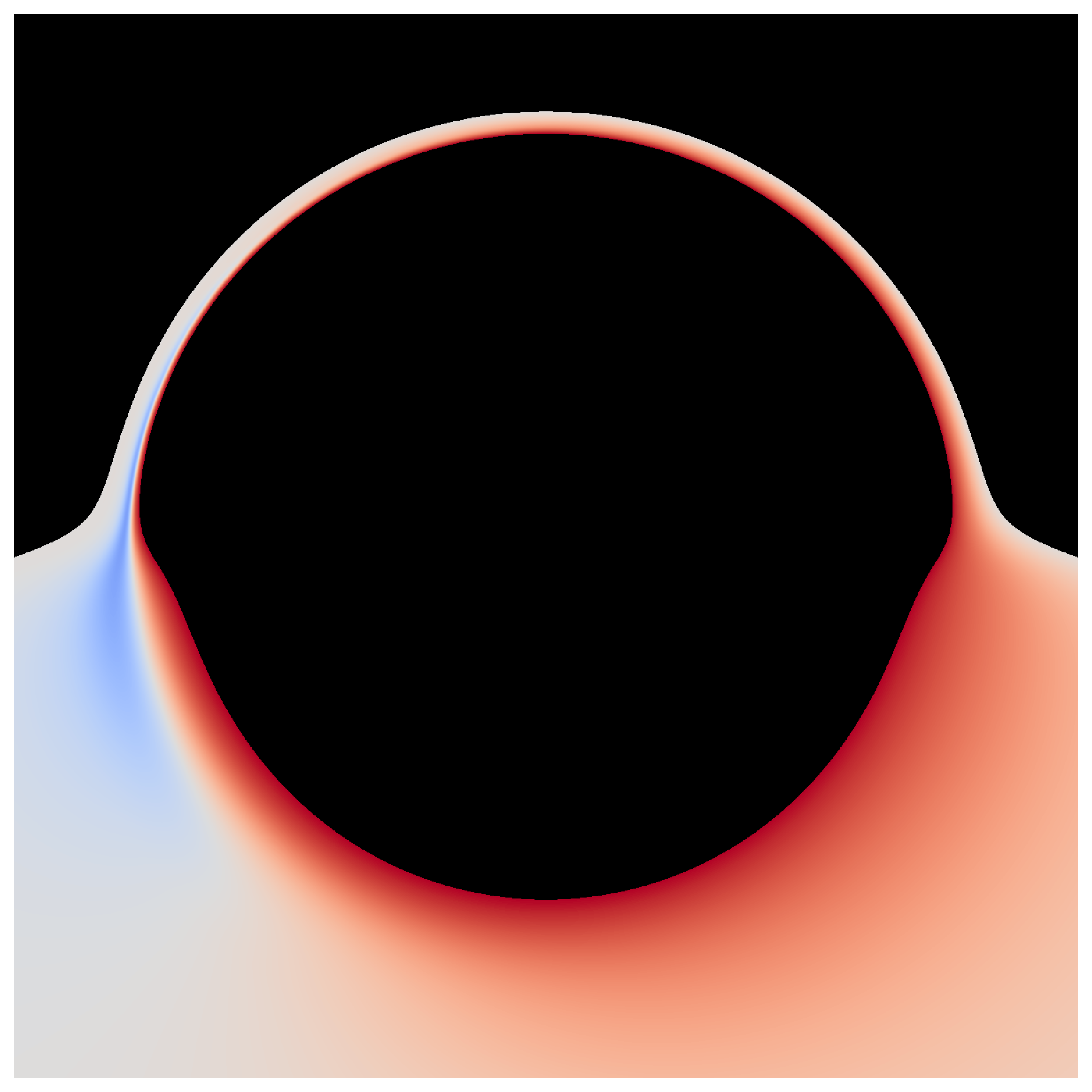}
\caption{Similar to figure 8, but showing the distributions of the second-order redshift factor.}}\label{fig9}
\end{figure*}

Compared to the first-order redshift factor, the second-order redshift factor occupies a significantly smaller region of the field of view, though its distribution still expands with increasing observation inclination. The boundary of the inner shadow becomes indistinguishable, as rays undergoing two disk crossings follow trajectories near the photon ring, causing the inner edge of the second-order redshift image to approach the critical curve. It is reasonable to predict that higher-order redshift factor images will converge more closely to the critical curve. For second-order redshift factors, distributions similar to those in figure 8 are observed: at high inclinations, the left side appears blueshifted and the right side redshifted, with an increase in $r_{\textrm{s}}$ expanding the image region and enhancing both redshift and blueshift effects. Moreover, based on our previous findings, it is reasonable to infer that varying $\rho_{\textrm{s}}$ would produce trends similar to those in figures 8 and 9, albeit with a much weaker magnitude of change.
\begin{figure*}%[tbph]
\center{
\includegraphics[width=5cm]{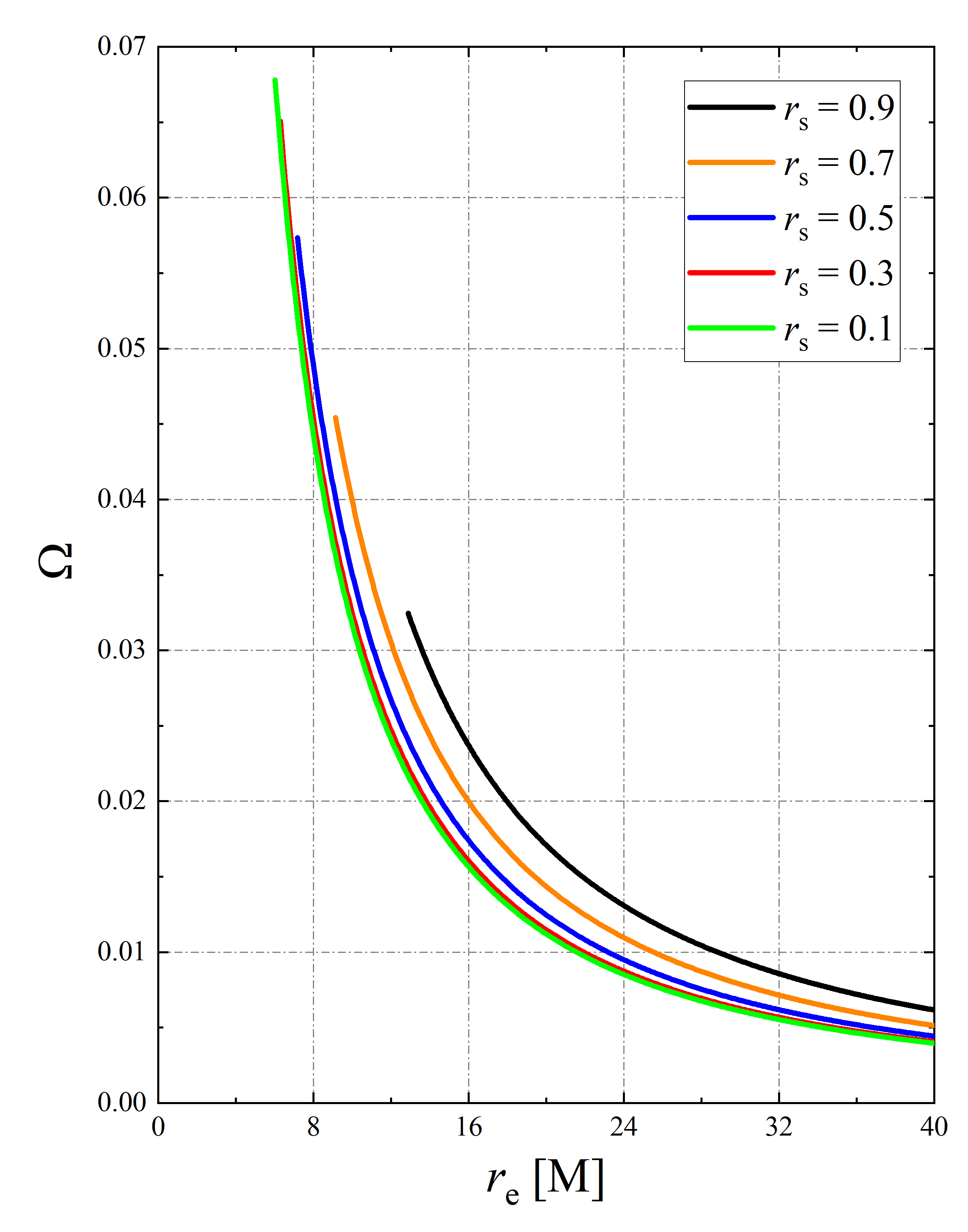}
\caption{Angular velocity $\Omega$ of the accretion disk as a function of radius for different values of the dark matter halo parameter $r_{\textrm{s}}$, with $\rho_{\textrm{s}}$ fixed at $0.5$. Each curve begins at the corresponding ISCO radius; thus, as $r_{\textrm{s}}$ decreases, the curves shift leftward. Additionally, the angular velocity increases with increasing $r_{\textrm{s}}$.}}\label{fig10}
\end{figure*}
\subsection{Black hole images}
First, we assume that the inner boundary of the accretion disk coincides with the black hole event horizon and that its emission follows the profile described in equation \eqref{22}. Under these assumptions, we simulate qualitative images at $86$ GHz and $230$ GHz for a Schwarzschild black hole surrounded by a dark matter halo, as presented in figures 11--13. The field of view is set to $x \in [-15,15]$ M and $y \in [-15,15]$ M with a resolution of $1500 \times 1500$ pixels. The specific intensity $I_{\textrm{obs}}$ is visualized using a continuous color map, ranging from black (indicating $I_{\textrm{obs}}=0$) through red and yellow to white.

\begin{figure*}%[tbph]
\center{
\includegraphics[width=3.5cm]{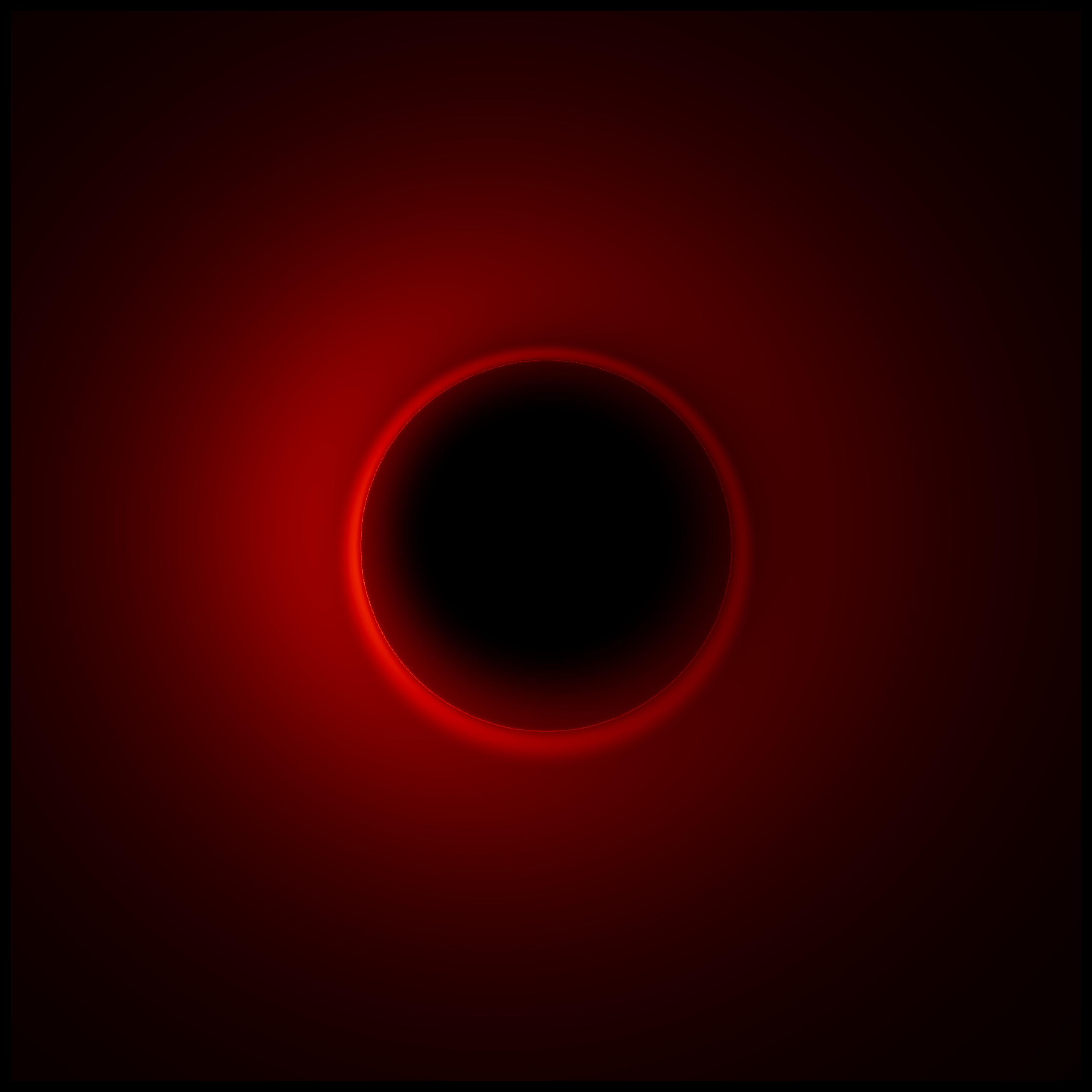}
\includegraphics[width=3.5cm]{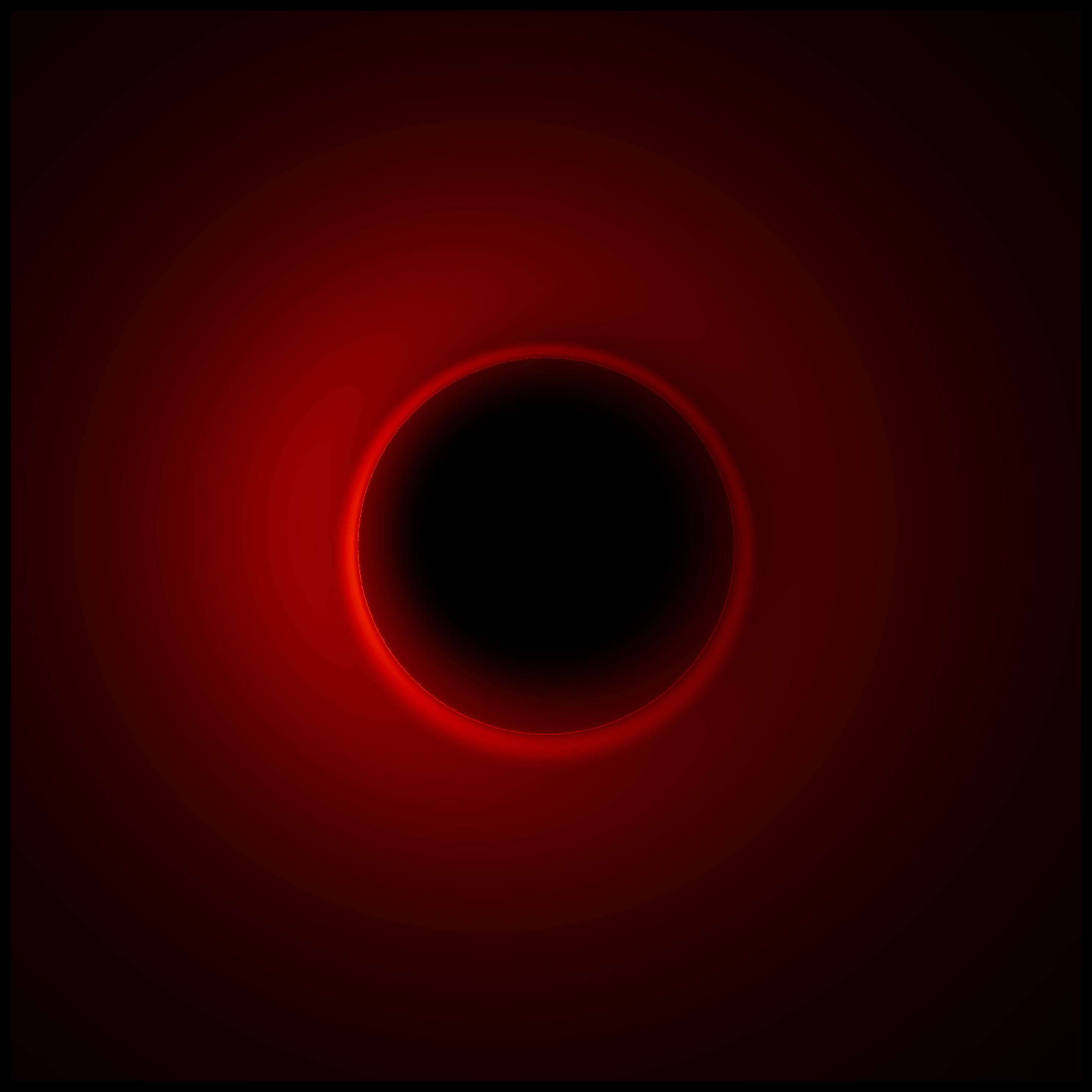}
\includegraphics[width=3.5cm]{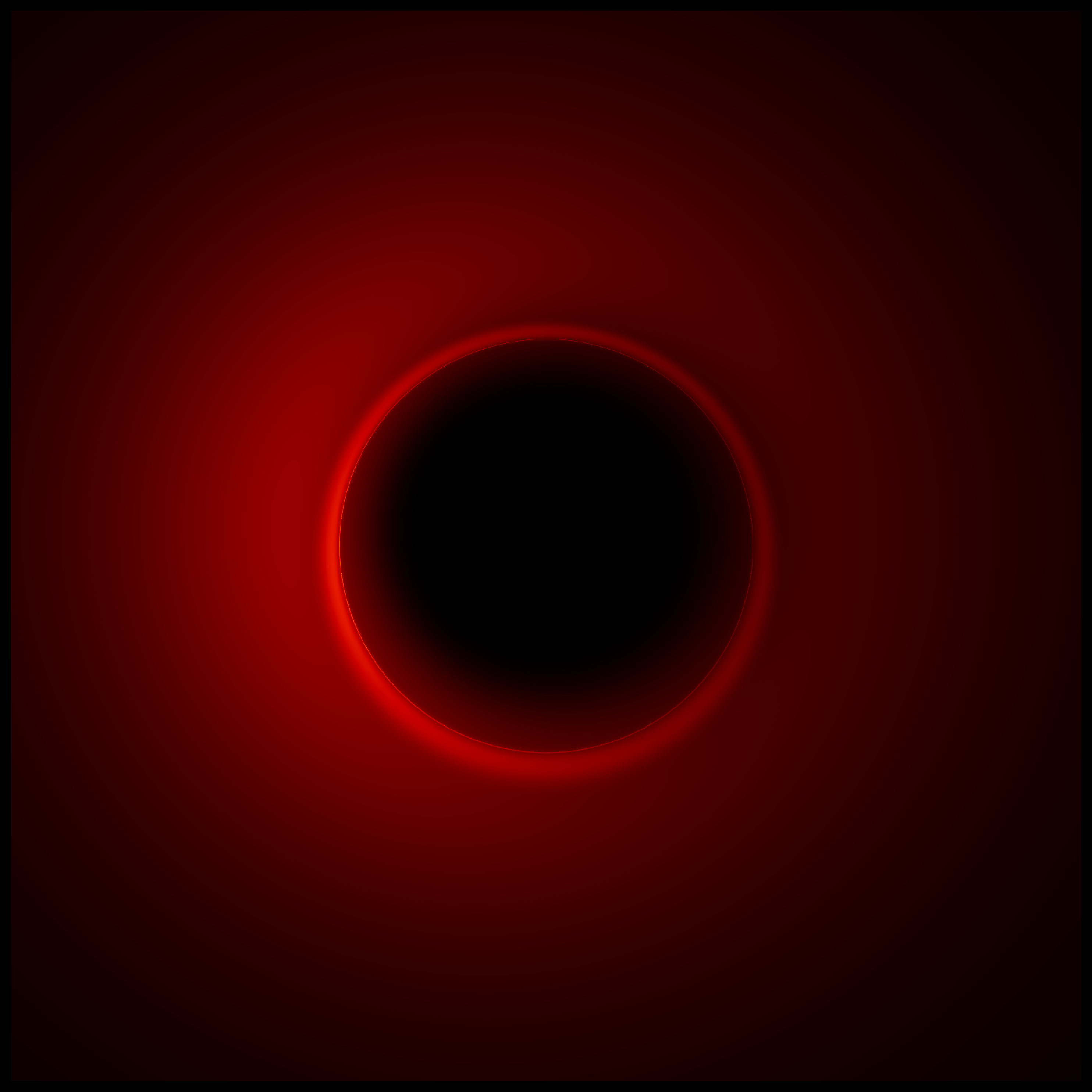}
\includegraphics[width=3.5cm]{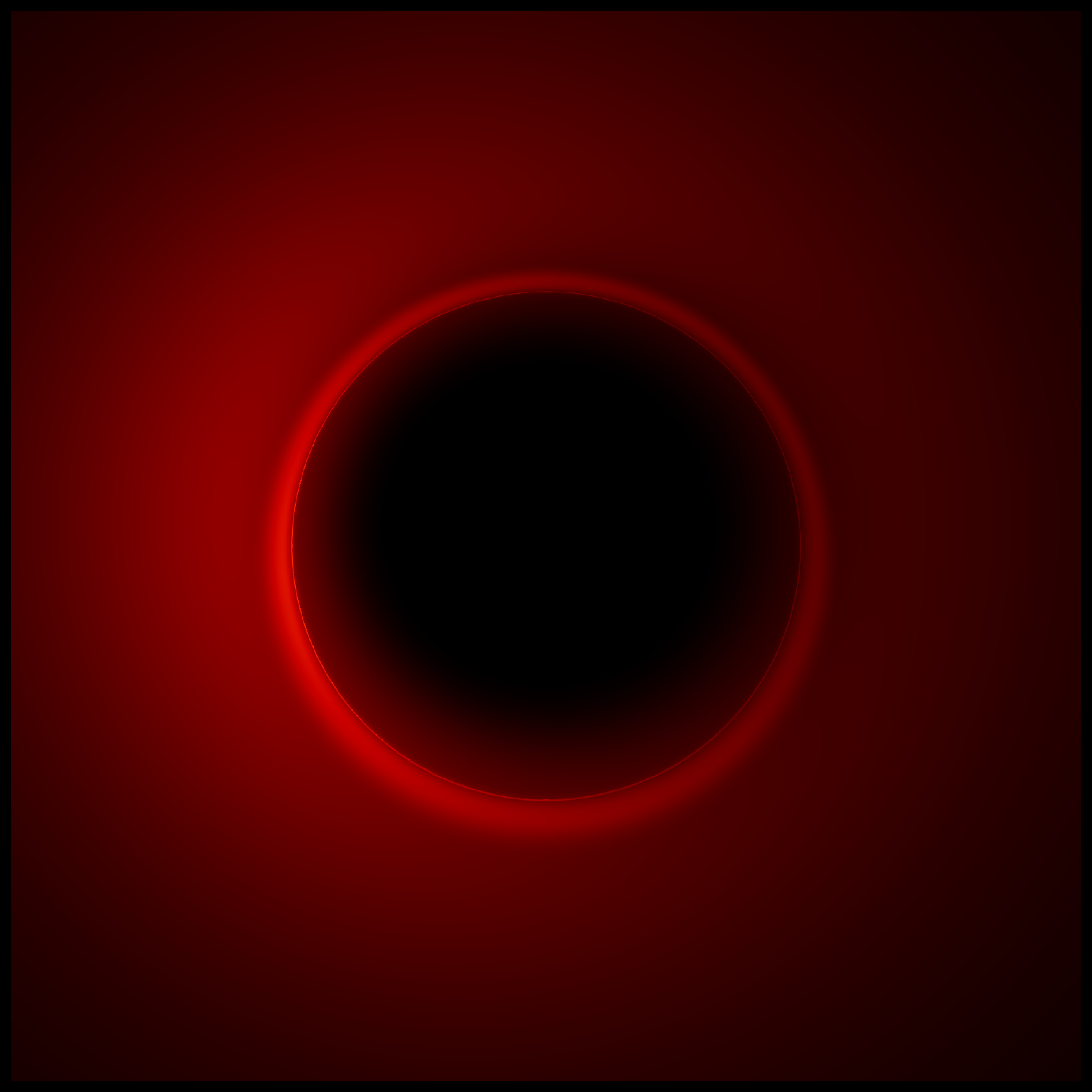}
\includegraphics[width=3.5cm]{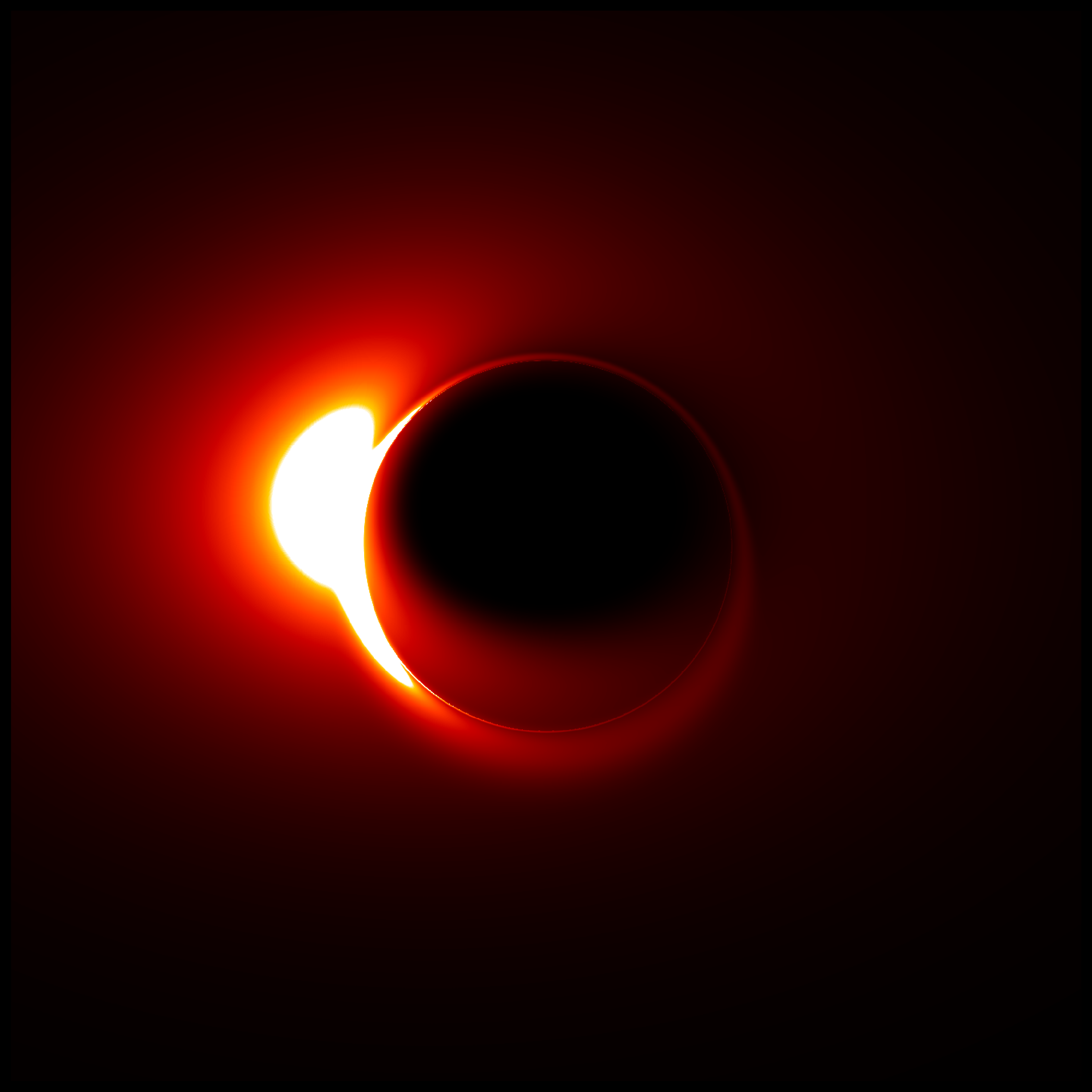}
\includegraphics[width=3.5cm]{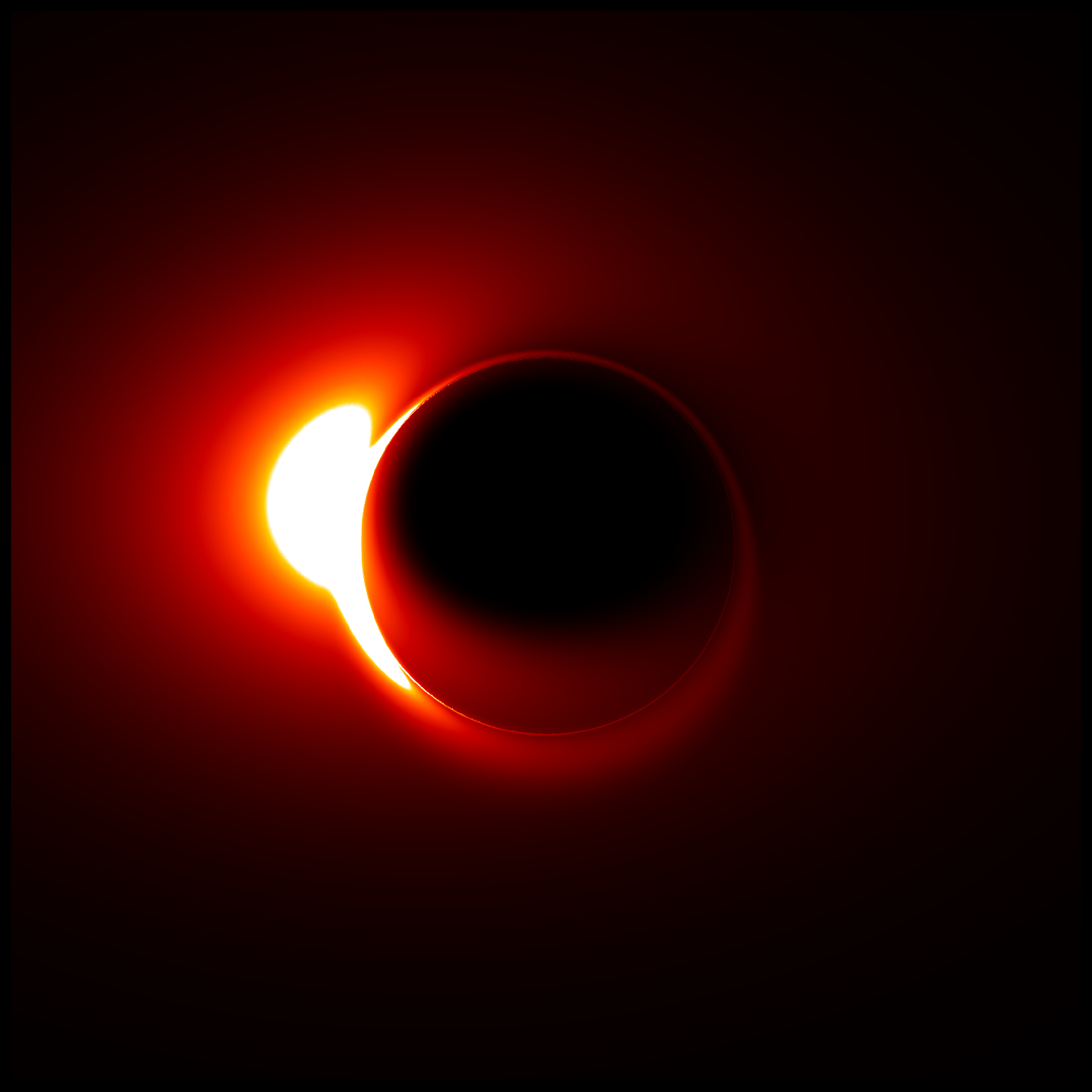}
\includegraphics[width=3.5cm]{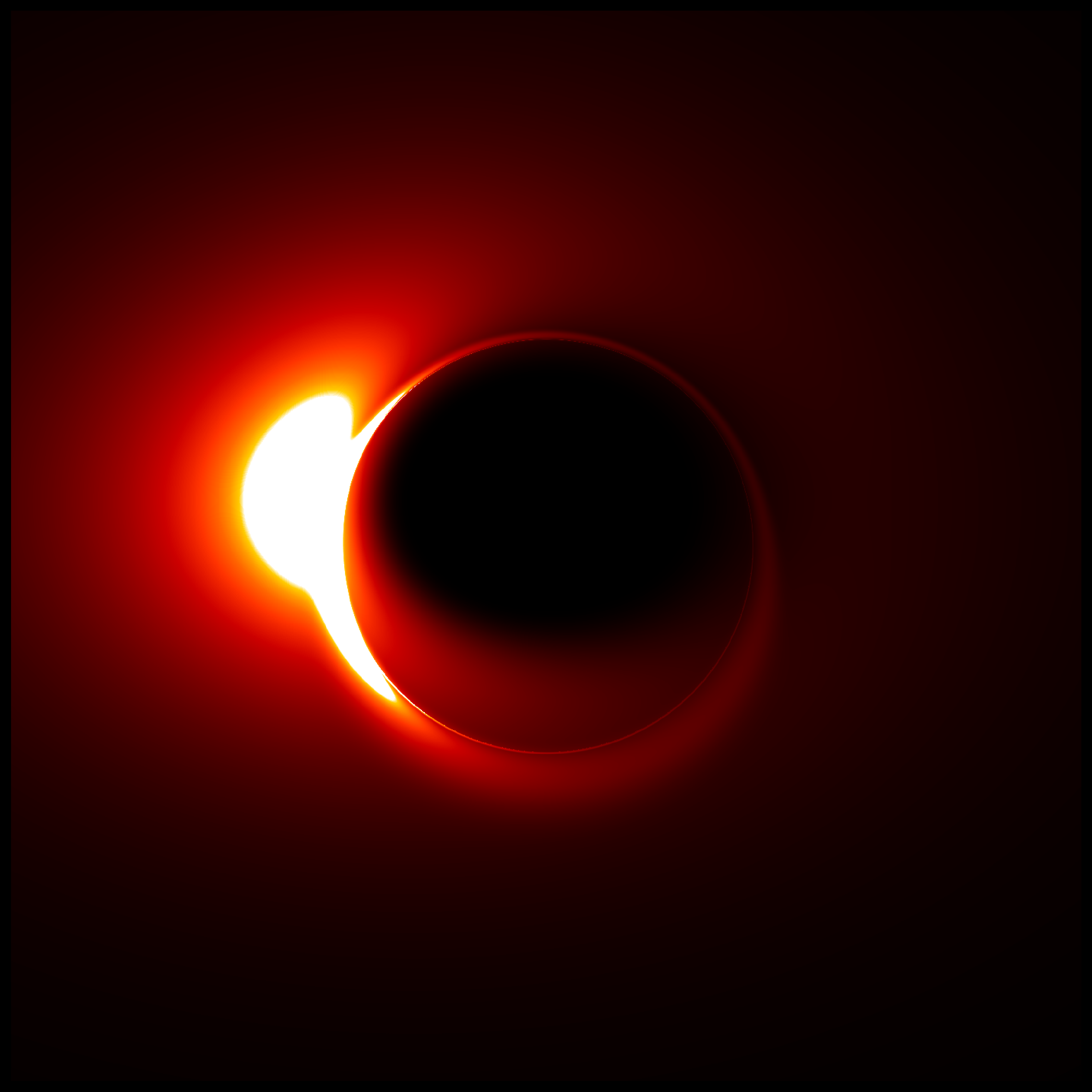}
\includegraphics[width=3.5cm]{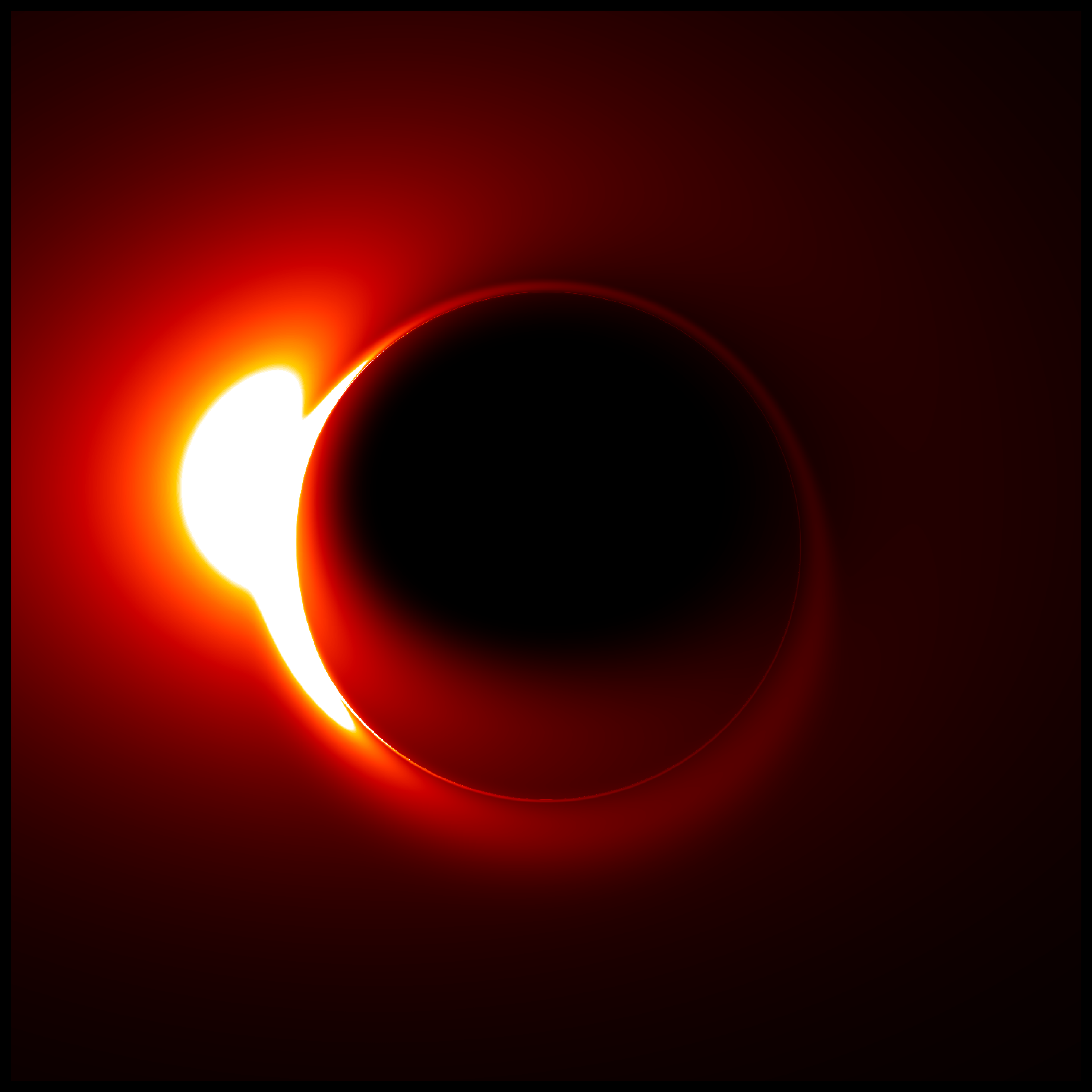}
\includegraphics[width=3.5cm]{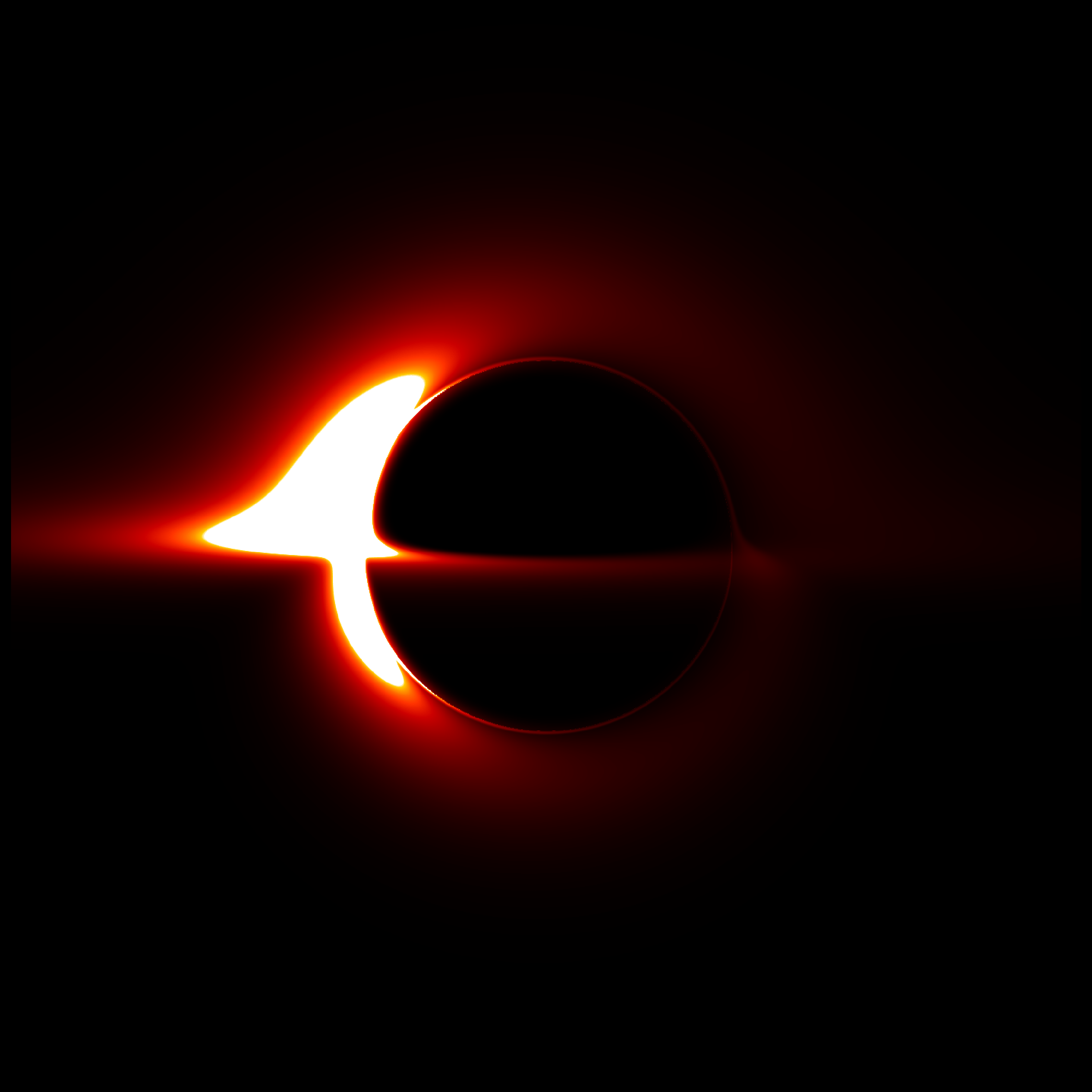}
\includegraphics[width=3.5cm]{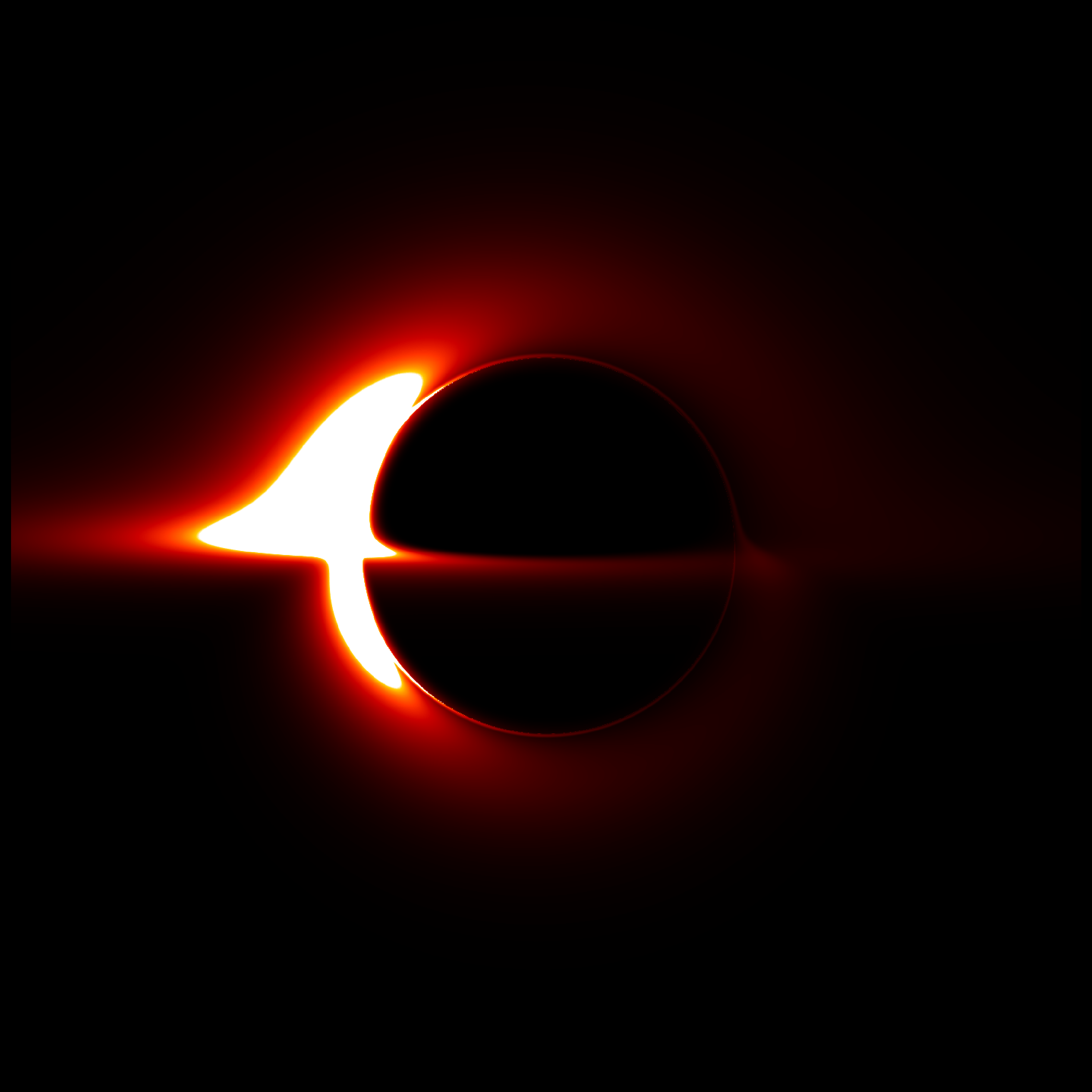}
\includegraphics[width=3.5cm]{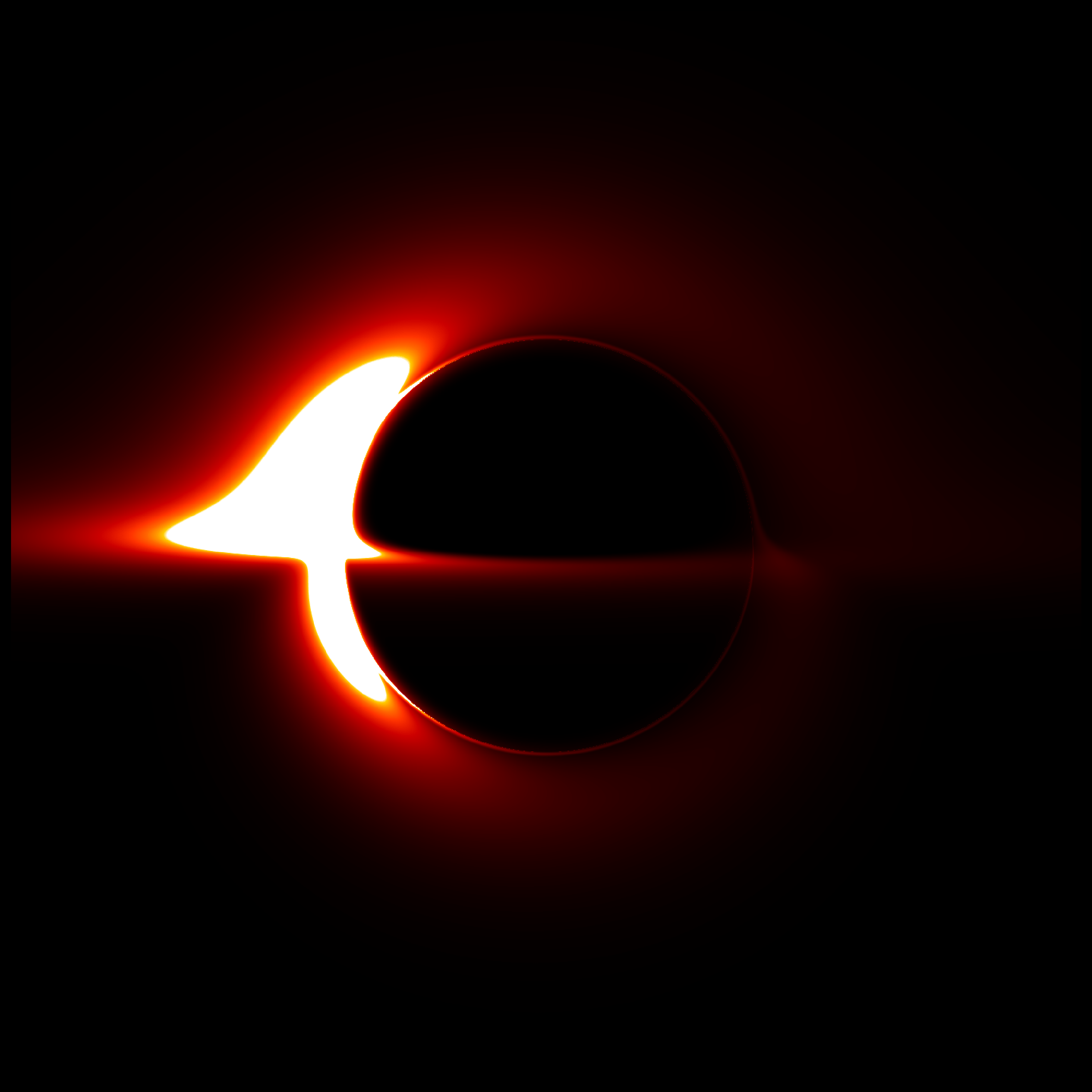}
\includegraphics[width=3.5cm]{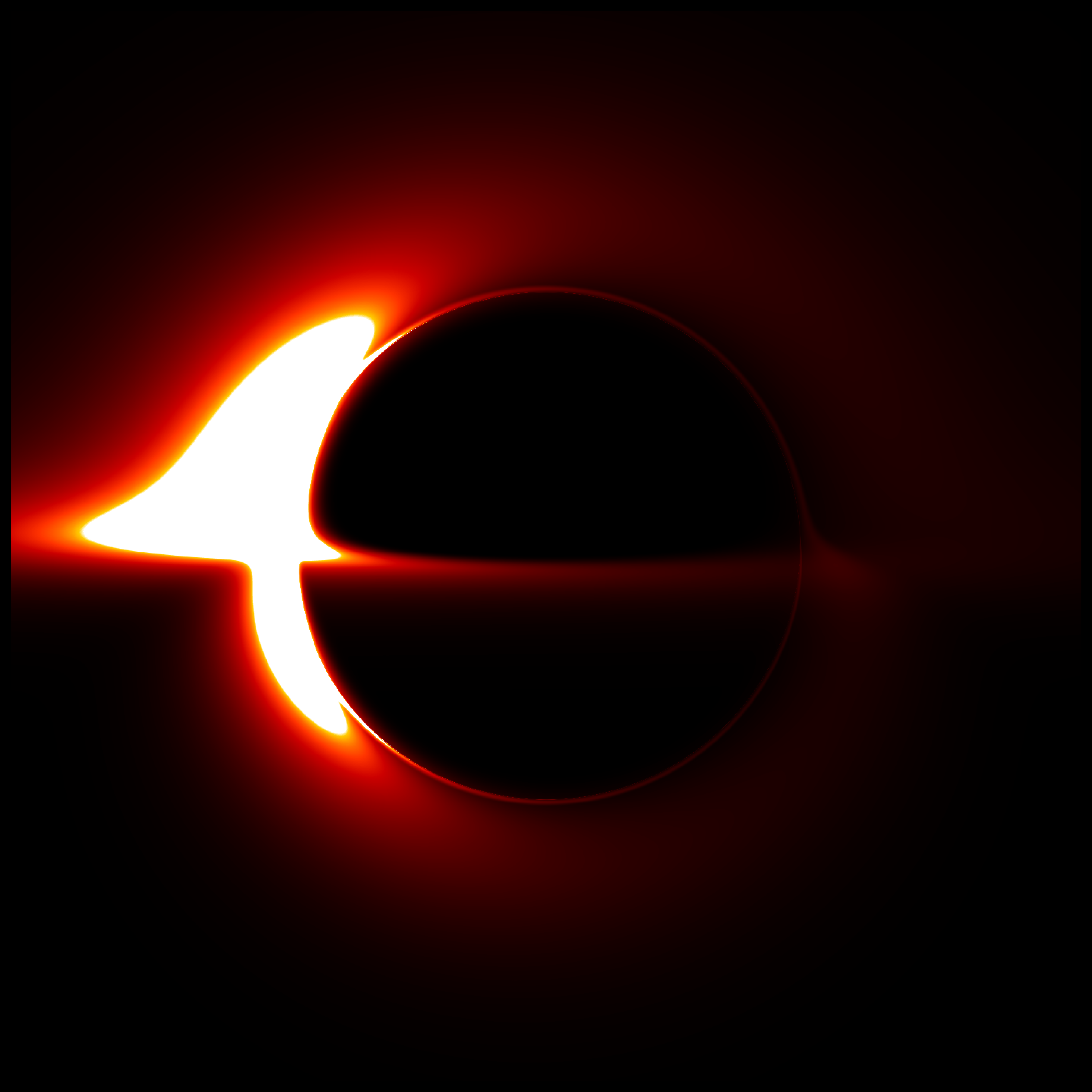}
\caption{86 GHz images of a Schwarzschild black hole surrounded by a dark matter halo across different parameter spaces. From left to right: scale parameter $r_{\textrm{s}}=$ $0$, $0.2$, $0.4$, $0.6$; from top to bottom: observation inclination $\omega=$ $17^{\circ}$, $50^{\circ}$, $85^{\circ}$. The resolution is fixed at $1500 \times 1500$, the density parameter at $\rho_{\textrm{s}}=0.5$, the specific intensity range at $I_{\textrm{obs}} \in [0,0.5]$, and the maximum number of disk crossings at $N_{\textrm{max}}=4$. The images are dominated by higher-order bright rings, inner shadows, and irregular bright spots, all of which expand with increasing $r_{\textrm{s}}$. This trend offers a potential diagnostic tool for distinguishing pure Schwarzschild black holes from those modified by a dark matter halo.}}\label{fig11}
\end{figure*}

In figure 11, we observe that, at small observation angles, the image consists of a central dim region surrounded by a bright ring near the critical curve. The central dark area is slightly larger than the inner shadow, which occurs because the accreting matter near the event horizon accelerates toward the black hole, causing extreme redshift in its emitted radiation, making it nearly undetectable. Consequently, the extremely low specific intensity from this region merges with the inner shadow, forming a dark zone that extends slightly beyond the inner shadow. Outside this central shadow, a distinct bright ring appears, formed by light rays that have crossed the accretion disk multiple times. As the observer's inclination increases, the inner shadow becomes distorted: it stretches laterally toward the edges of the field of view while being compressed vertically. Along with this distortion, an irregular bright patch appears on the left side, confirming the left-right brightness asymmetry in the image---an effect attributable to Doppler beaming. When the observation inclination reaches $85^{\circ}$, this asymmetry becomes highly pronounced. Notably, these intensity distributions are fully consistent with the corresponding redshift factor patterns. Furthermore, we emphasize that increasing the core size of the dark matter halo enlarges all image features---such as the bright ring, inner shadow, and bright patches---without significantly altering their shapes. This behavior helps establish a direct connection between image characteristics and dark matter halo parameters.
\begin{figure*}%[tbph]
\center{
\includegraphics[width=3.5cm]{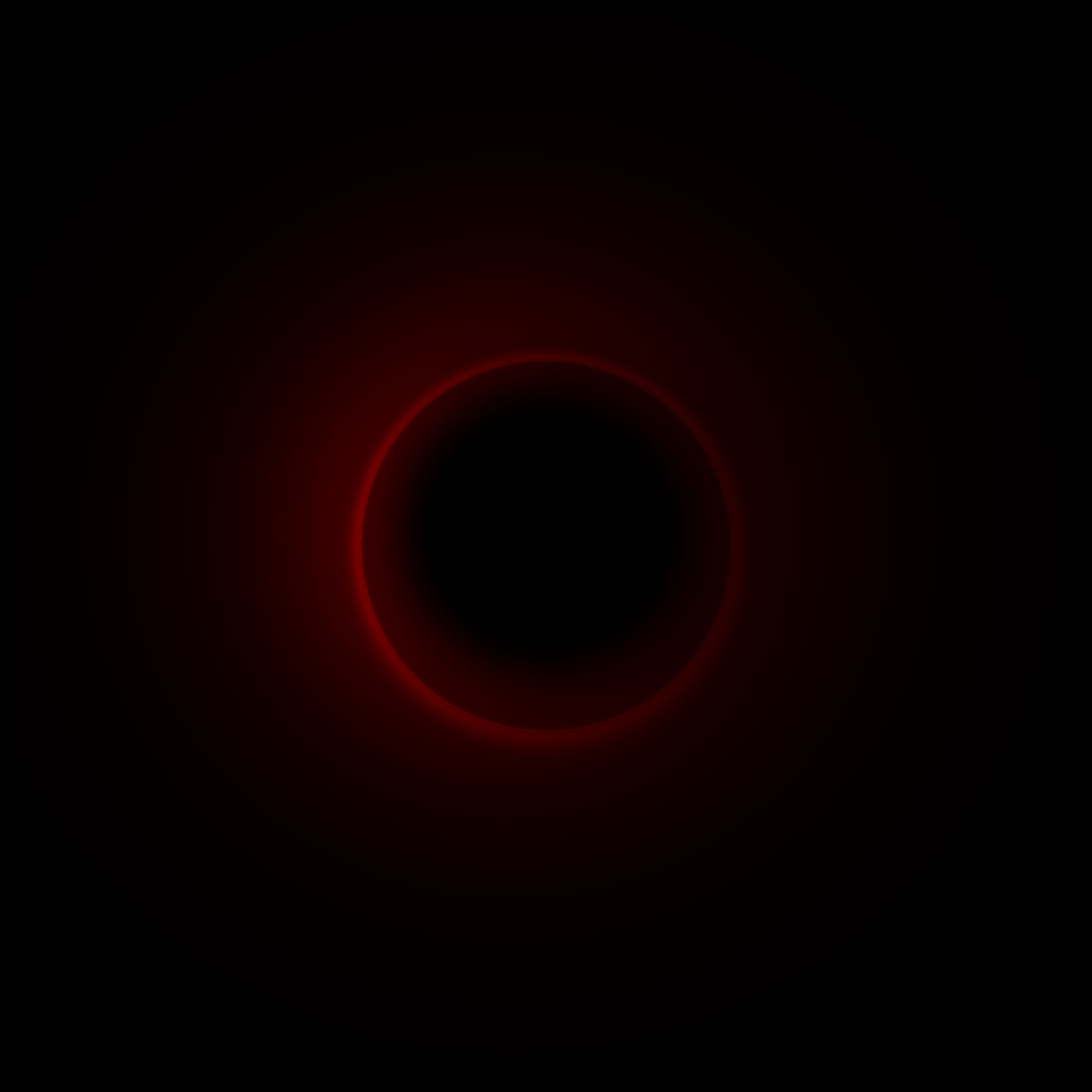}
\includegraphics[width=3.5cm]{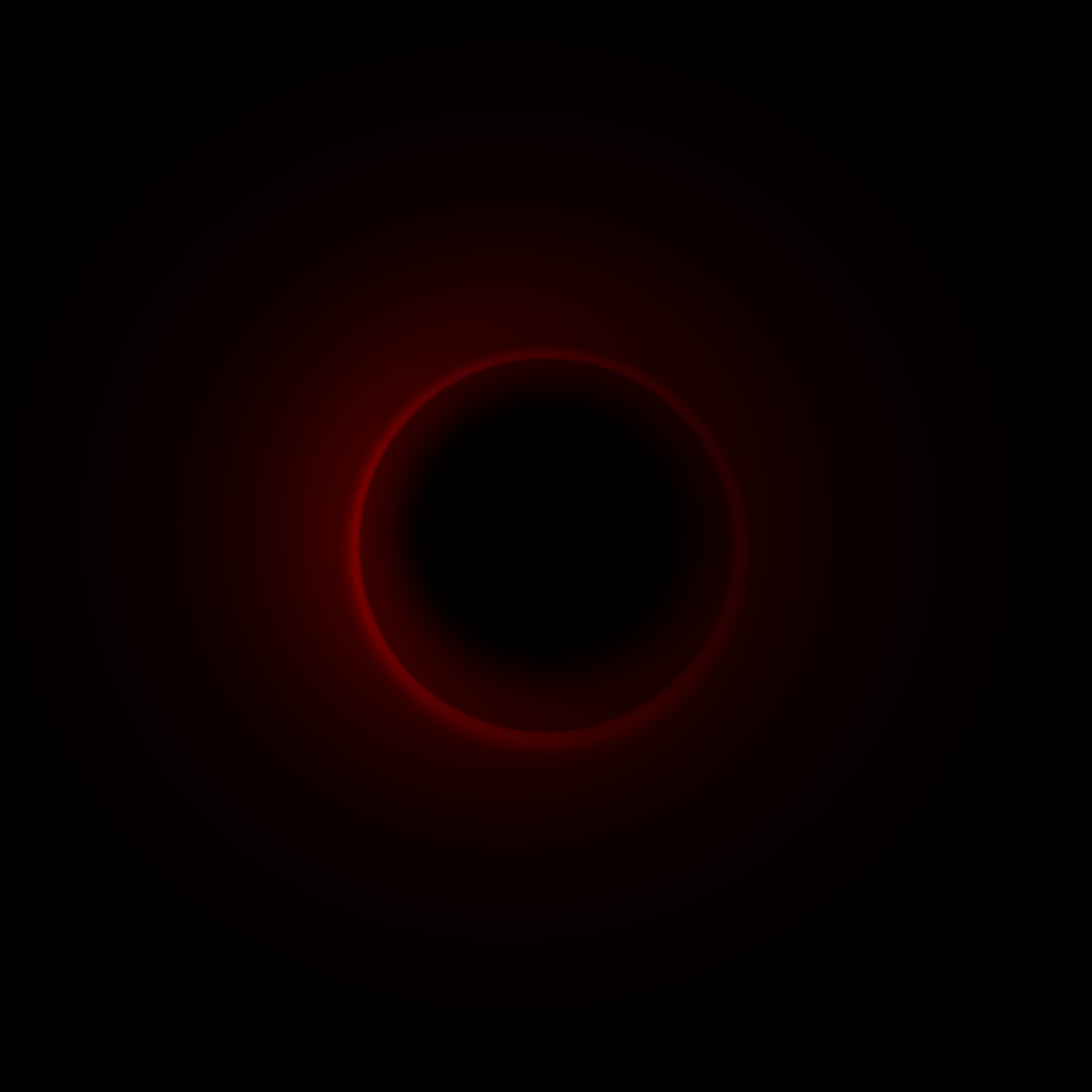}
\includegraphics[width=3.5cm]{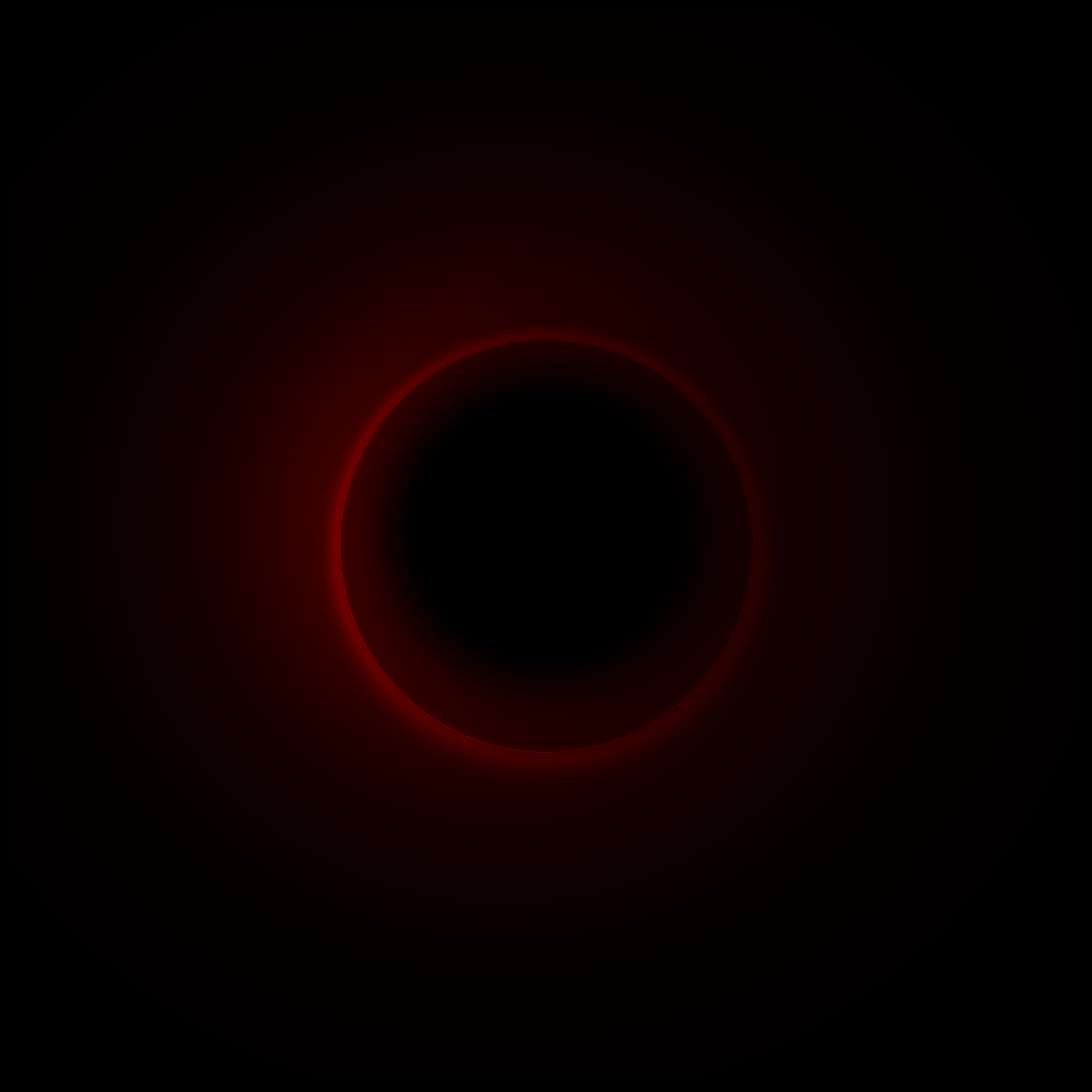}
\includegraphics[width=3.5cm]{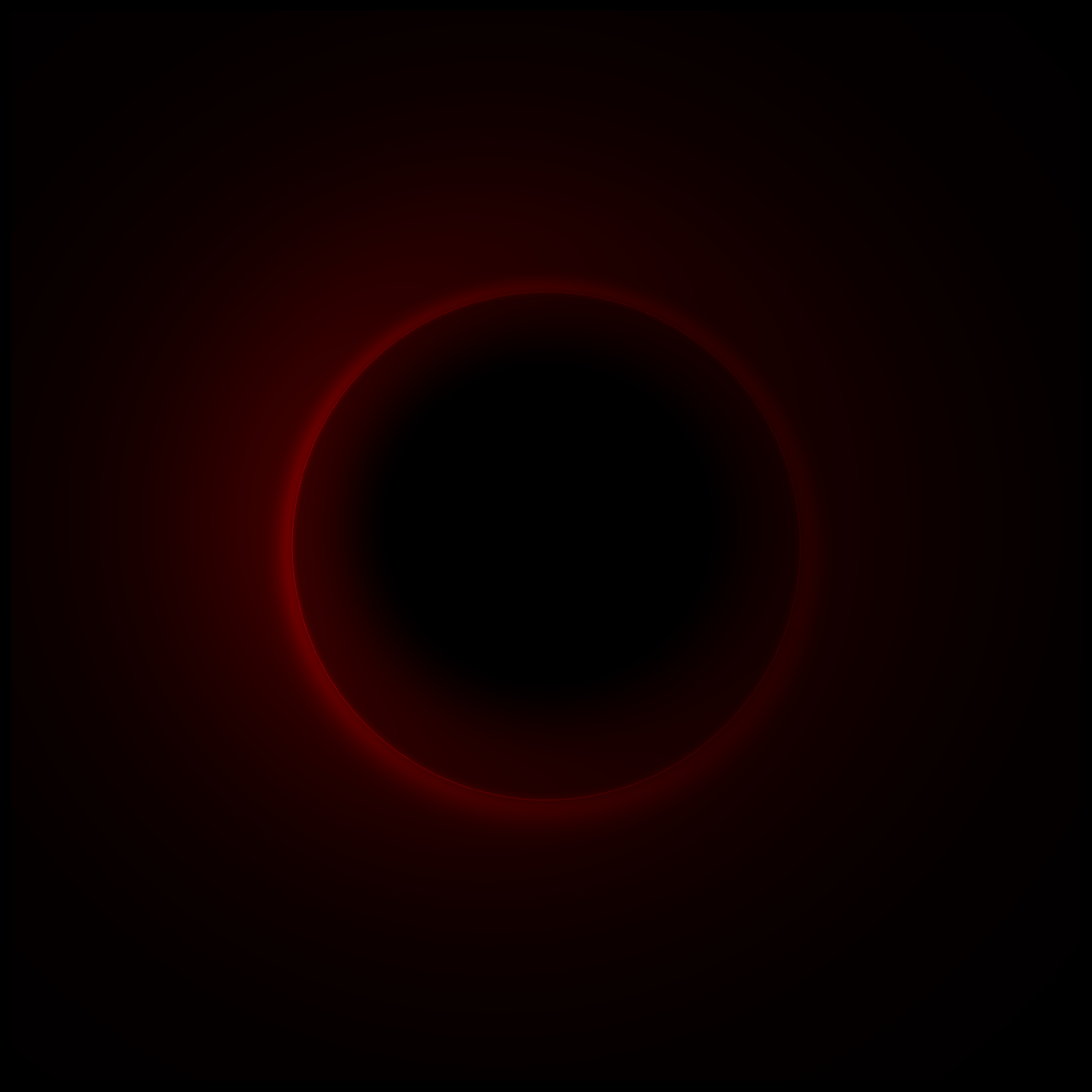}
\includegraphics[width=3.5cm]{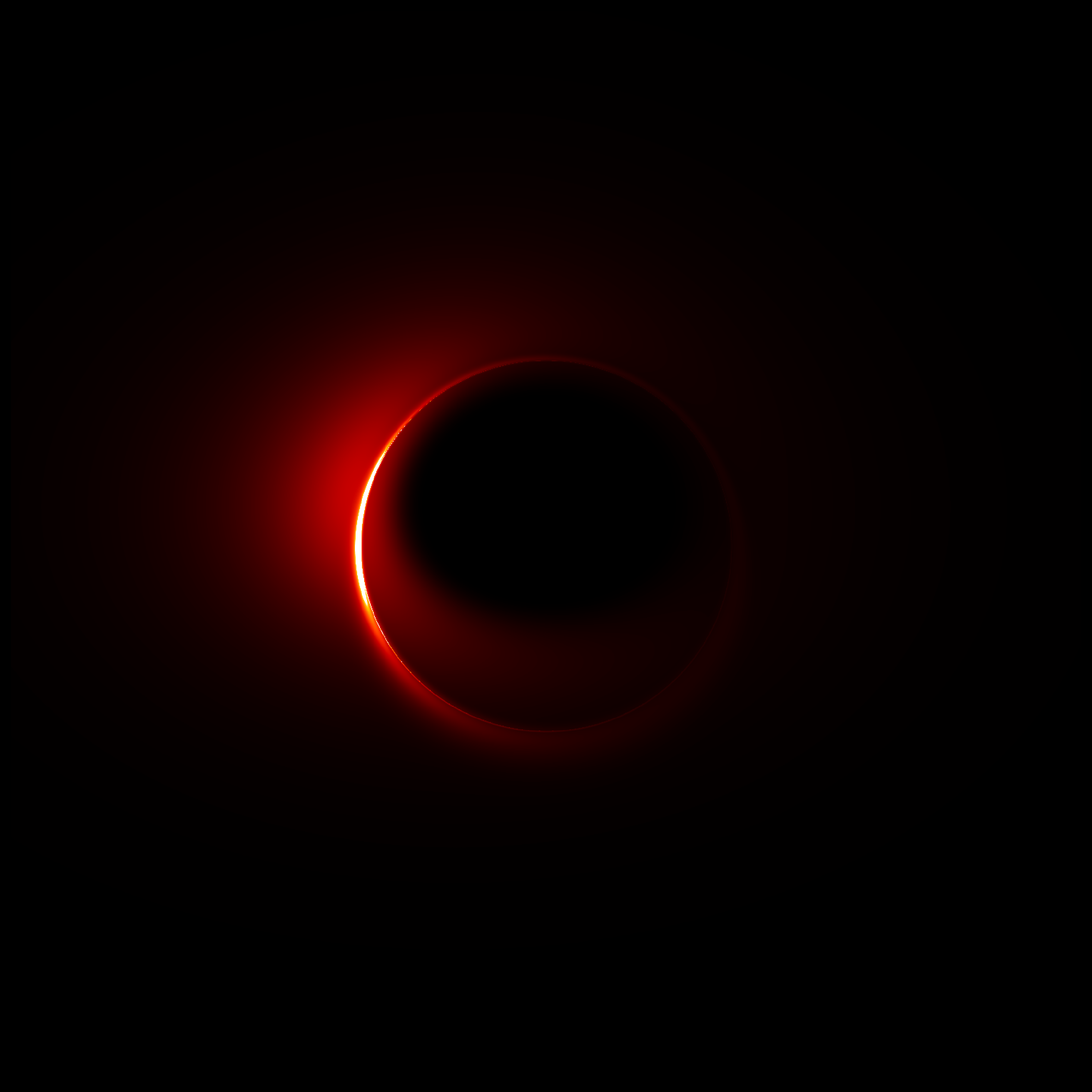}
\includegraphics[width=3.5cm]{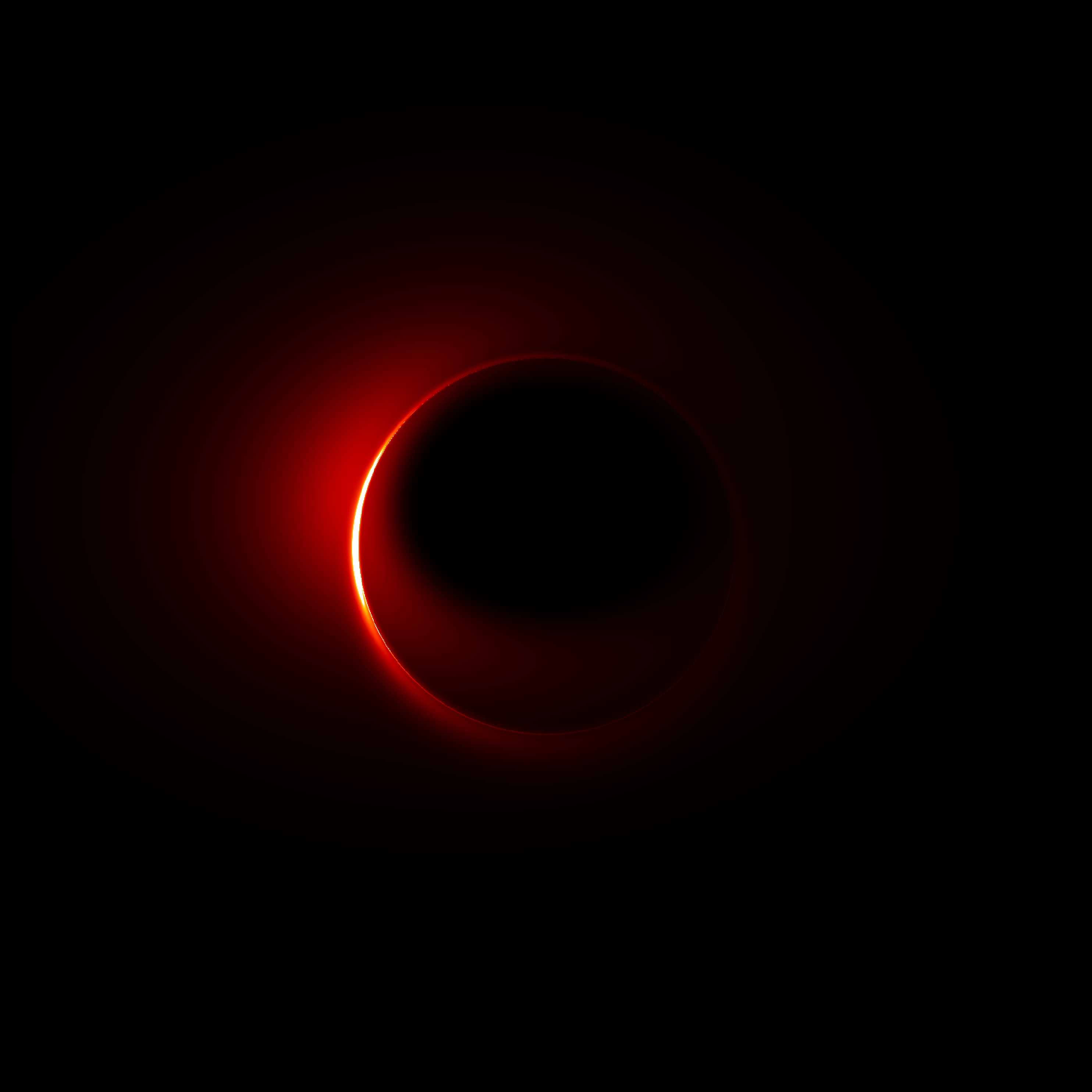}
\includegraphics[width=3.5cm]{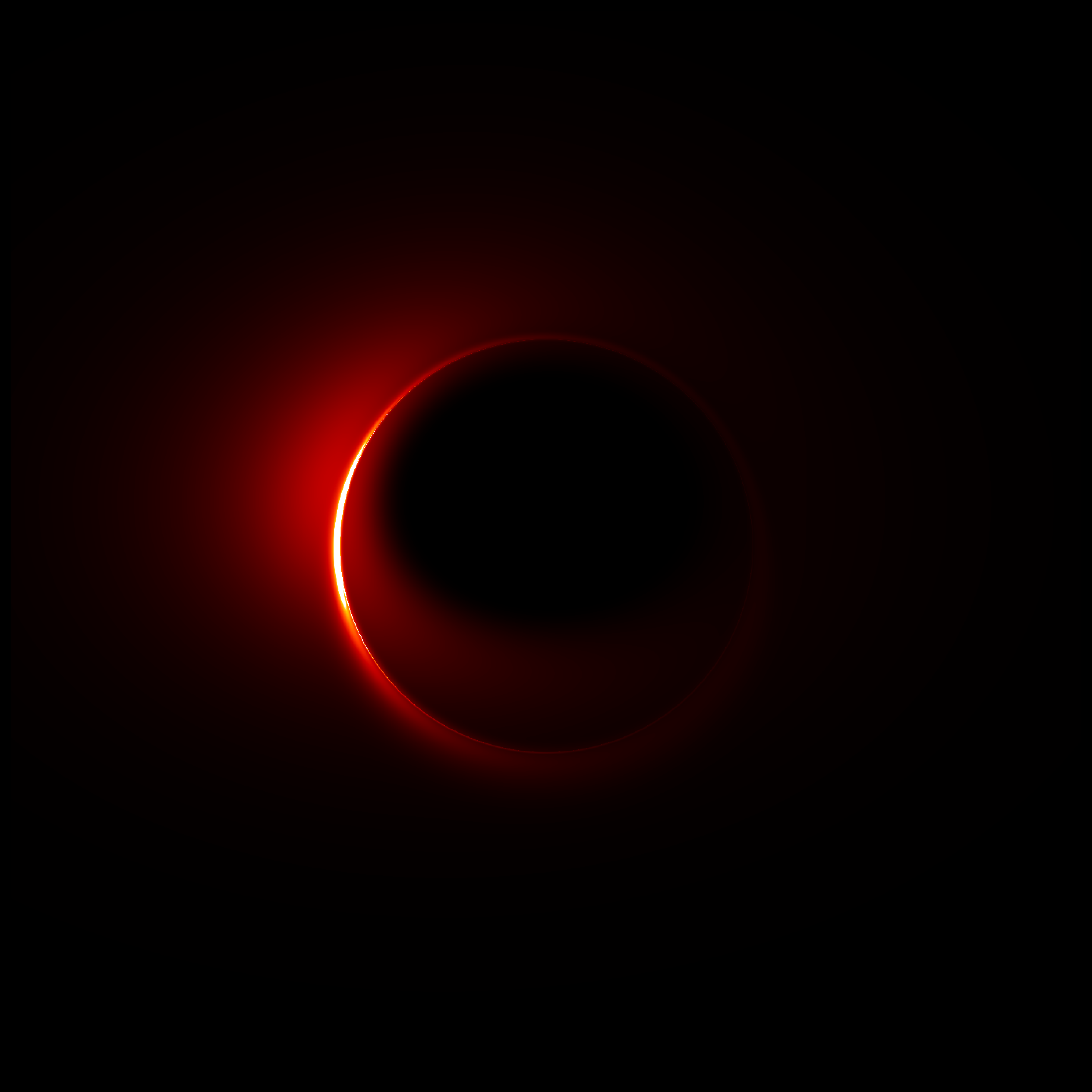}
\includegraphics[width=3.5cm]{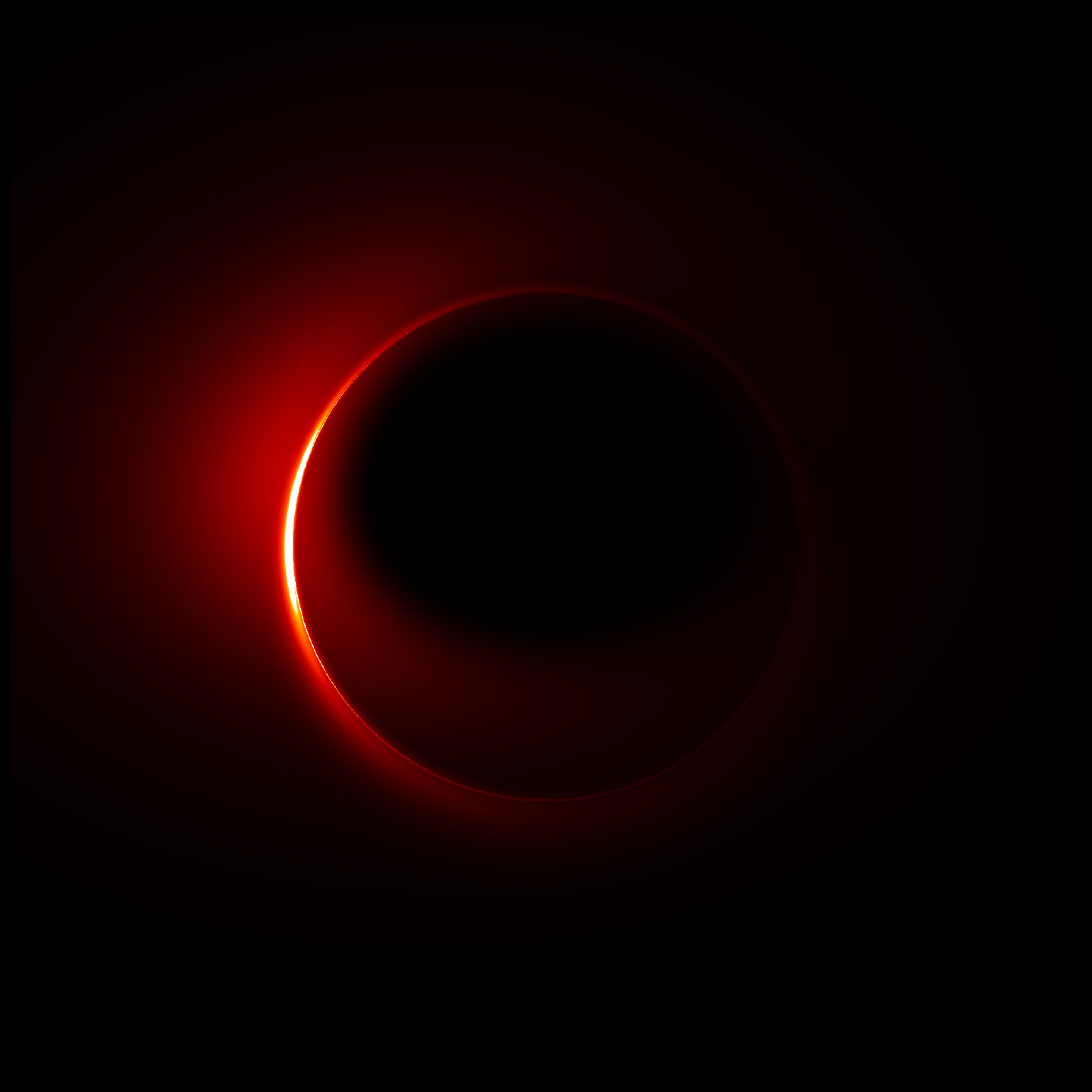}
\includegraphics[width=3.5cm]{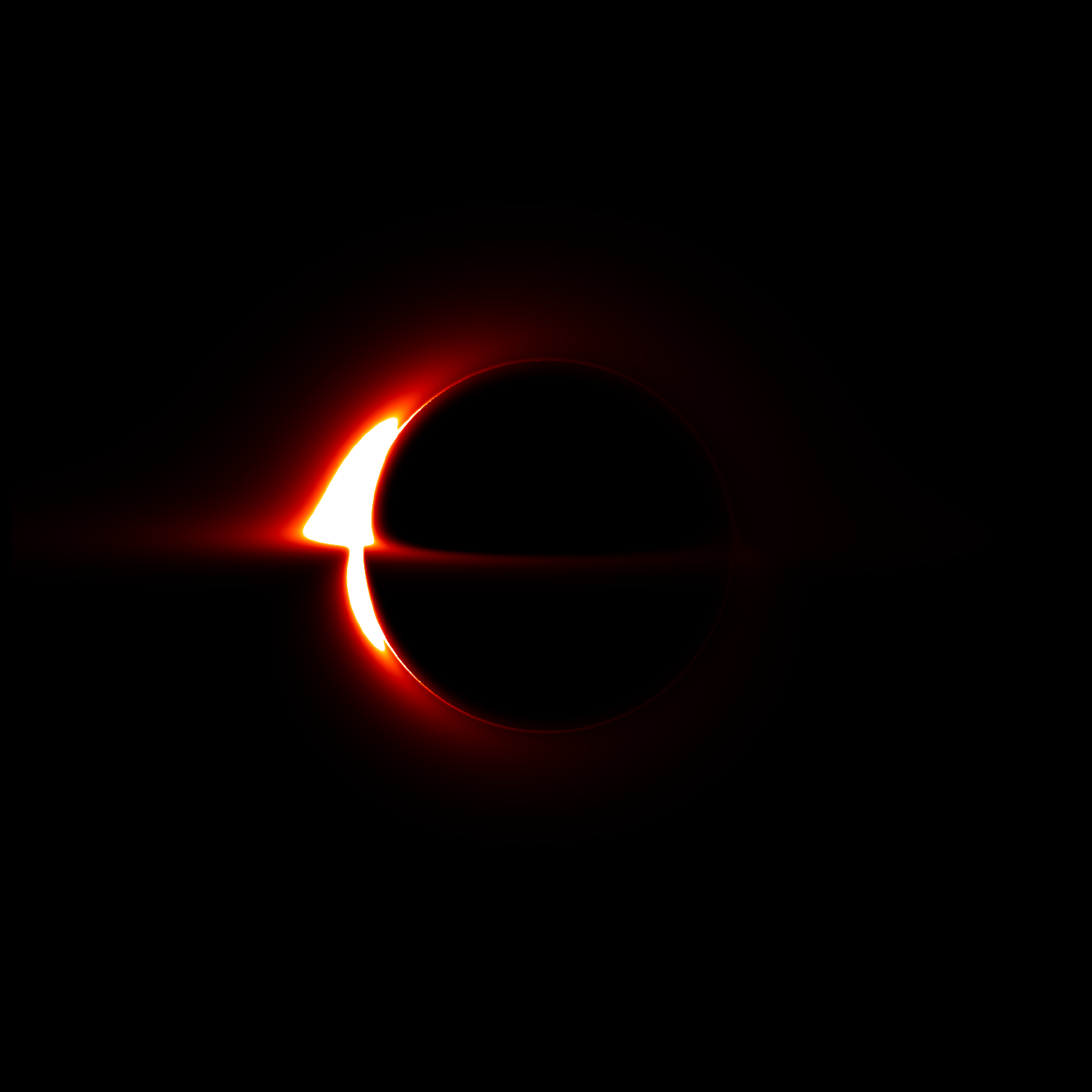}
\includegraphics[width=3.5cm]{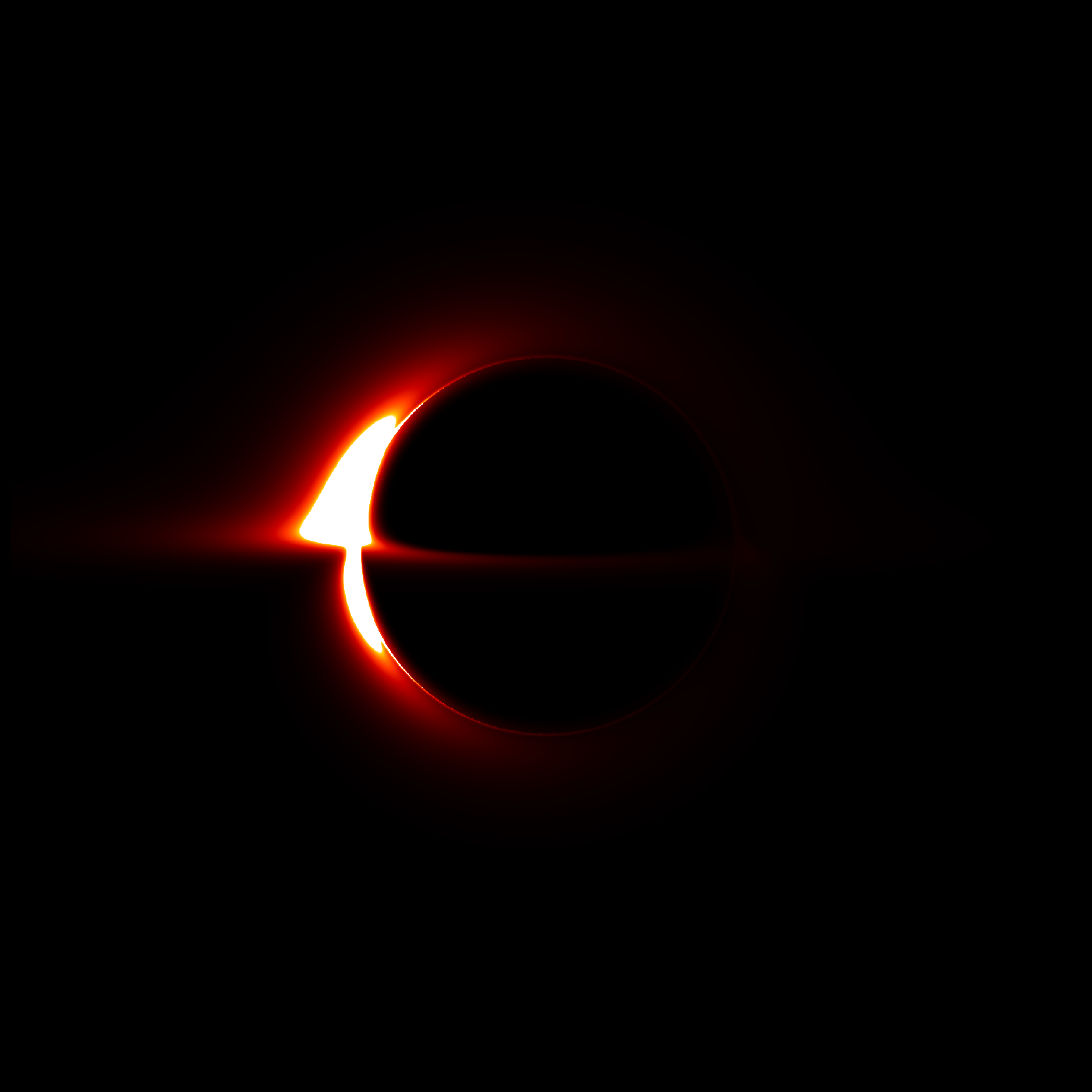}
\includegraphics[width=3.5cm]{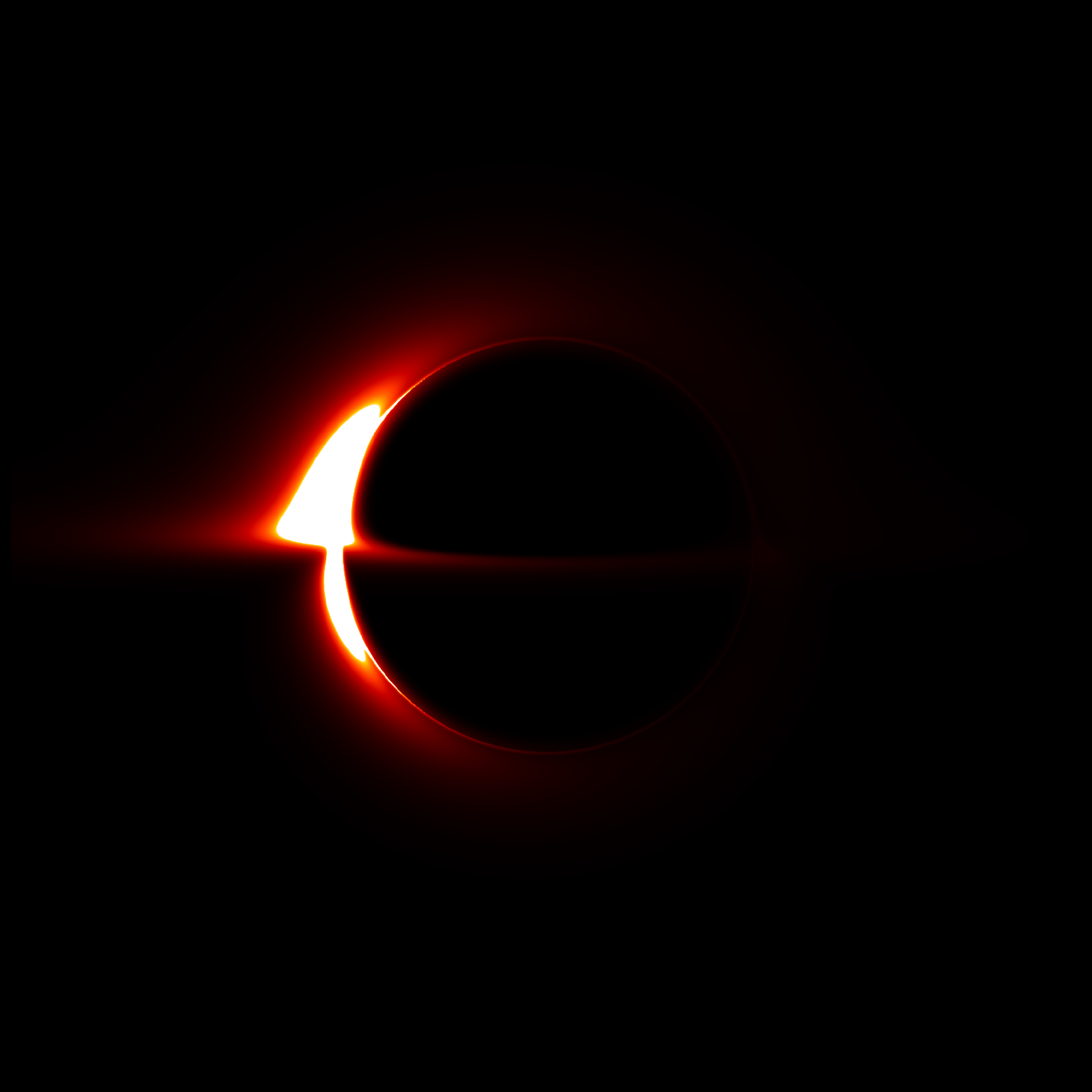}
\includegraphics[width=3.5cm]{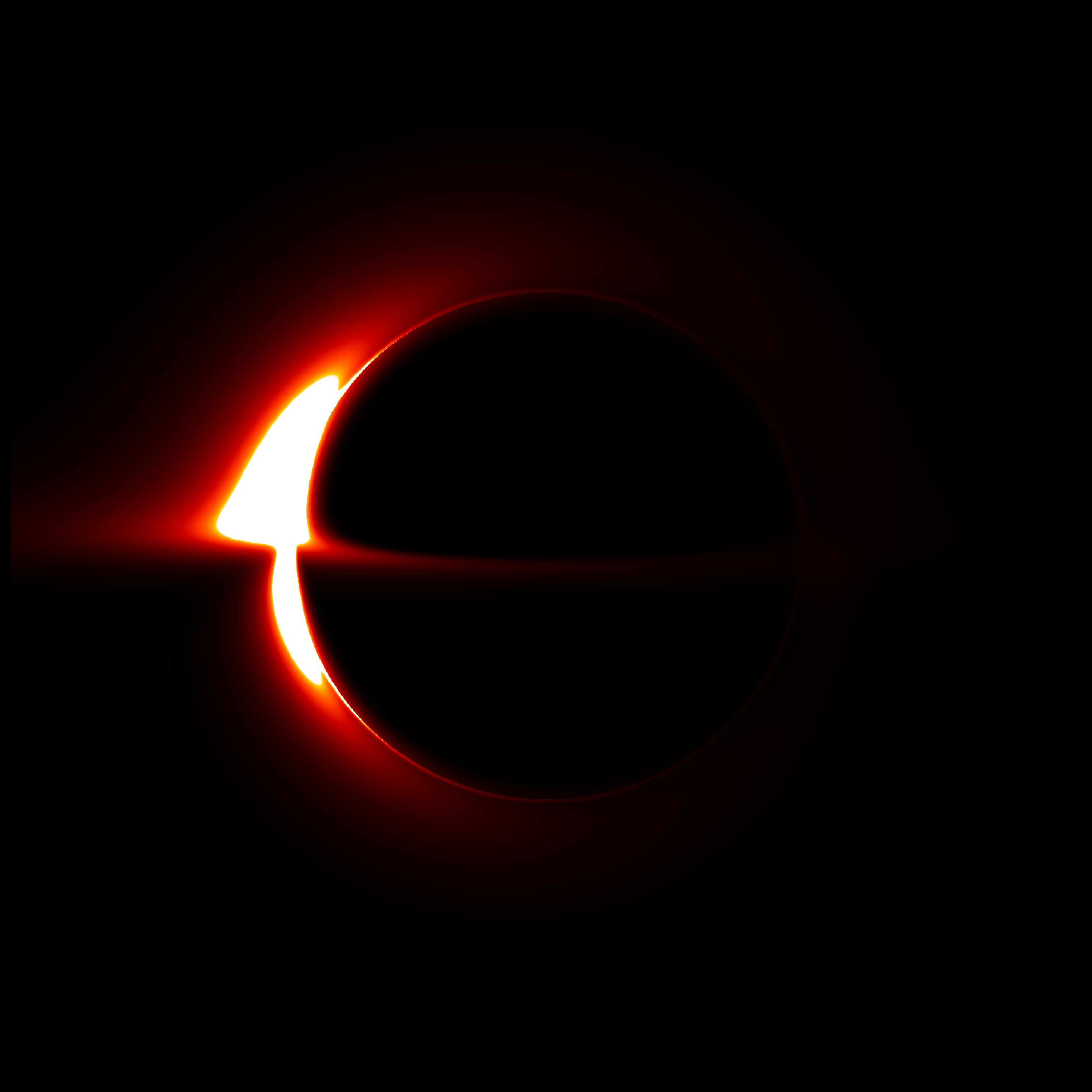}
\caption{Similar to figure 11, but for 230 GHz images, with the specific intensity range adjusted to $I_{\textrm{obs}} \in [0,0.25]$.}}\label{fig12}
\end{figure*}

The image features at 230 GHz are qualitatively similar to those at 86 GHz, with the inner shadow and bright patches clearly visible within specific parameter ranges. However, the overall brightness at the higher frequency is noticeably lower, making it difficult to identify complete higher-order bright rings. Instead, these rings appear as crescent-shaped luminous bands. Additionally, the bright patches are significantly smaller in extent compared to the 86 GHz case. Similarly, variations in $r_{\textrm{s}}$ do not alter the shapes of the image features but only affect their scale.
\begin{figure*}%[tbph]
\center{
\includegraphics[width=3.5cm]{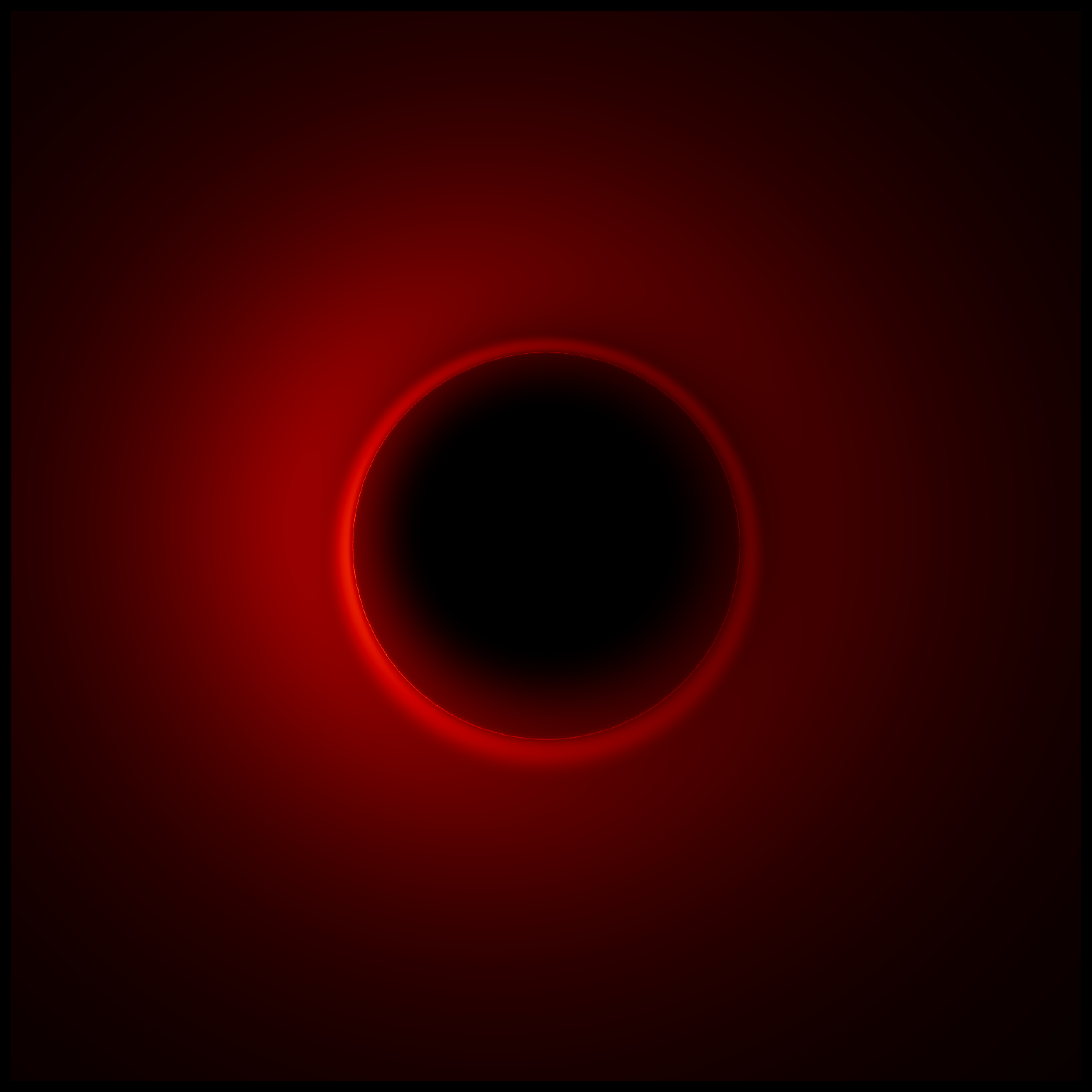}
\includegraphics[width=3.5cm]{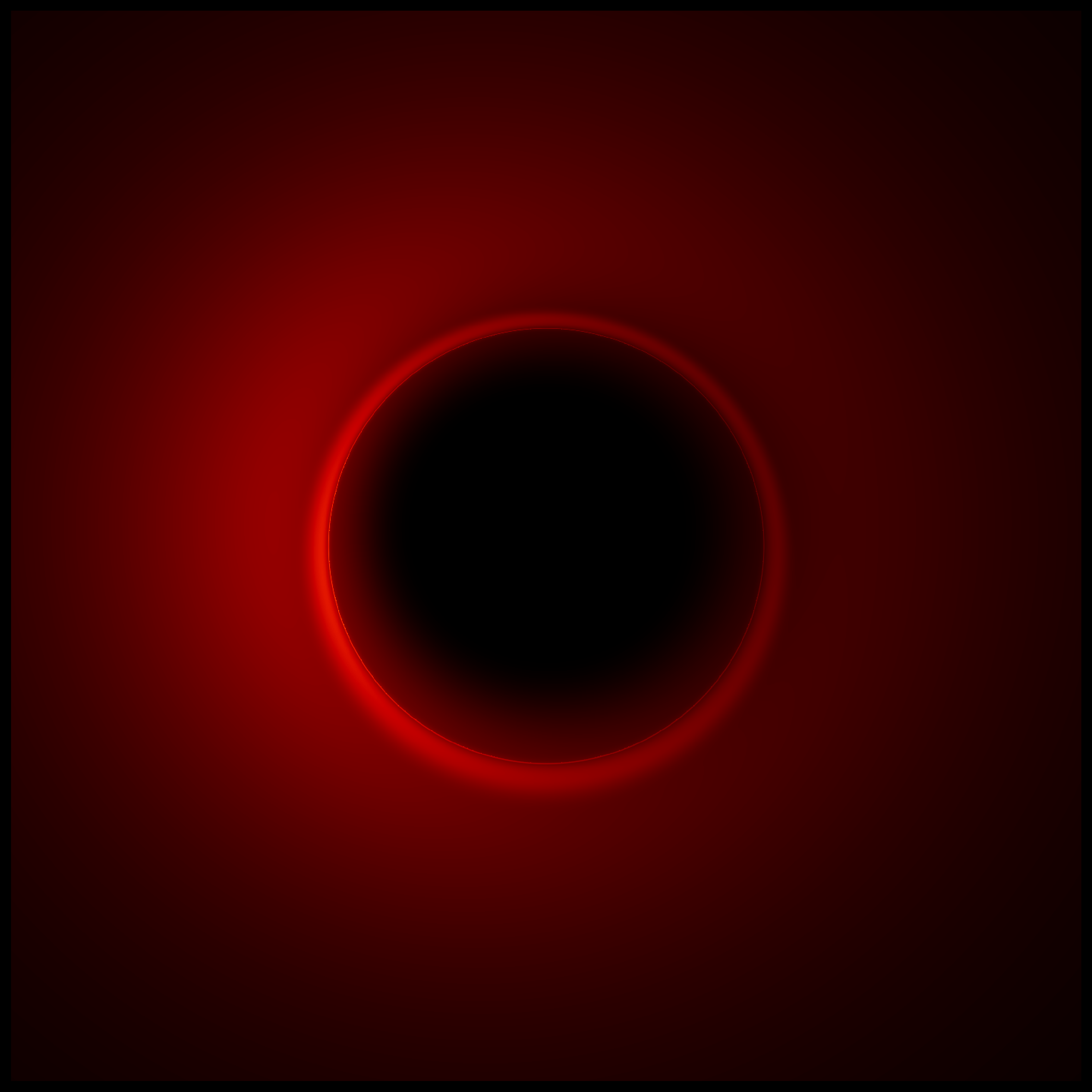}
\includegraphics[width=3.5cm]{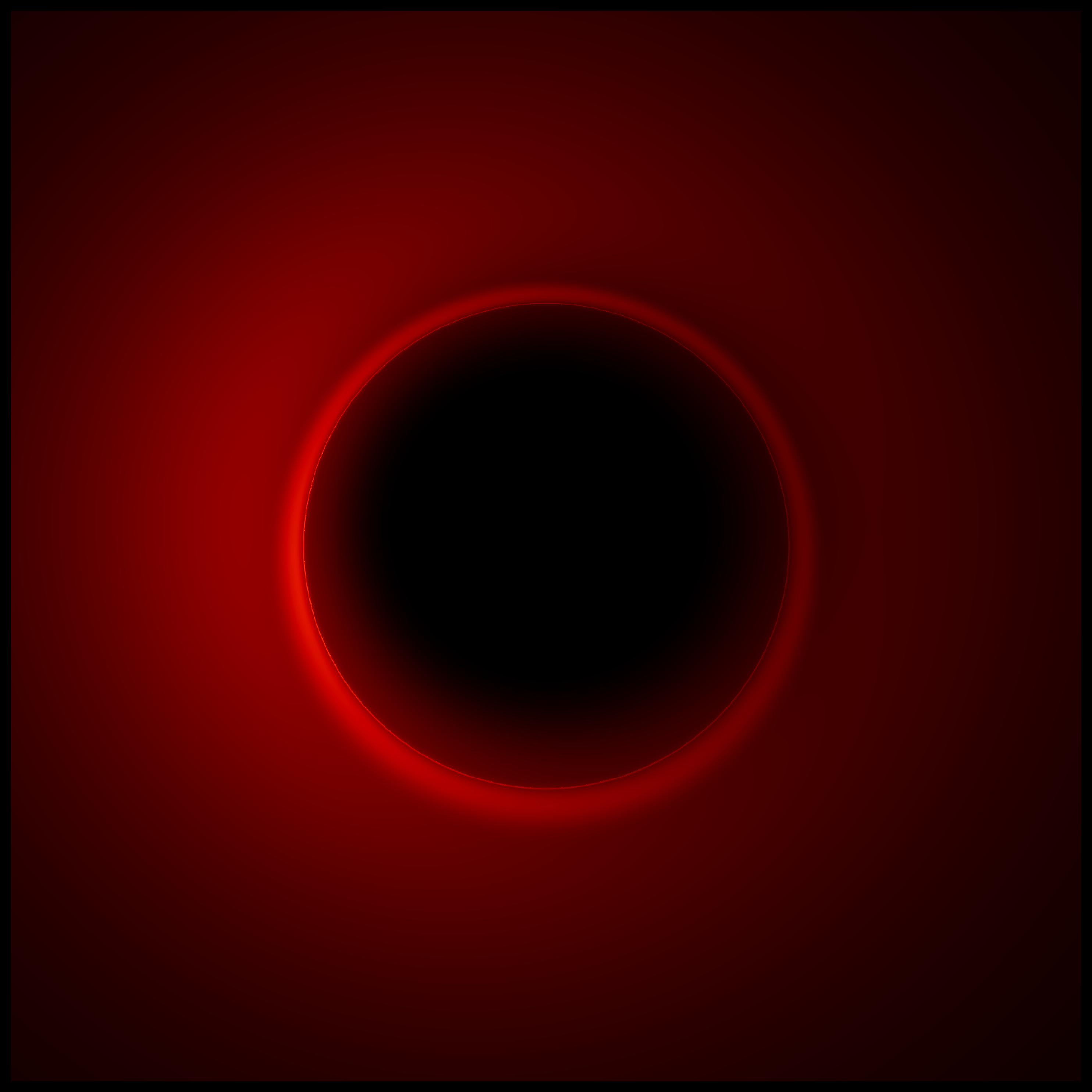}
\includegraphics[width=3.5cm]{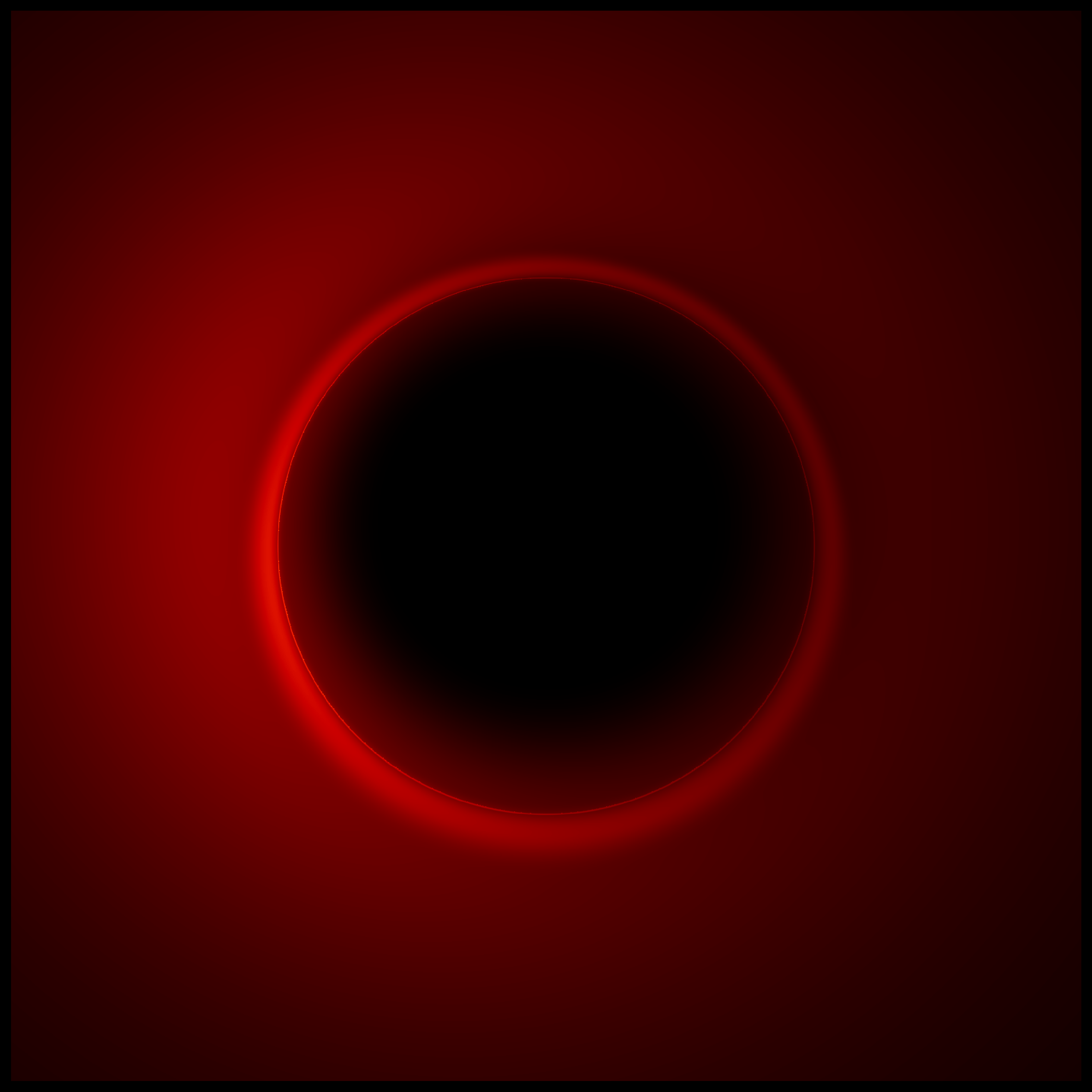}
\includegraphics[width=3.5cm]{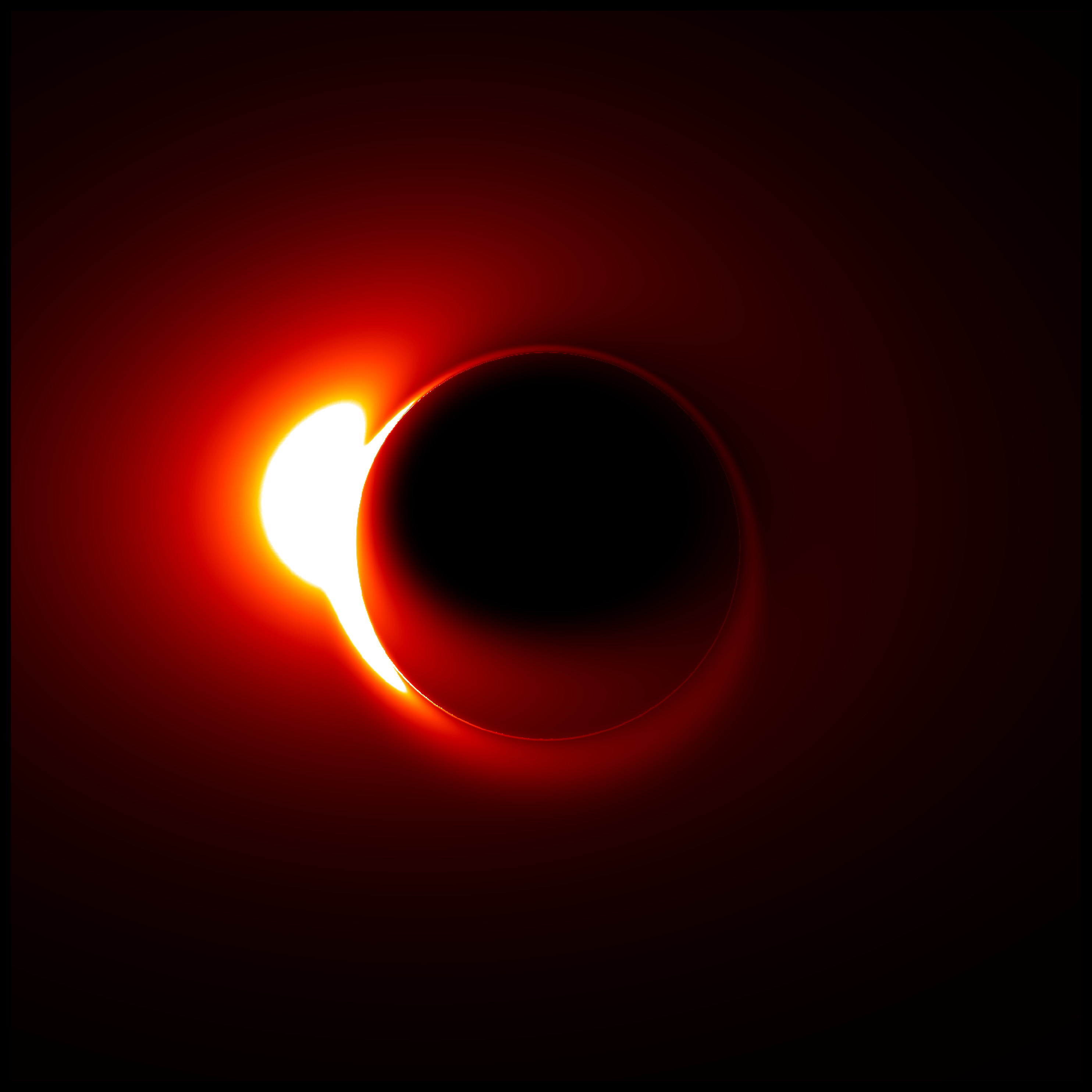}
\includegraphics[width=3.5cm]{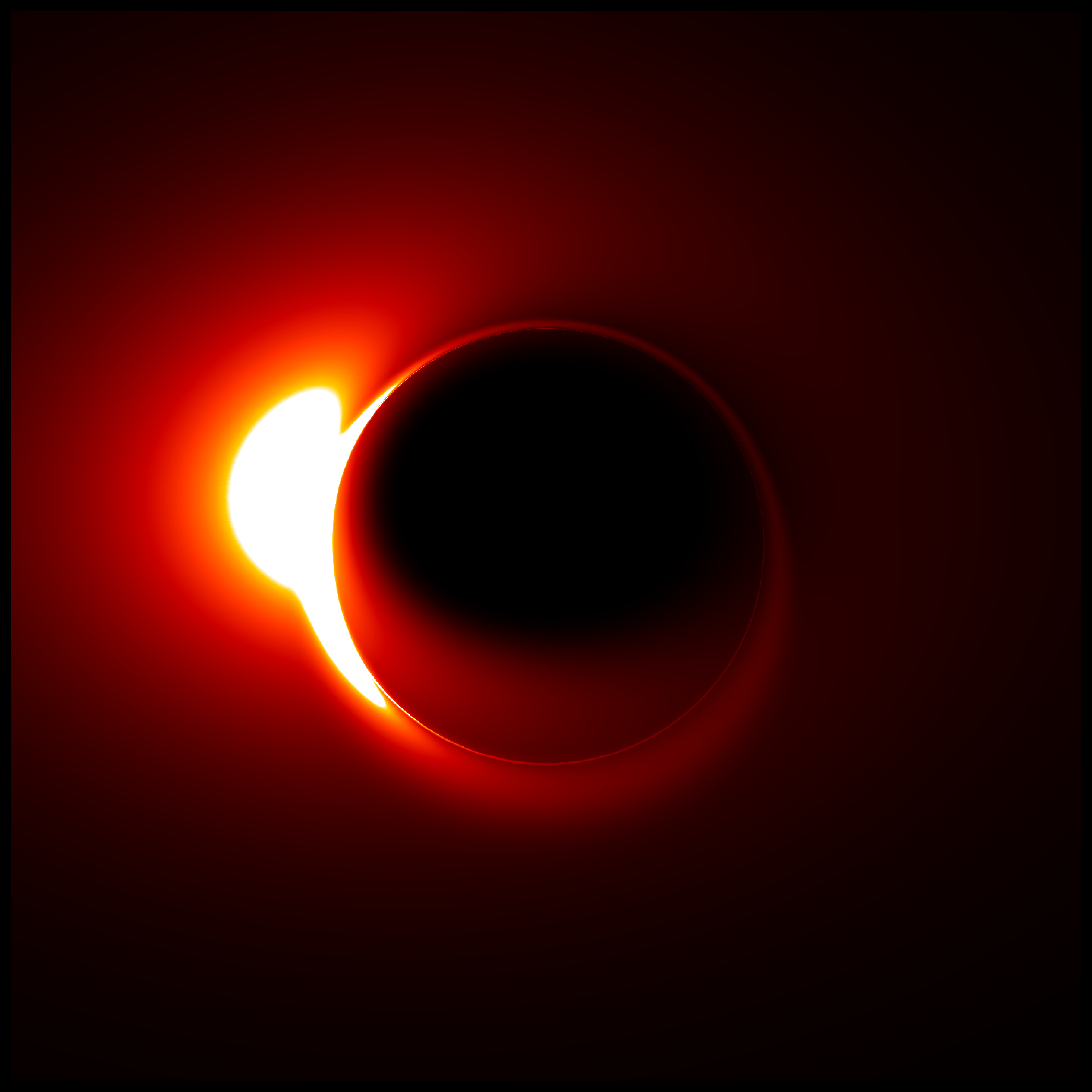}
\includegraphics[width=3.5cm]{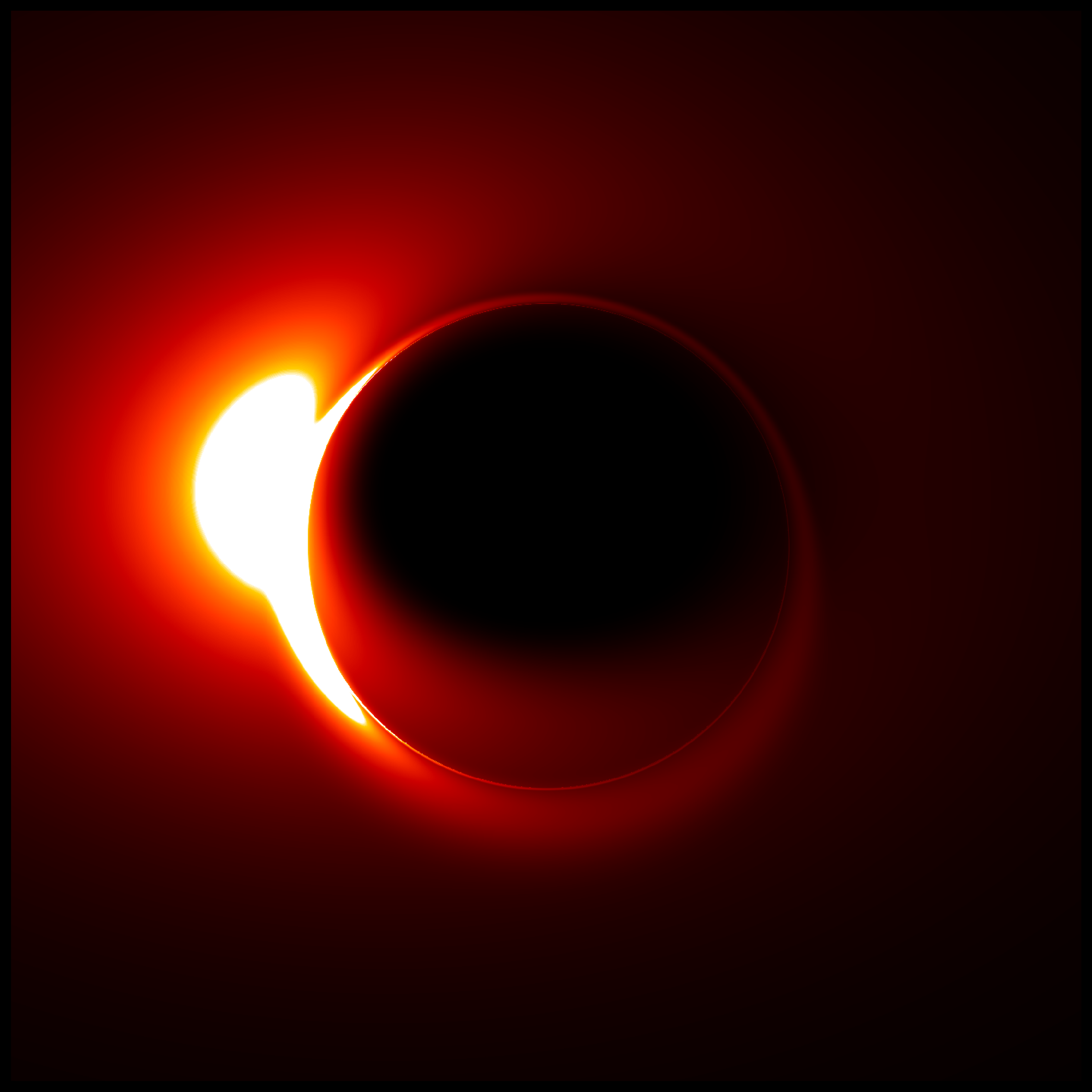}
\includegraphics[width=3.5cm]{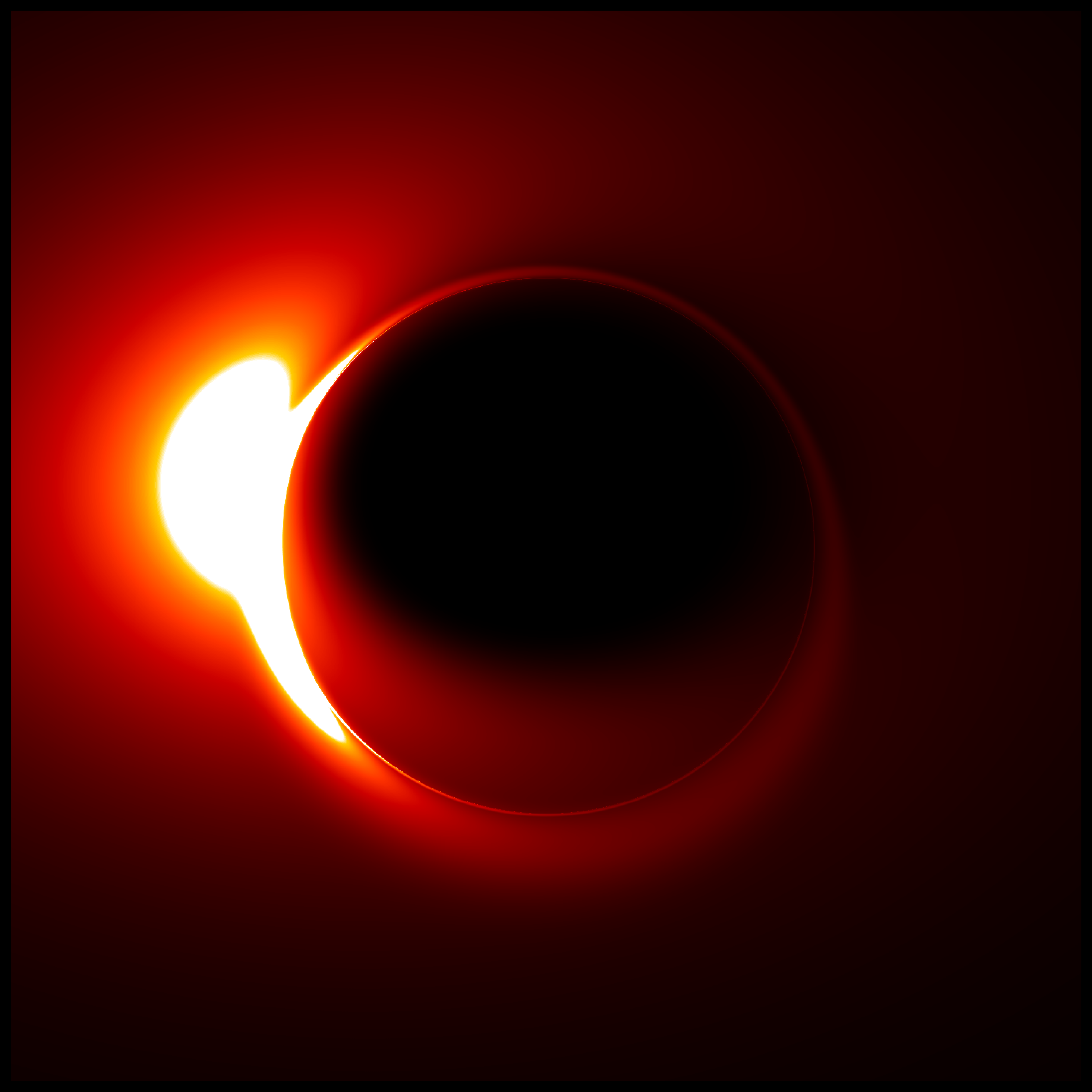}
\includegraphics[width=3.5cm]{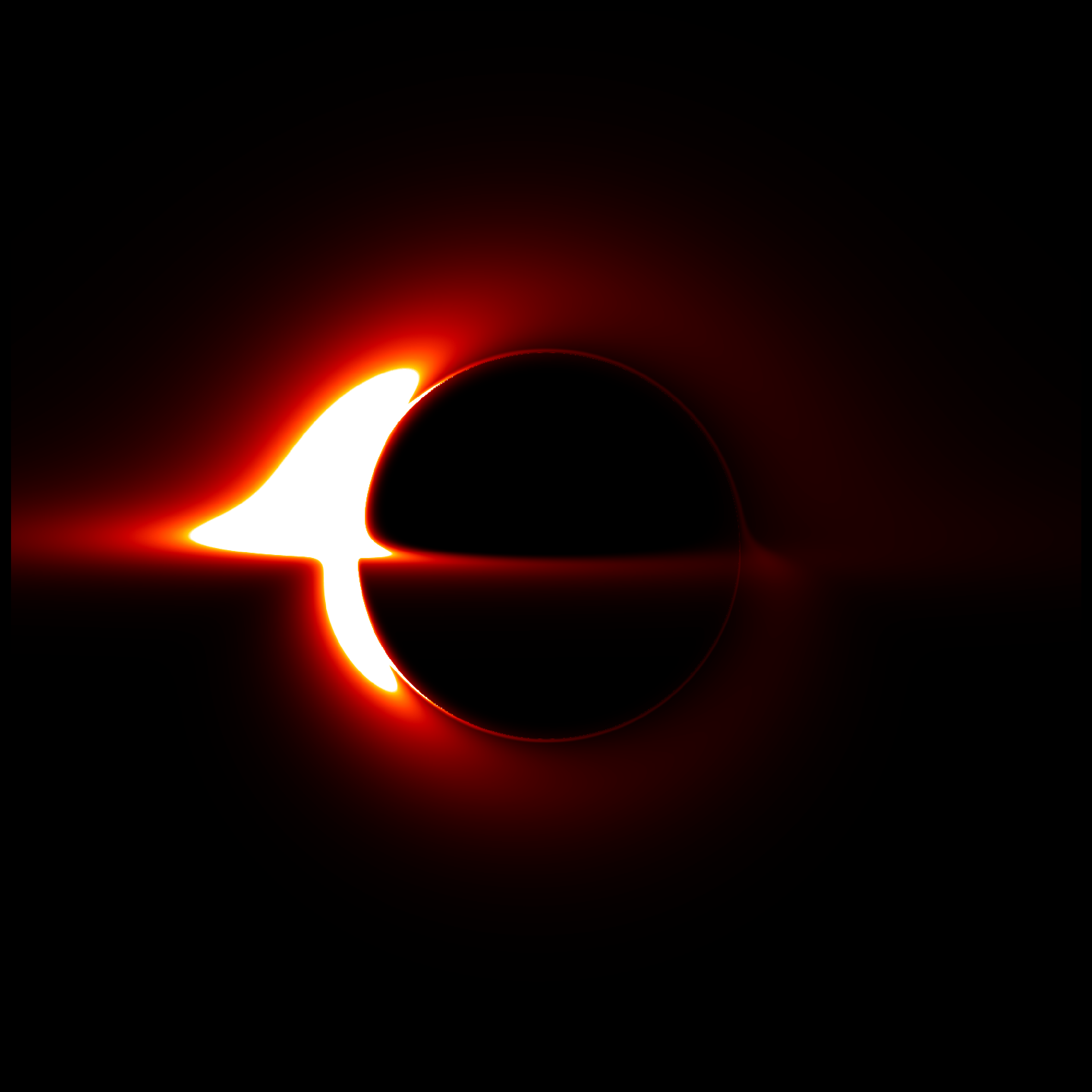}
\includegraphics[width=3.5cm]{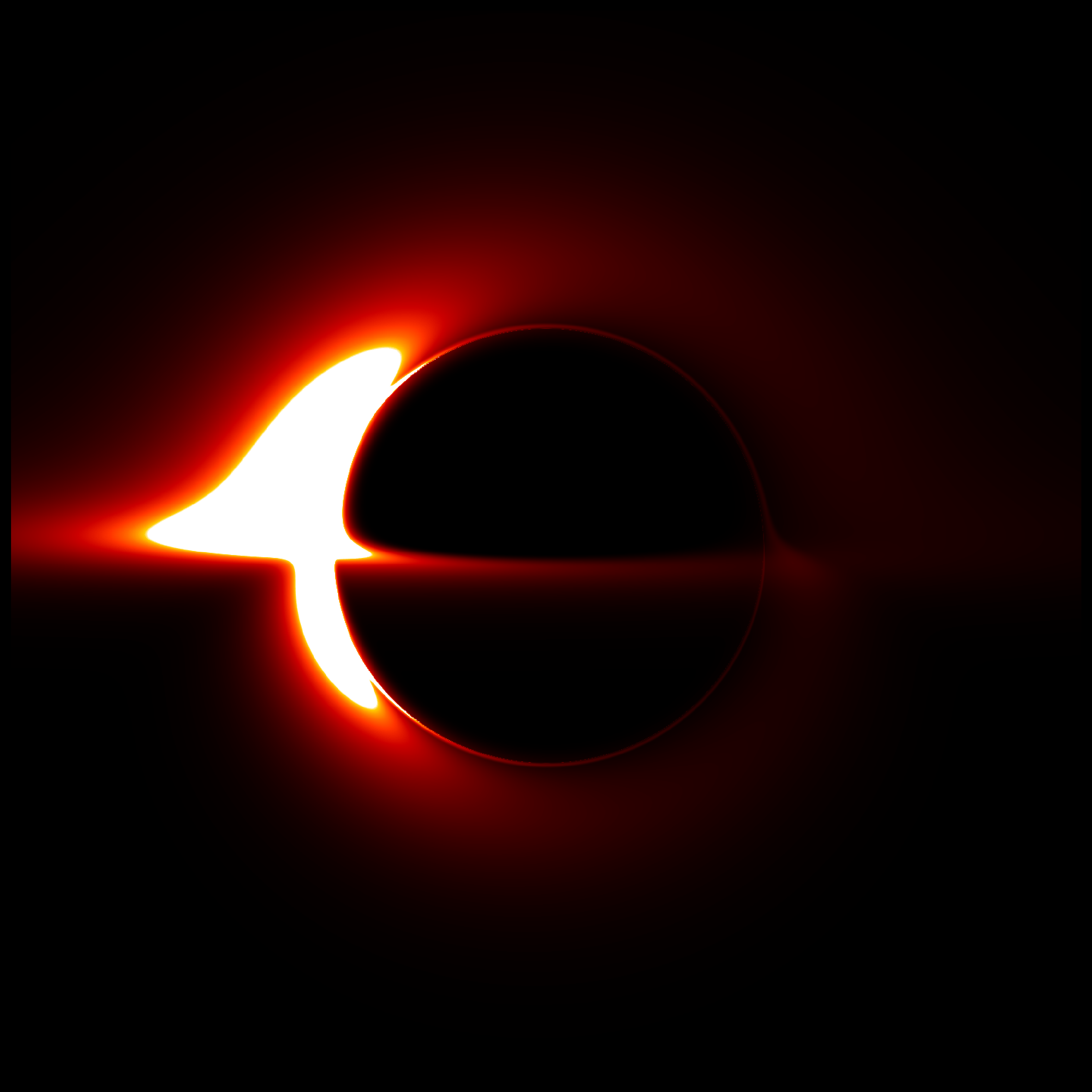}
\includegraphics[width=3.5cm]{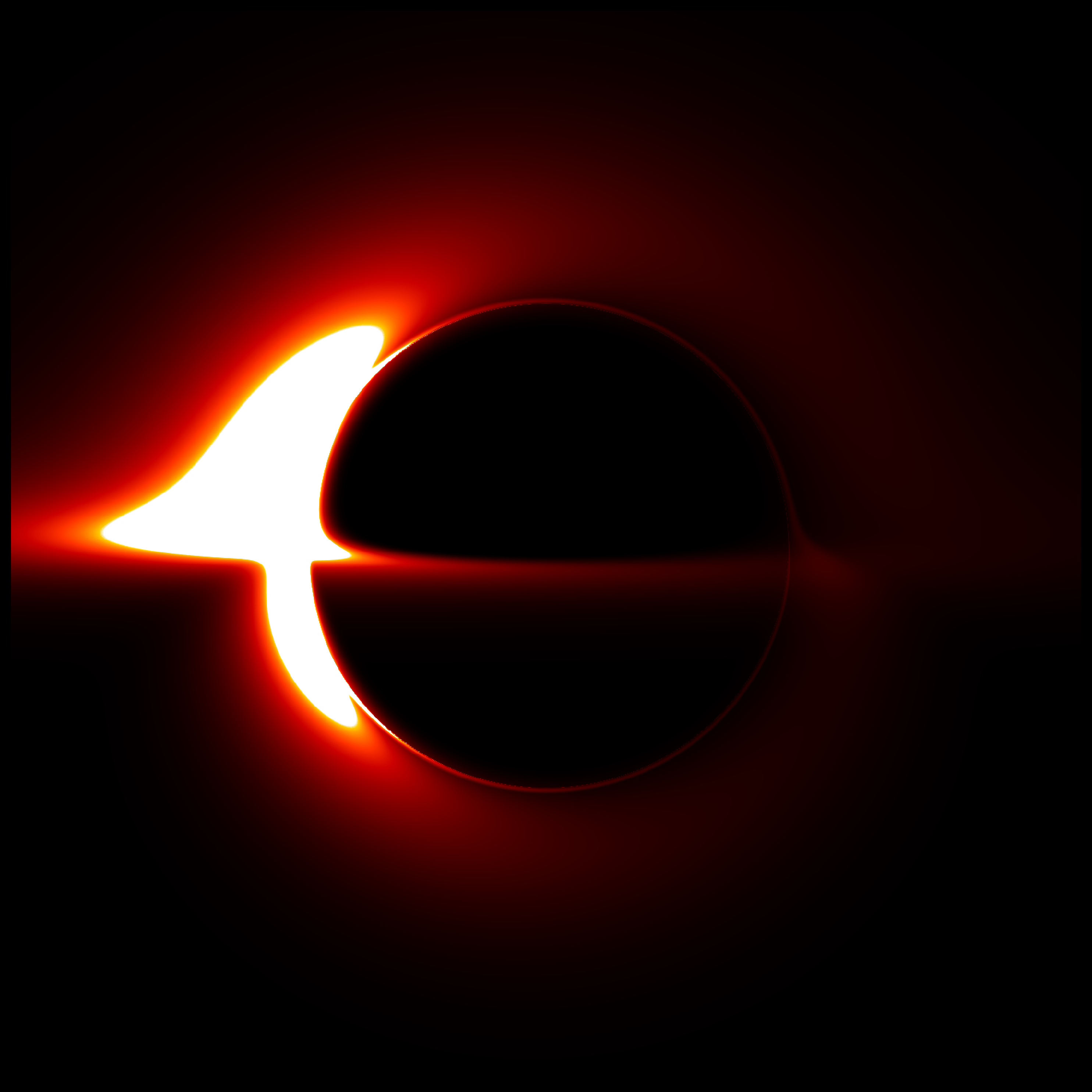}
\includegraphics[width=3.5cm]{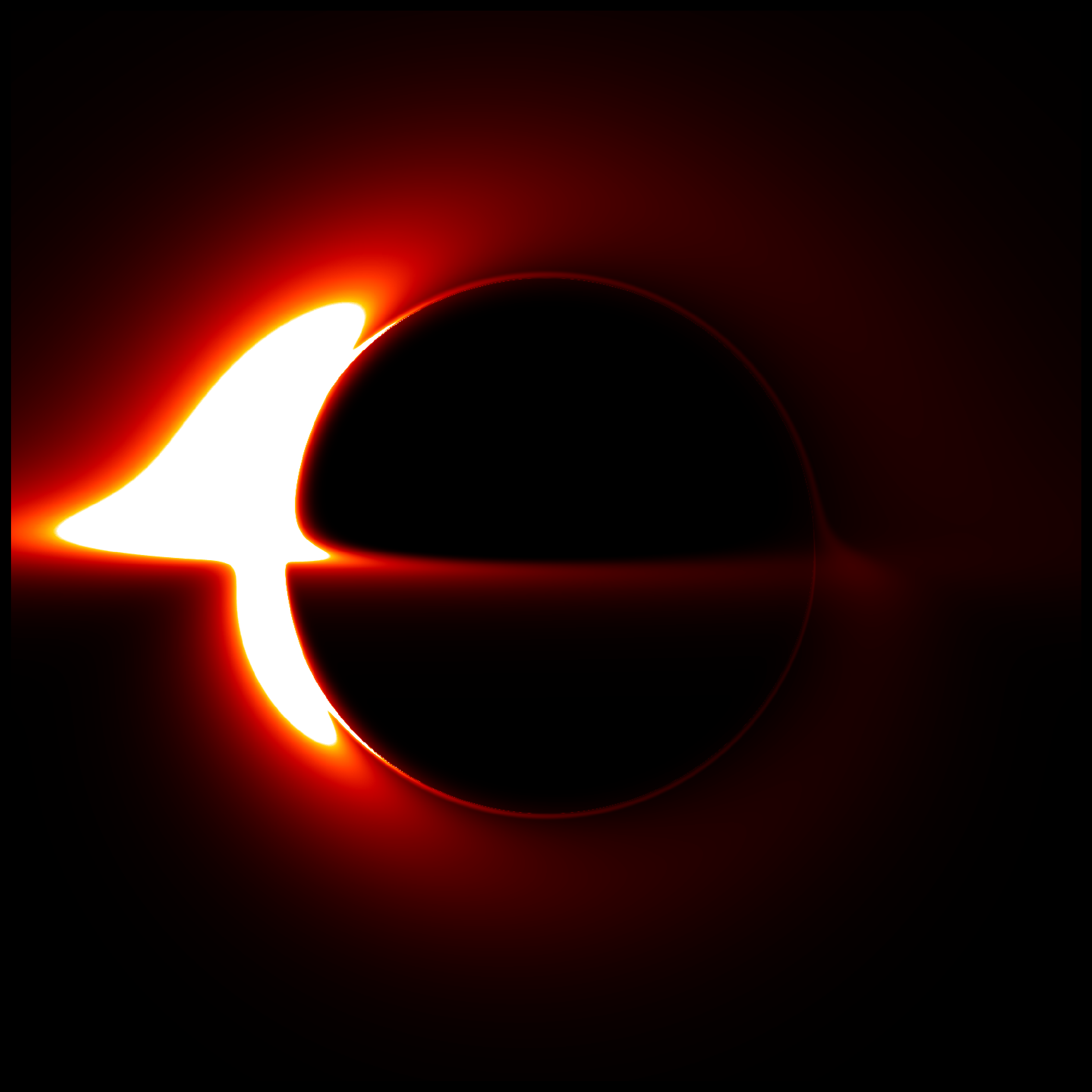}
\caption{86 GHz images of the target black hole for different values of the density parameter $\rho_{\textrm{s}}$ and observation inclination $\omega$, with $r_{\textrm{s}}$ fixed at $0.5$. From left to right: $\rho_{\textrm{s}}=$ $0.1$, $0.4$, $0.7$, $1$; from top to bottom: $\omega=$ $17^{\circ}$, $50^{\circ}$, $85^{\circ}$. All other simulation parameters are consistent with those in figure 11.}}\label{fig13}
\end{figure*}

Figure 13 presents the 86 GHz images for different values of the observation inclination $\omega$ and the dark matter halo density $\rho_{\textrm{s}}$, with $r_{\textrm{s}}$ fixed at $0.5$. As expected, the trends observed here align with those in figures 11 and 12: an increase in $\rho_{\textrm{s}}$ enlarges the bright patches, higher-order rings, and inner shadow. However, the effect of $\rho_{\textrm{s}}$ on the image features exhibits a nearly linear scaling, while the influence of $r_{\textrm{s}}$ follows an exponential growth pattern.

In addition to setting the accretion disk's inner boundary at the event horizon, OCTOPUS also allows for specifying an arbitrary cutoff radius $r_{\textrm{in}}$ for disk emission, such as $r_{\textrm{in}}=r_{\textrm{p}}$ or $r_{\textrm{in}}=r_{\textrm{isco}}$. Furthermore, the outer boundary $r_{\textrm{out}}$ of the accretion disk can be freely defined. 

We assume the presence of a plasma torus orbiting within the black hole's equatorial plane, with the inner and outer boundaries denoted by $r_{\textrm{in}}$ and $r_{\textrm{out}}$, respectively, and adopting the emission model given in equation \eqref{22}. Figure 14 presents images of tori placed at various locations. It is observed that when a torus is positioned near the critical photon orbit ($r_{\textrm{p}}=3.6093$ for $r_{\textrm{s}}=\rho_{\textrm{s}}=0.5$), its emission appears exceedingly faint and nearly undetectable, as shown in panel (a). Increasing the observation inclination enhances Doppler blueshift, which partially mitigates this effect. When the torus is placed outside the photon ring, its image becomes clearly visible, although the luminosity decreases as the torus moves further outward. Increasing the observation inclination not only distorts the torus shape but also generates secondary images---a consequence of gravitational lensing. At high inclinations, emission from the back side of the black hole is bent toward the upper and lower regions of the image. Moreover, in near-edge-on views, cap-like structures appear attached to both sides of the torus image. These features gradually diminish as the torus moves closer to the black hole and may serve as potential probes for inferring accretion environments.
\begin{figure*}%[tbph]
\center{
\includegraphics[width=3.5cm]{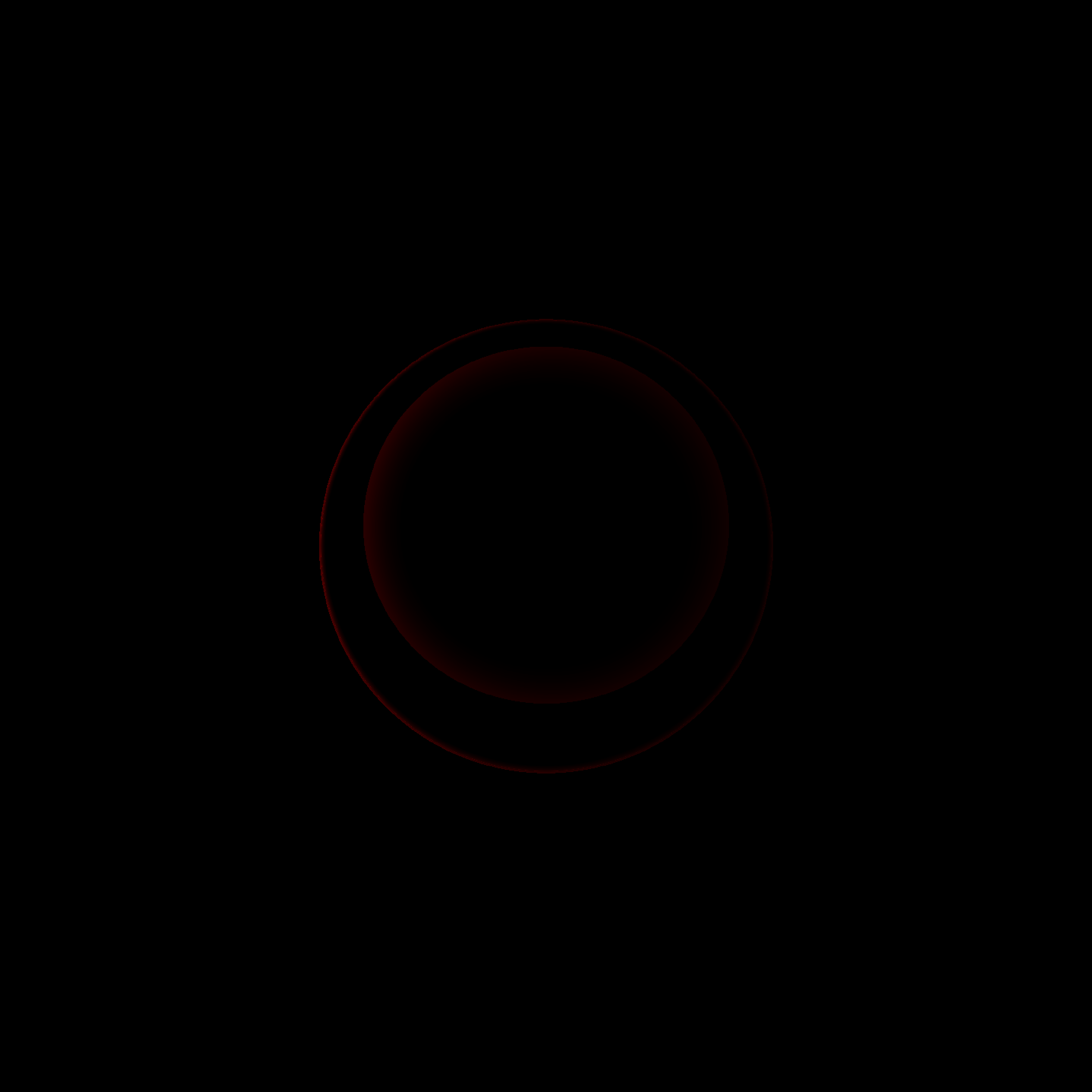}
\includegraphics[width=3.5cm]{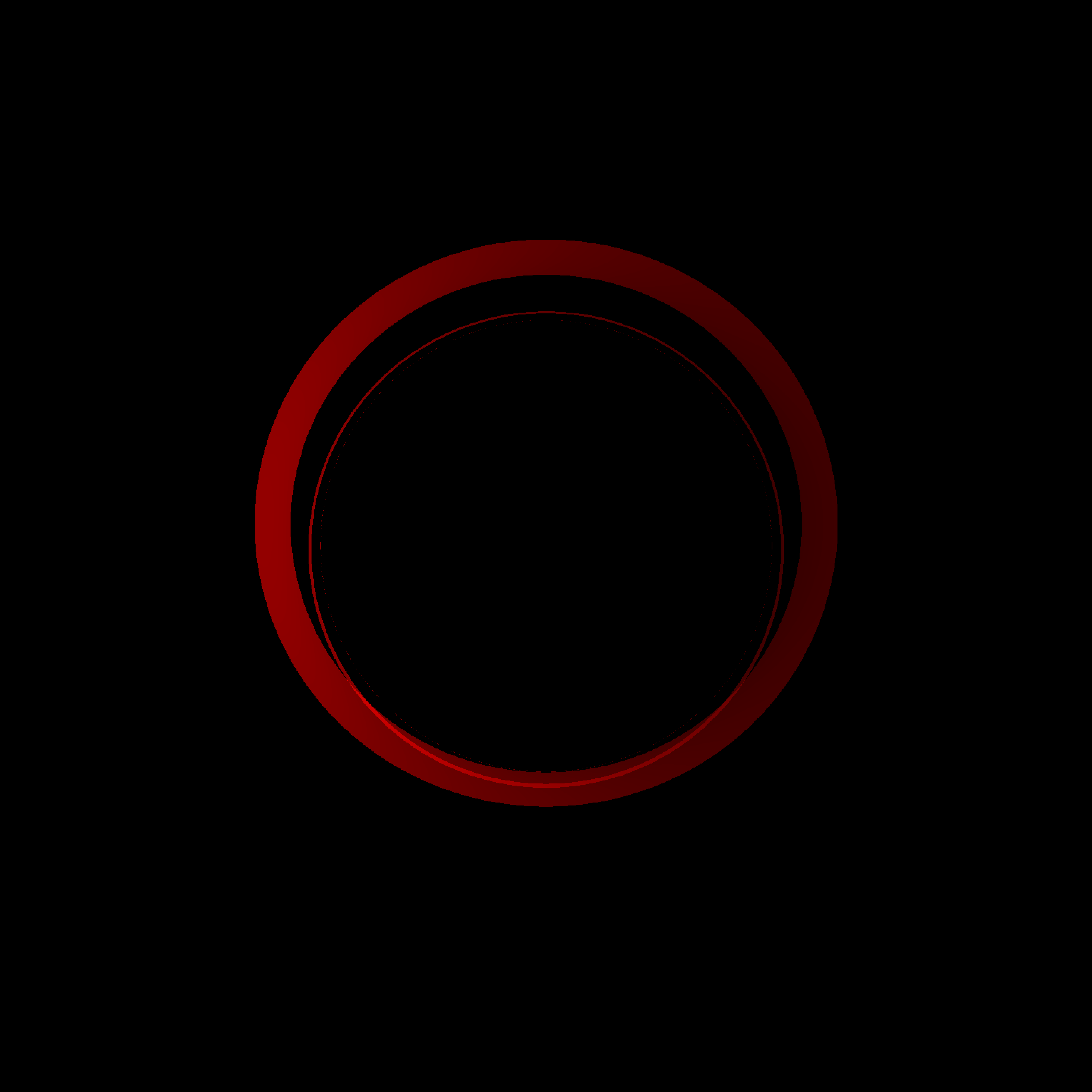}
\includegraphics[width=3.5cm]{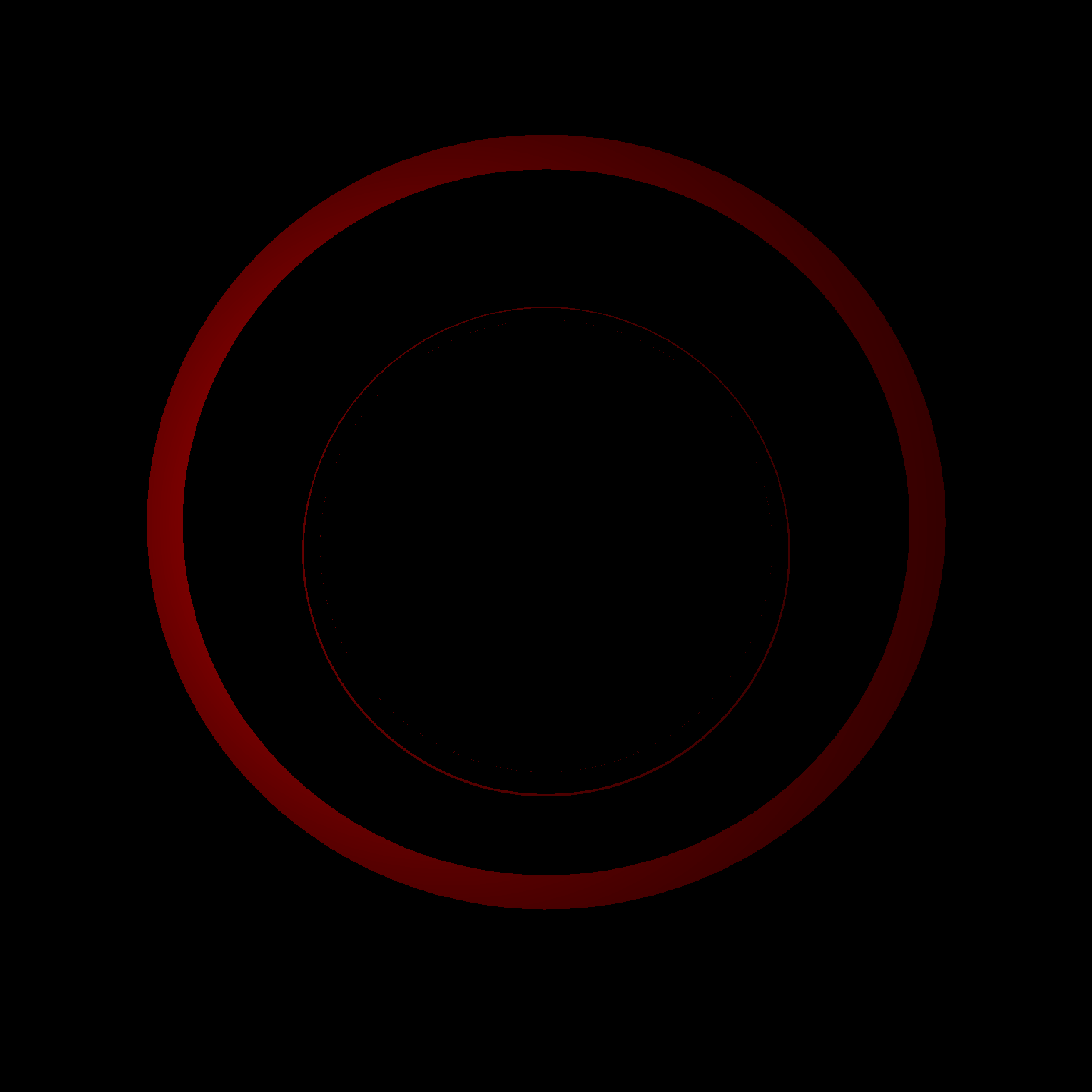}
\includegraphics[width=3.5cm]{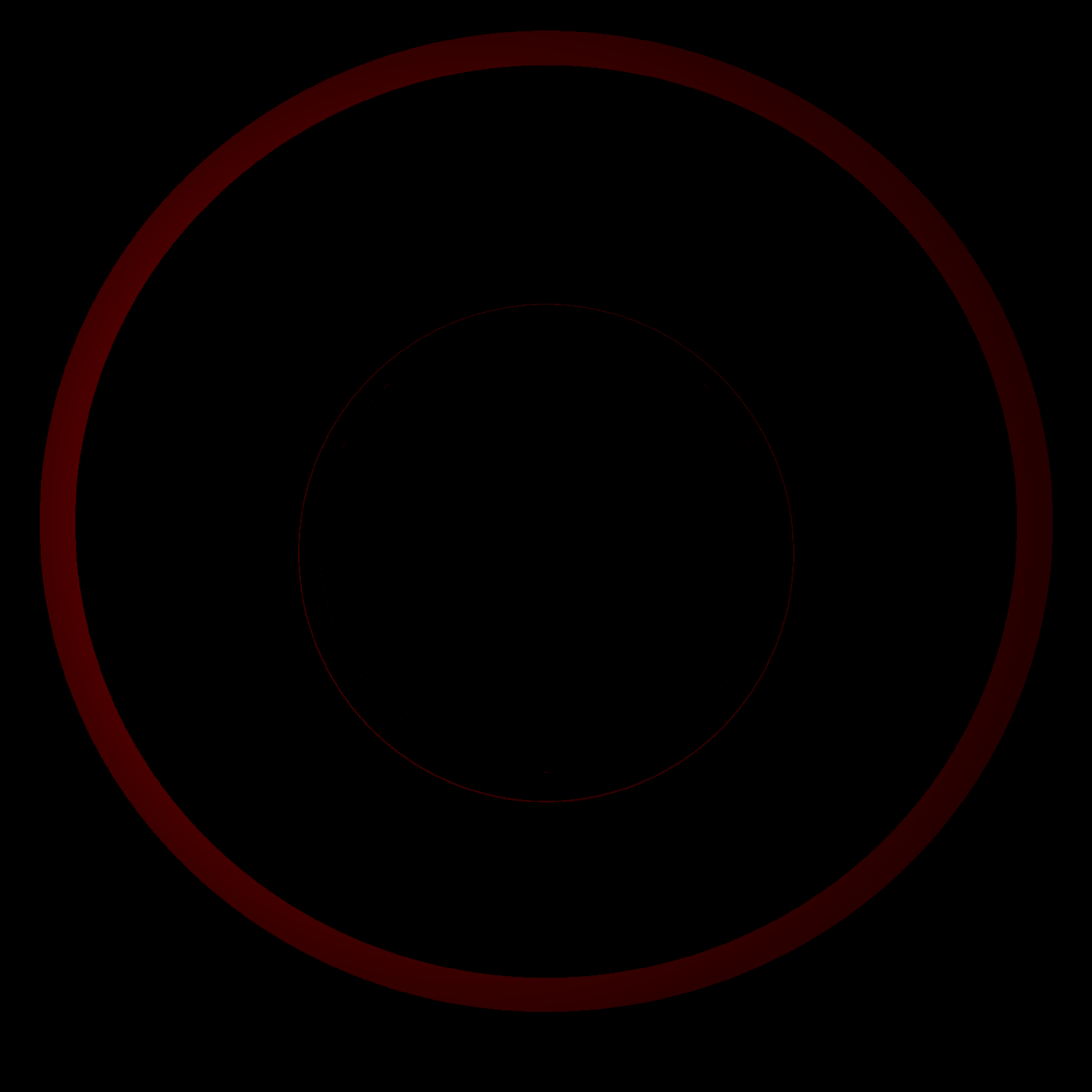}
\includegraphics[width=3.5cm]{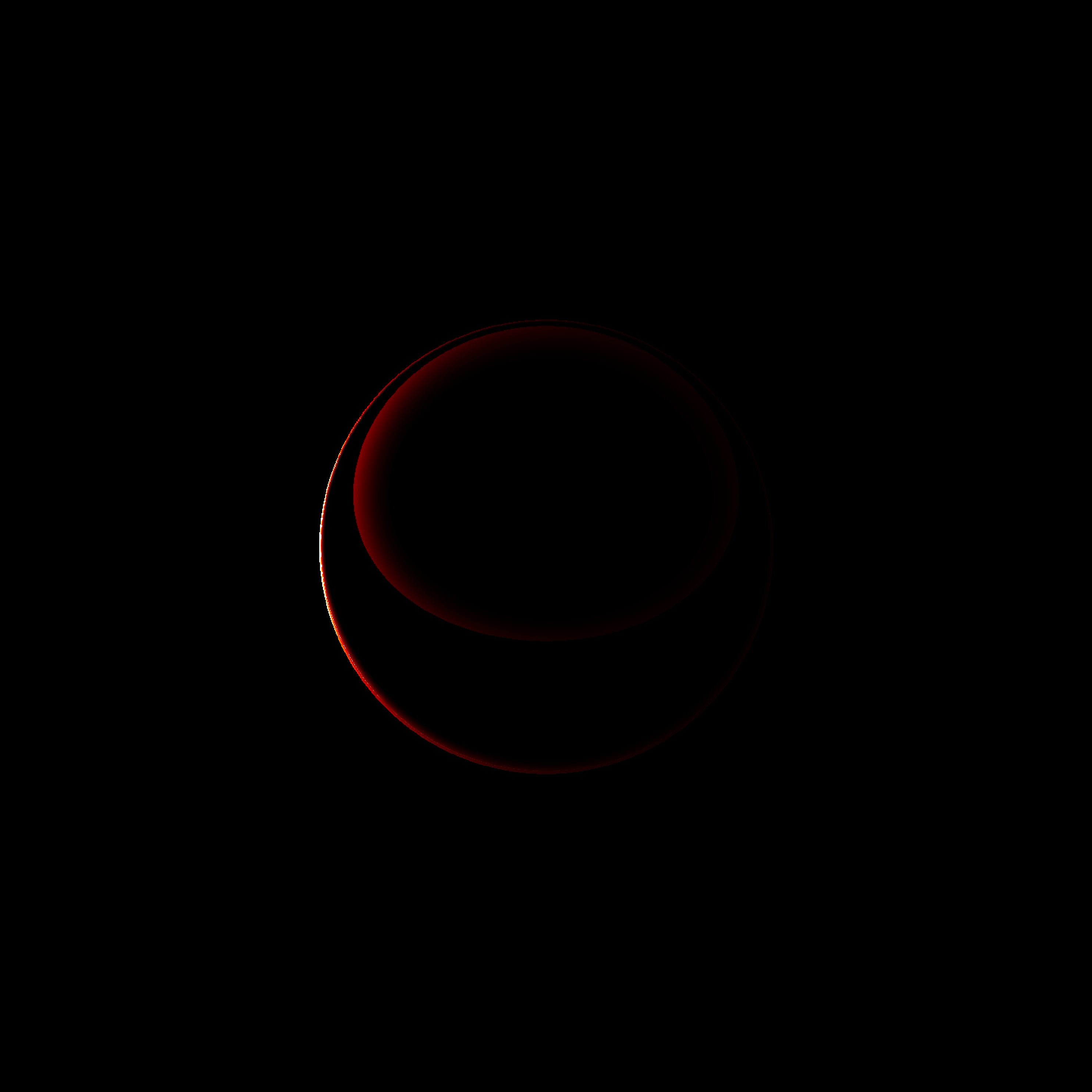}
\includegraphics[width=3.5cm]{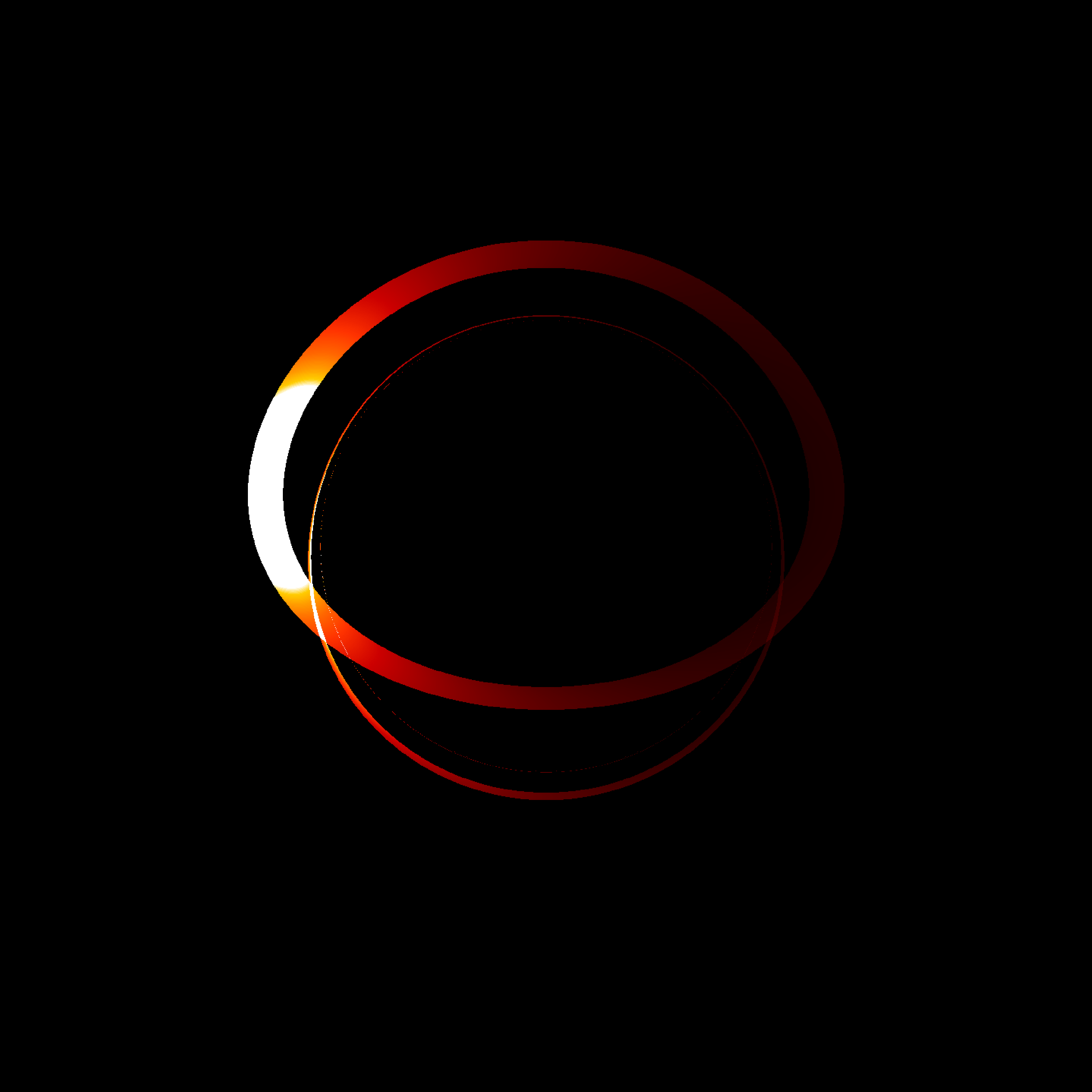}
\includegraphics[width=3.5cm]{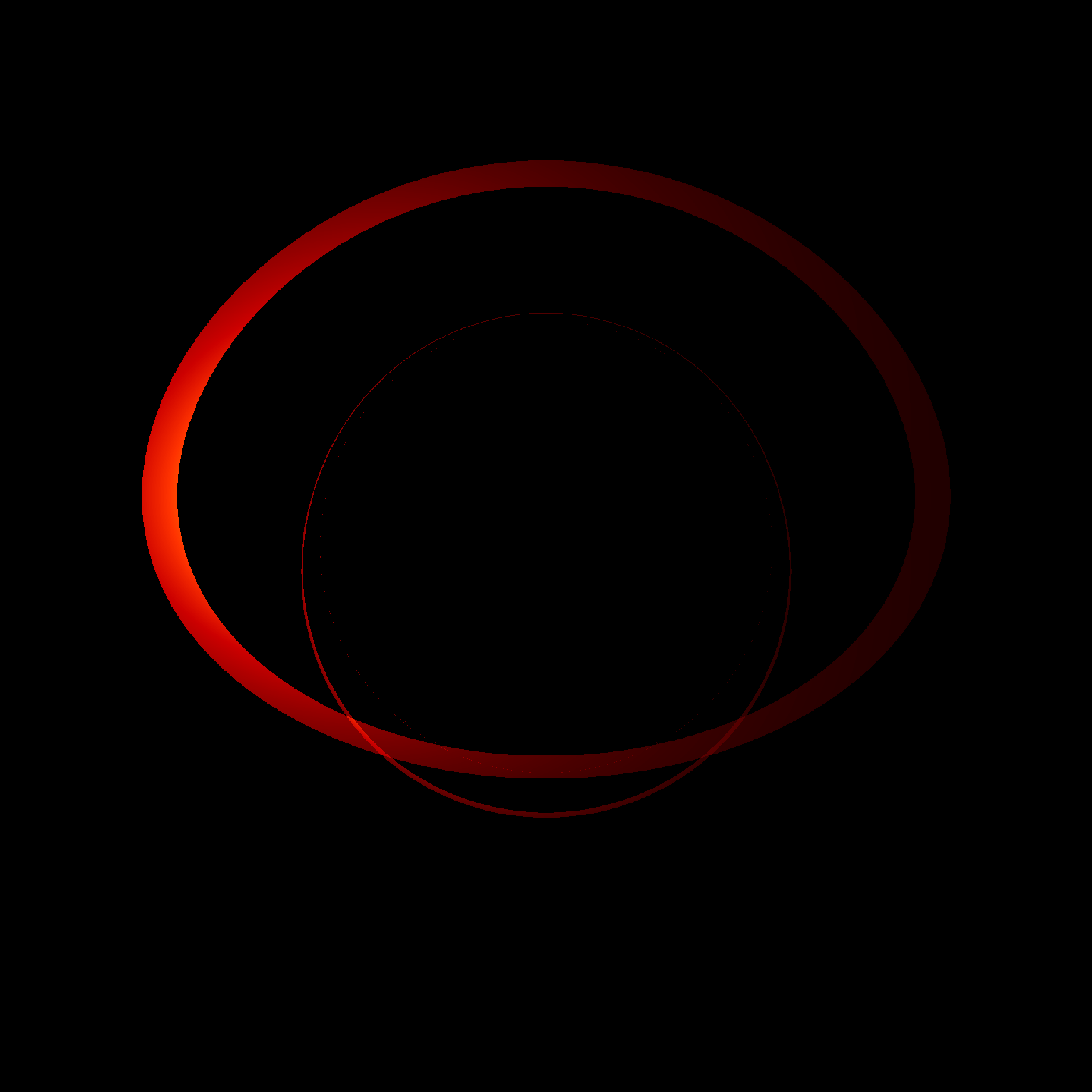}
\includegraphics[width=3.5cm]{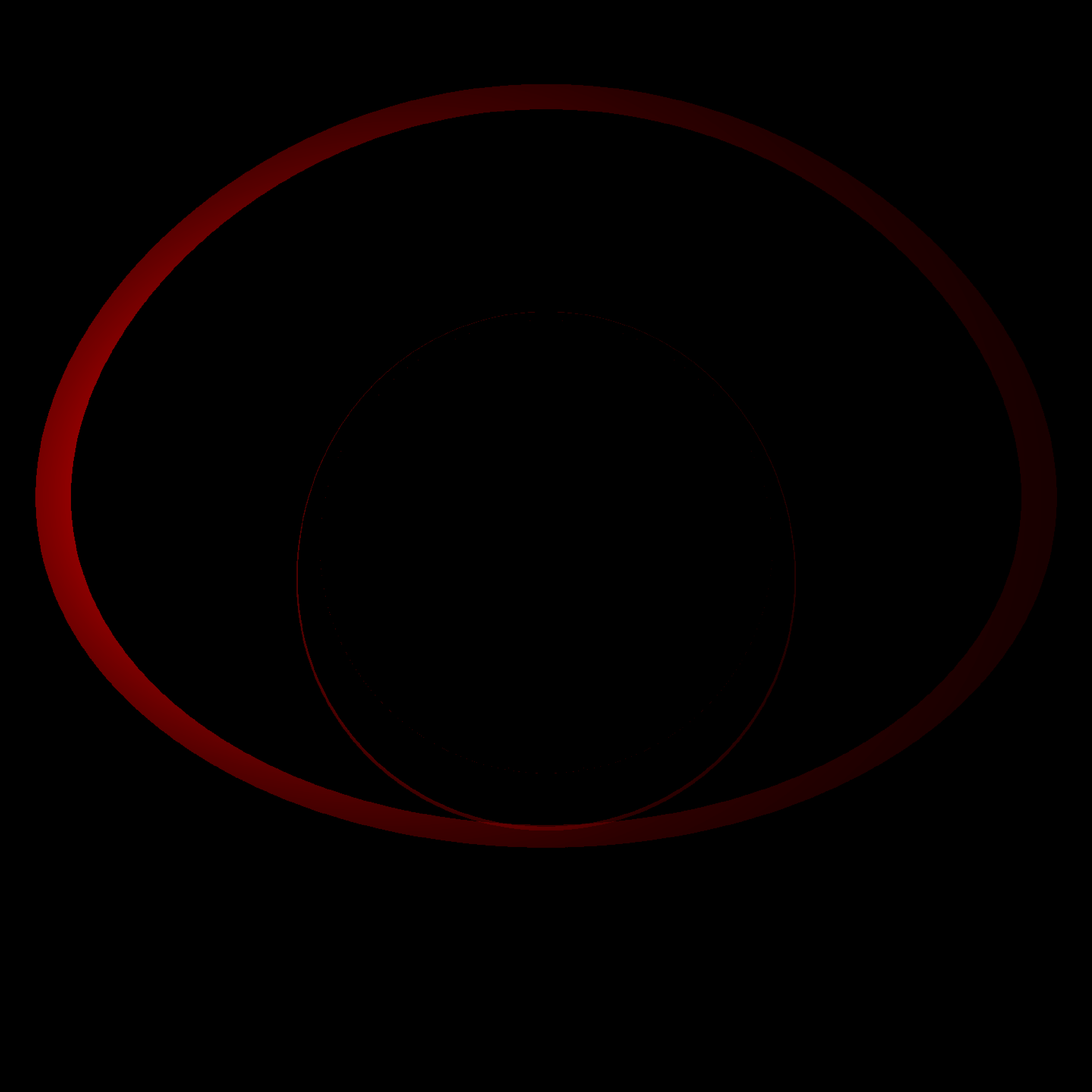}
\includegraphics[width=3.5cm]{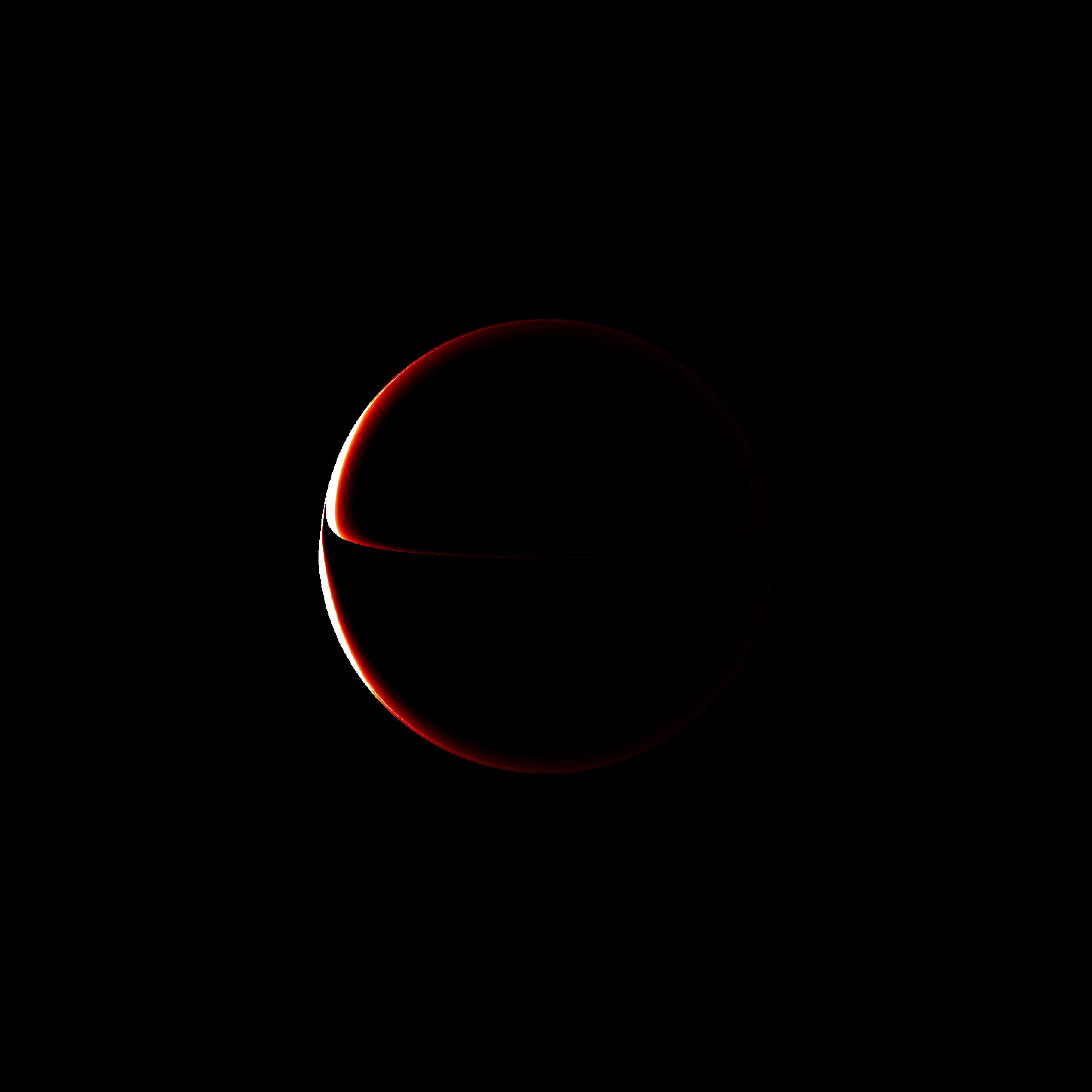}
\includegraphics[width=3.5cm]{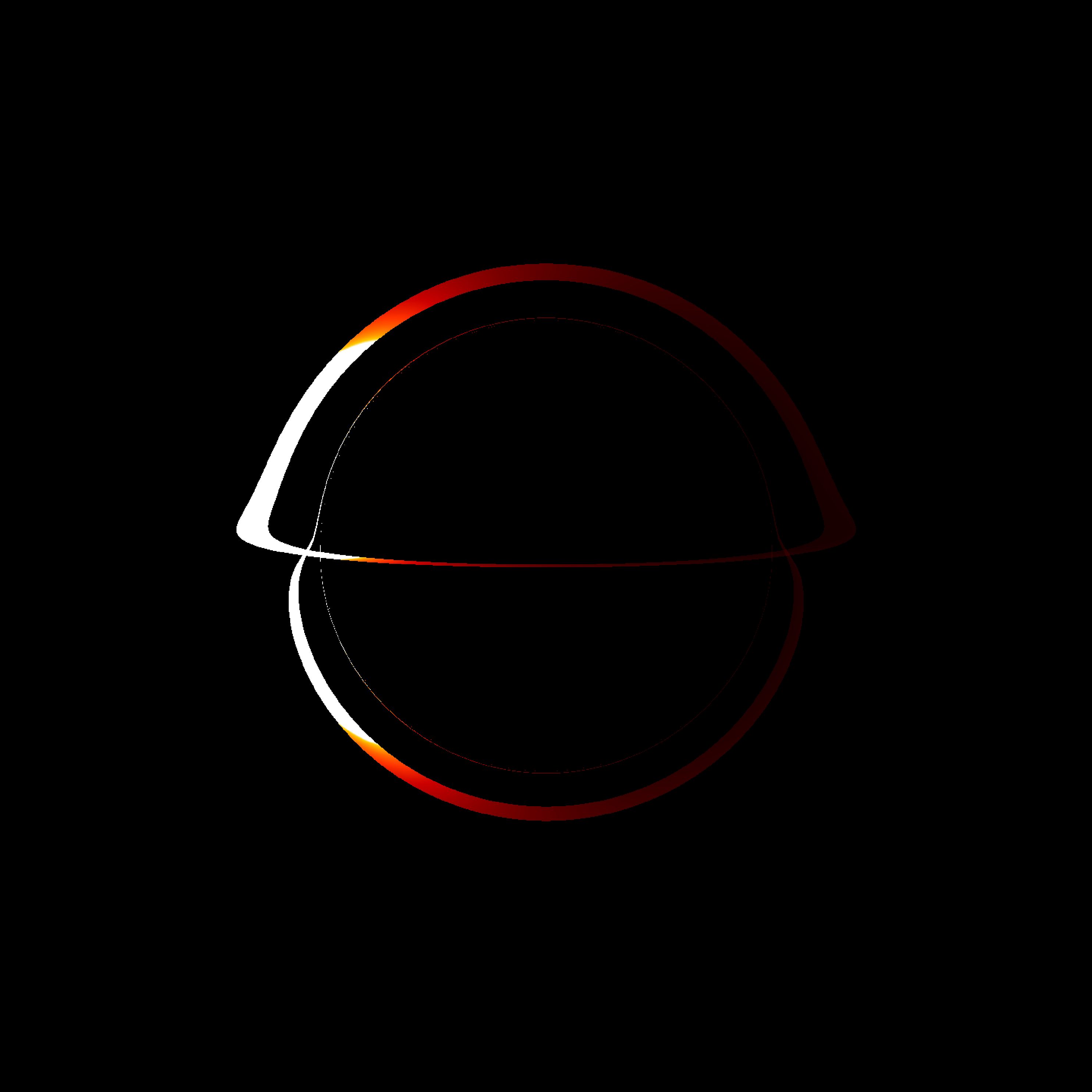}
\includegraphics[width=3.5cm]{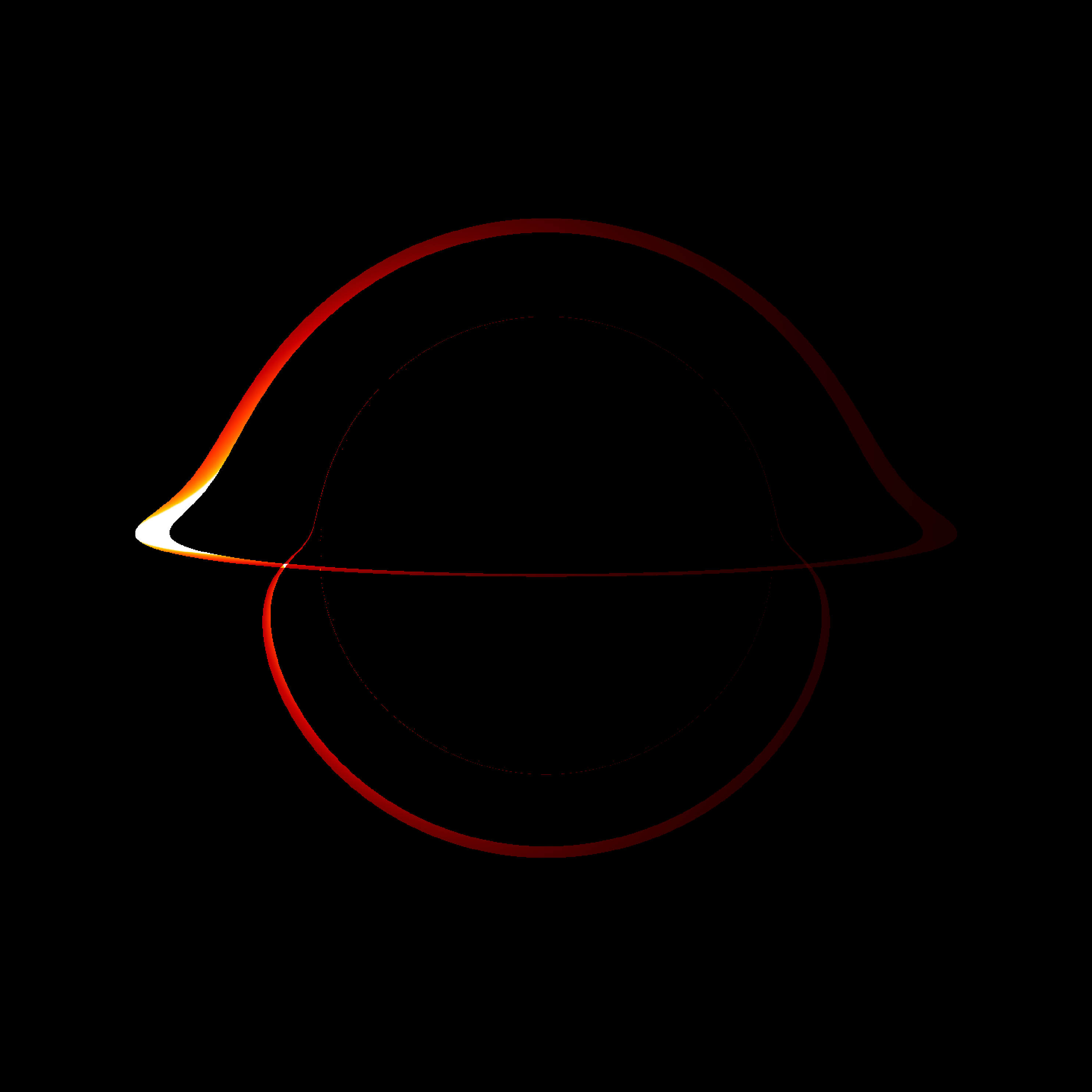}
\includegraphics[width=3.5cm]{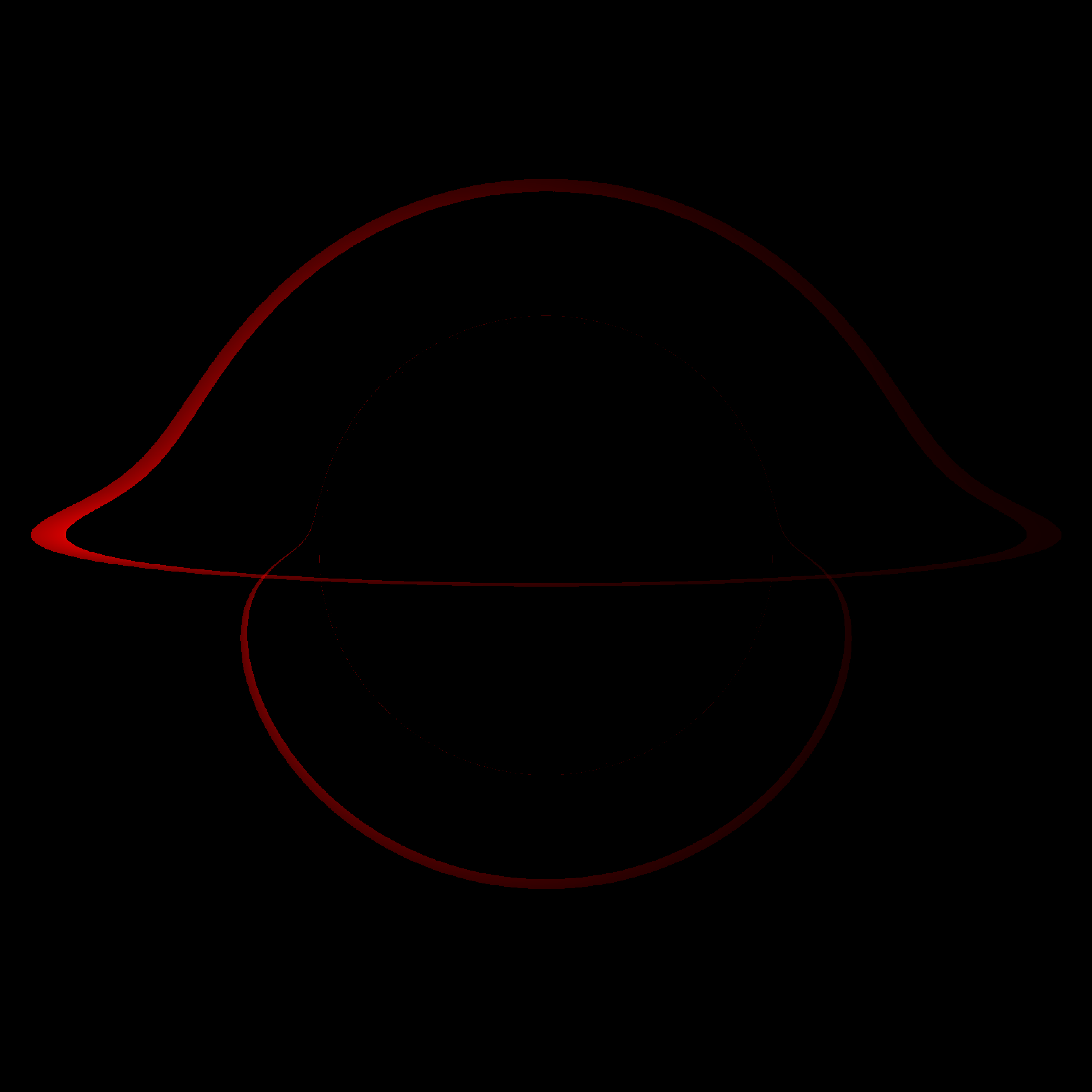}
\caption{Images of tori at different locations under varying observation inclinations. From top to bottom: observation inclinations of $17^{\circ}$, $50^{\circ}$, and $85^{\circ}$; from left to right: torus radial ranges of $3$--$4$, $6$--$7$, $9$--$10$, $12$--$13$ M. Here, we fix the dark matter halo parameters at $r_{\textrm{s}}=\rho_{\textrm{s}}=0.5$, the specific intensity range at $I_{\textrm{obs}} \in [0,0.5]$, and the resolution at $1500 \times 1500$ pixels.Note that the morphology of the torus image depends not only on the observation inclination but also on the torus position relative to the critical photon orbit.}}\label{fig14}
\end{figure*}

With $\rho_{\textrm{s}}=0.5$ and the torus positioned between 6--7 M, we examine the combined effects of $r_{\textrm{s}}$ and the observation inclination on the torus image, as summarized in figure 15. We find that increasing $r_{\textrm{s}}$ enlarges the torus image, with this effect becoming more pronounced at higher inclination angles. Simultaneously, the expansion of the event horizon, critical curve, and photon ring due to a larger dark matter halo scale brings the torus closer to the black hole. This enhances gravitational redshift, resulting in a dimmer image, as shown in the right column of the figure.
\begin{figure*}%[tbph]
\center{
\includegraphics[width=2.8cm]{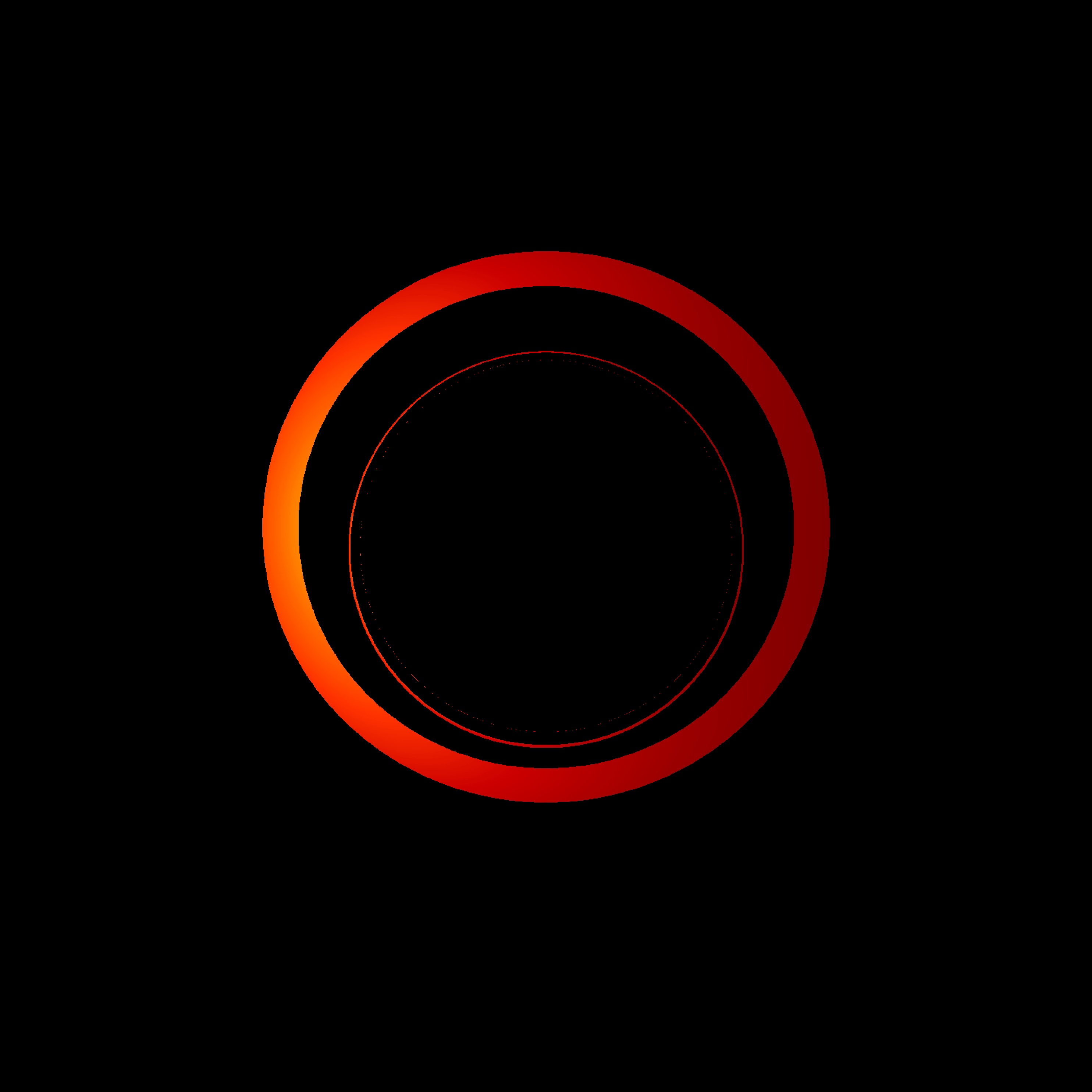}
\includegraphics[width=2.8cm]{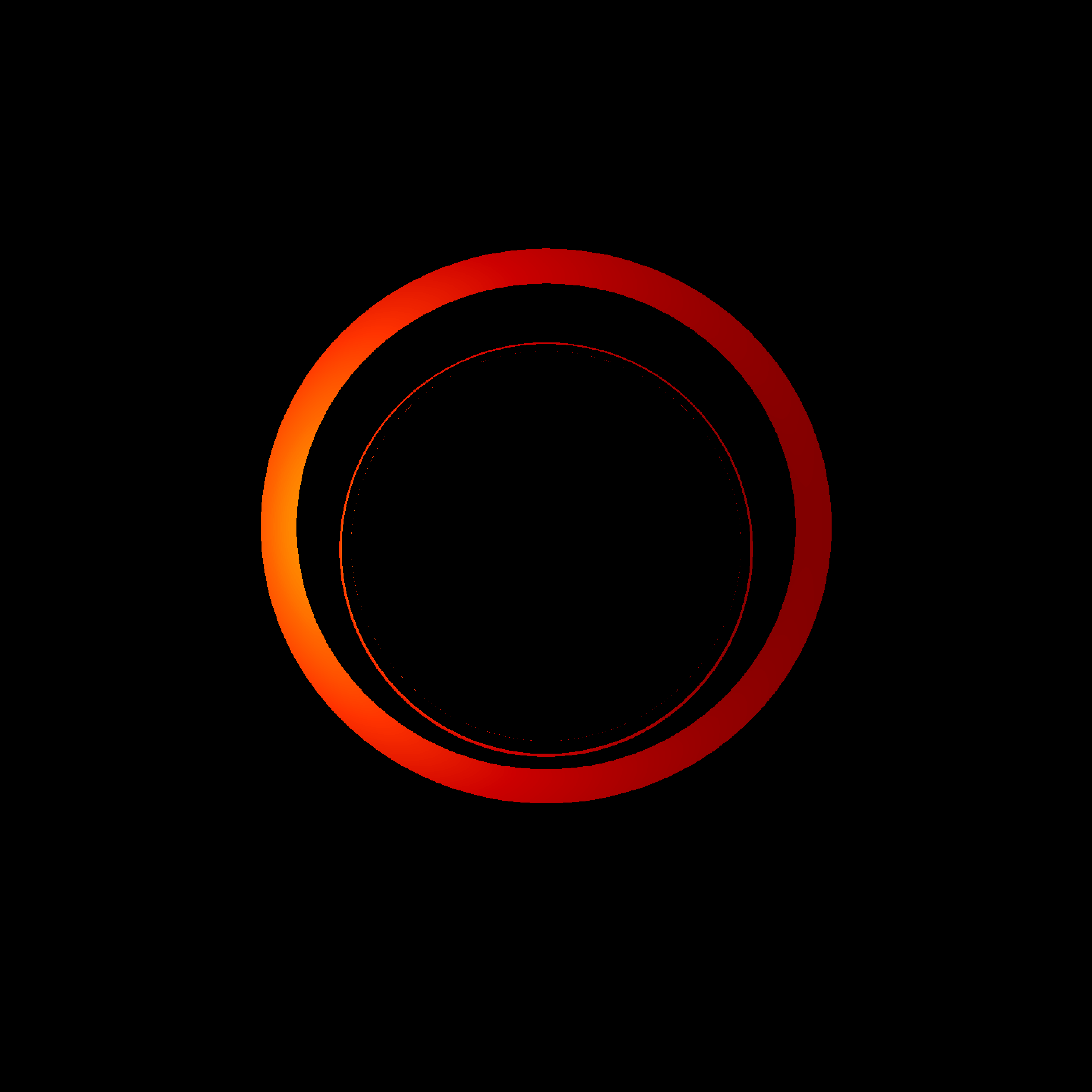}
\includegraphics[width=2.8cm]{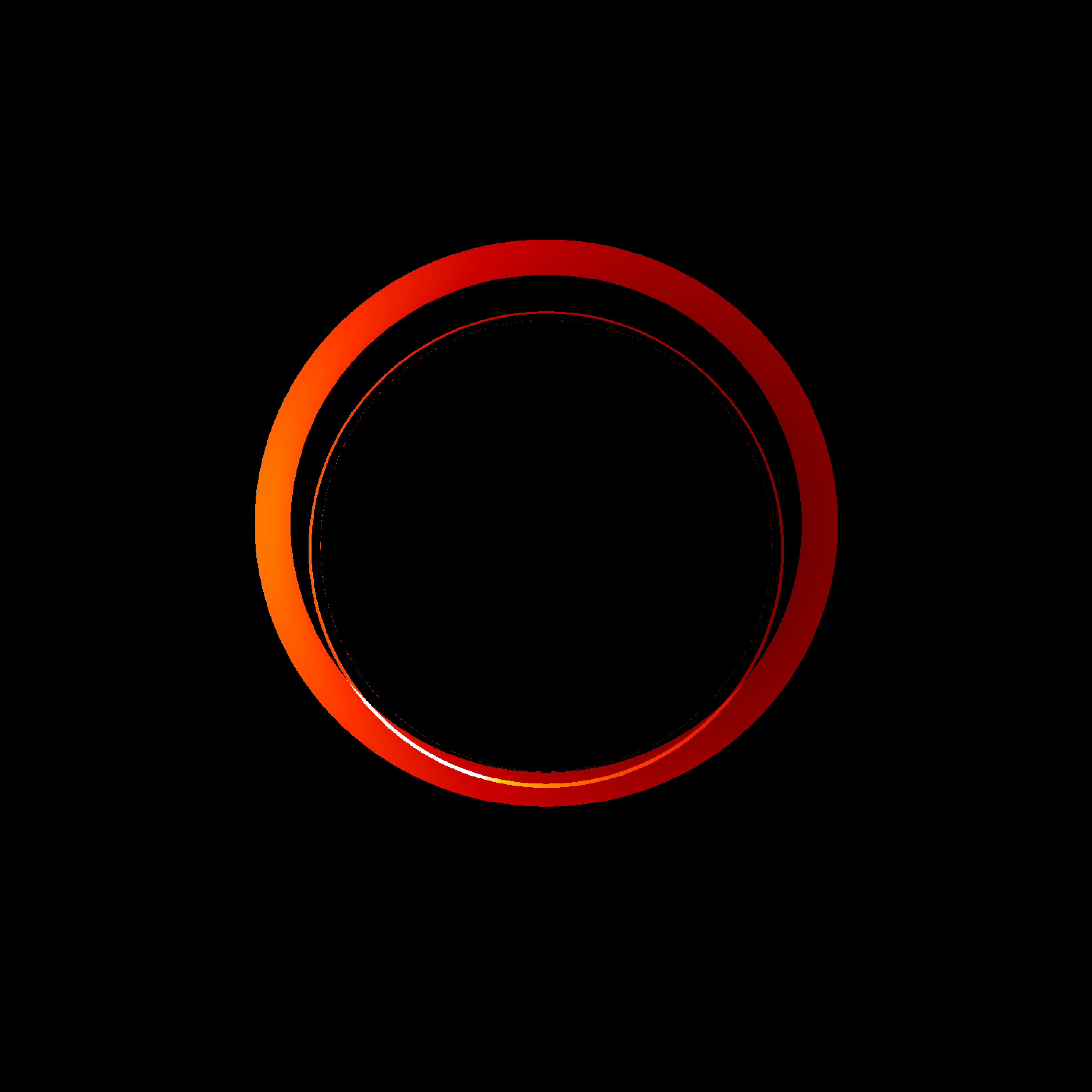}
\includegraphics[width=2.8cm]{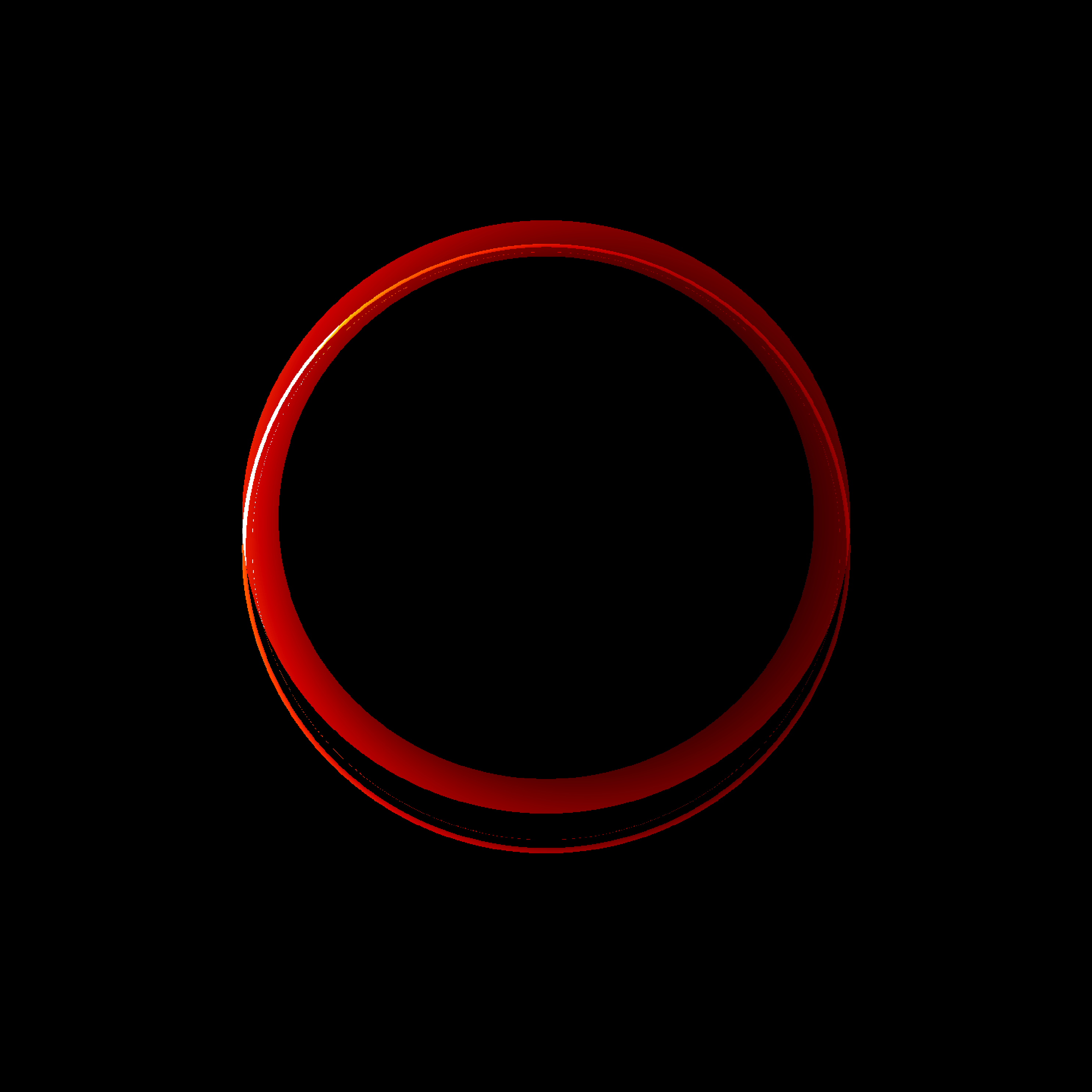}
\includegraphics[width=2.8cm]{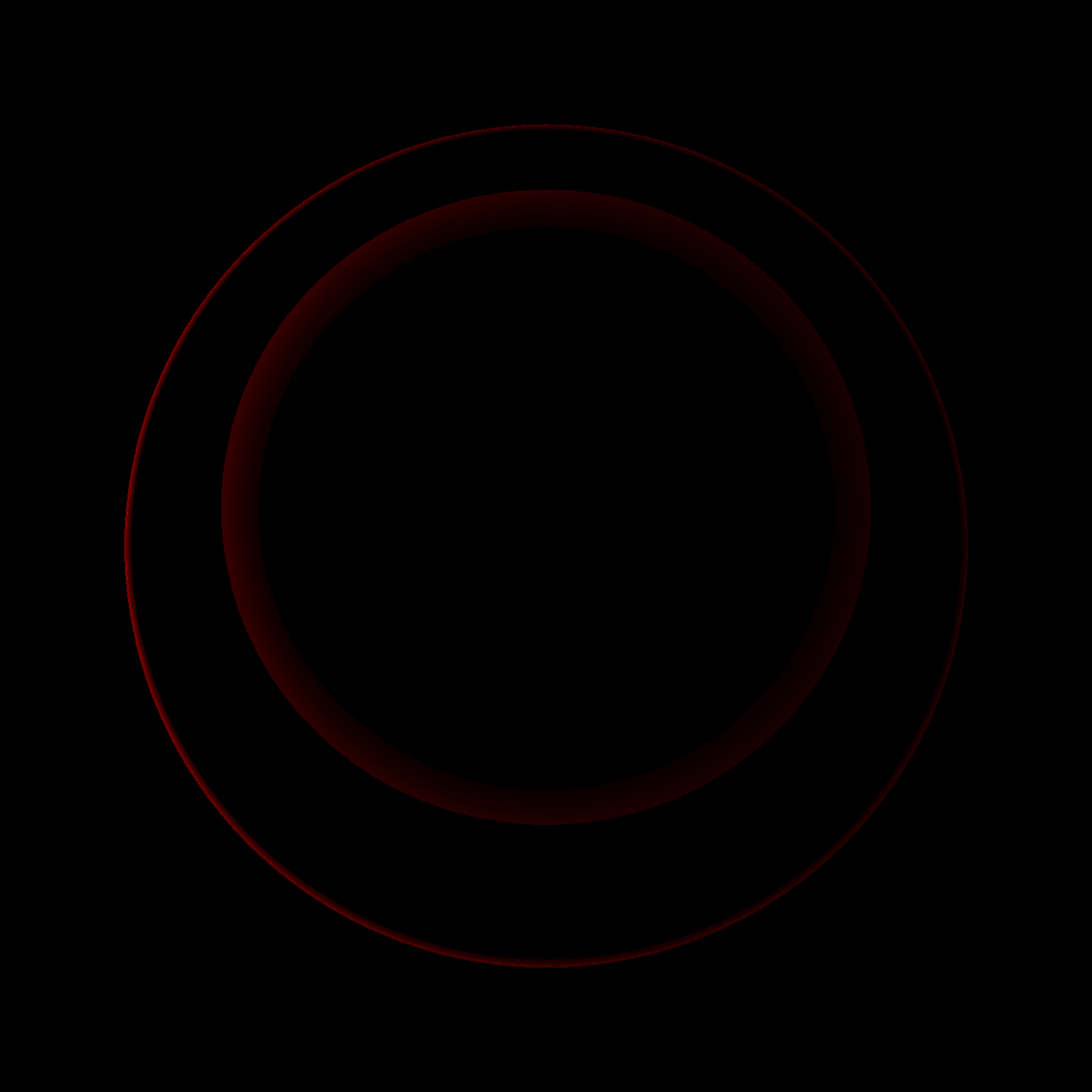}
\includegraphics[width=2.8cm]{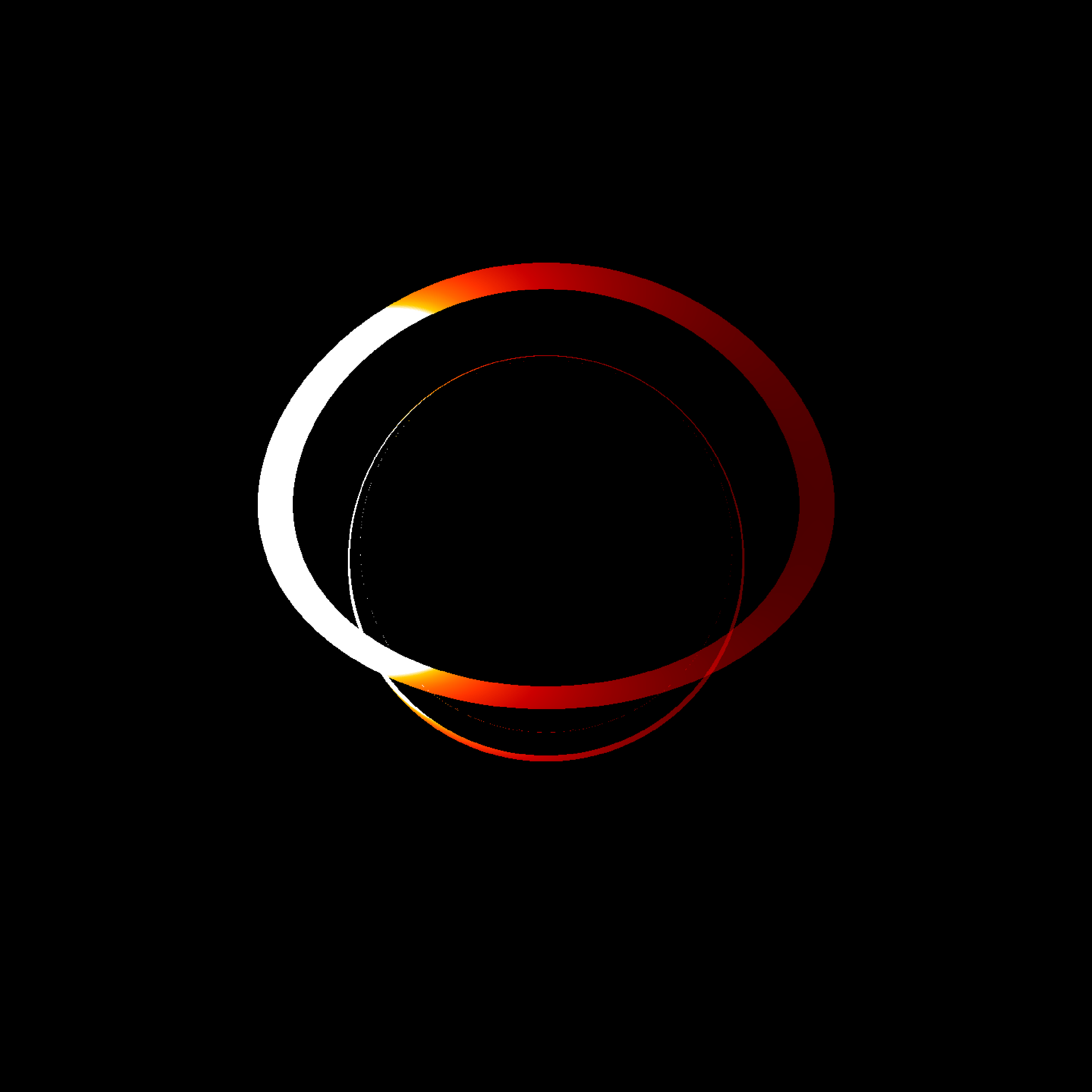}
\includegraphics[width=2.8cm]{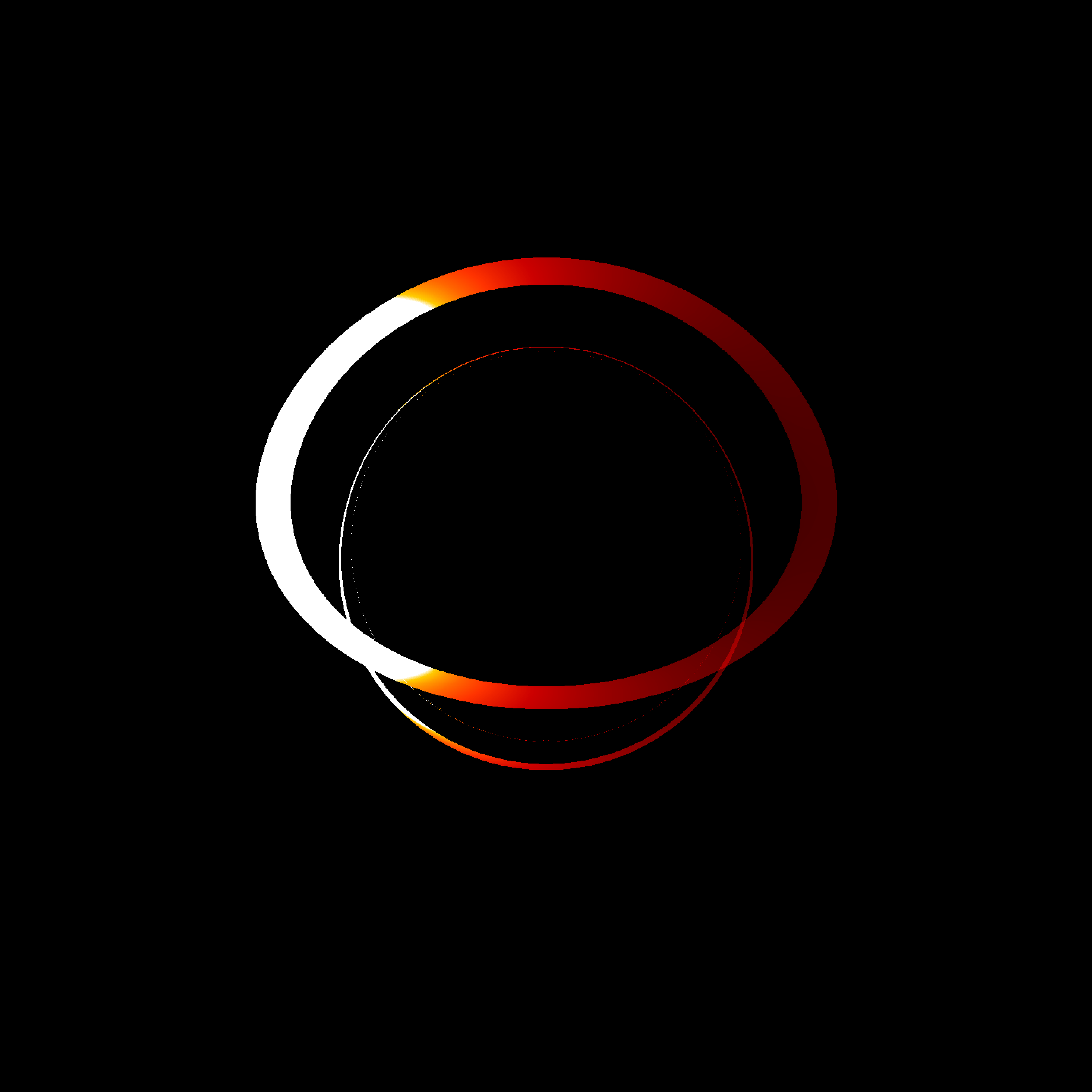}
\includegraphics[width=2.8cm]{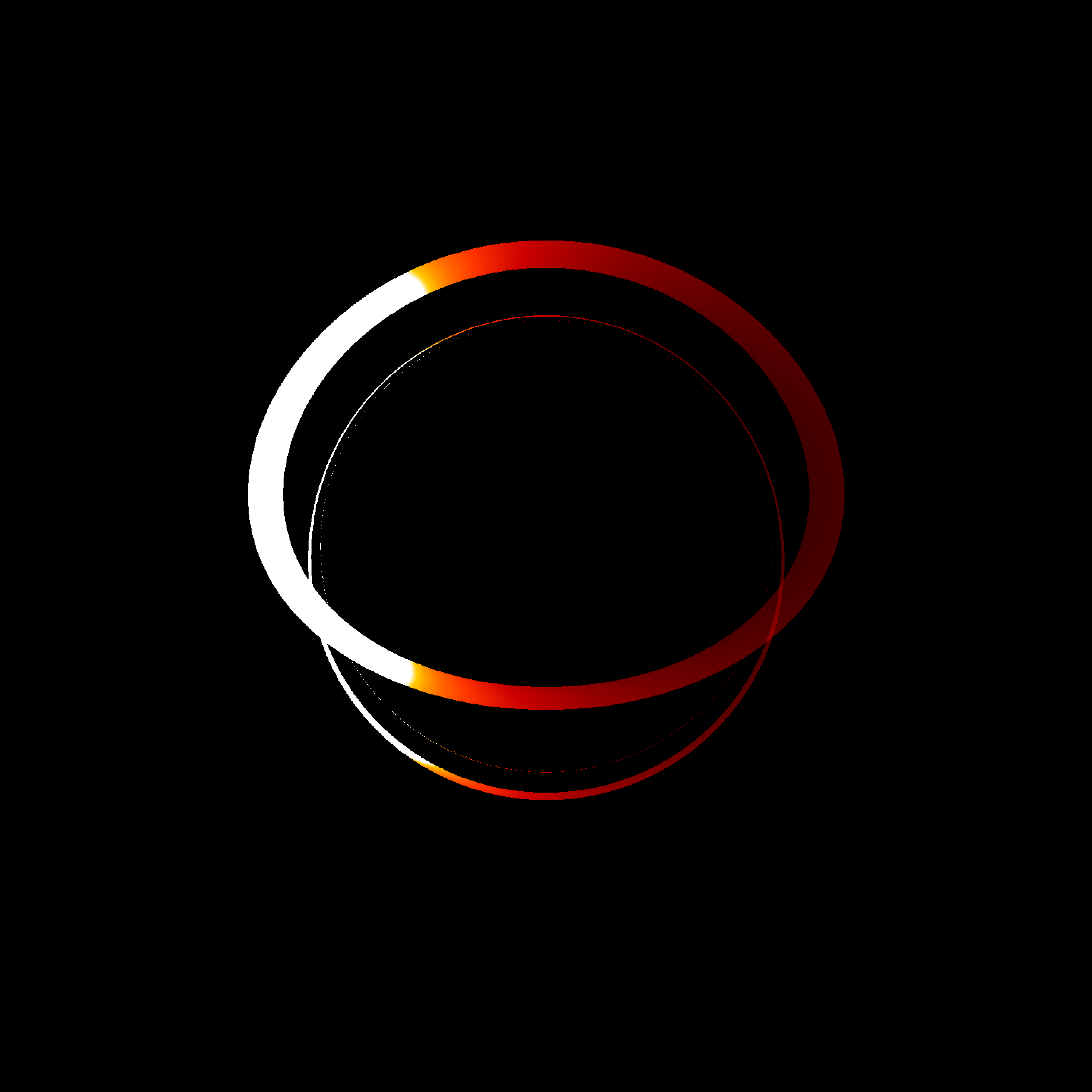}
\includegraphics[width=2.8cm]{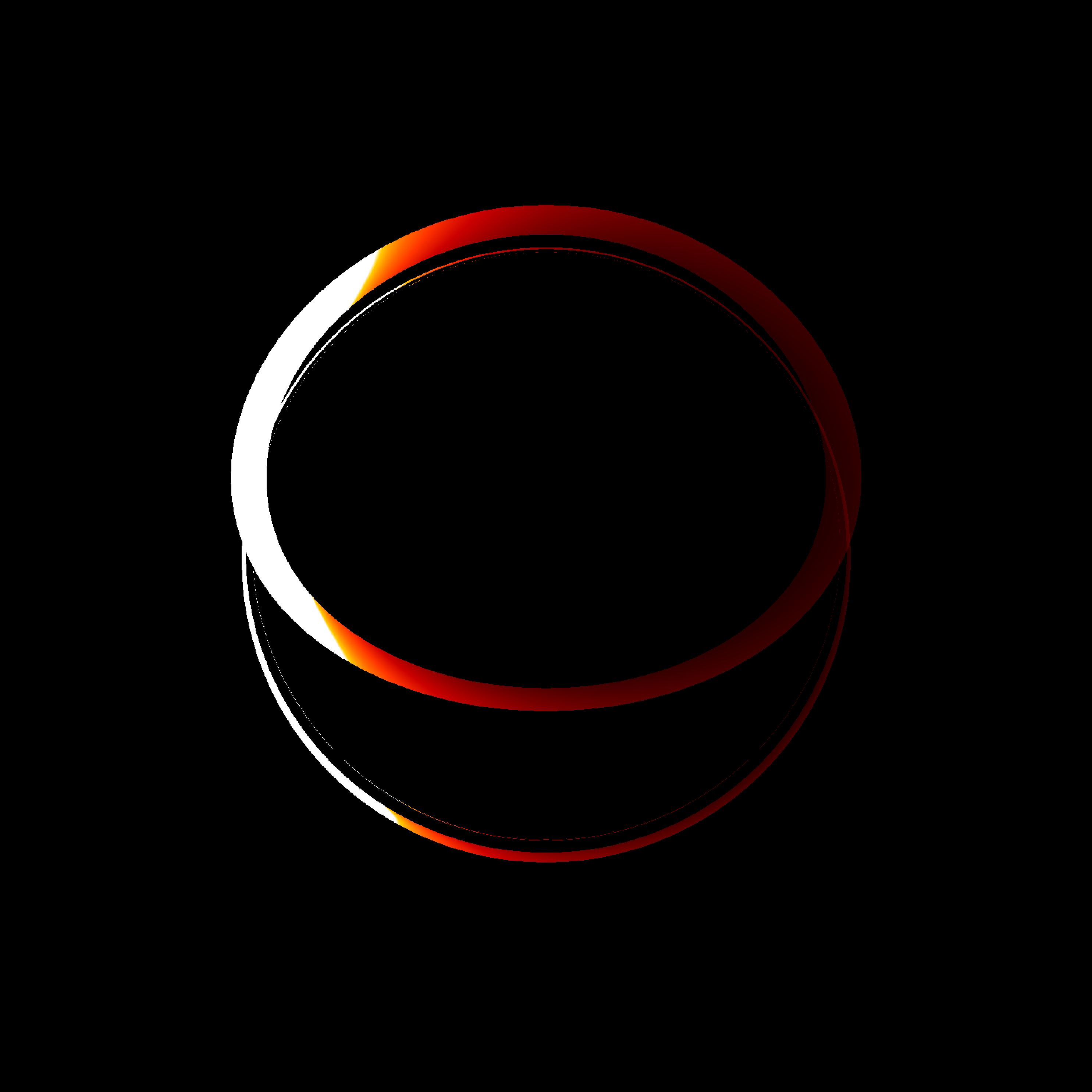}
\includegraphics[width=2.8cm]{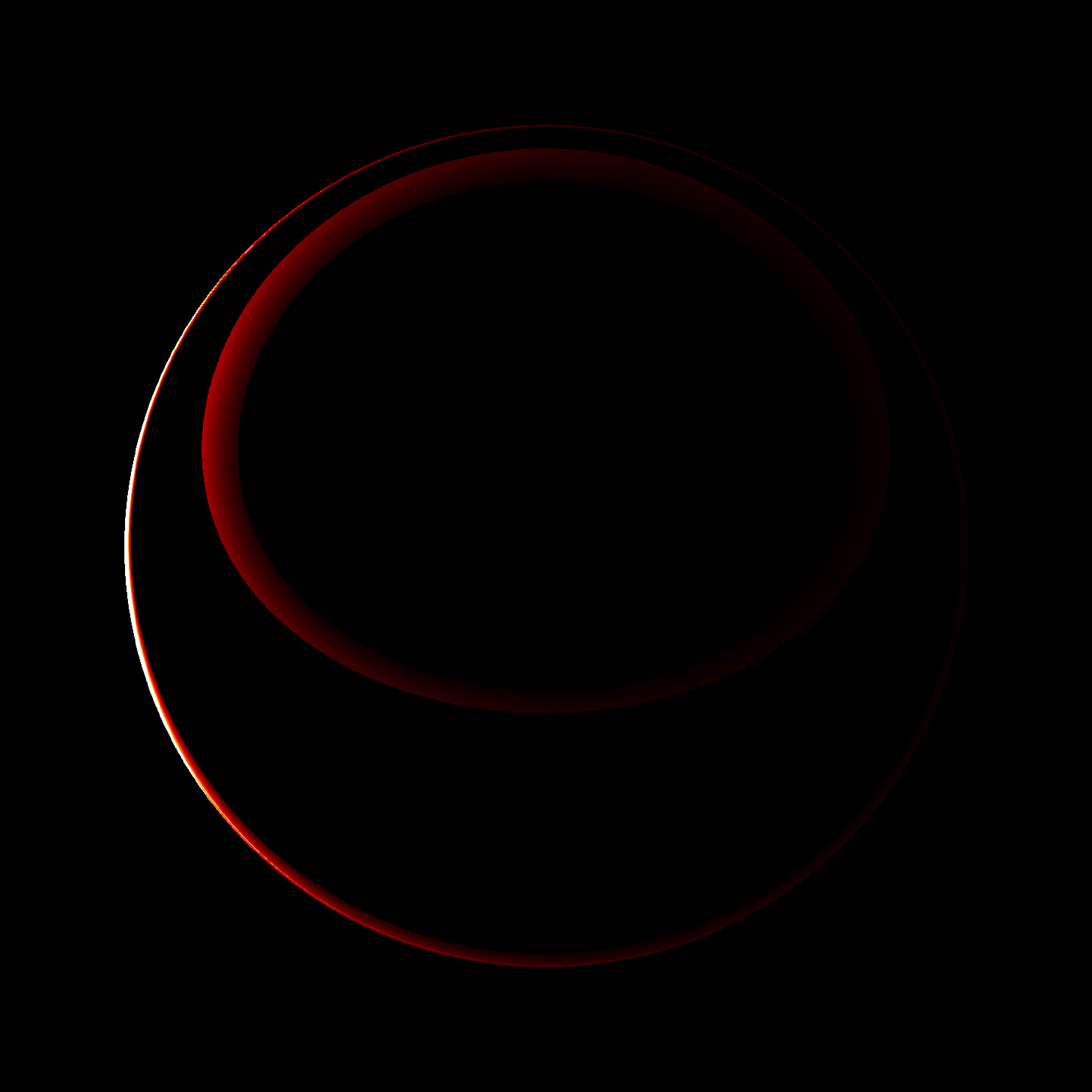}
\includegraphics[width=2.8cm]{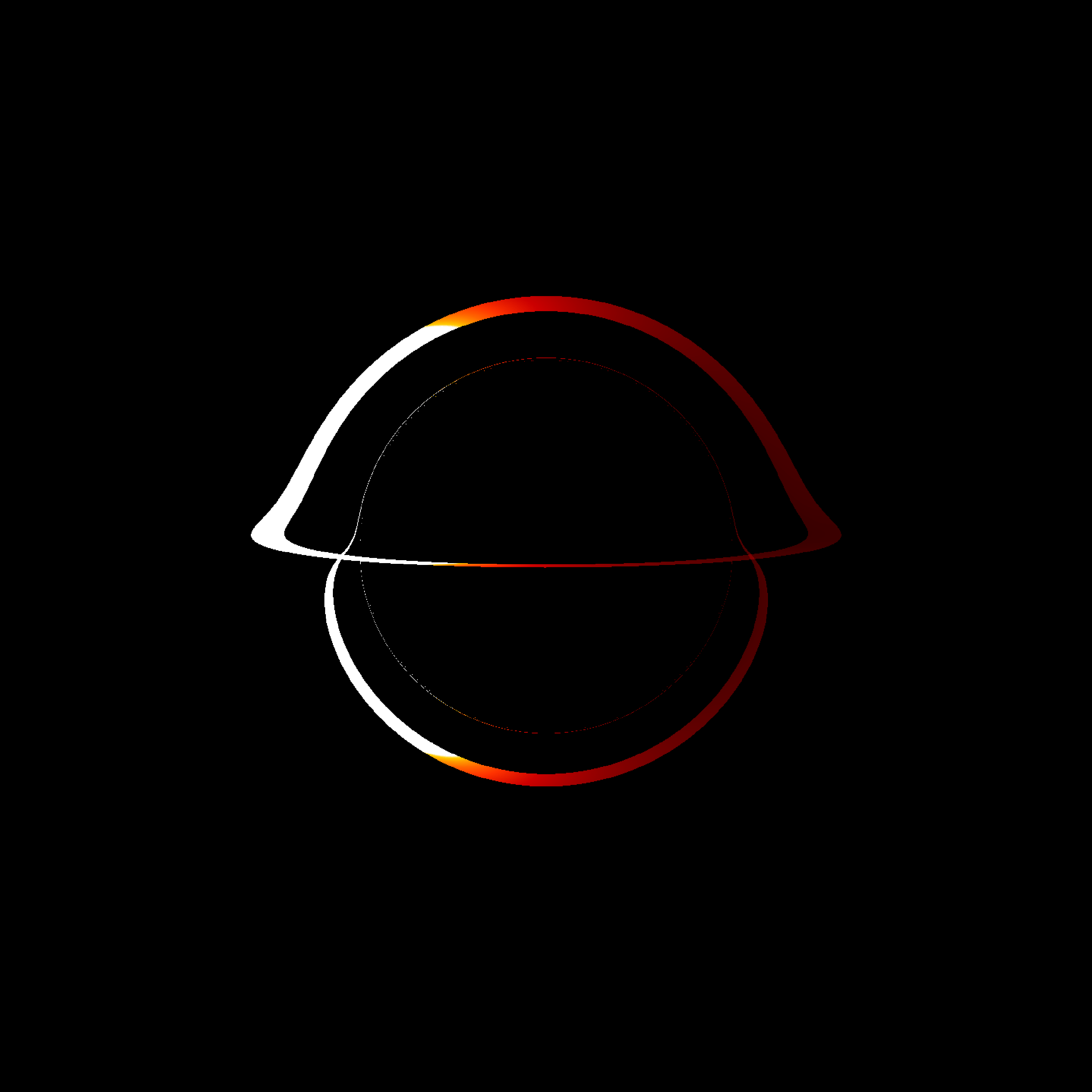}
\includegraphics[width=2.8cm]{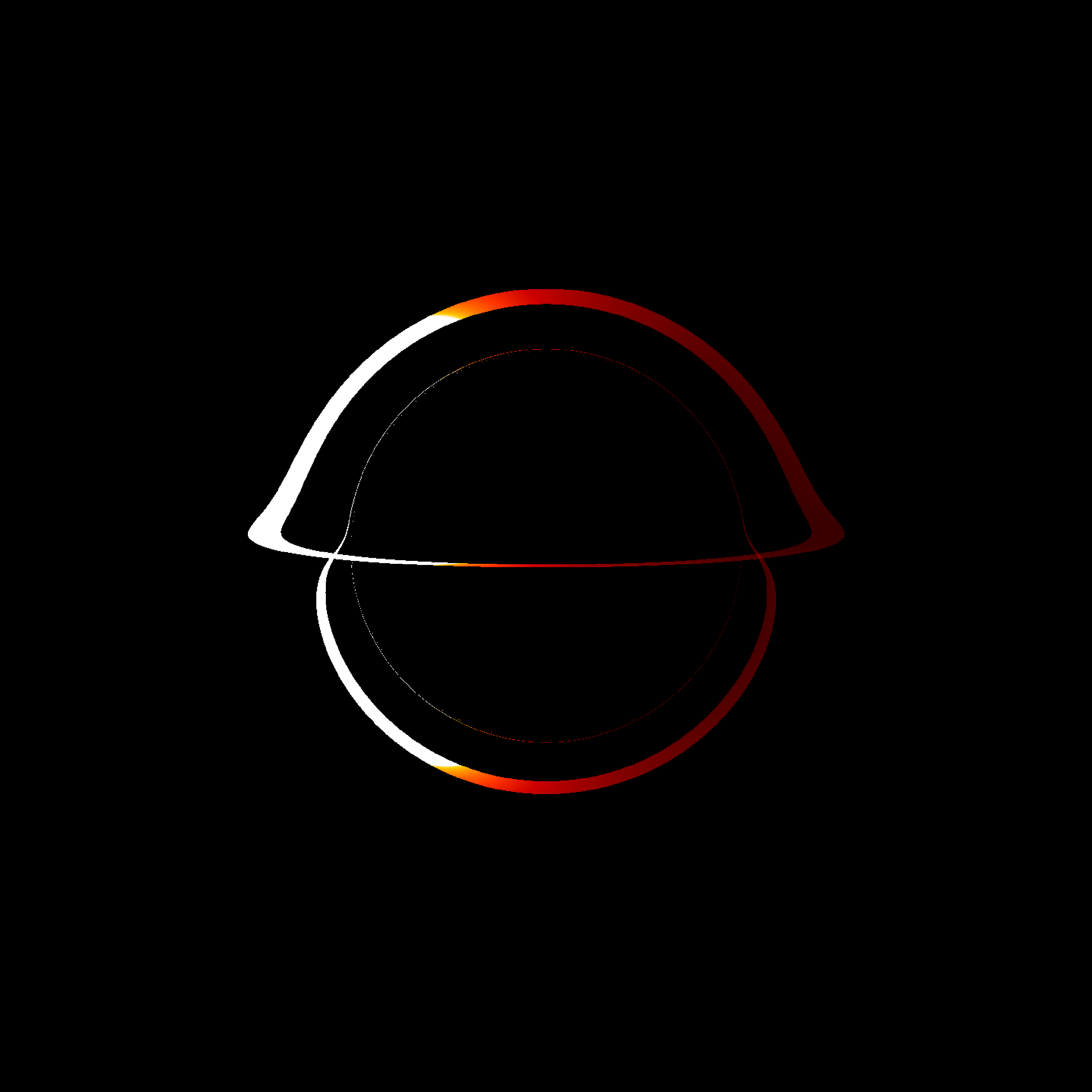}
\includegraphics[width=2.8cm]{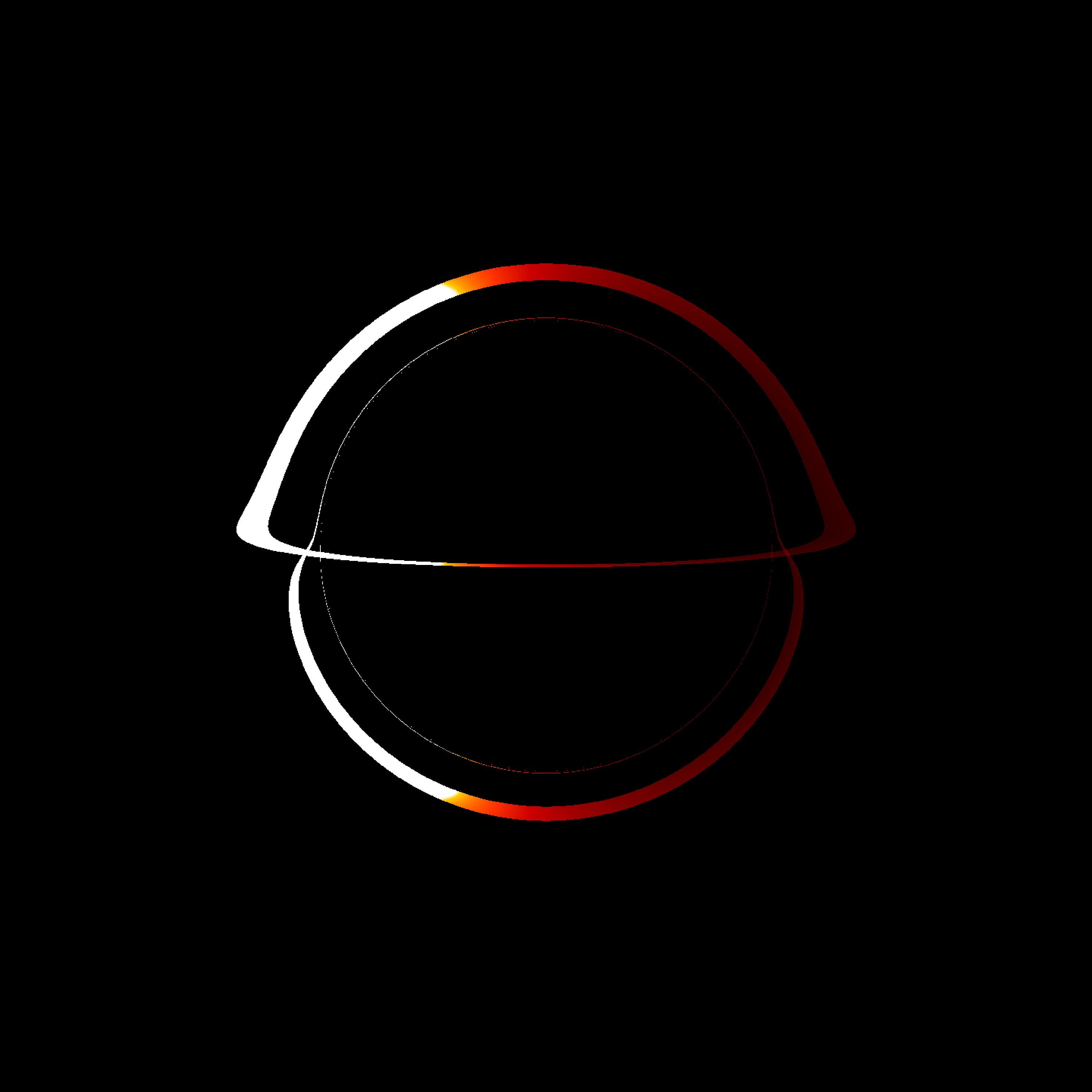}
\includegraphics[width=2.8cm]{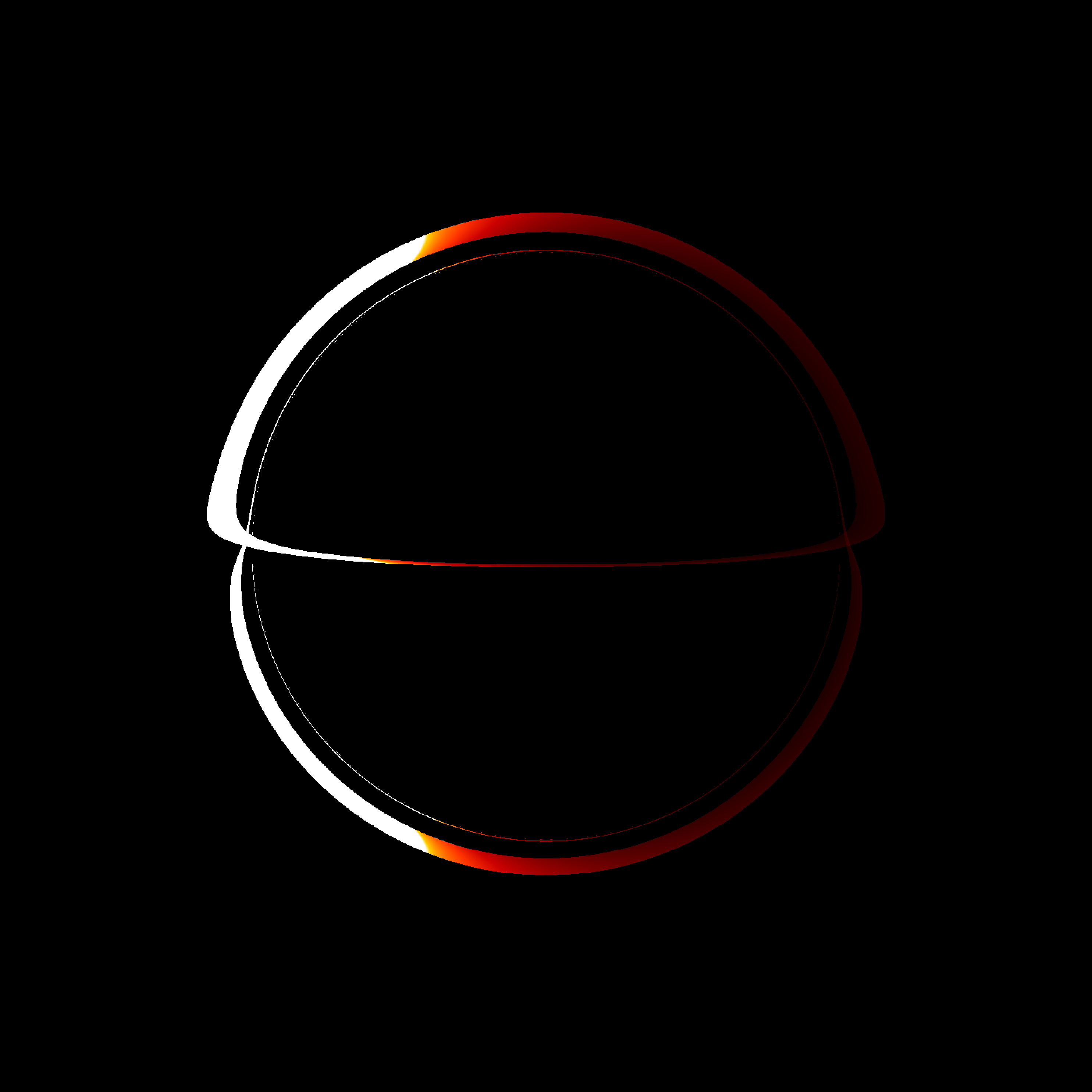}
\includegraphics[width=2.8cm]{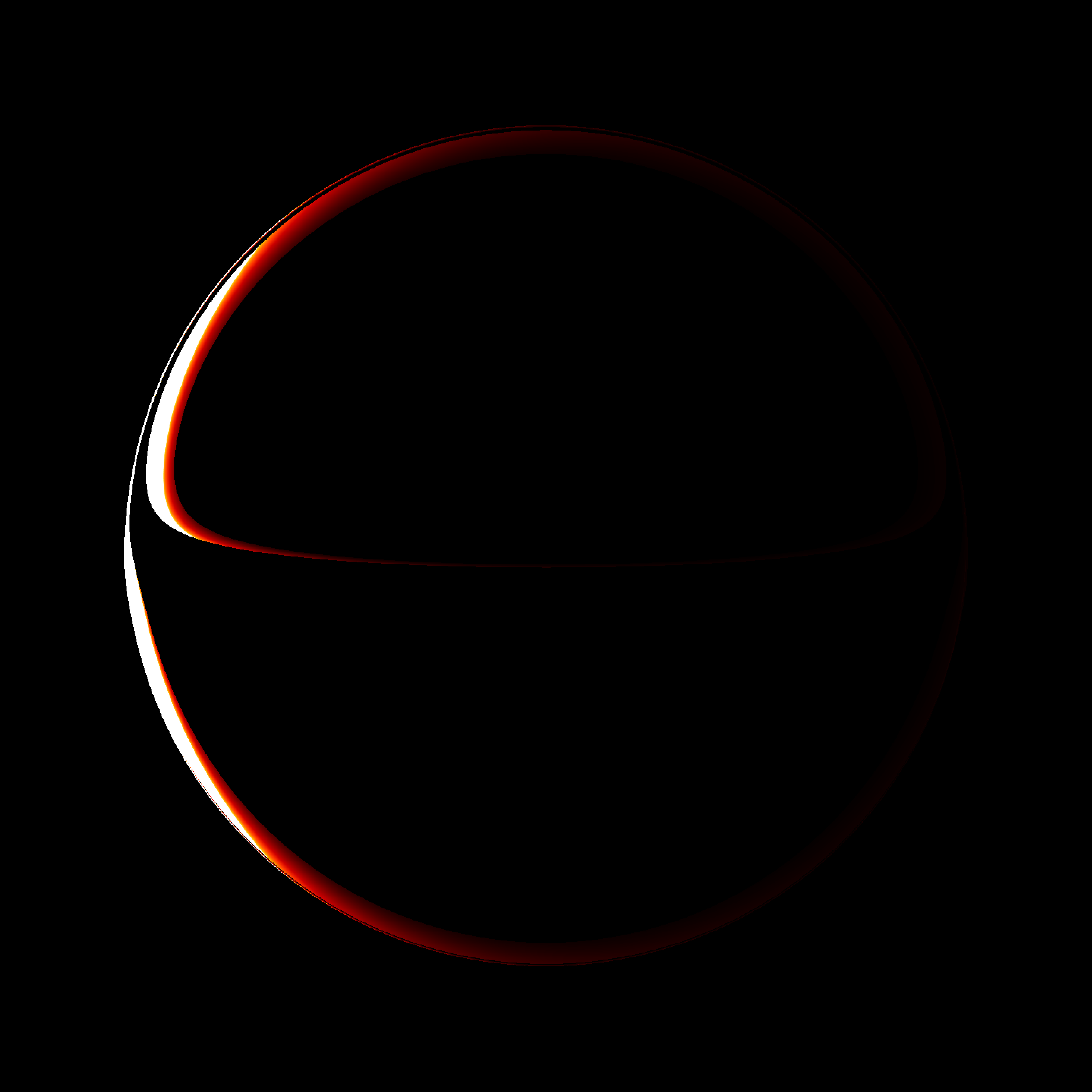}
\caption{Images of the torus across different parameter spaces. From left to right: $r_{\textrm{s}}=$ $0.1$, $0.3$, $0.5$, $0.7$, $0.9$; from top to bottom: observation inclination $\omega=$ $17^{\circ}$, $50^{\circ}$, $85^{\circ}$. Here, we fix $\rho_{\textrm{s}}=0.5$, the torus radial range at $6$--$7$ M, and the specific intensity range at $I_{\textrm{obs}} \in [0,0.25]$. It is evident that although the torus remains at a fixed location, its apparent image expands with increasing $r_{\textrm{s}}$, an effect attributable to gravitational lensing.}}\label{fig15}
\end{figure*}
\subsection{Gravitational lensing}
We consider a stationary spherical point source with radius $r_{\textrm{source}}=0.5$ M, located at $(x^{\prime},y^{\prime},z^{\prime})=(-8,0,0)$. Figure 16 shows its projection onto the observer's screen for different inclination angles. When the line of sight is nearly aligned with the black hole's polar axis, the point source appears as a quasi-circular spot (orange) in the upper part of the field of view---this is the direct image, formed by rays that reach the source directly. A striped orange image appears near the lower shadow boundary, corresponding to the secondary image, formed by gravitationally lensed rays. As the inclination $\omega$ increases, both the direct and secondary images evolve into crescent shapes and eventually merge into an Einstein ring when $\omega = 90^{\circ}$, representing a strong manifestation of gravitational lensing. It is noteworthy that the size of the Einstein ring depends on black hole parameters, as shown in figure 17, where the ring radius increases with $r_{\textrm{s}}$. Thus, in principle, Einstein rings offer a potential means to infer the scale and density of the dark matter halo.
\begin{figure*}%[tbph]
\center{
\includegraphics[width=4cm]{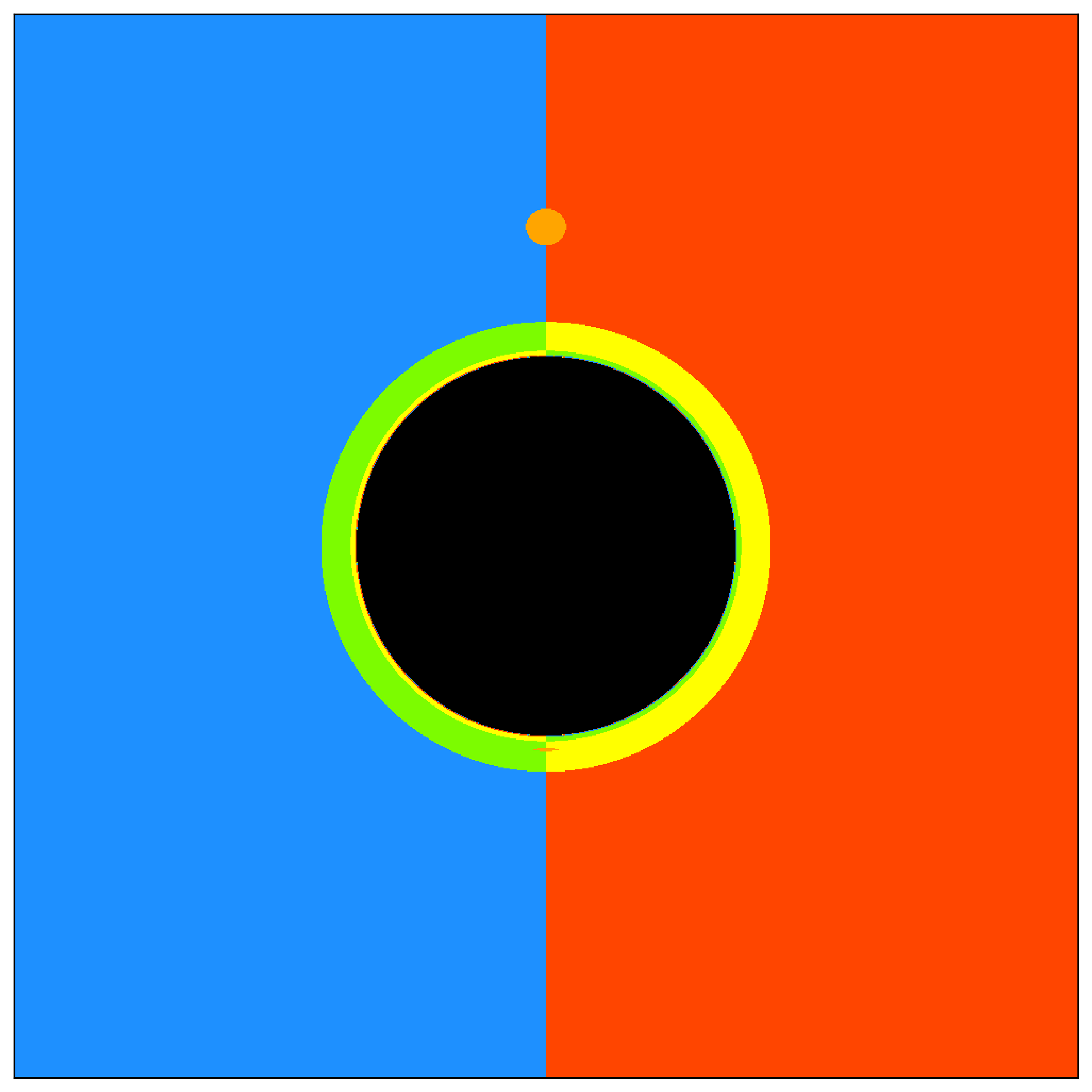}
\includegraphics[width=4cm]{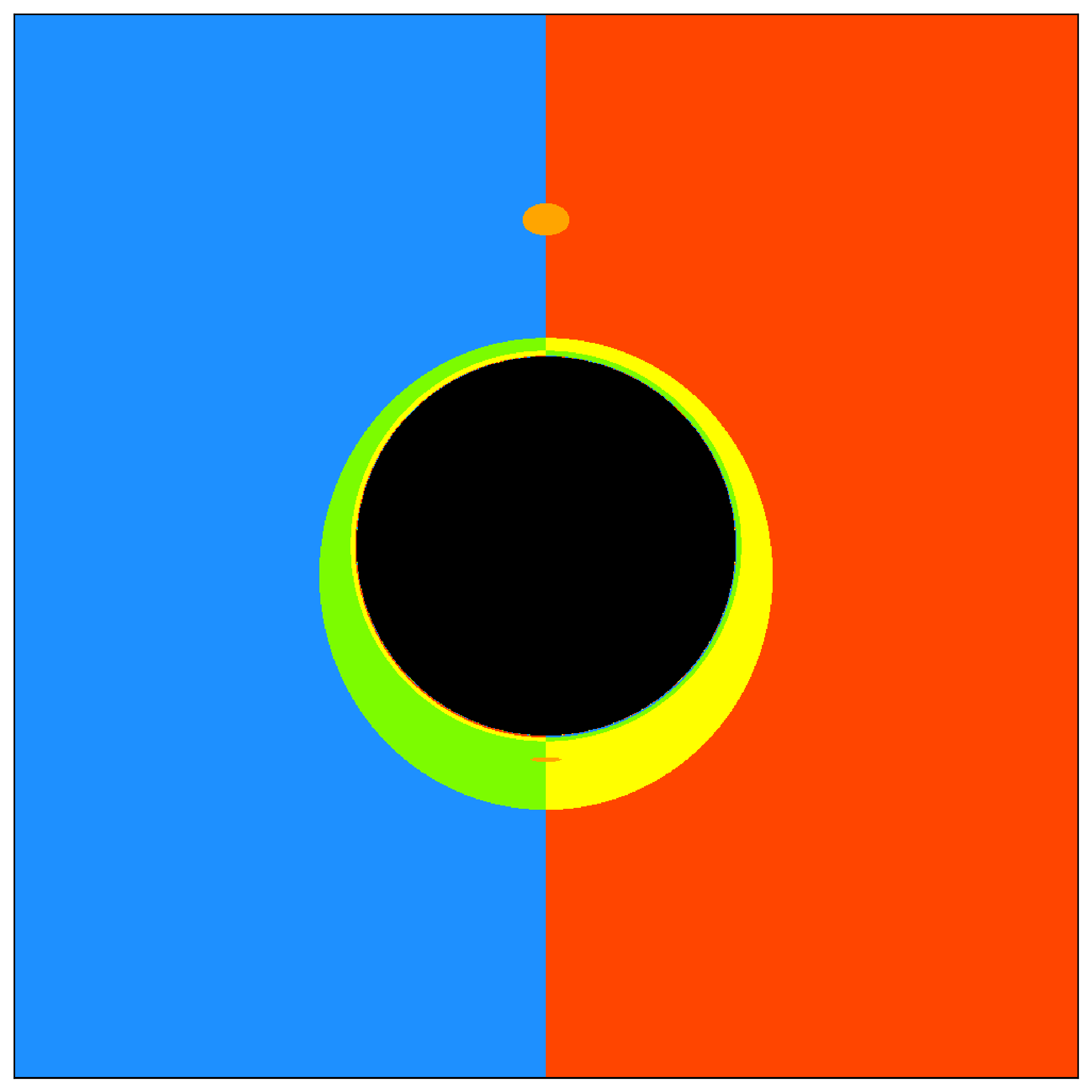}
\includegraphics[width=4cm]{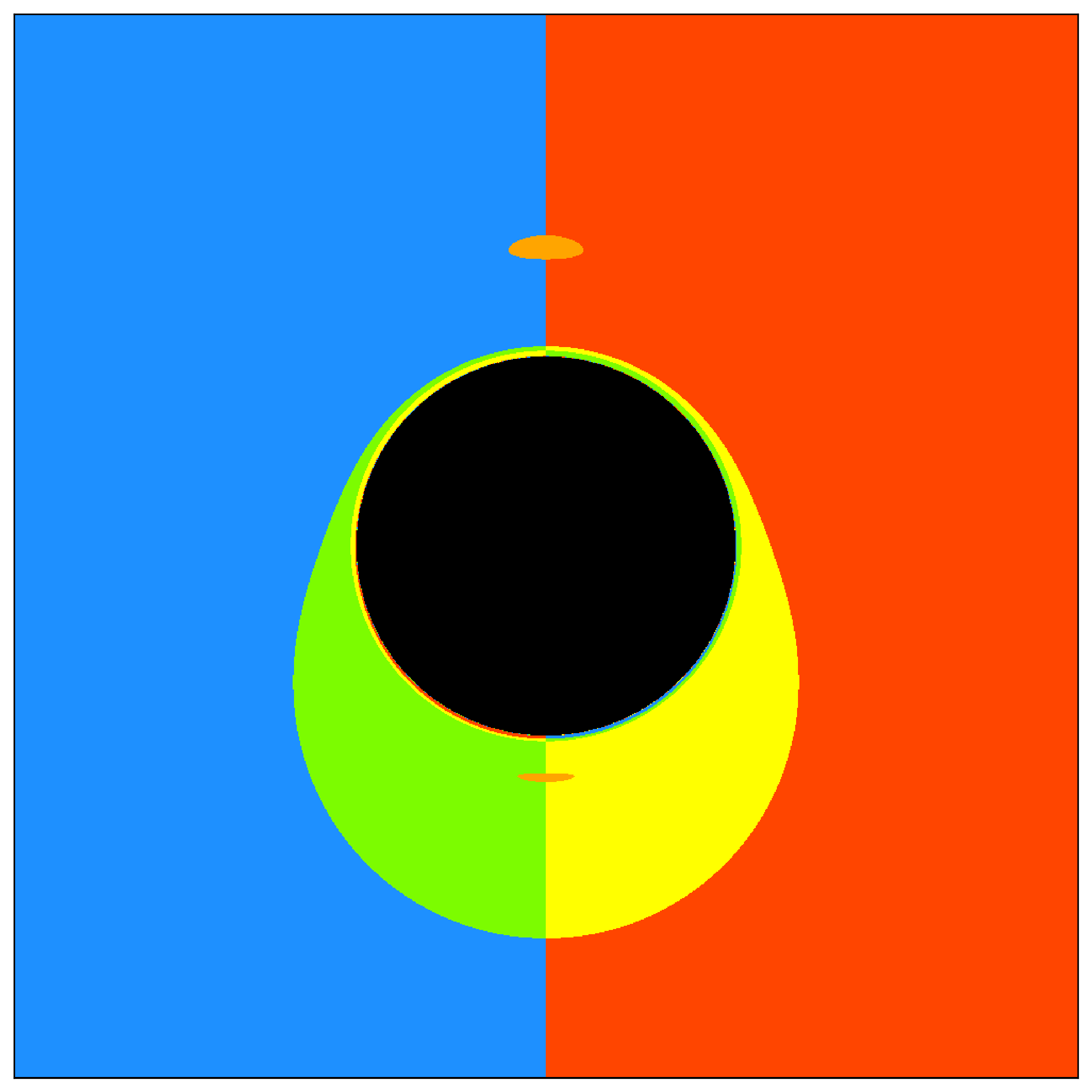}
\includegraphics[width=4cm]{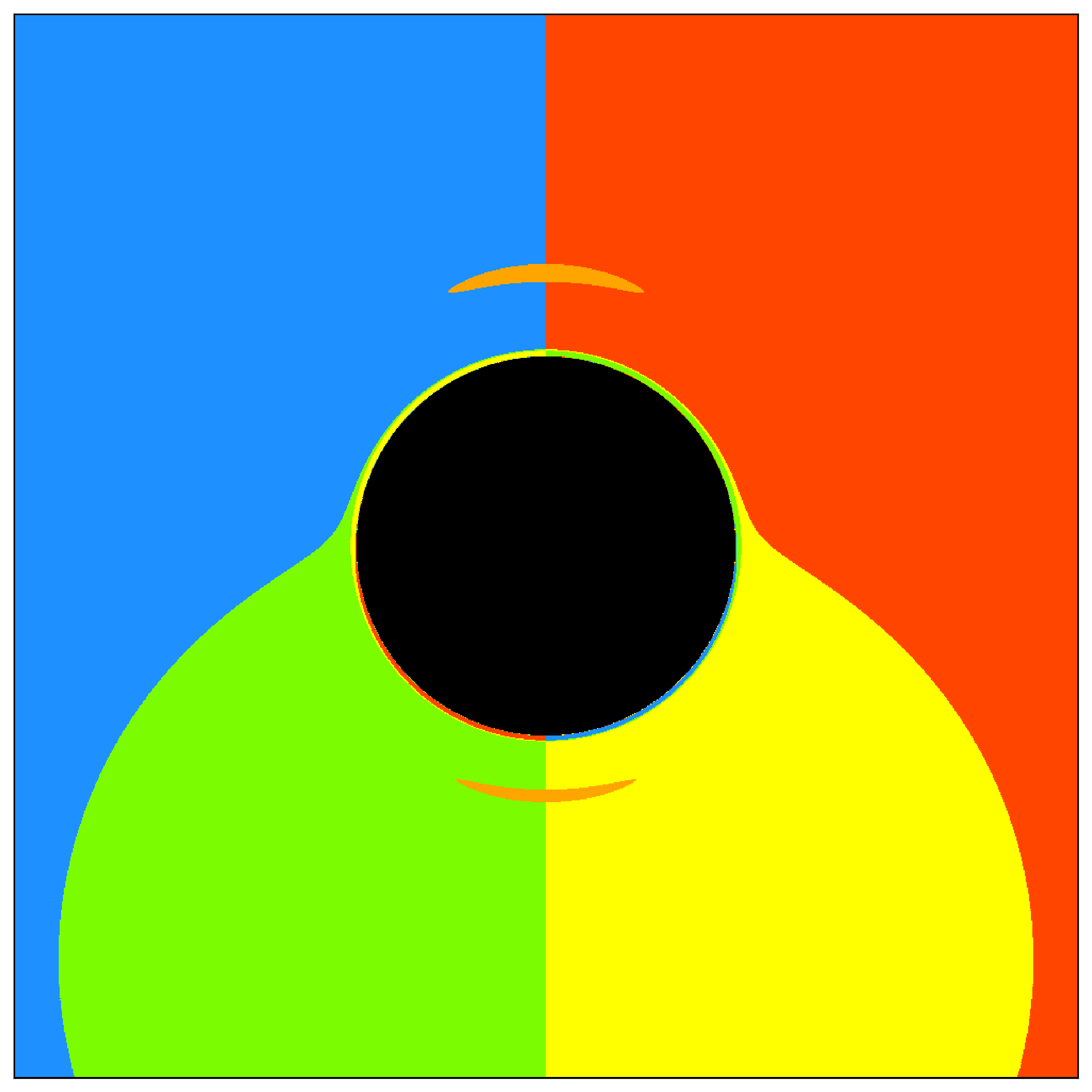}
\includegraphics[width=4cm]{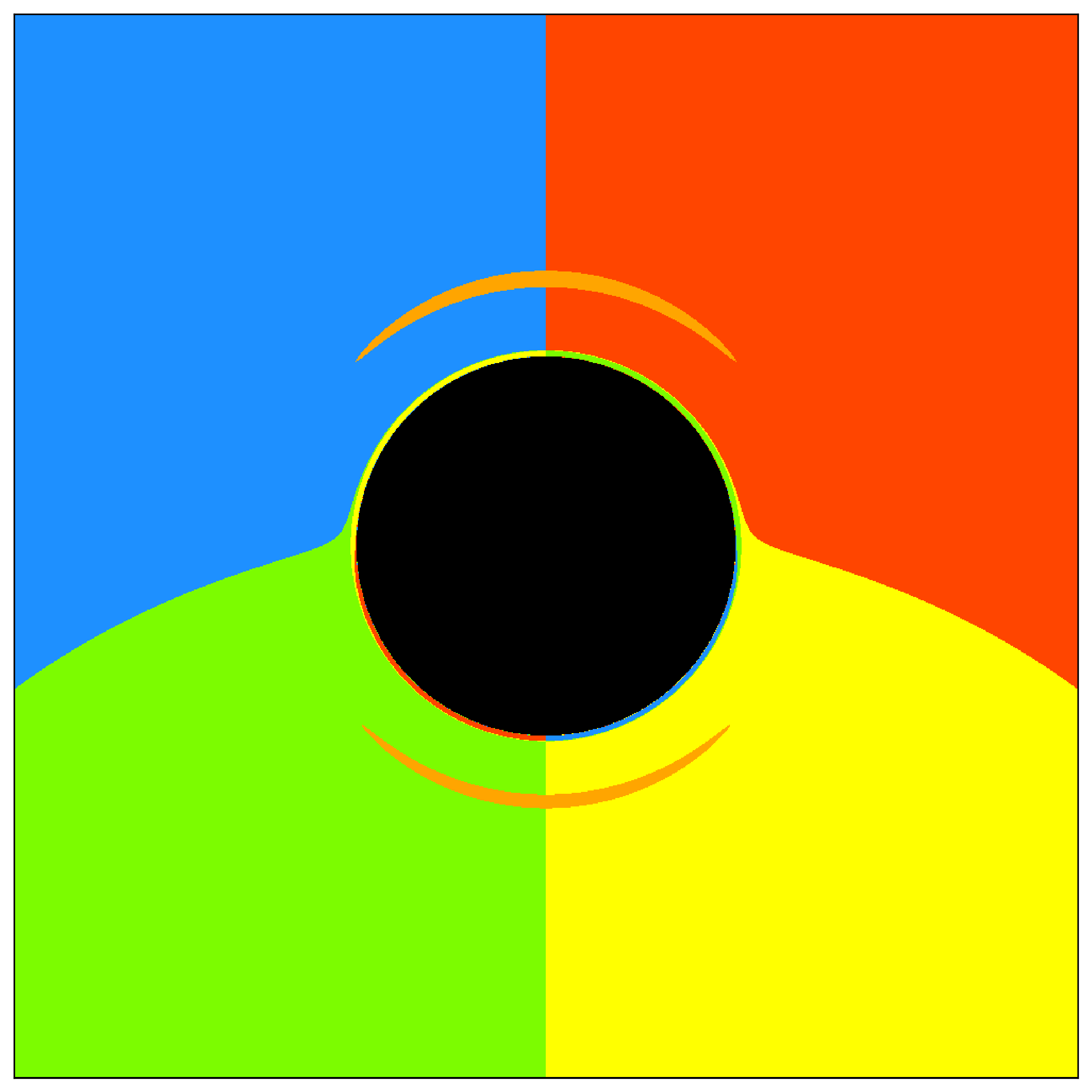}
\includegraphics[width=4cm]{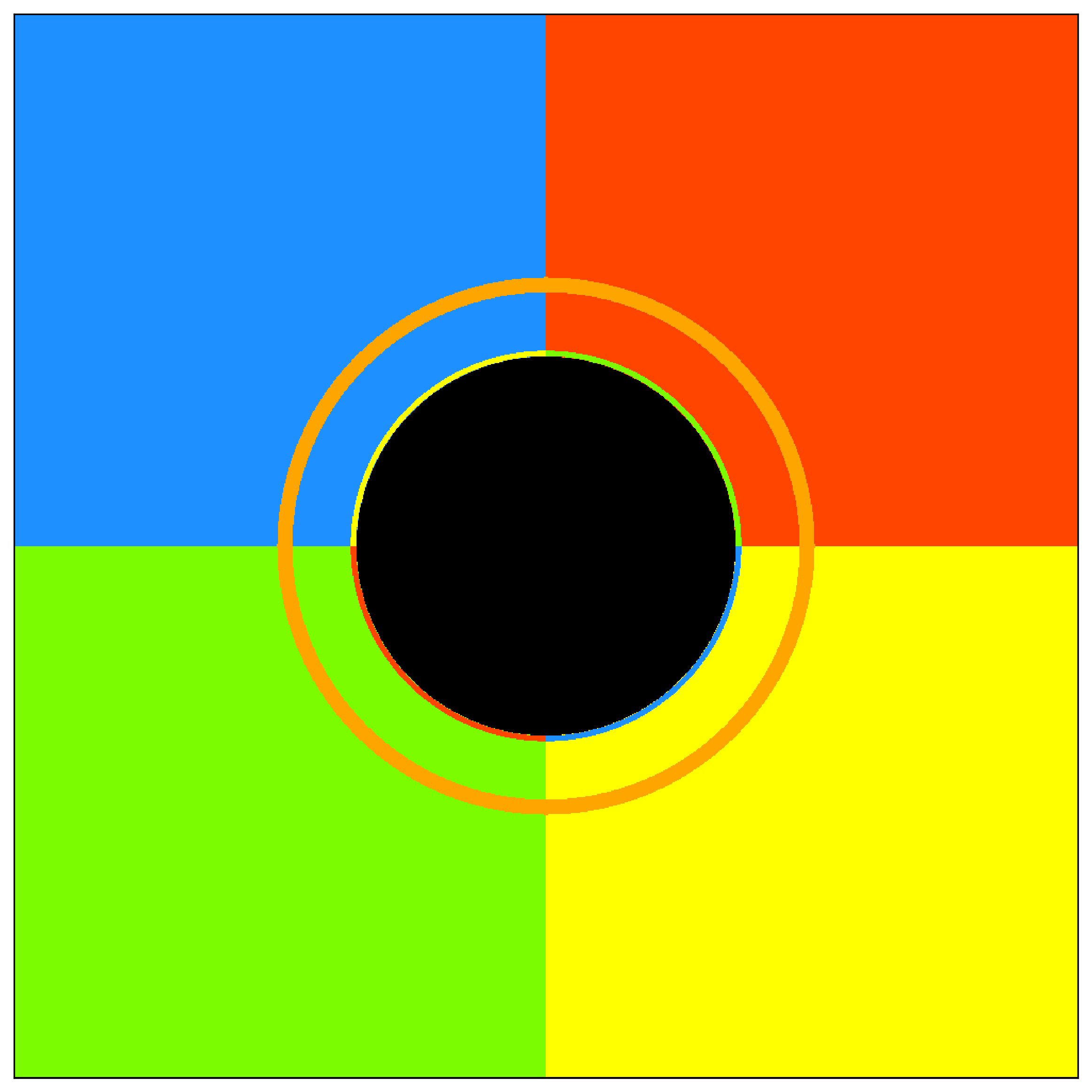}
\caption{Gravitational lensing images of a point source under different observation inclinations. From top-left to bottom-right: observation inclinations of $1^{\circ}$, $30^{\circ}$, $60^{\circ}$, $80^{\circ}$, $85^{\circ}$, and $90^{\circ}$. We fix the dark matter halo parameters at $r_{\textrm{s}}=\rho_{\textrm{s}}=0.3$, the simulation resolution at $1000 \times 1000$ pixels, and the point source position at $(-8,0,0)$ with a radius of $r_{\textrm{source}} = 0.5$. As the inclination increases, the image morphology evolves from a spot-like shape to a crescent, with the appearance of secondary images. When the observer, black hole, and point source are aligned, an Einstein ring forms.}}\label{fig16}
\end{figure*}

\begin{figure*}%[tbph]
\center{
\includegraphics[width=3.5cm]{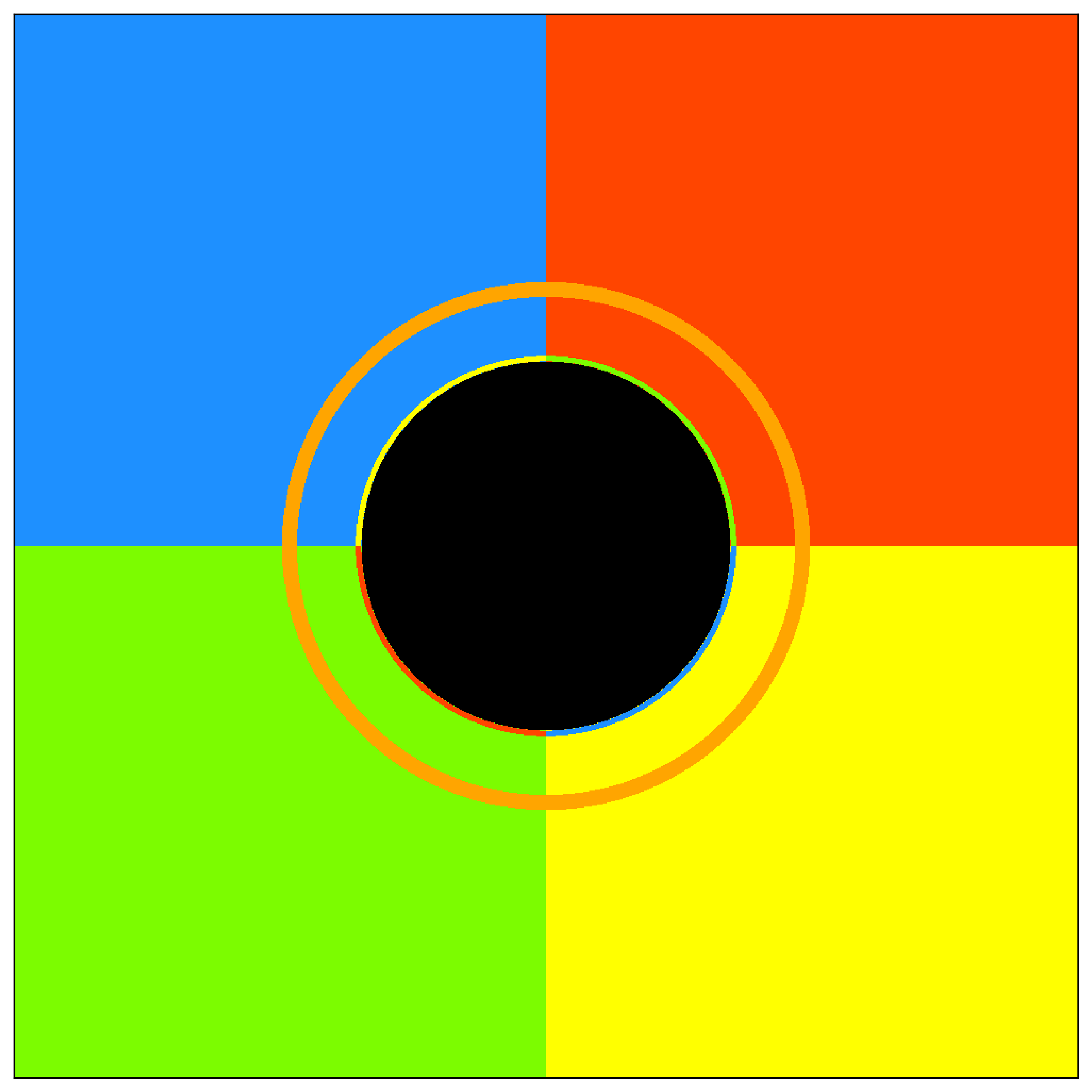}
\includegraphics[width=3.5cm]{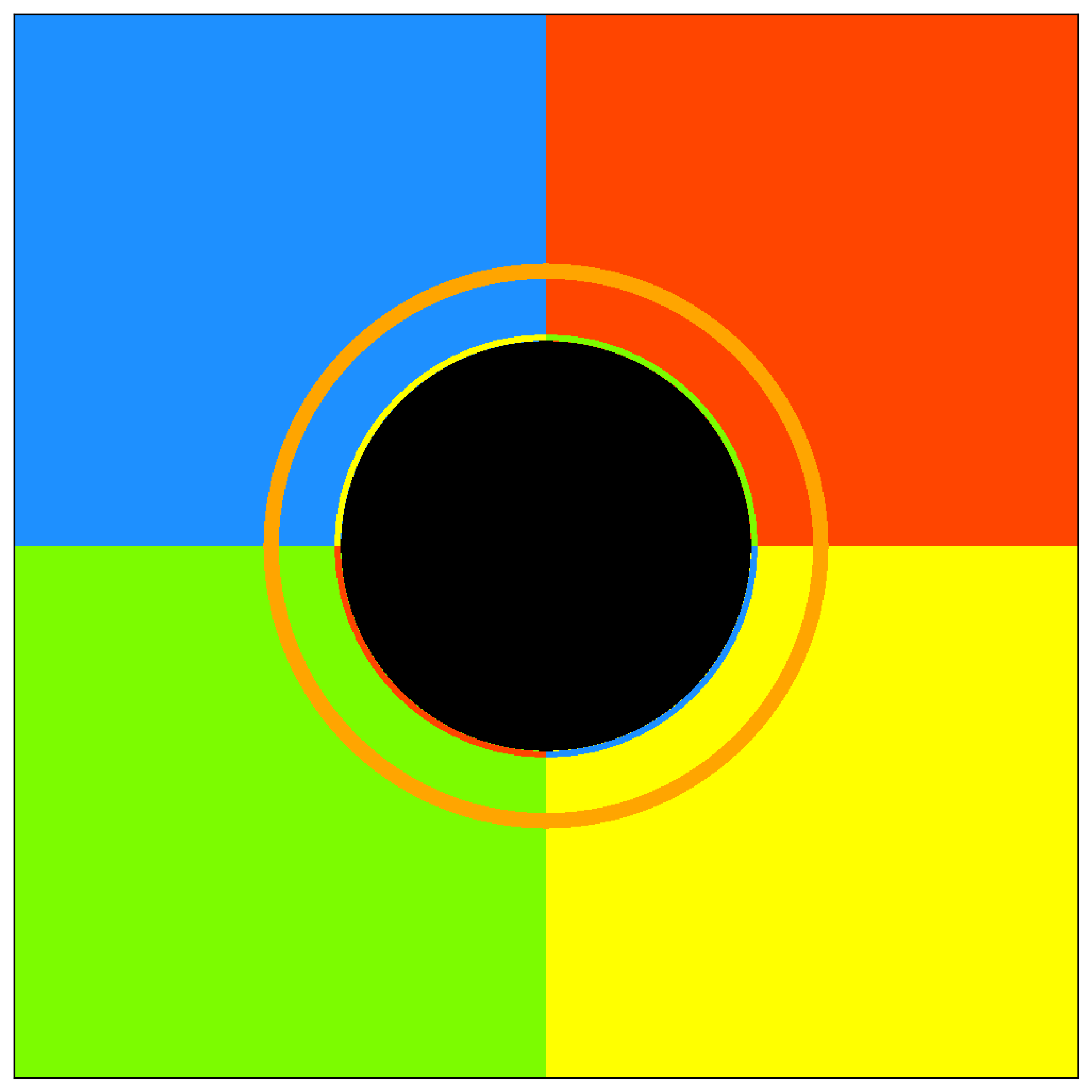}
\includegraphics[width=3.5cm]{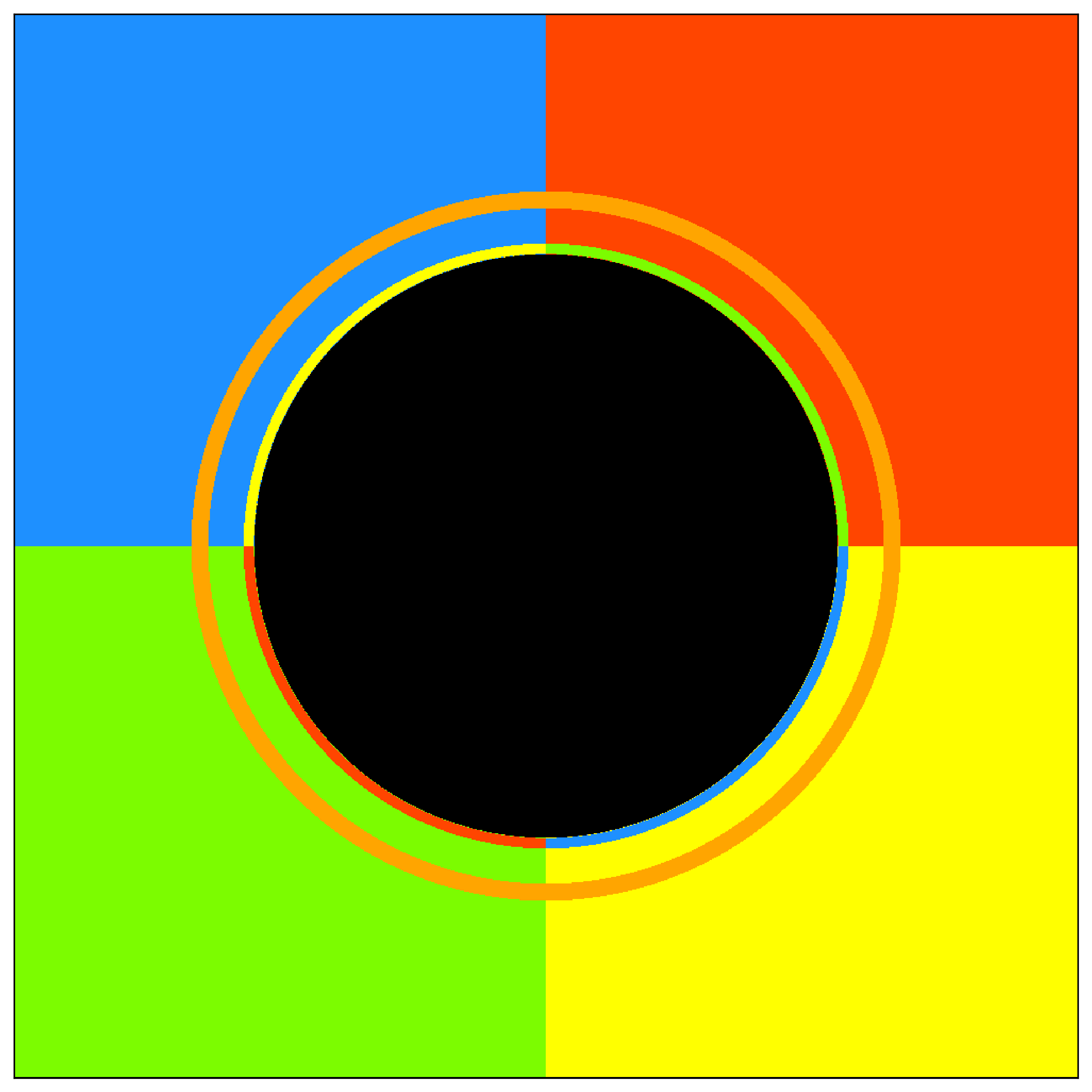}
\includegraphics[width=3.5cm]{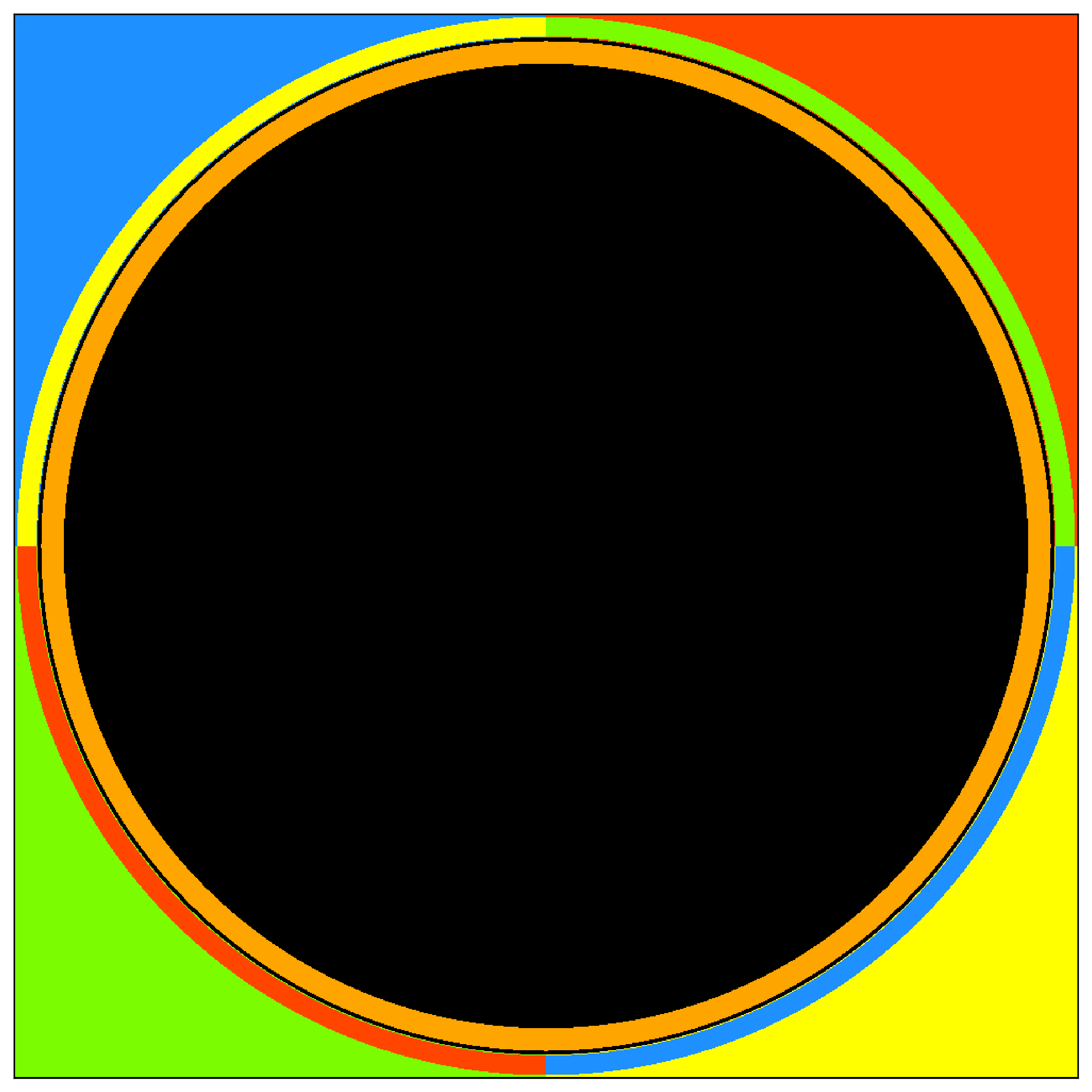}
\caption{Dependence of the gravitational lensing image of a fixed static point source on the dark matter halo scale parameter $r_{\textrm{s}}$. From left to right: $r_{\textrm{s}}=$ $0.1$, $0.4$, $0.7$, $1$. The point source is fixed at $(x^{\prime},y^{\prime},z^{\prime})=(-8,0,0)$, the dark matter halo density at $\rho_{\textrm{s}}=0.5$, and the observation inclination at $90^{\circ}$. We observe that the Einstein ring expands with increasing $r_{\textrm{s}}$, although at a slower rate than the critical curve. Consequently, the ring eventually falls inside the shadow region. Furthermore, using the Einstein ring radius to infer spacetime parameters holds promising potential.}}\label{fig17}
\end{figure*}

Theoretically, a point source can be treated as an emitting medium, allowing light rays propagating through it and accumulating specific intensity. We idealize its emission and absorption properties using equations \eqref{61} and \eqref{62}, and numerically simulate the specific intensity distribution of the point source corresponding to each panel in figure 16. The results are presented in figure 18, offering a more intuitive and physically realistic approach for investigating gravitational lensing phenomena.
\begin{figure*}%[tbph]
\center{
\includegraphics[width=4cm]{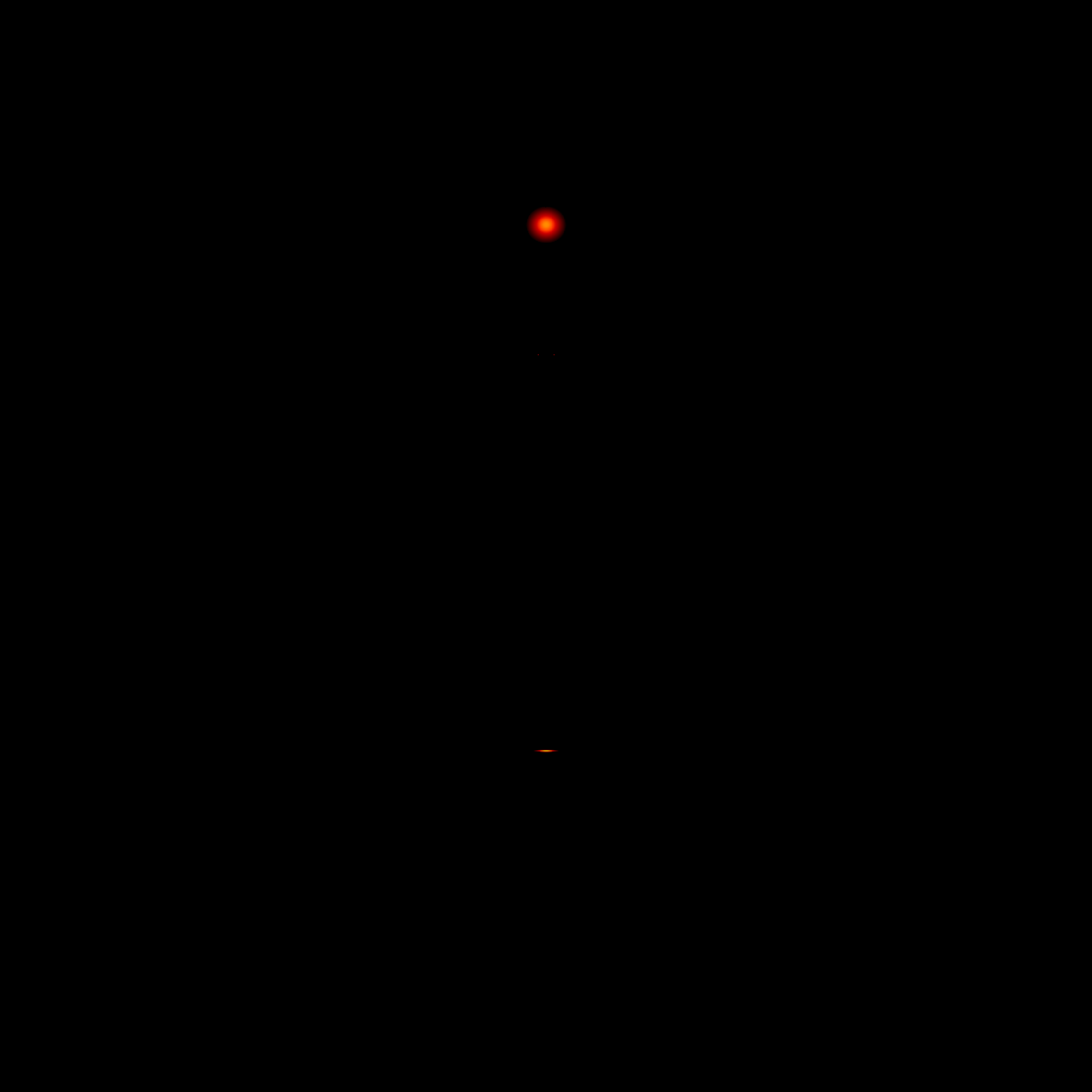}
\includegraphics[width=4cm]{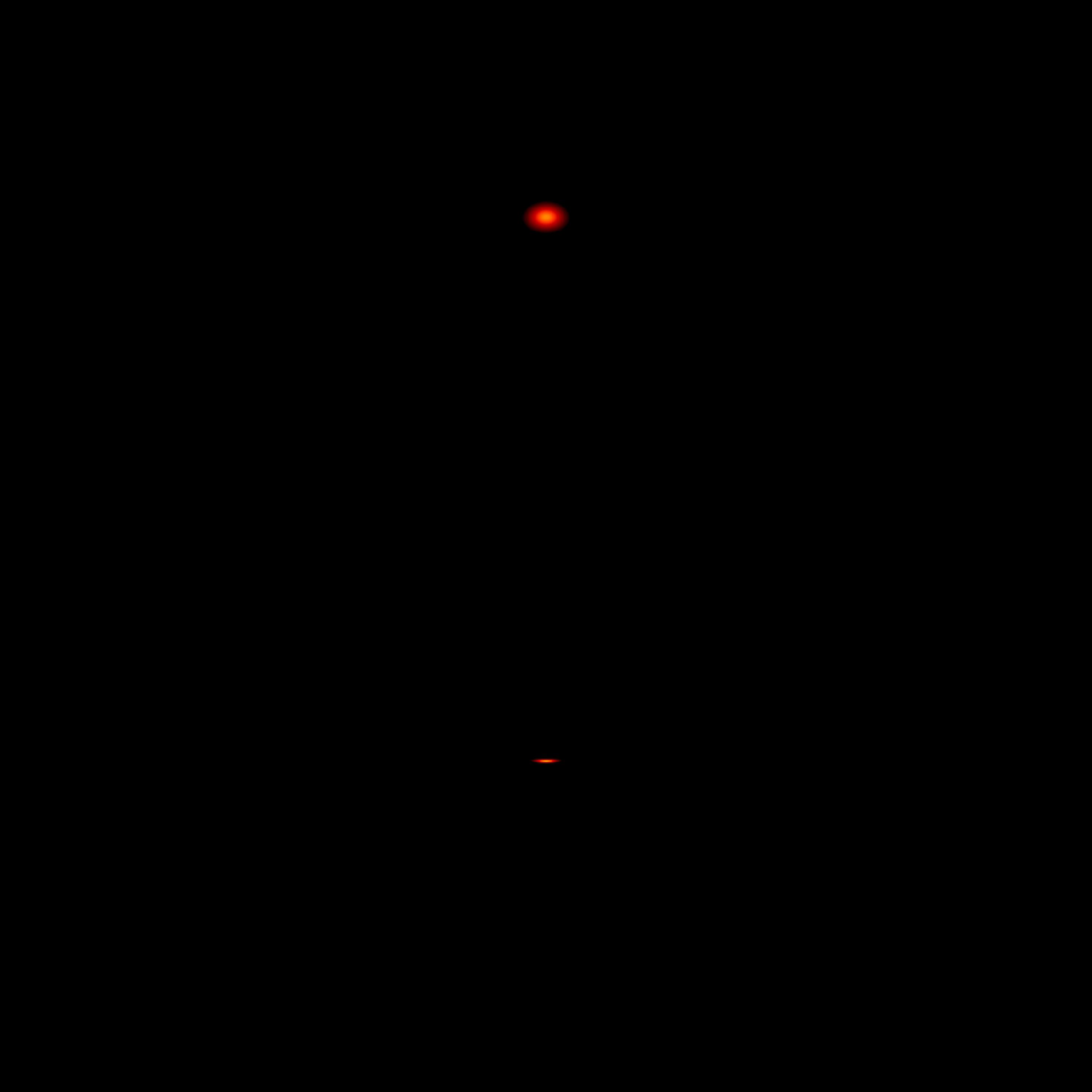}
\includegraphics[width=4cm]{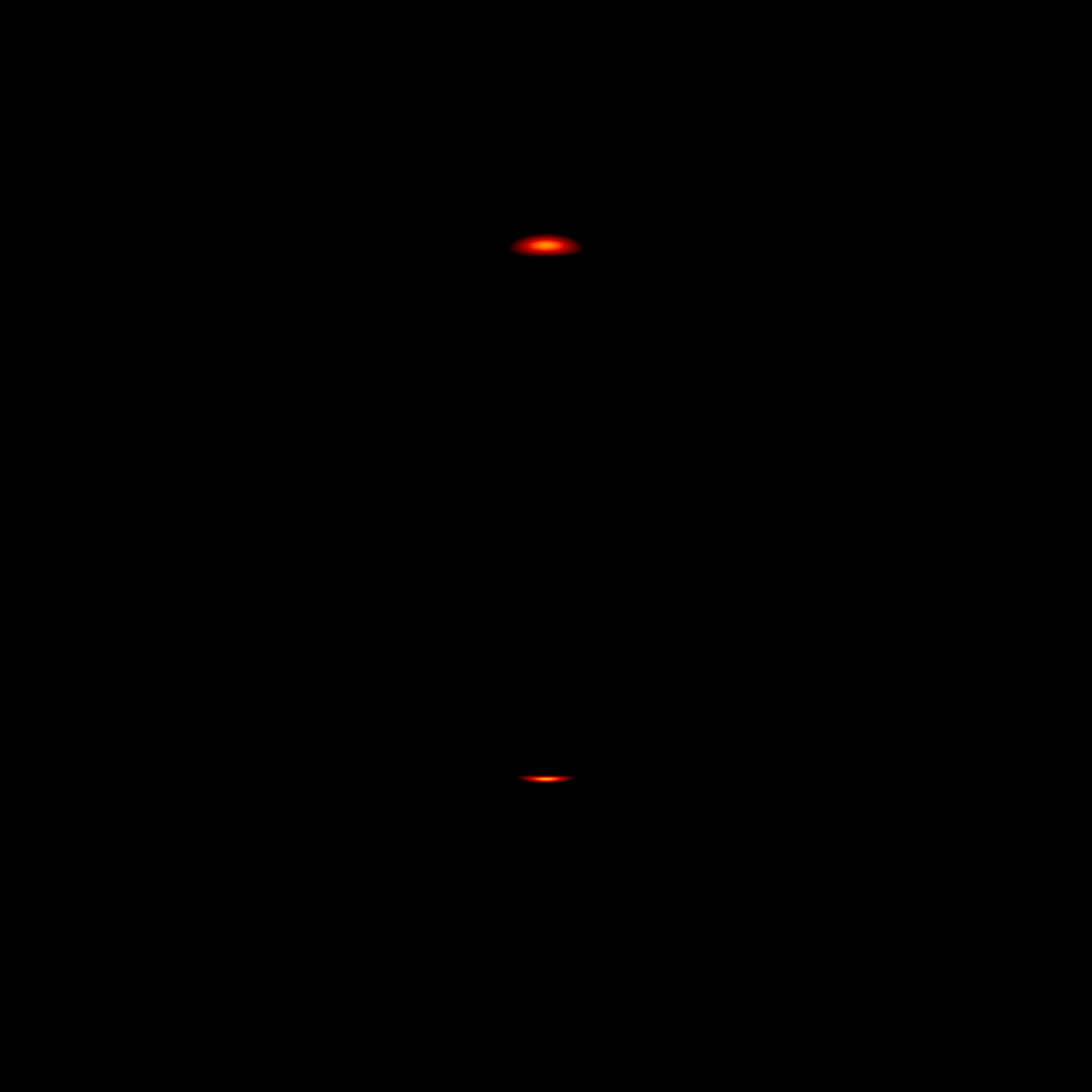}
\includegraphics[width=4cm]{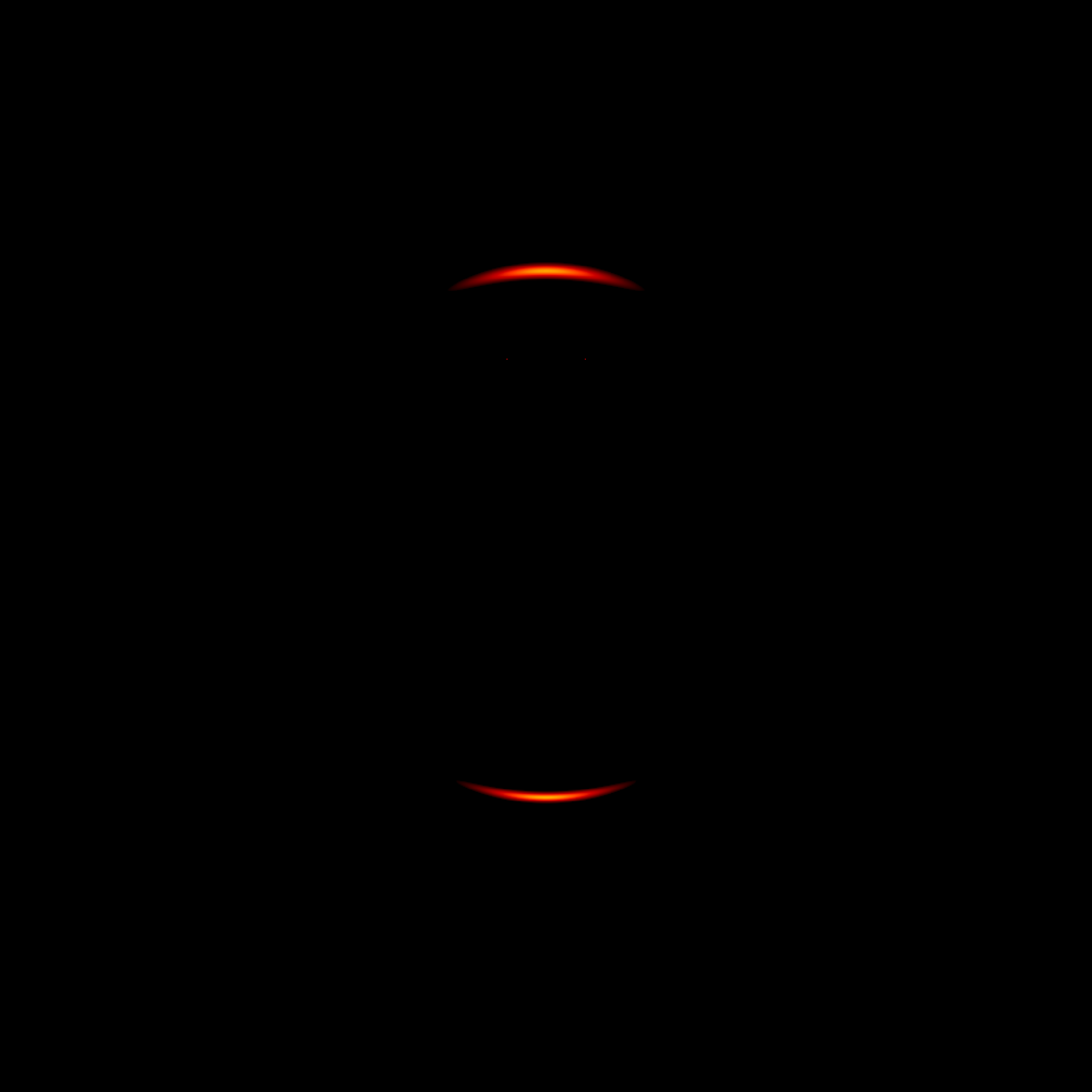}
\includegraphics[width=4cm]{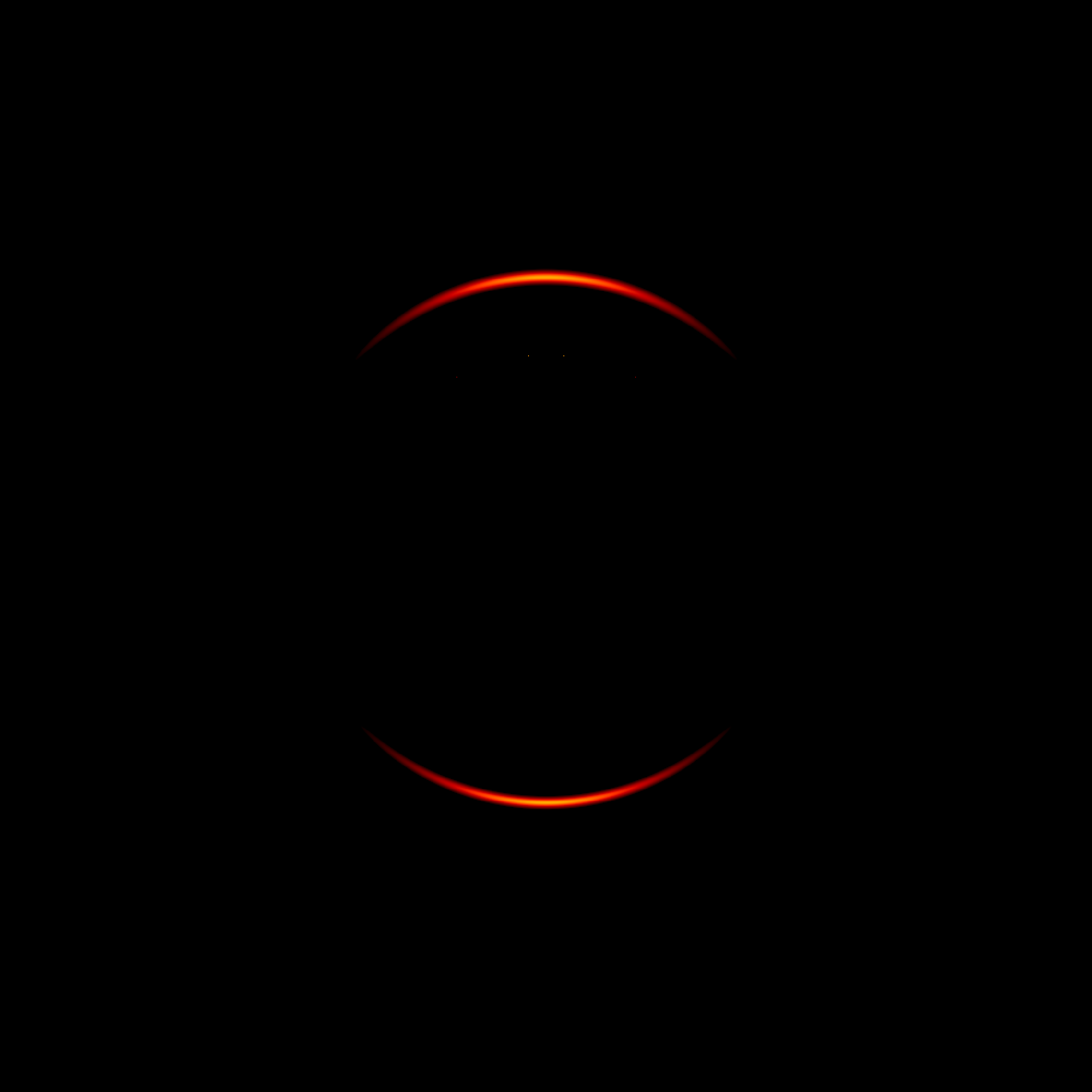}
\includegraphics[width=4cm]{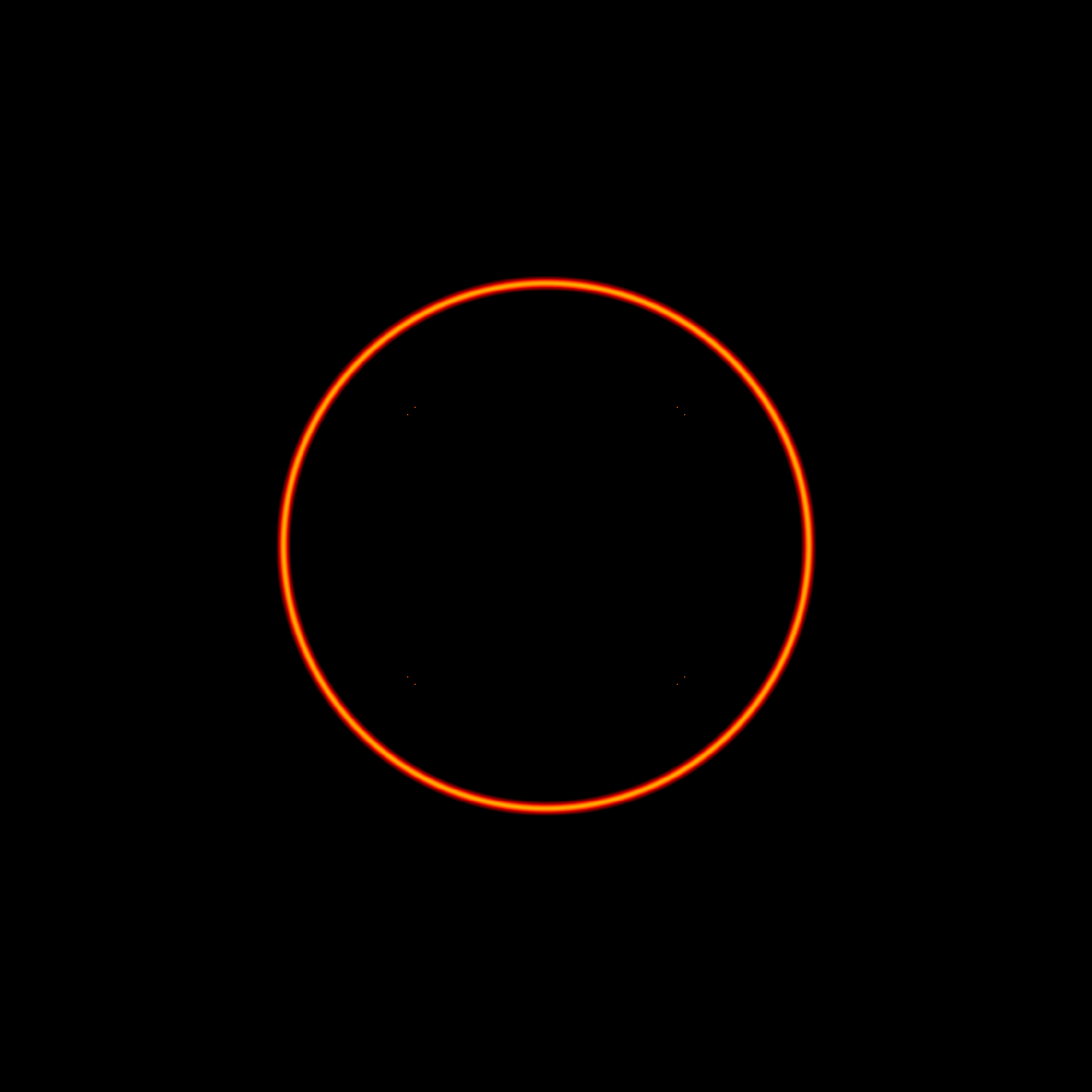}
\caption{Similar to the configuration in figure 16, but showing the lensing image of a stationary point source incorporating a simple radiation model. The specific intensity is plotted over the range $I_{\textrm{obs}} \in [0,0.5]$.}}\label{fig23}
\end{figure*}
\subsection{Orbits, gravitational emissions, and light curves}
In the previous section, we examined gravitational lensing images from stationary point sources. However, in realistic curved spacetime, such sources are, in fact, hot-spots moving along timelike geodesics and emitting electromagnetic radiation. It is evident that the observational signatures of hot-spots are both more prominent and scientifically valuable.

We selected eight hot-spot orbits, with their initial conditions listed in table 2. Orbits 1--4 are standard circular orbits with a fixed radius of 15 M, though the specific energy and angular momentum required to maintain circular motion differ due to variations in the dark matter halo parameters. Figure 19 shows the light curves produced by hot-spots moving along these four circular orbits under different observational configurations. We observe that all light curves exhibit periodic behavior across the parameter space. As the observation inclination increases (from top to bottom in each panel), photons tend to concentrate within specific time intervals, leading to higher peaks and distinct pulses---an effect attributed to gravitational lensing focusing. Moreover, an increase in $r_{\textrm{s}}$ extends the duration of the light curves, as the stronger gravitational field induced by the dark matter halo increase the time delay of light signals. Additionally, larger $r_{\textrm{s}}$ sharpen the peaks of the light curves, particularly at high observation inclinations. In other words, extracting spacetime parameters from light curves appears to be a promising approach.
\begin{table*}
\begin{center}
\small \caption{Initial conditions and spacetime parameters for the eight hot-spot orbits. Other parameters are fixed as $\theta=\pi/2$, $\varphi=0$, $t=0$, and $p_{r}=p_{\theta}=0$. Orbits 1--4 are standard circular orbits, while Orbits 5--8 are precessing quasi-periodic orbits, classified according to the scheme in \cite{Levin and Perez-Giz (2008),Liu et al. (2019),Deng (2020a),Deng (2020b),Wang et al. (2022),Lin and Deng (2023),Tu et al. (2023),Huang and Deng (2024)}.} \label{t2}
\begin{tabular}{cccccc}\hline
Orbit & $r_{\textrm{s}}$ & $\rho_{\textrm{s}}$ & $r$ & $p_{t}$ & $p_{\varphi}$ \\
\hline
$1$ & $0.3$ & $0.5$ & $15$ & $-0.967375116667521$ & $4.47450786056199$ \\
\hline
$2$ & $0.5$ & $0.5$ & $15$ & $-0.961756732028602$ & $4.97134509967810$ \\
\hline
$3$ & $0.7$ & $0.5$ & $15$ & $-0.950373267529097$ & $6.04002654765628$ \\
\hline
$4$ & $0.9$ & $0.5$ & $15$ & $-0.935958487675813$ & $8.05218427187439$ \\
\hline
$5$ & $0.2$ & $0.4$ & $20.702903$ & $-0.965488$ & $3.77841$ \\
\hline
$6$ & $0.2$ & $0.4$ & $23.274039$ & $-0.967983$ & $3.77841$ \\
\hline
$7$ & $0.2$ & $0.4$ & $23.784133$ & $-0.969139$ & $3.887002$ \\
\hline
$8$ & $0.2$ & $0.4$ & $28.449032$ & $-0.972679$ & $3.887002$ \\
\hline
\end{tabular}
\end{center}
\end{table*}

\begin{figure*}%[tbph]
\center{
\includegraphics[width=7cm]{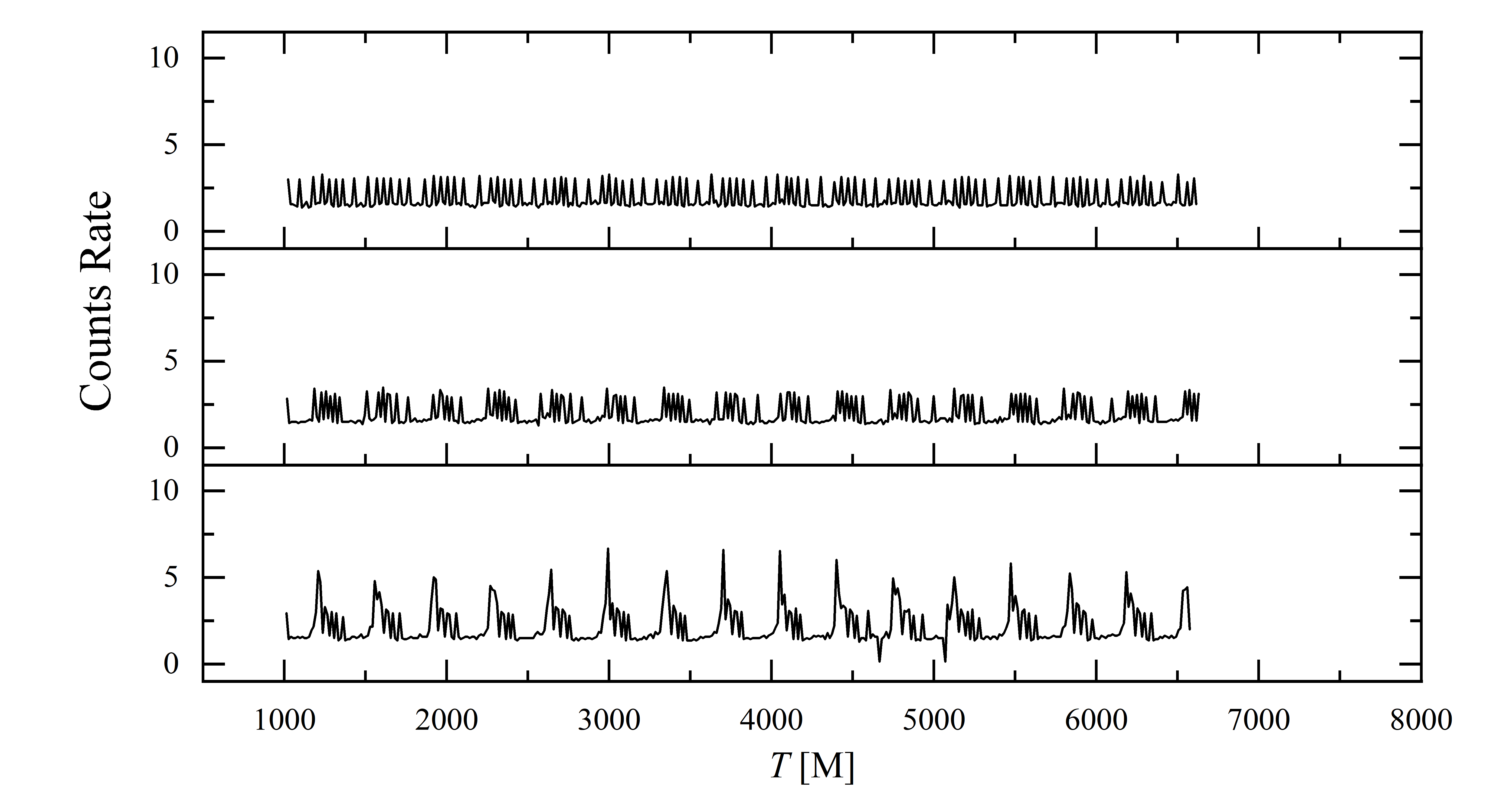}
\includegraphics[width=7cm]{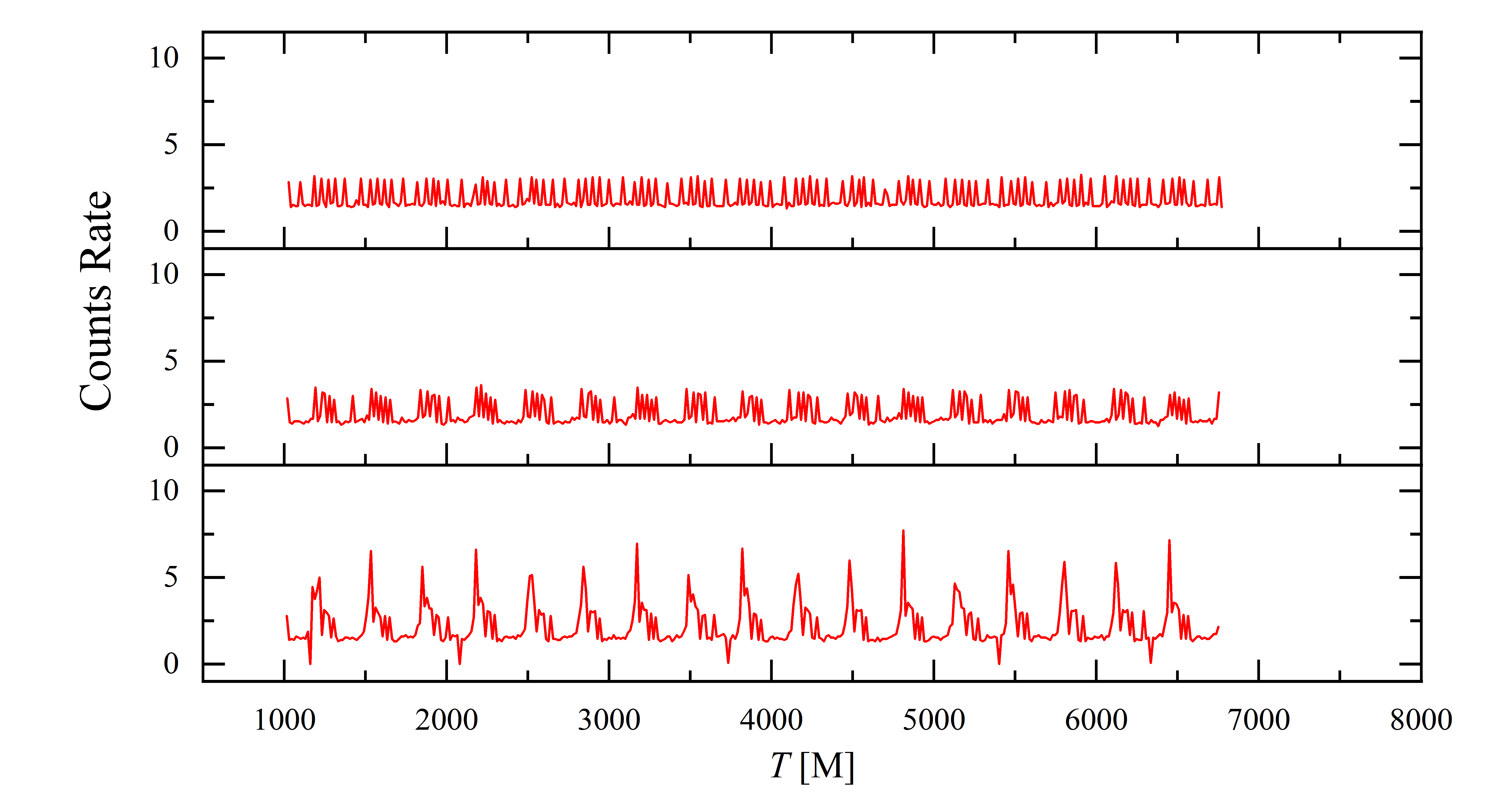}
\includegraphics[width=7cm]{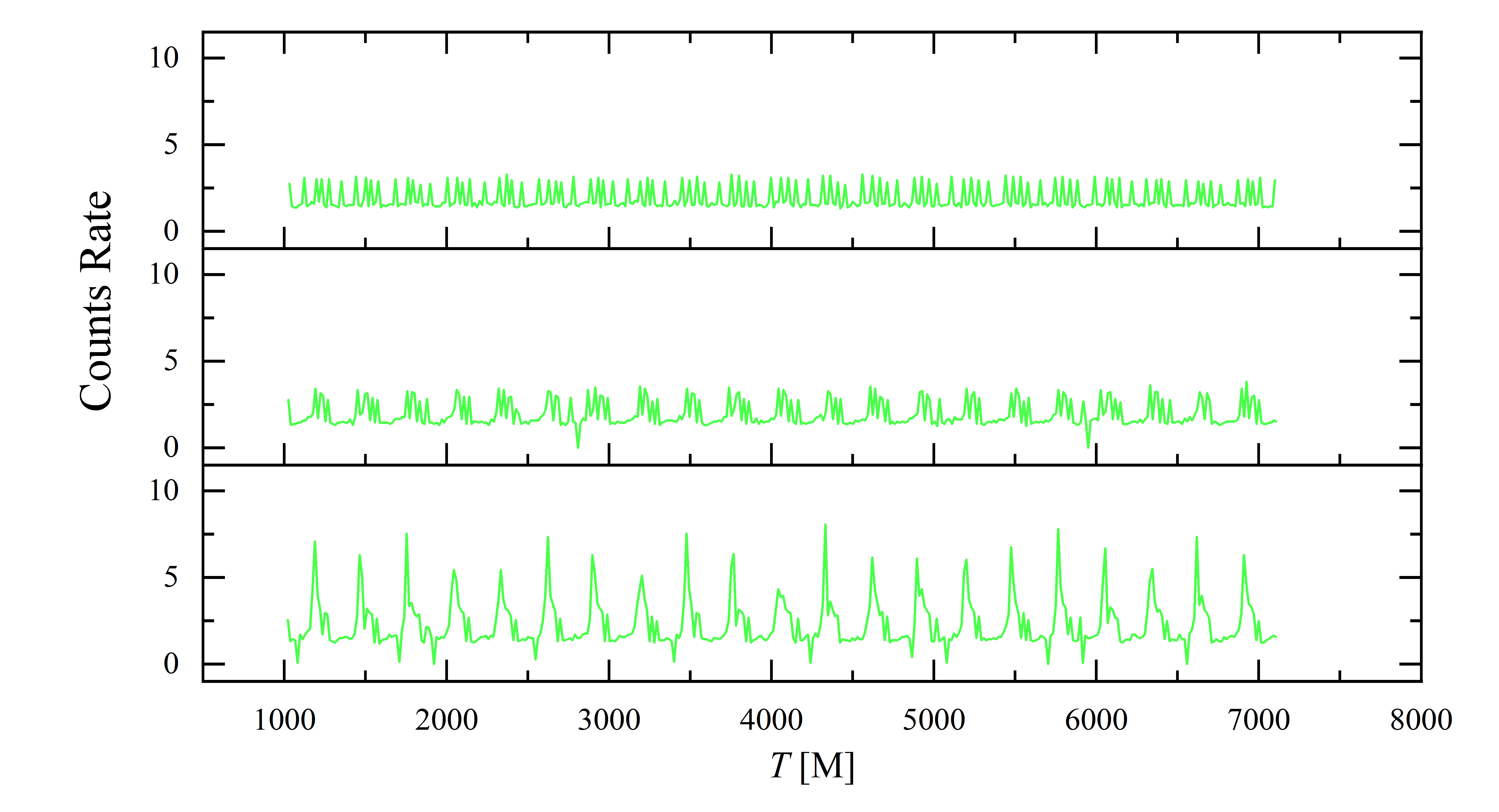}
\includegraphics[width=7cm]{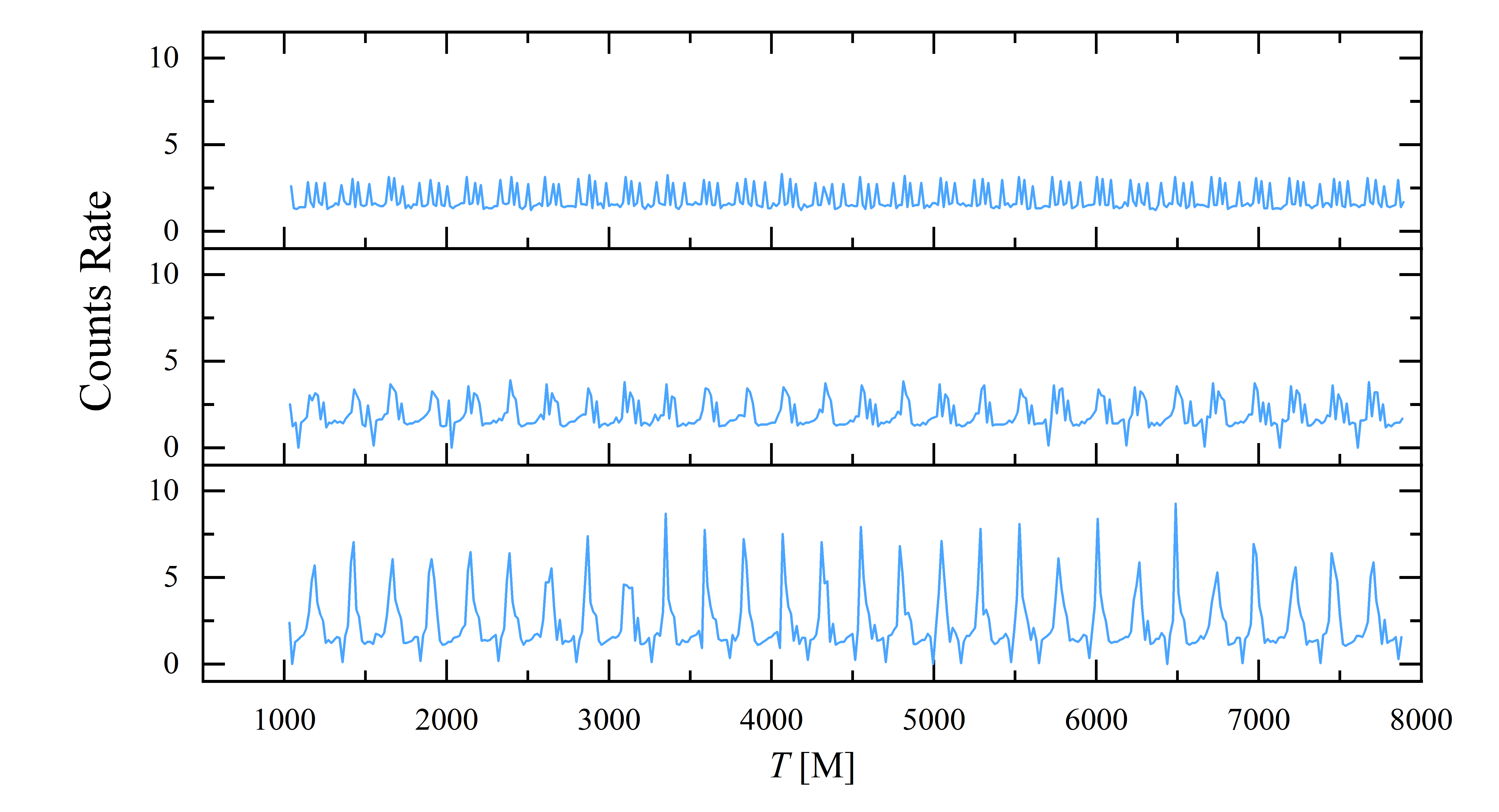}
\caption{Light curves of Orbits 1--4 (shown in black, red, green, and blue, respectively) under different observational configurations. In each panel, from top to bottom: observation inclinations of $17^{\circ}$, $50^{\circ}$, and $80^{\circ}$. The field of view is set to $x \in [-20,20]$ M and $y \in [-20,20]$ M, with a resolution of $2000 \times 2000$ pixels, a hot-spot radius of $r_{\textrm{source}}=0.05$, and a simulation duration of 5000 M. We find that while the four circular orbits exhibit similar trends in their light curves, their durations increase with $r_{\textrm{s}}$ due to enhanced gravitational lensing time delays. Moreover, despite sharing the same orbital radius, subtle differences in the light curves---manifested in peak height, width, and density---become discernible at high observation inclinations.}}\label{fig19}
\end{figure*}

Figure 20 displays the trajectories of Orbits 5--8. Unlike the first four circular orbits, these are not strictly periodic or closed; instead, they exhibit precessing quasi-periodic motion, demonstrating different numbers of leaves, whirls, and vertices, as similarly defined in \cite{Levin and Perez-Giz (2008),Liu et al. (2019),Deng (2020a),Deng (2020b),Wang et al. (2022),Lin and Deng (2023),Tu et al. (2023),Huang and Deng (2024)}. The light curves of the hot-spot moving along these four orbits under different observation inclinations are shown in figure 21. We find that, at low viewing angles, it is difficult to infer the orbital structure from the light curve. This is because gravitational lensing effects are weak under such conditions, allowing photons to arrive at the observer in a nearly continuous manner, resulting in smooth, mountain-like light curves. As the inclination increases, time delays caused by lensing become significant, superimposing this effect onto the ``mountain'' features of the light curves and producing both broad and narrow peaks. The broad peaks generally originate from the outer sections of the orbit, while the narrow peaks are predominantly contributed by the inner orbital segments. At an observation inclination of $80^ {\circ}$, the peak features become more pronounced. Nevertheless, it remains challenging to establish a clear correlation between the characteristics of the light curves and the underlying orbital configurations. It is worth noting that since figure 21 is based on photon counts without incorporating redshift factors, this may limit the amount of orbital information encoded in the light curves. In subsequent studies, we plan to supplement this approach by simulating light curves from a flux-based perspective.
\begin{figure*}%[tbph]
\center{
\includegraphics[width=6cm]{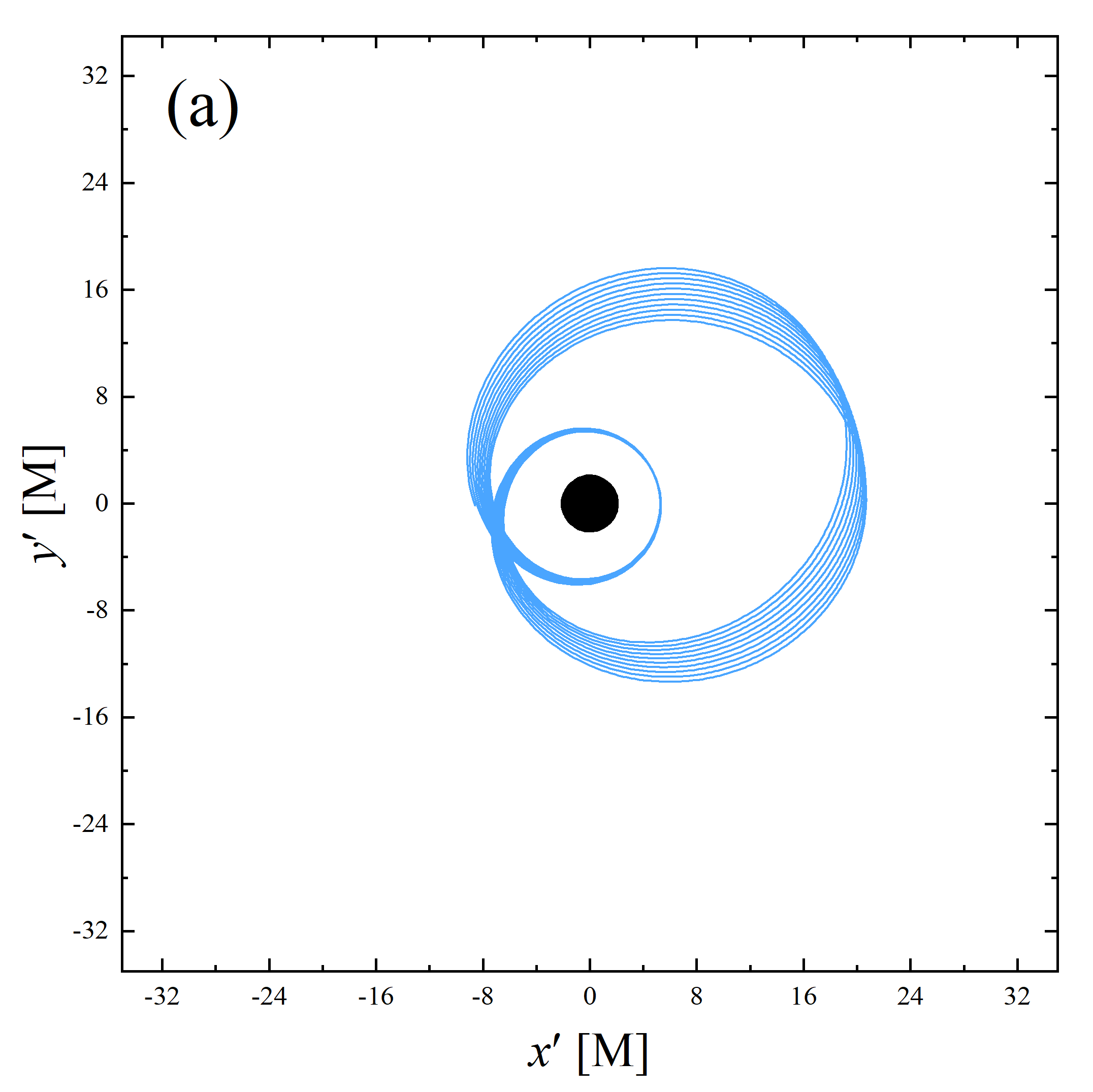}
\includegraphics[width=6cm]{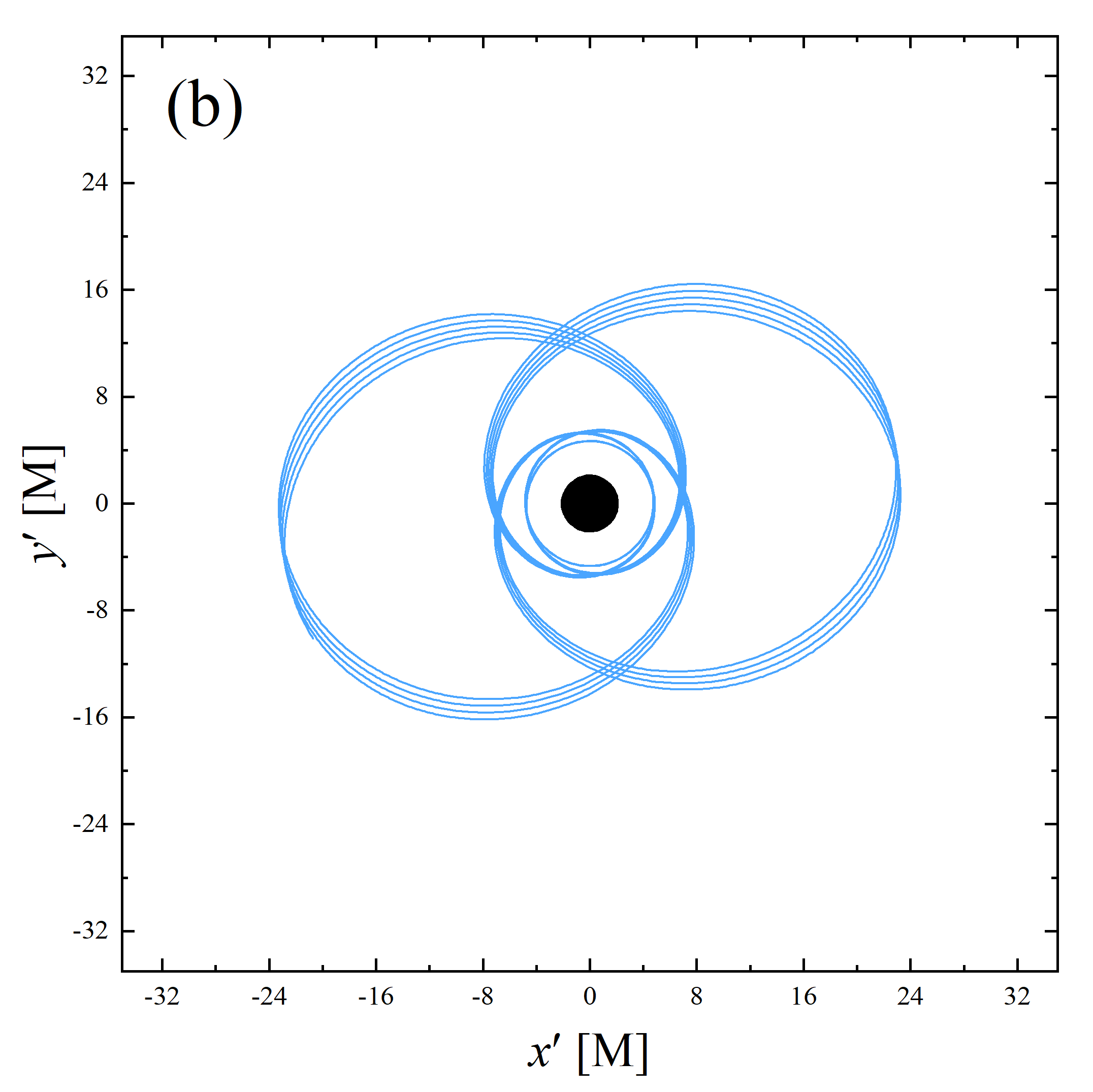}
\includegraphics[width=6cm]{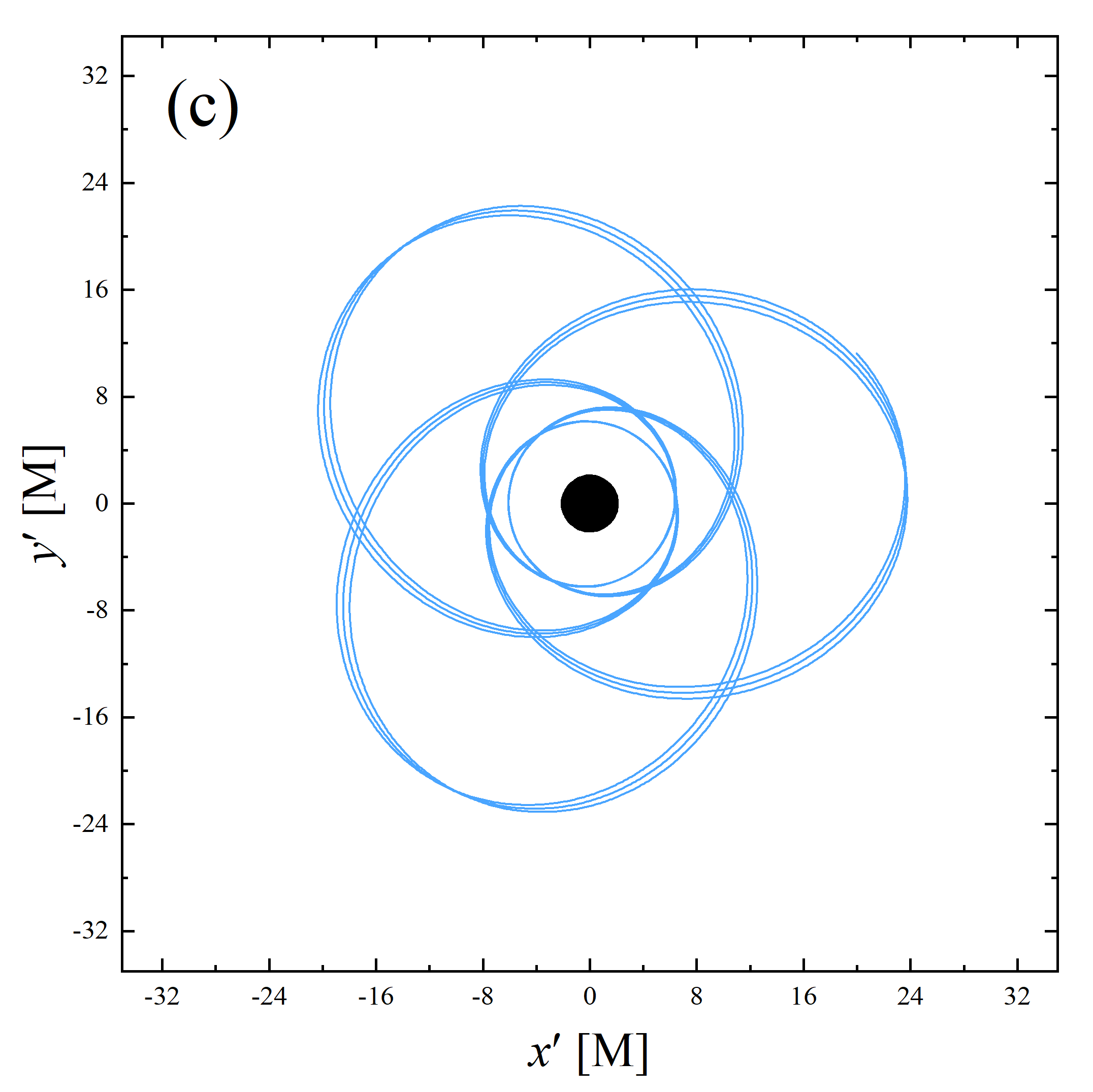}
\includegraphics[width=6cm]{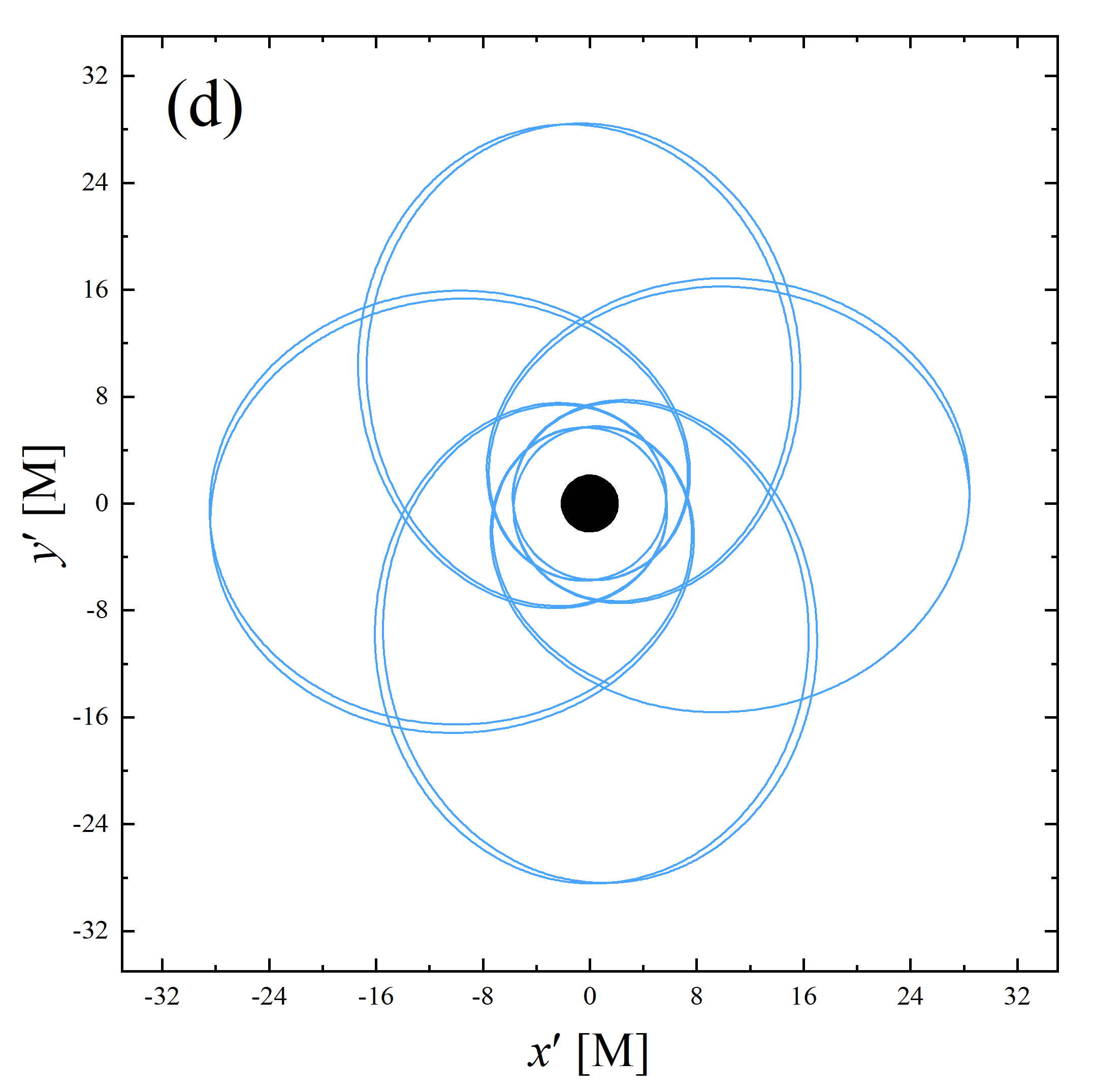}
\caption{Panels (a) to (d) display the trajectories of Orbits 5--8 from table 2, respectively. The black dot marks the location of the black hole. All orbits are integrated over a duration of $5000$ M. These four trajectories successively exhibit 1, 2, 3, and 4 petals.}}\label{fig20}
\end{figure*}
\begin{figure*}%[tbph]
\center{
\includegraphics[width=7cm]{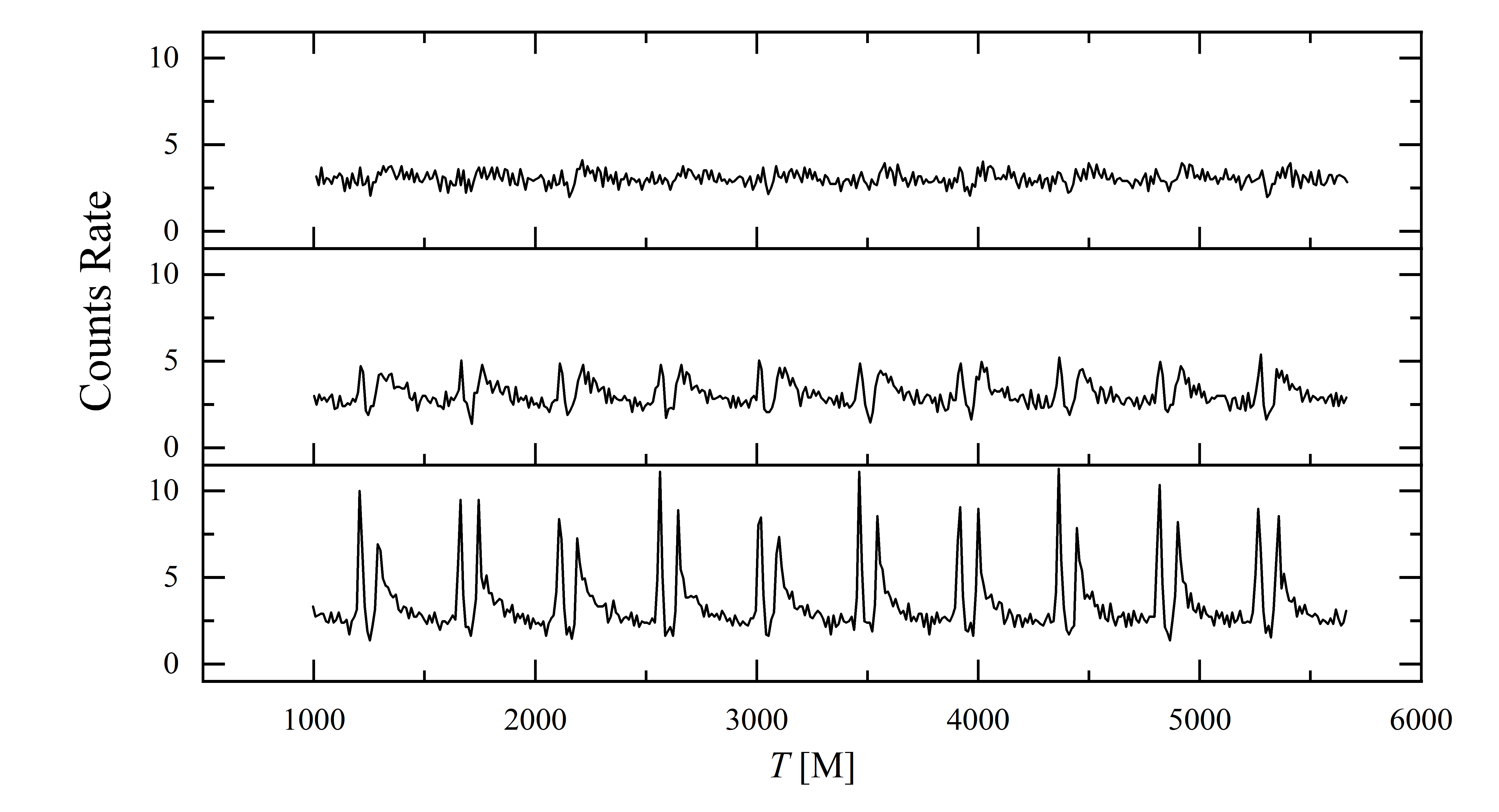}
\includegraphics[width=7cm]{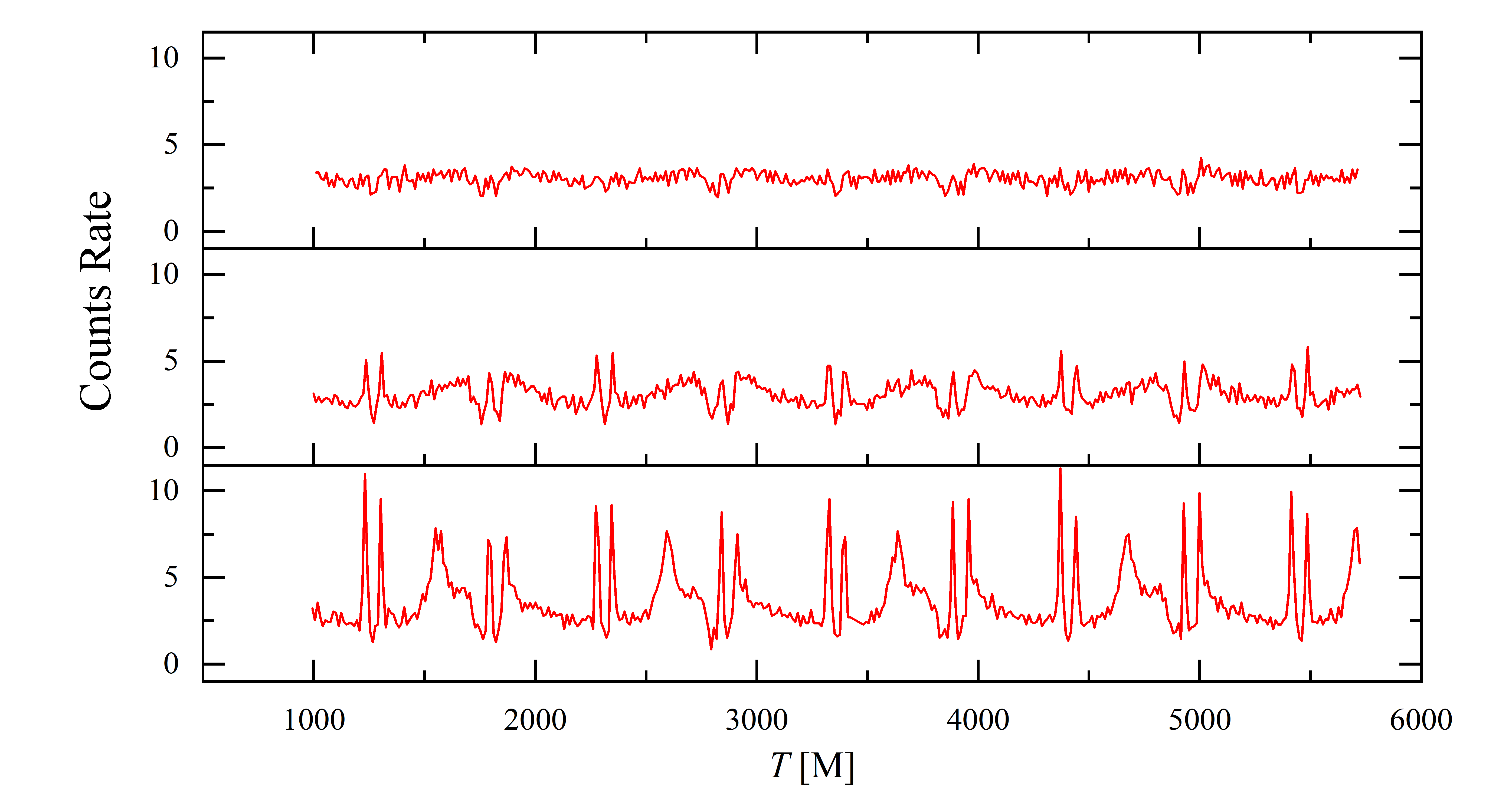}
\includegraphics[width=7cm]{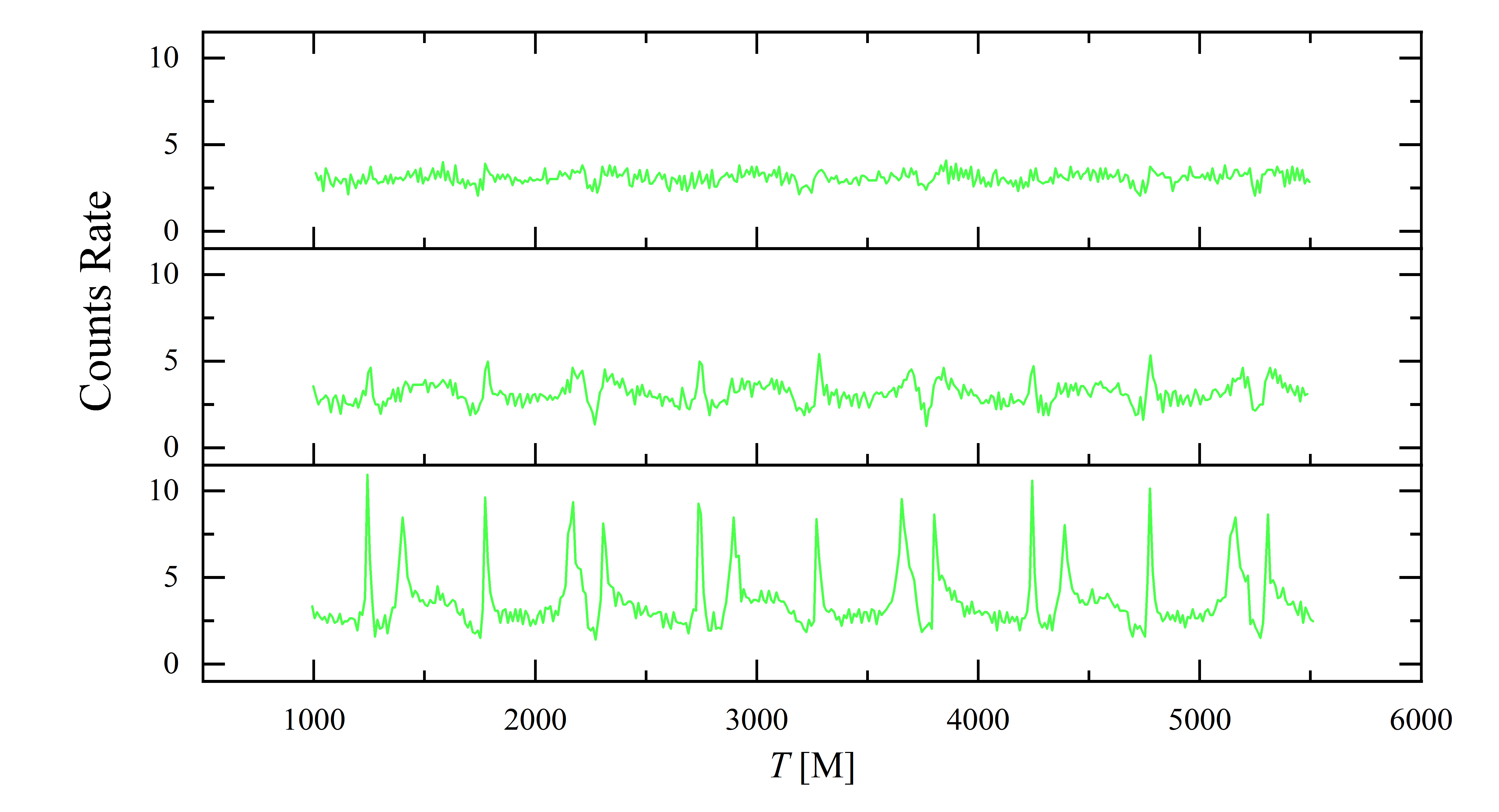}
\includegraphics[width=7cm]{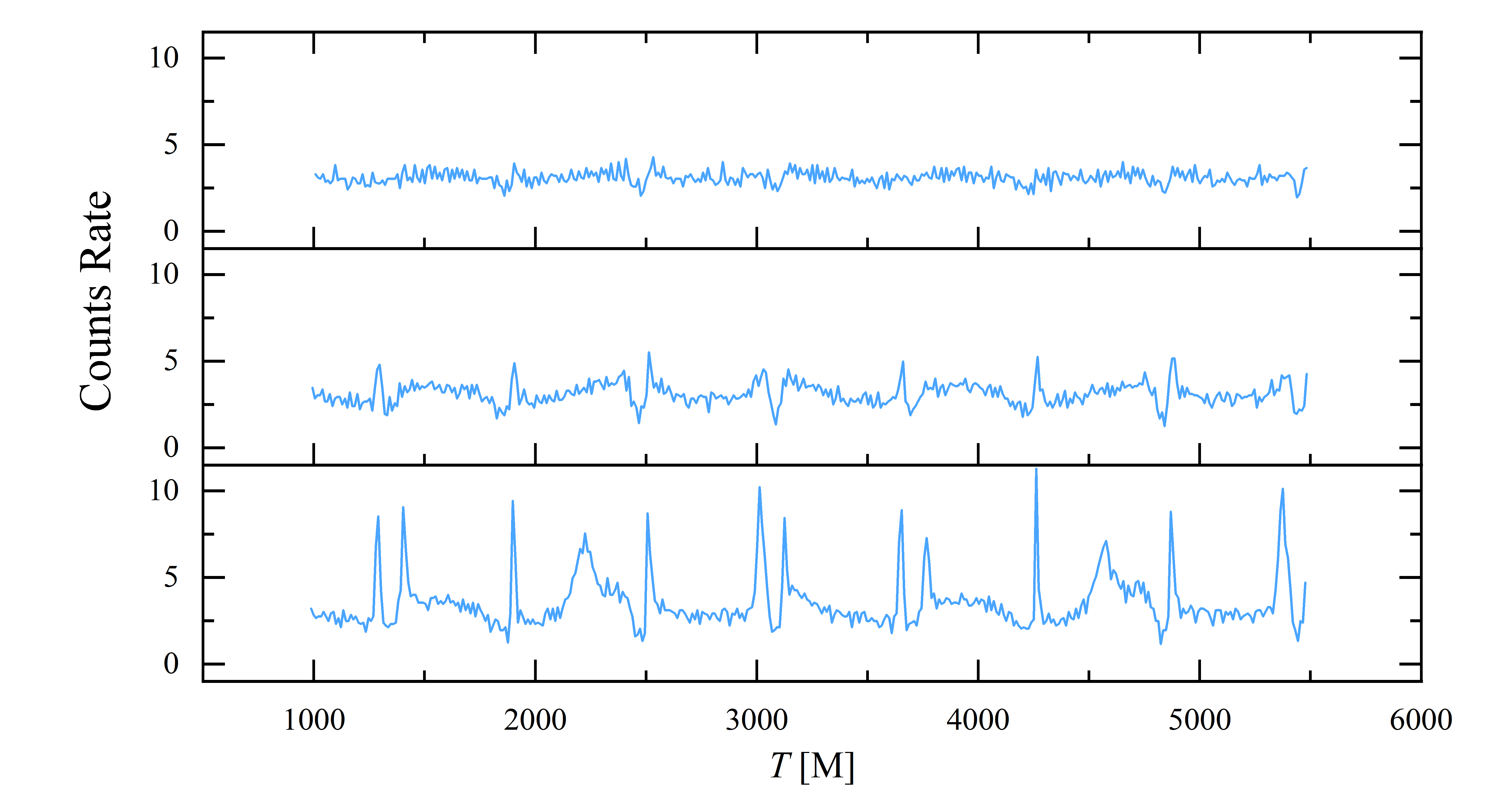}
\caption{Similar to figure 19, but showing the light curves of Orbits 5--8, with the field of view changed to $x \in [-35,35]$ M and $y \in [-35,35]$ M, and the hot-spot evolution time set to $4000$ M. Although these four orbits exhibit distinctly different structures, such differences are not readily apparent in the global characteristics of their light curves.}}\label{fig21}
\end{figure*}

According to the kludge waveform model, figure 22 displays, from top to bottom, the gravitational wave emissions for Orbits 1--8 over a fixed integration period, where the left and right columns correspond to the ``plus'' and ``cross'' polarizations, respectively. For Orbits 1--4, both polarizations exhibit regular sinusoidal or cosinusoidal profiles. However, the gravitational wave frequency also rises due to the enhancement of the circular orbital angular velocity---and thus the orbital frequency---by larger values of $r_{\textrm{s}}$. For Orbits 5--8, the gravitational waveforms consist of two distinct components: sinusoidal or cosinusoidal envelopes and abrupt variations in amplitude. The former arises from orbital segments with larger curvature radii, such as the petal-shaped trajectories shown in figure 20, where the radial coordinate and azimuthal angle evolve gradually, producing broad peaks in the gravitational radiation. The latter is primarily contributed by orbital segments close to the black hole, where the timelike object passes rapidly through turning points multiple times within short intervals, generating tightly spaced, high-amplitude narrow features in the waveform. 
\begin{figure*}%[tbph]
\center{
\includegraphics[width=7cm]{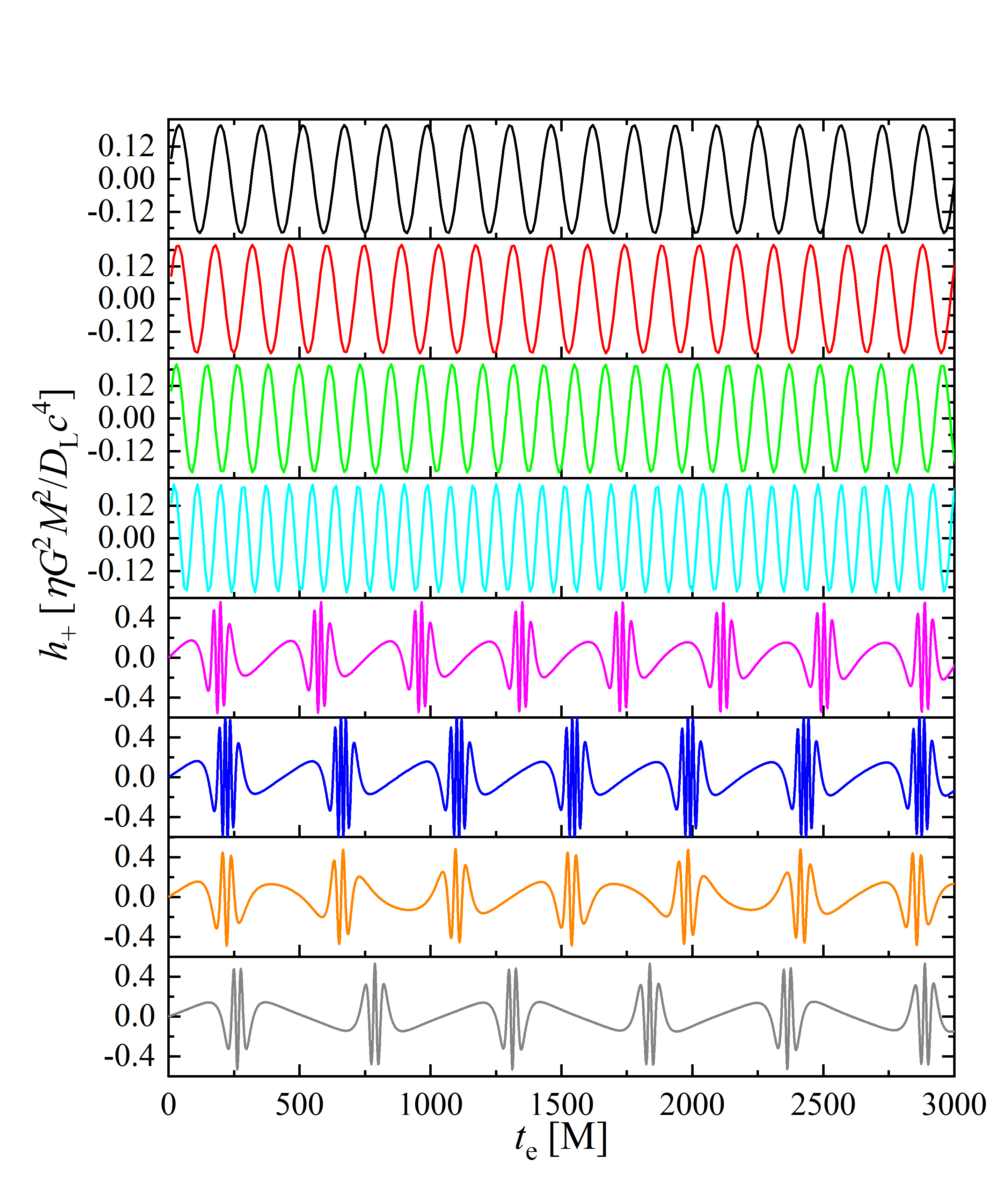}
\includegraphics[width=7cm]{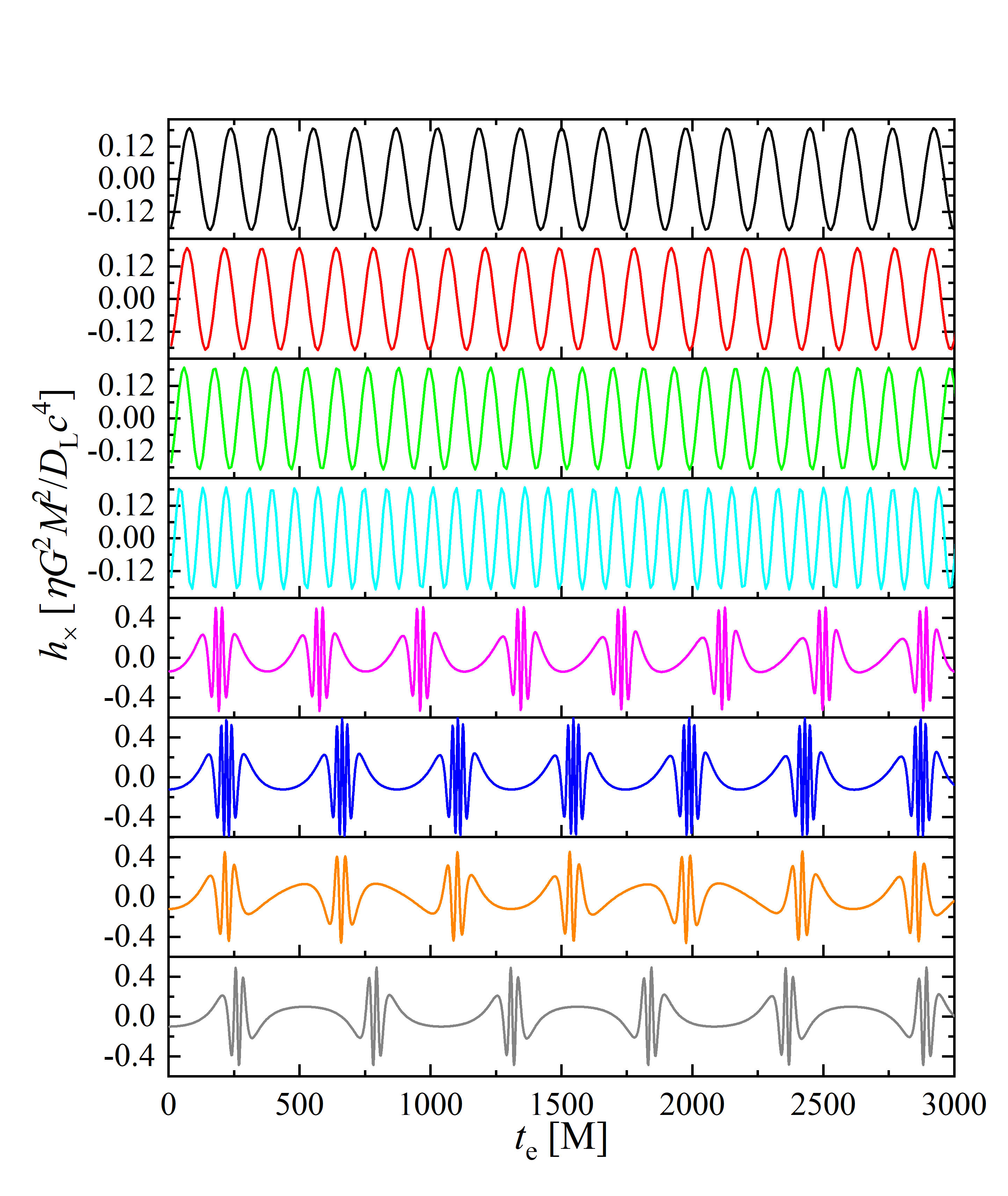}
\caption{Gravitational waveforms in the plus (left) and cross (right) polarizations for different orbits. From top to bottom: Orbits 1--8. The waveforms of circular orbits (Orbits 1--4) exhibit distinct periodic patterns, with their frequencies increasing as $r_{\textrm{s}}$ increases. In contrast, waveforms from quasi-periodic orbits display abrupt variations, primarily due to the inner orbital segments. Furthermore, gravitational wave signals can effectively distinguish between circular and quasi-periodic orbits.}}\label{fig22}
\end{figure*}

We note that while gravitational waves can distinguish between standard circular orbits and precessing trajectories, extracting detailed parameters of precessing orbits---such as the number of leaves, whirls, and vertices---from the waveform remains challenging. Figures 19--22 demonstrate that OCTOPUS is capable of simulating both light curves and gravitational wave emissions, highlighting the algorithm's promising potential for studying multi-messenger correlations between gravitational radiation and electromagnetic signals.
\begin{figure*}%[tbph]
\center{
\includegraphics[width=2.8cm]{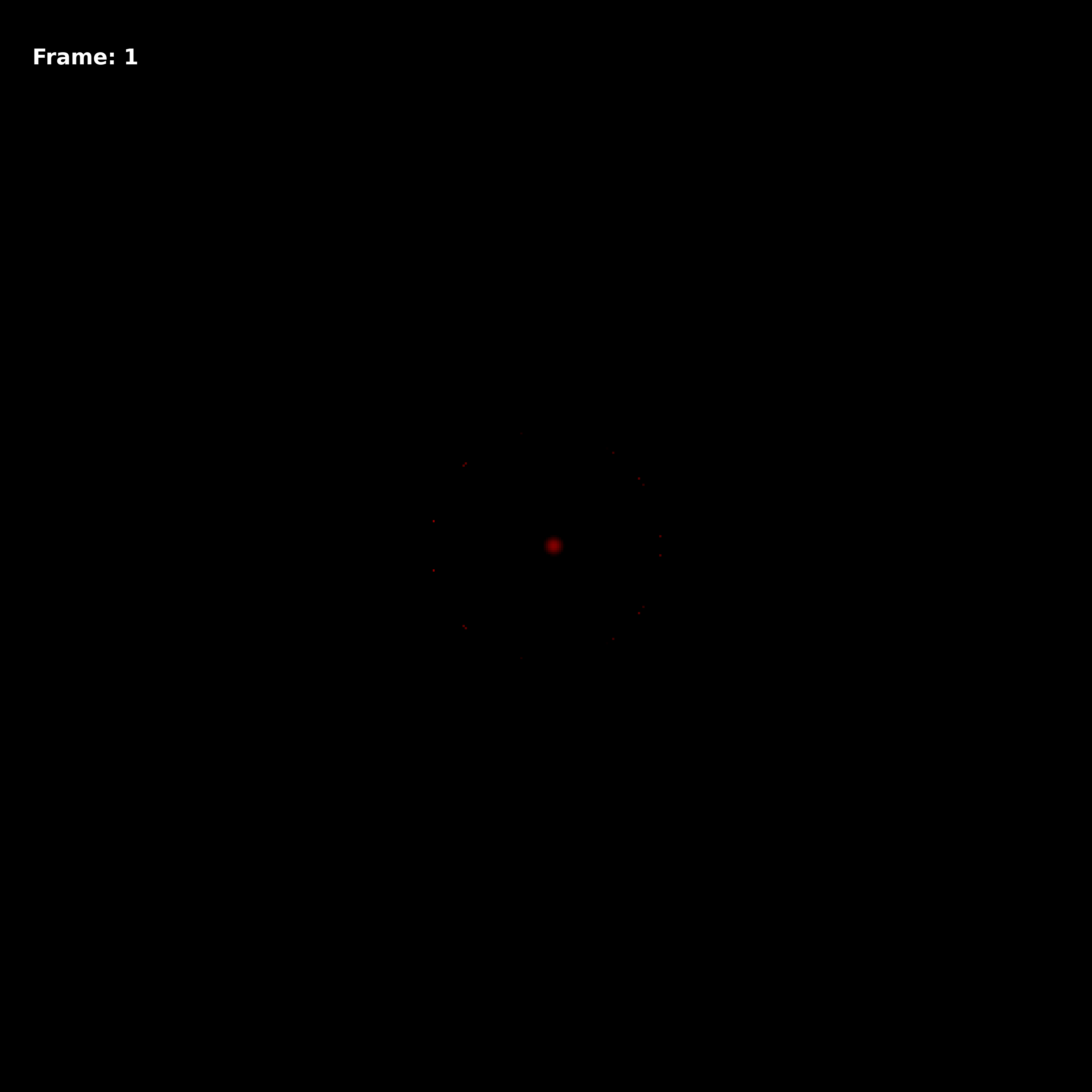}
\includegraphics[width=2.8cm]{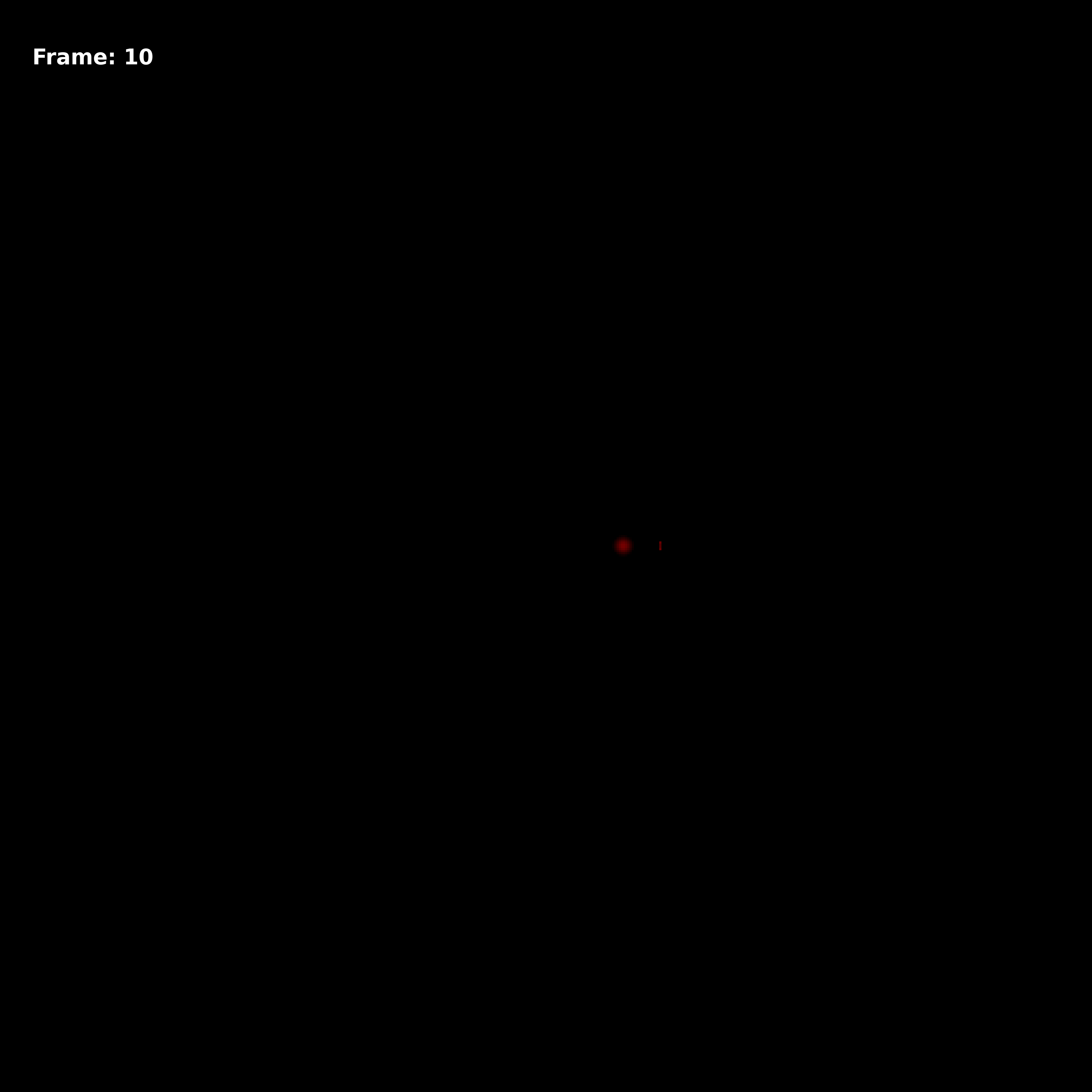}
\includegraphics[width=2.8cm]{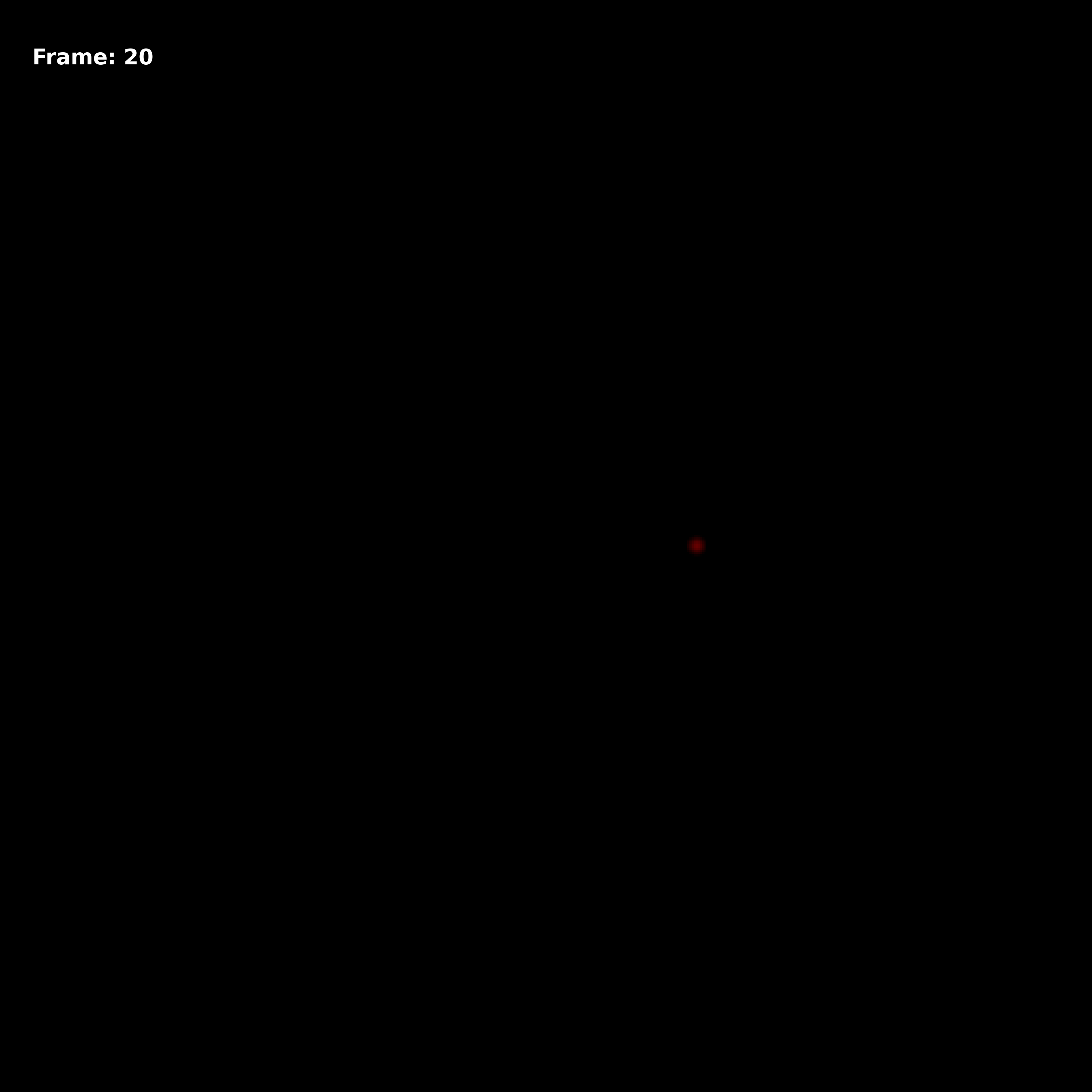}
\includegraphics[width=2.8cm]{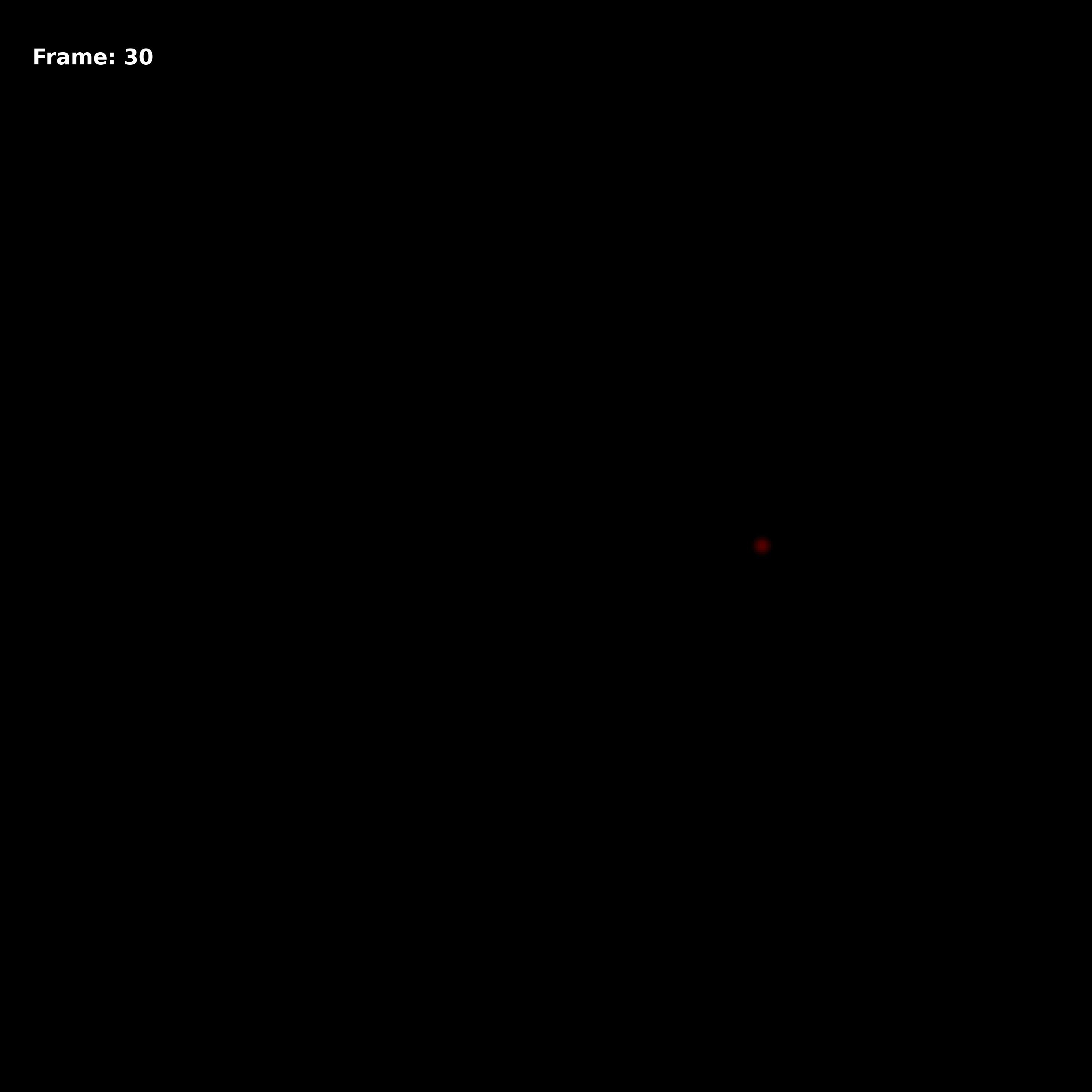}
\includegraphics[width=2.8cm]{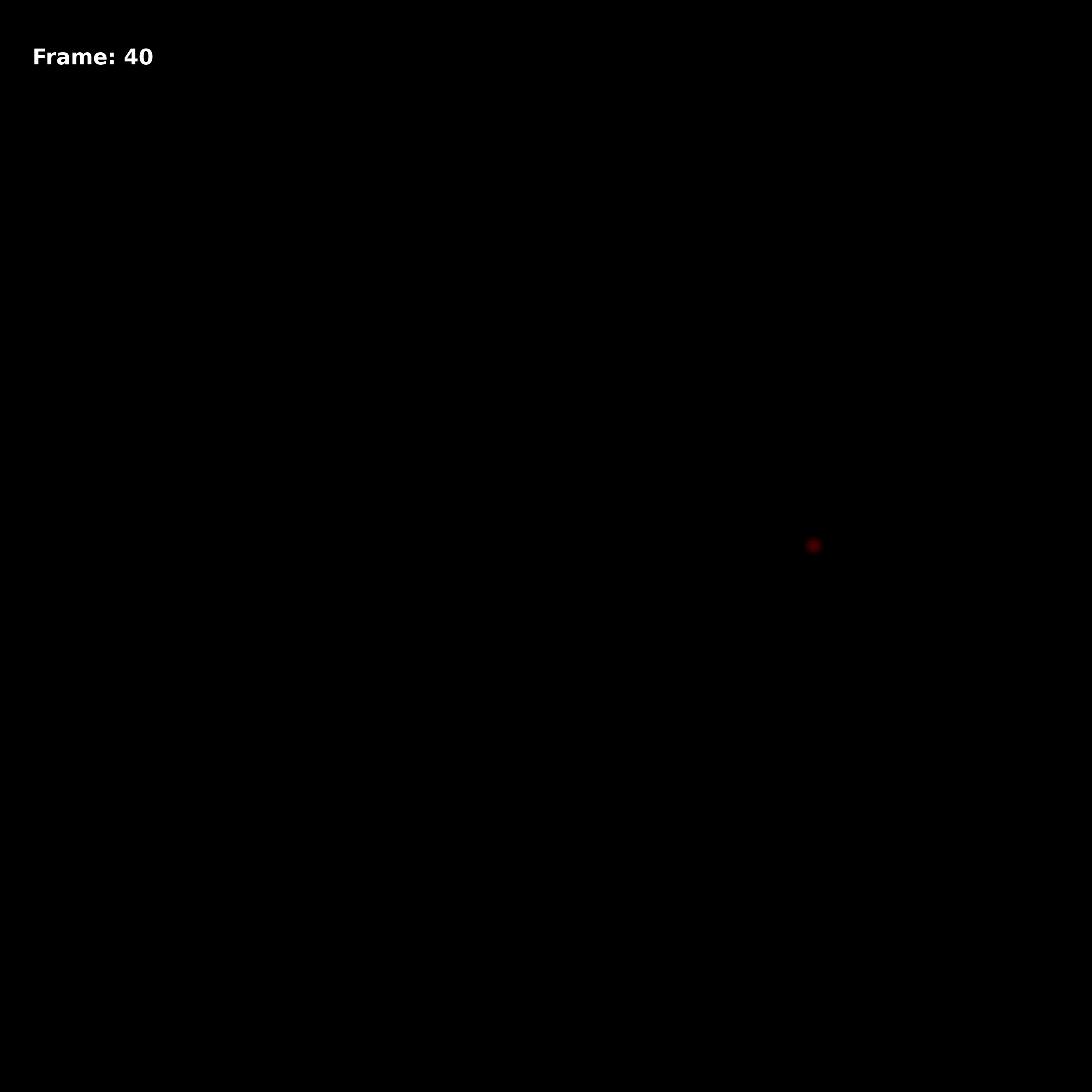}
\includegraphics[width=2.8cm]{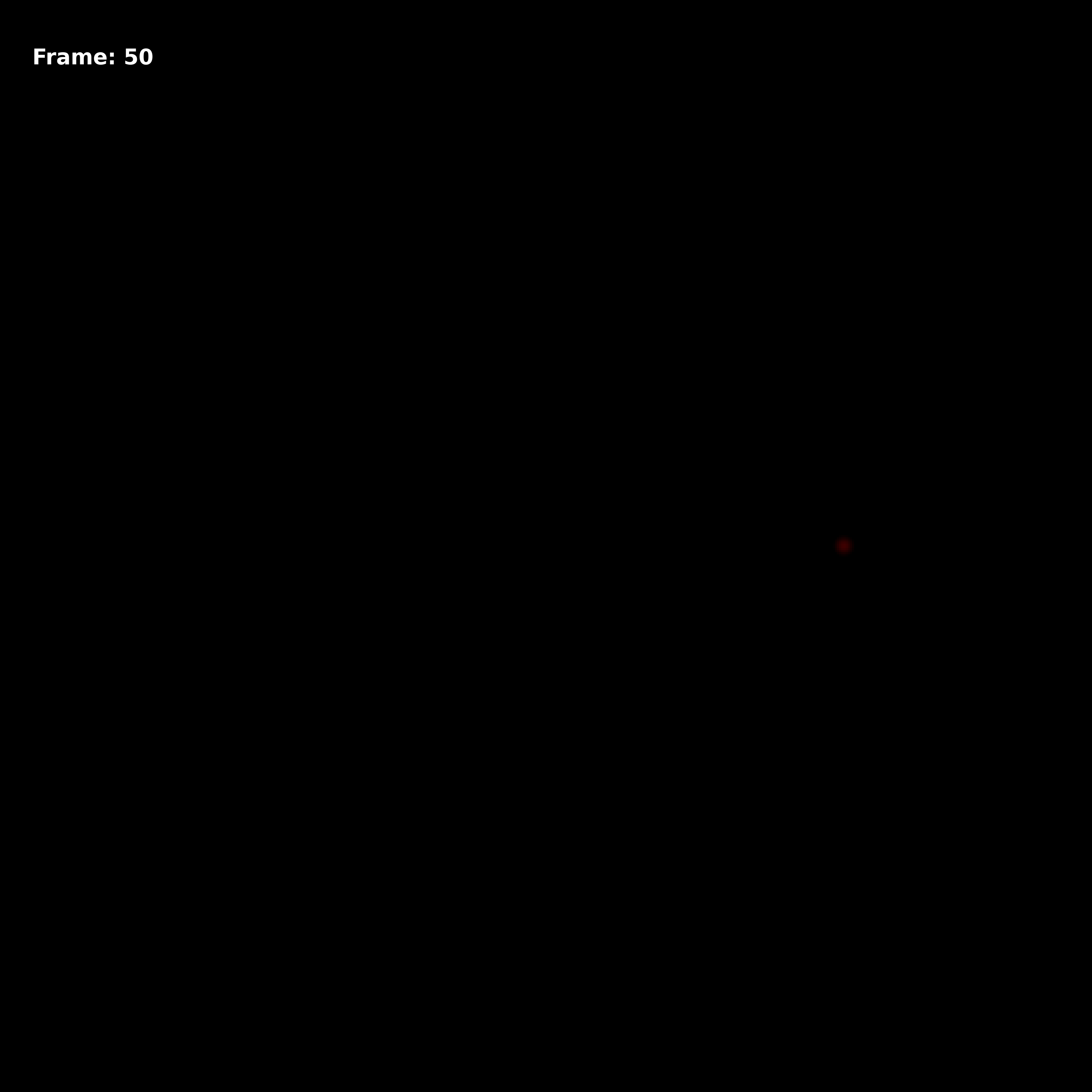}
\includegraphics[width=2.8cm]{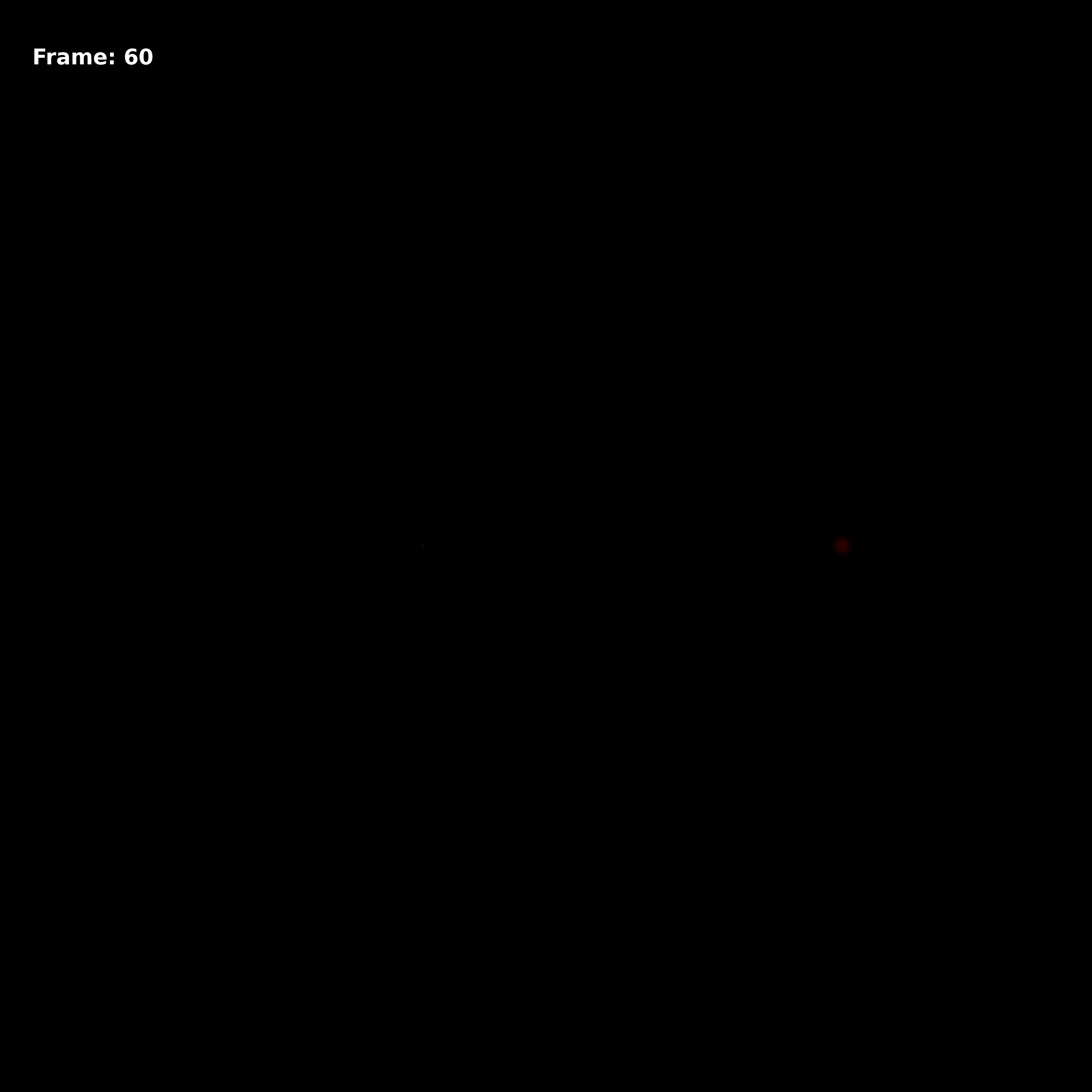}
\includegraphics[width=2.8cm]{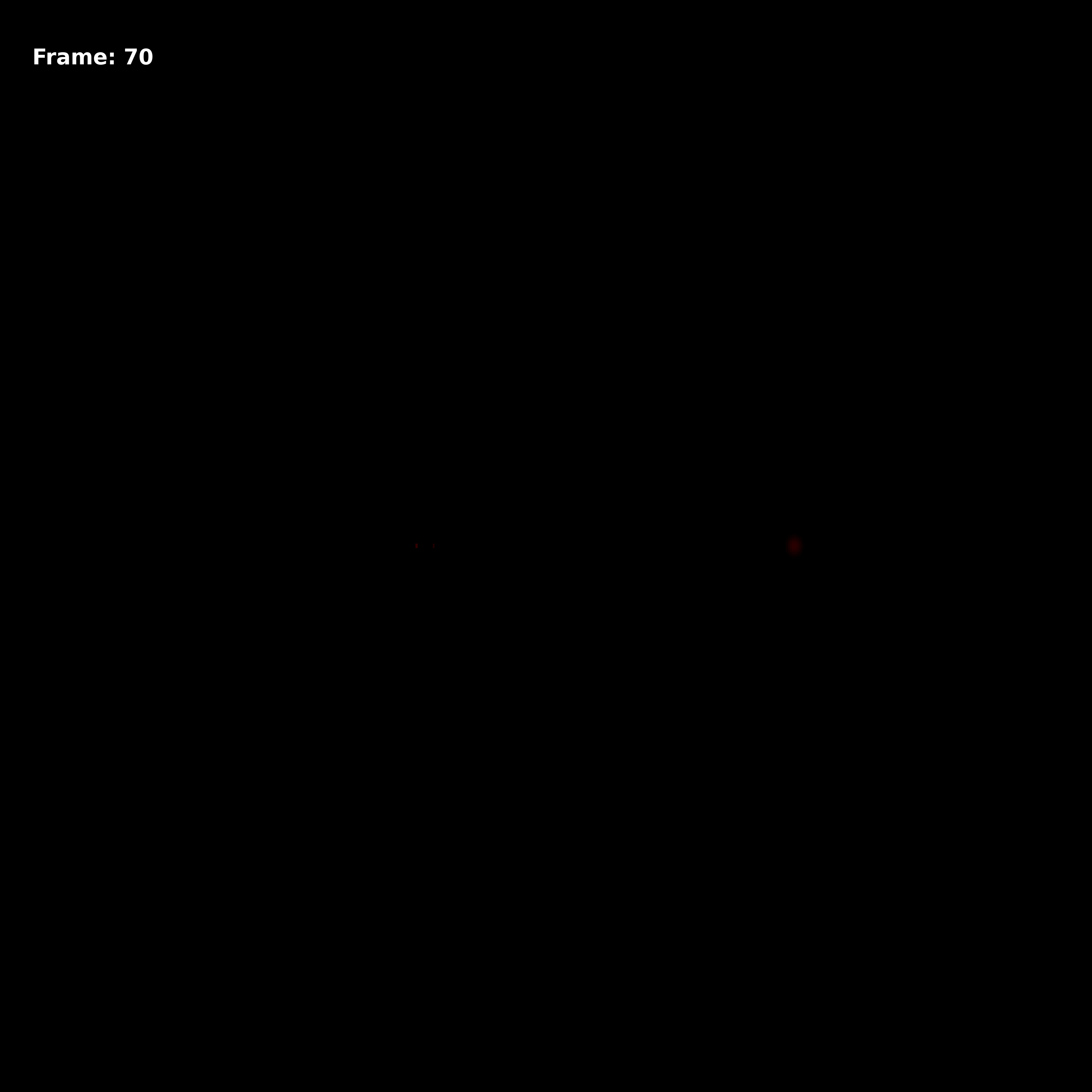}
\includegraphics[width=2.8cm]{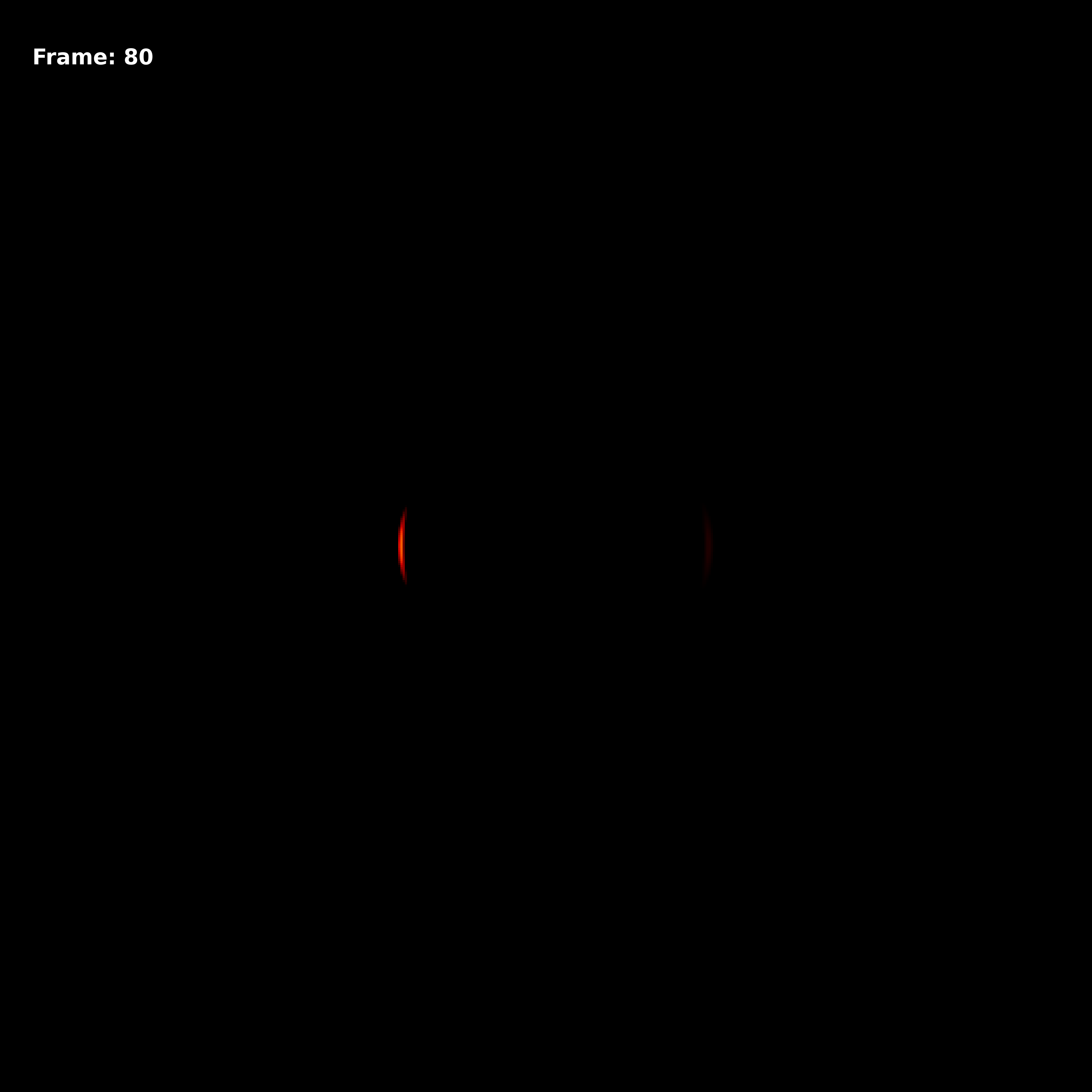}
\includegraphics[width=2.8cm]{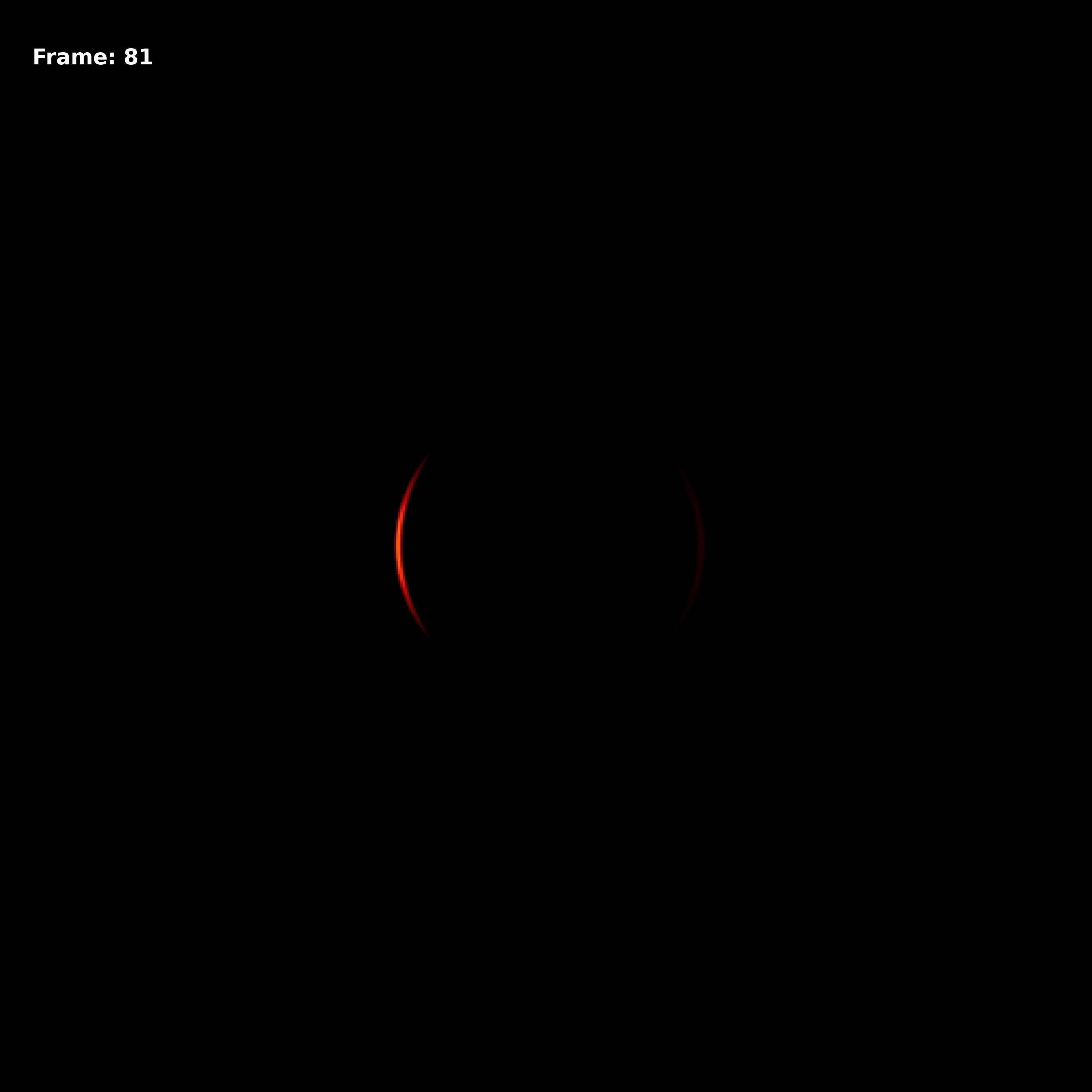}
\includegraphics[width=2.8cm]{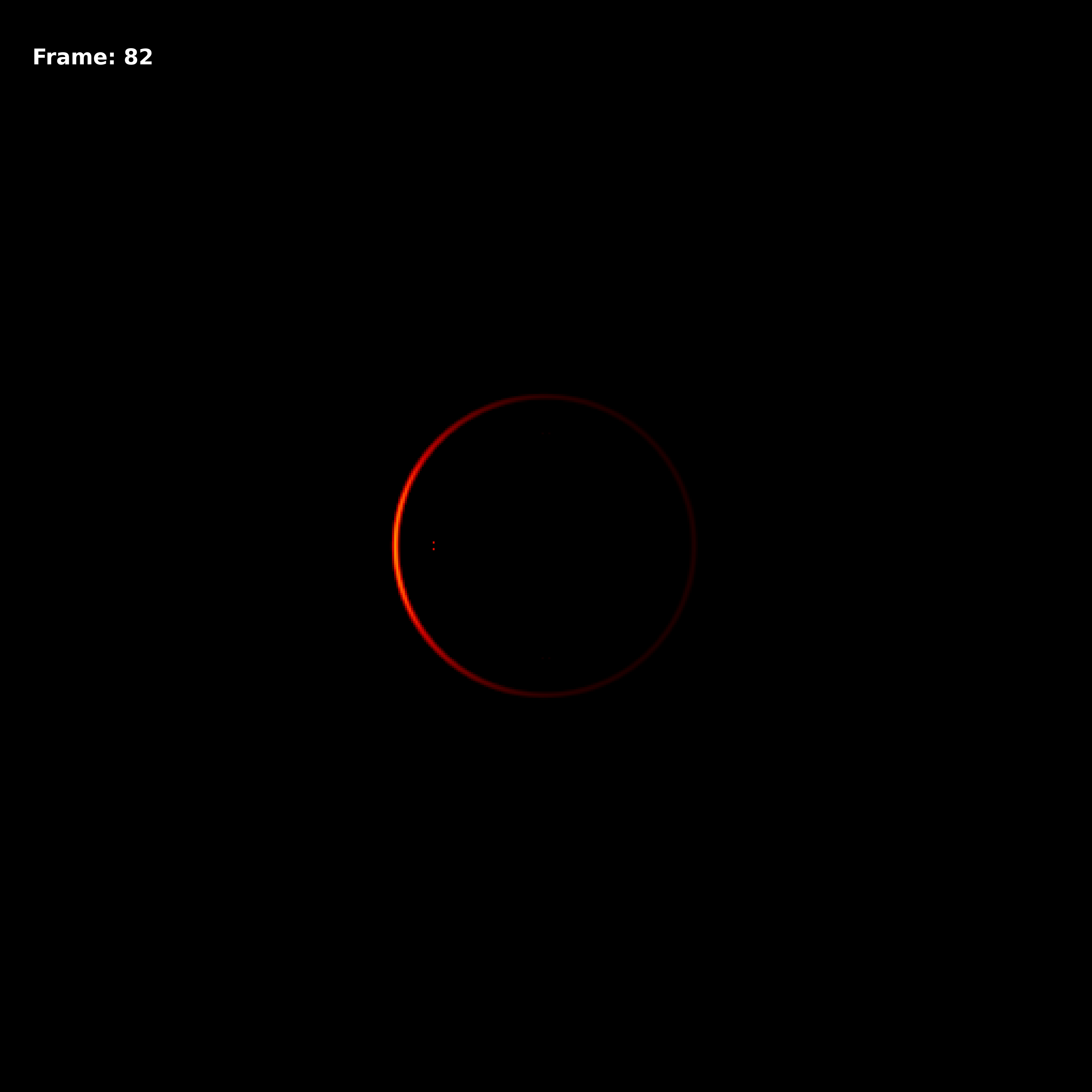}
\includegraphics[width=2.8cm]{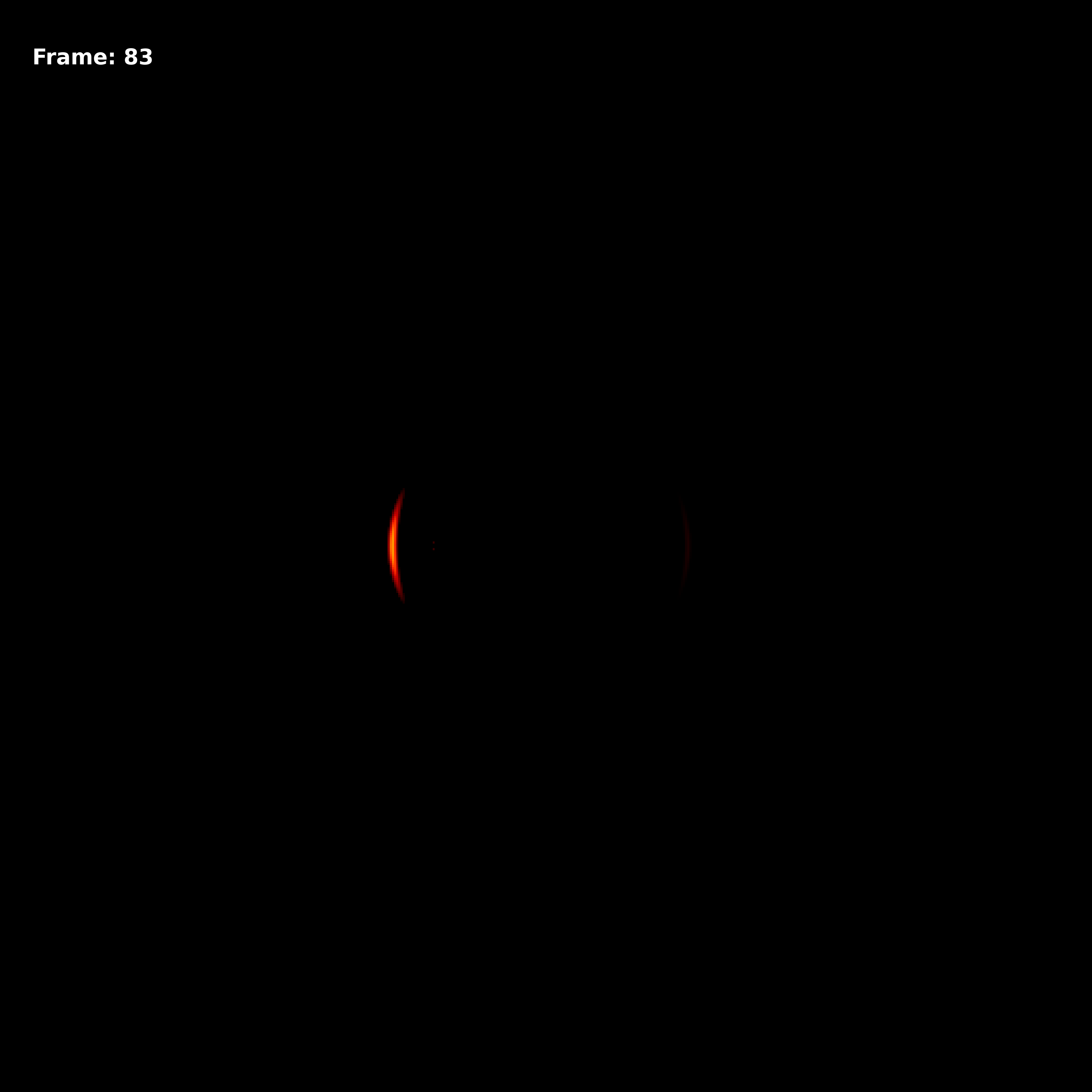}
\includegraphics[width=2.8cm]{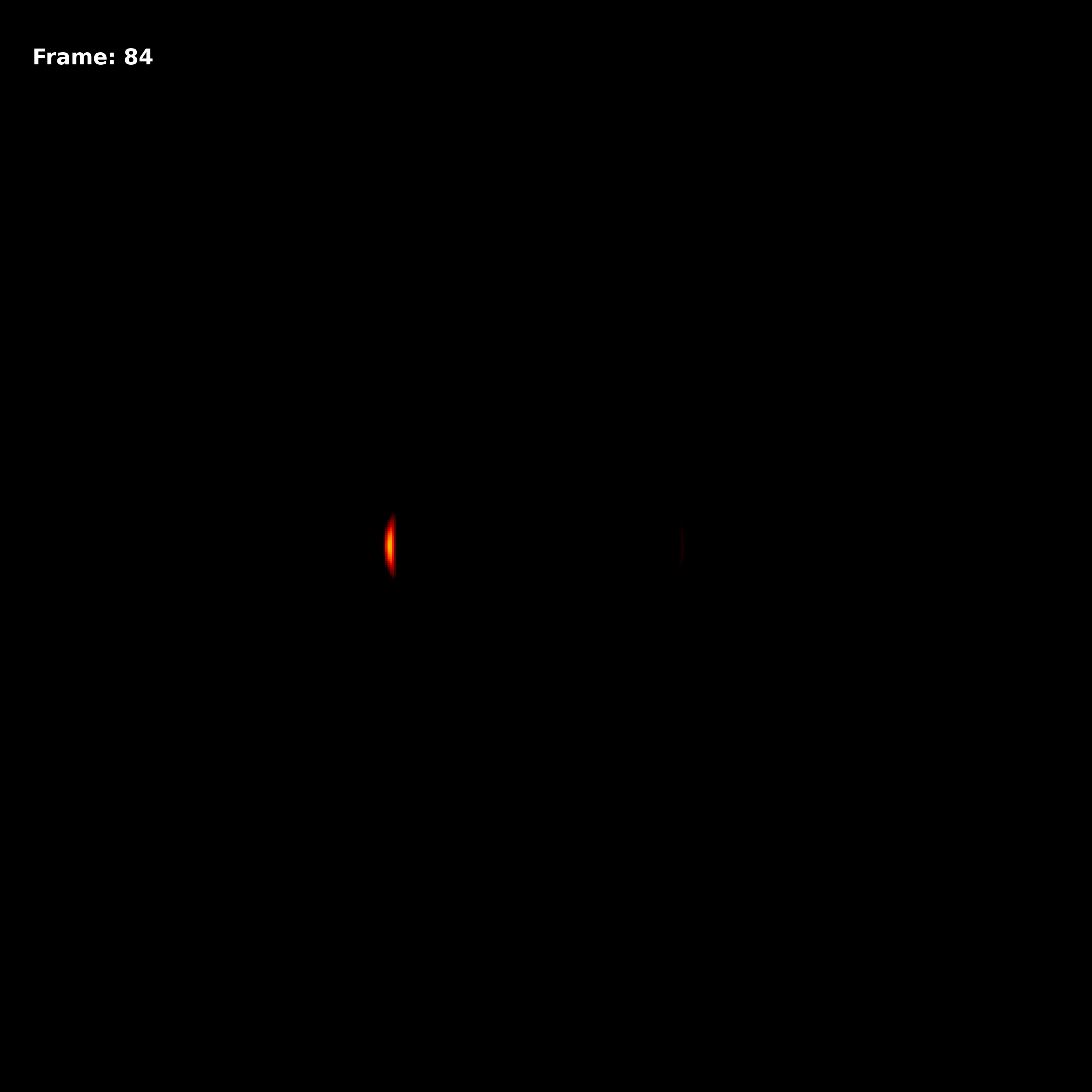}
\includegraphics[width=2.8cm]{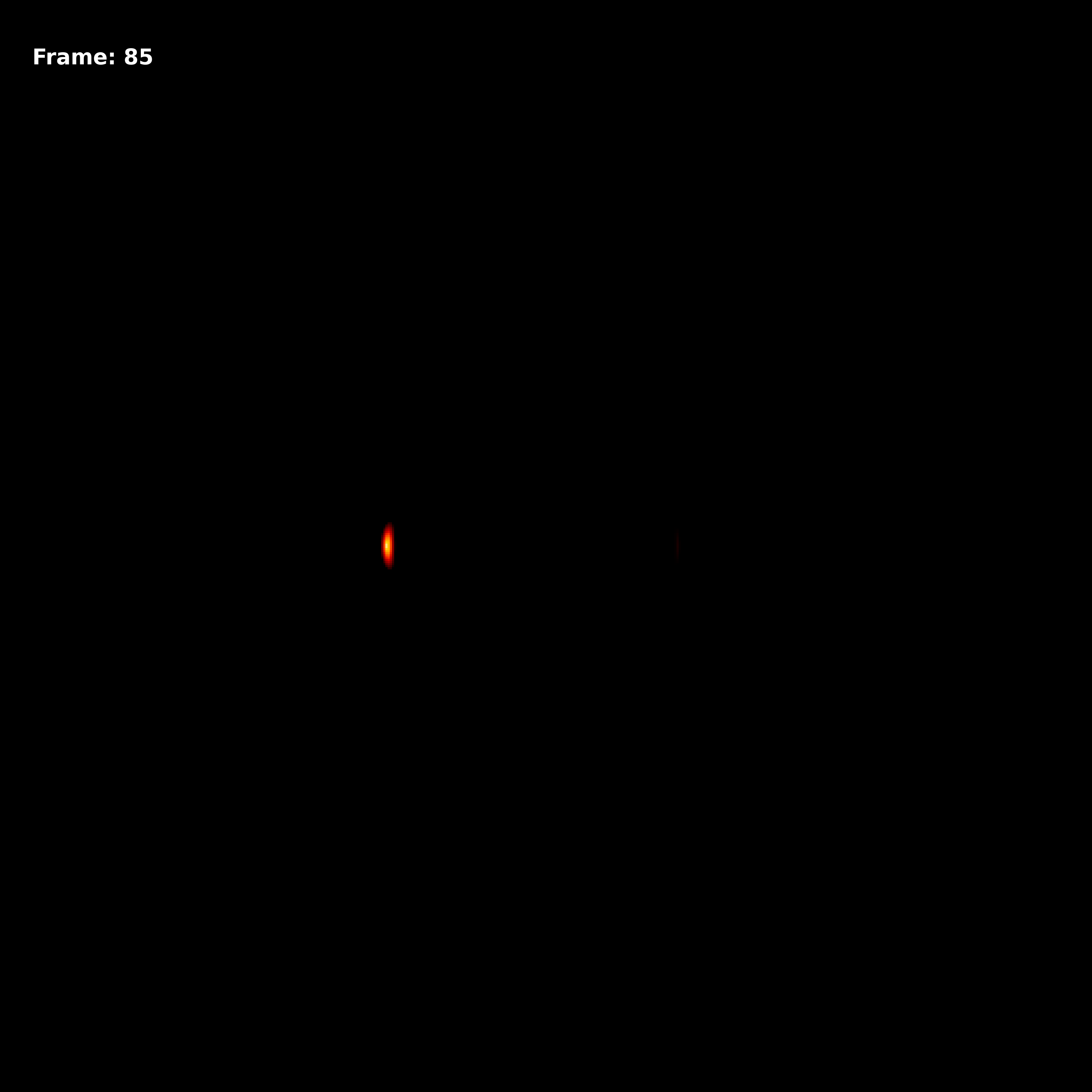}
\includegraphics[width=2.8cm]{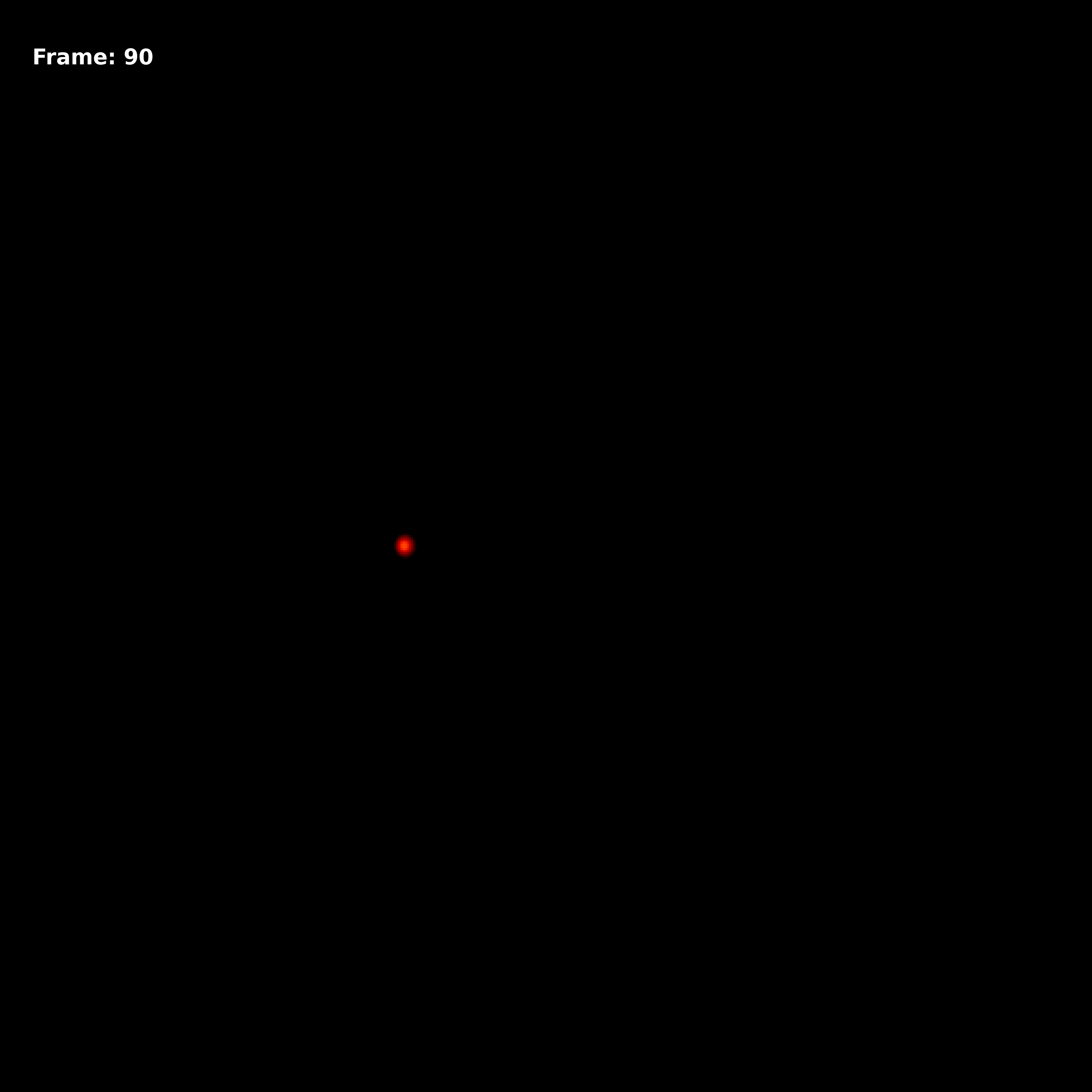}
\includegraphics[width=2.8cm]{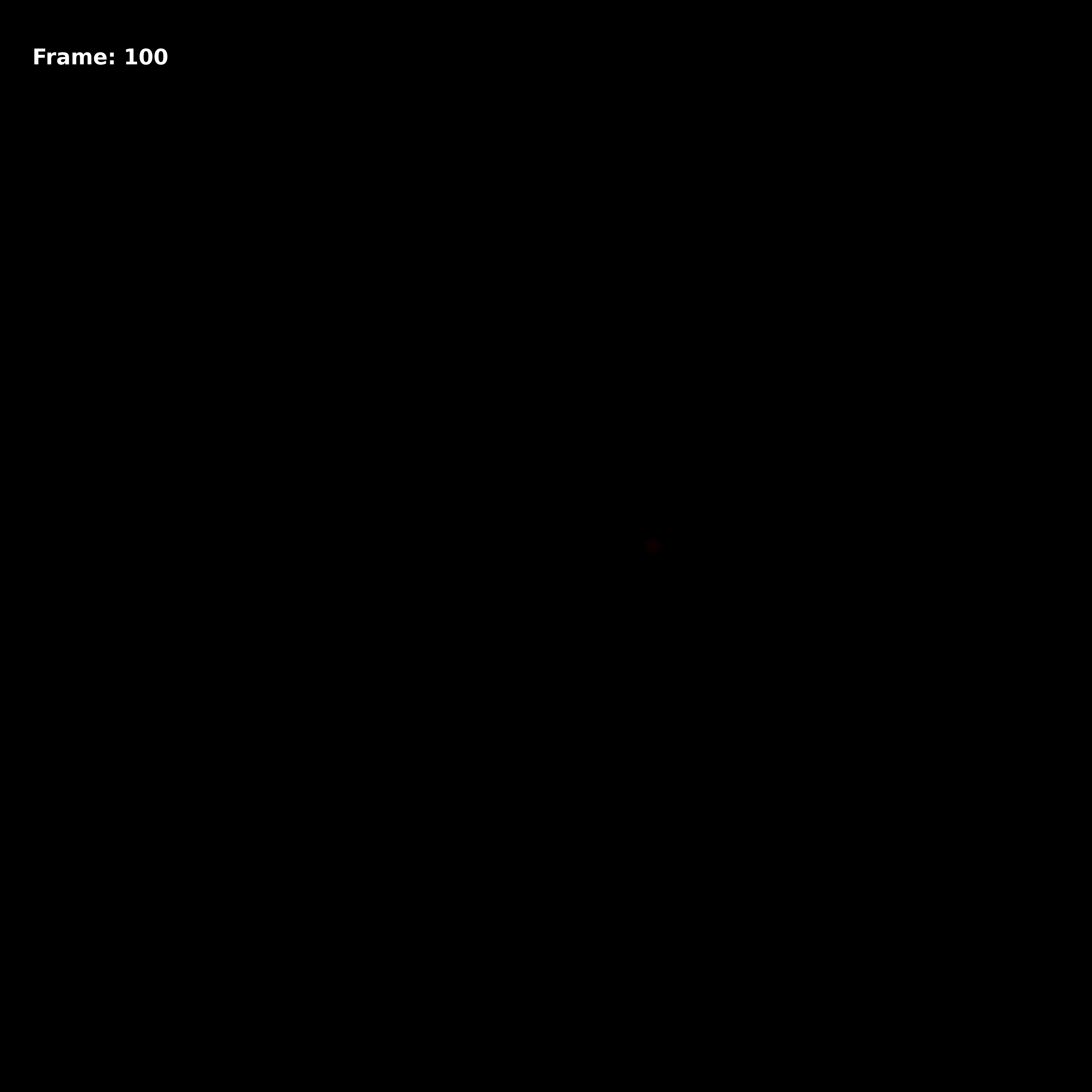}
\includegraphics[width=2.8cm]{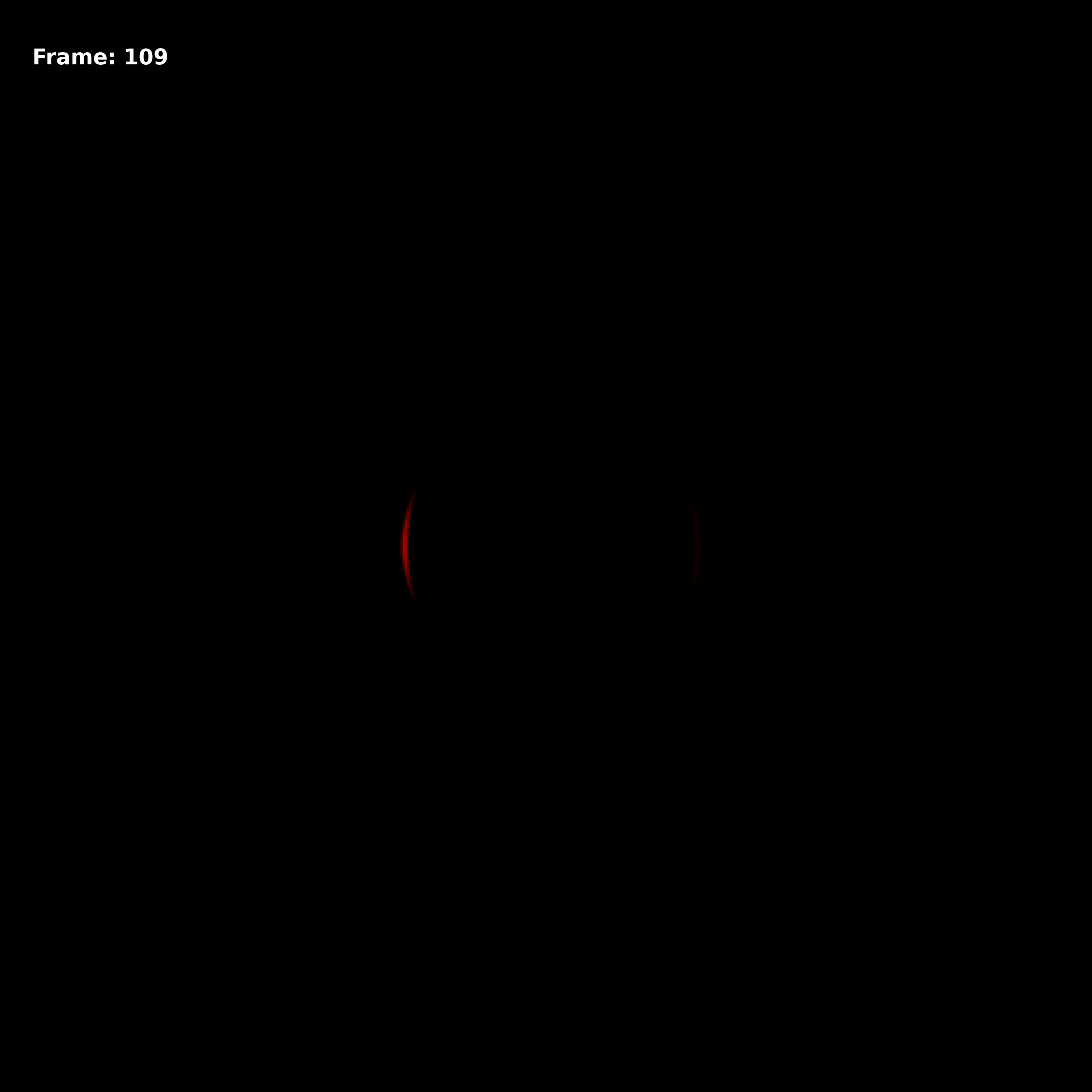}
\includegraphics[width=2.8cm]{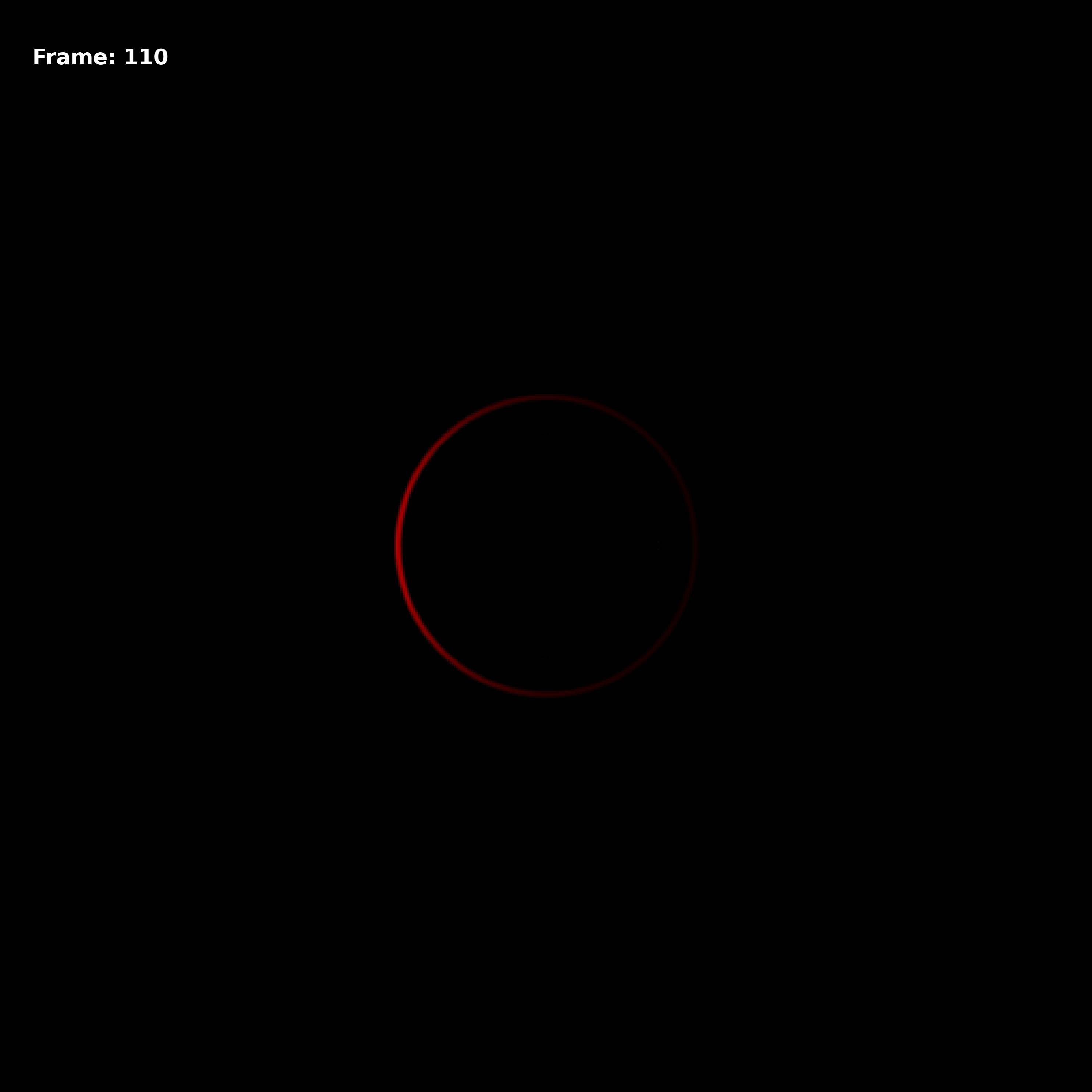}
\includegraphics[width=2.8cm]{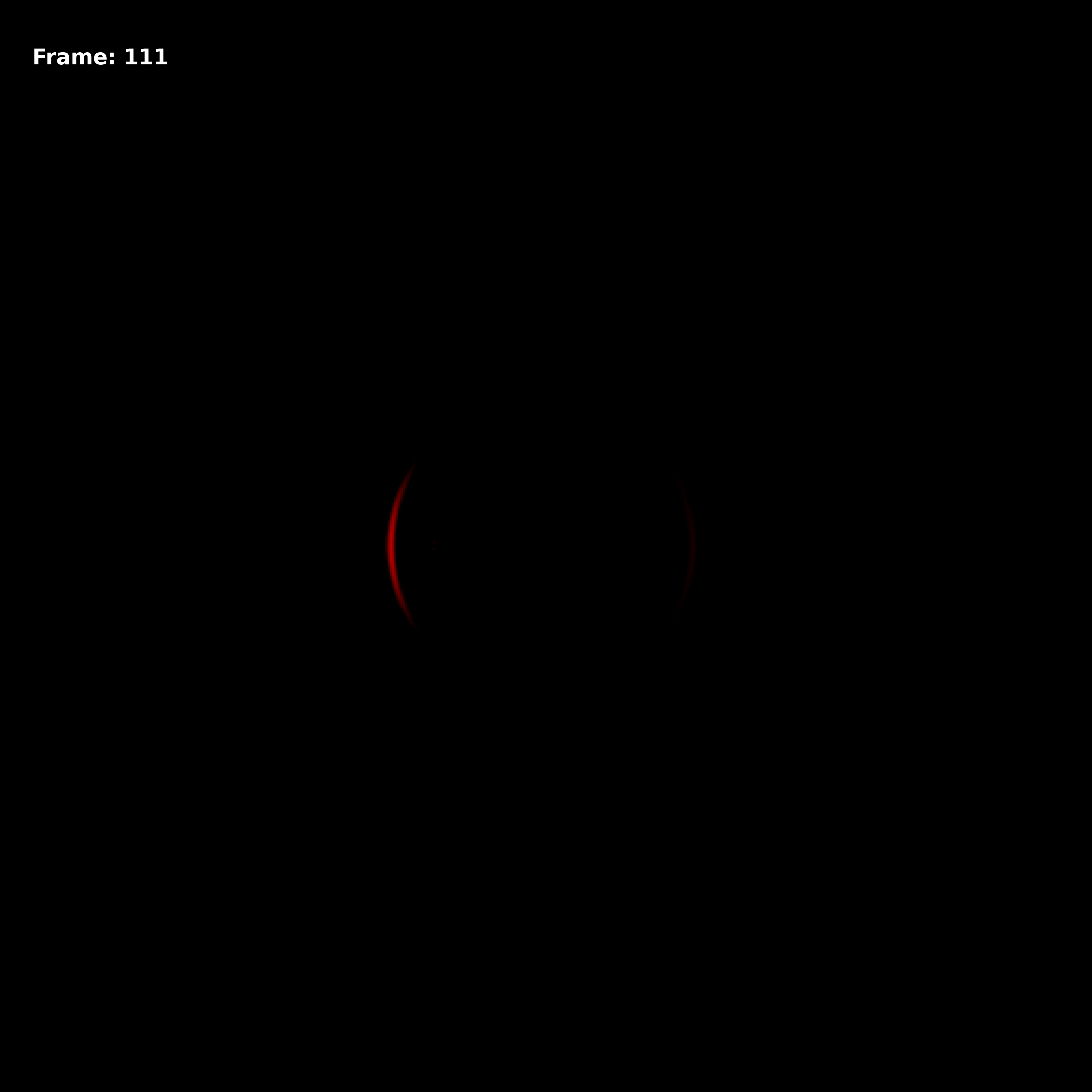}
\includegraphics[width=2.8cm]{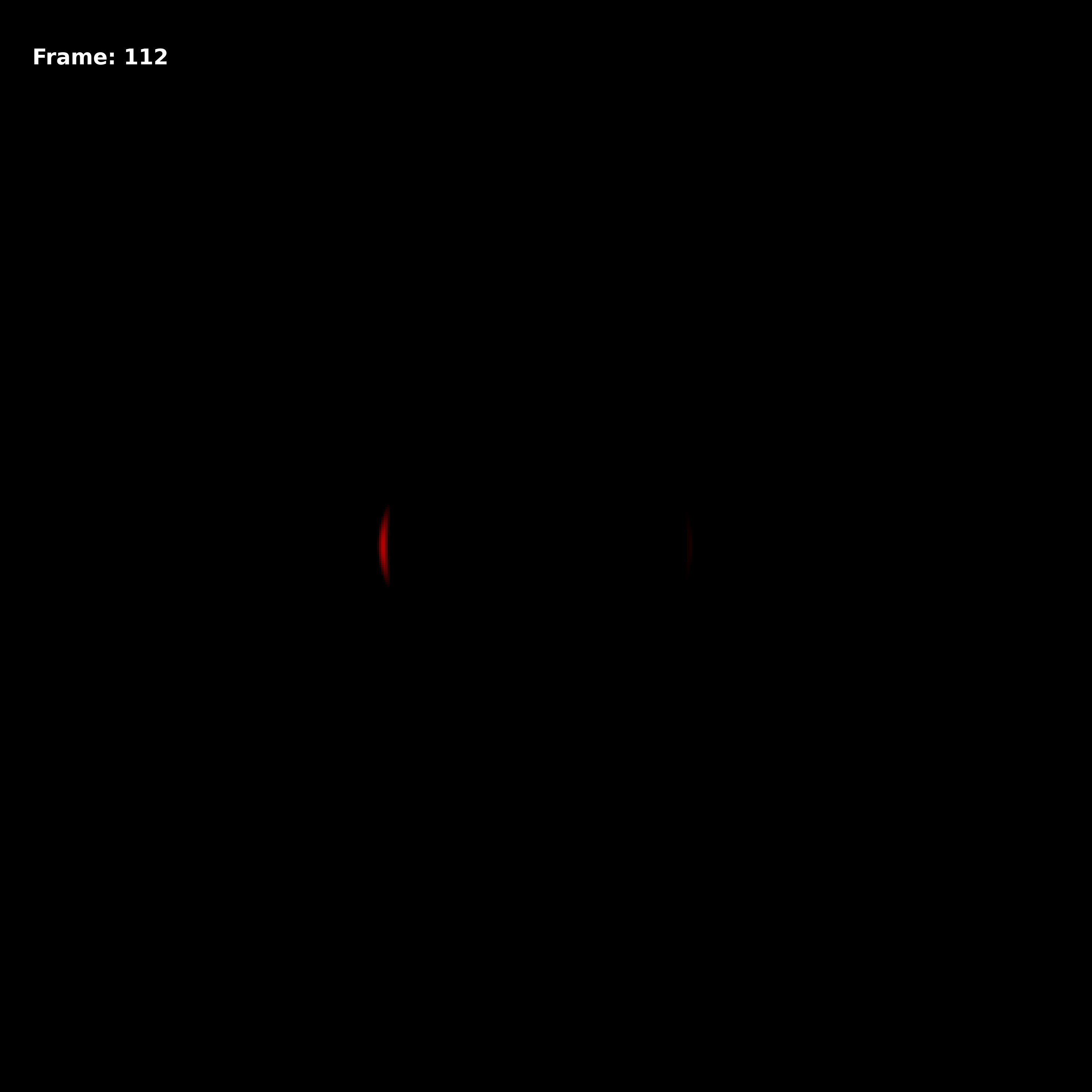}
\includegraphics[width=2.8cm]{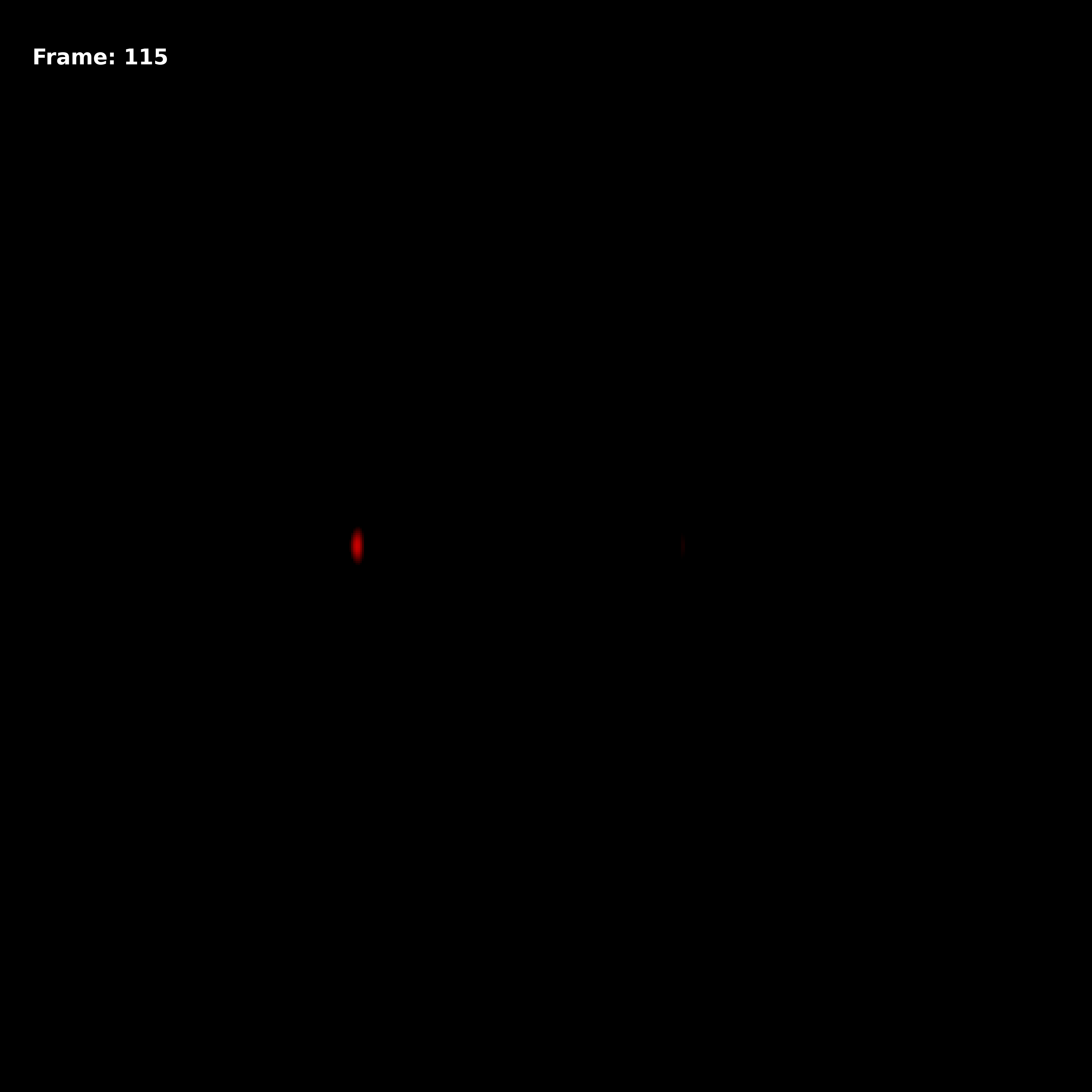}
\includegraphics[width=2.8cm]{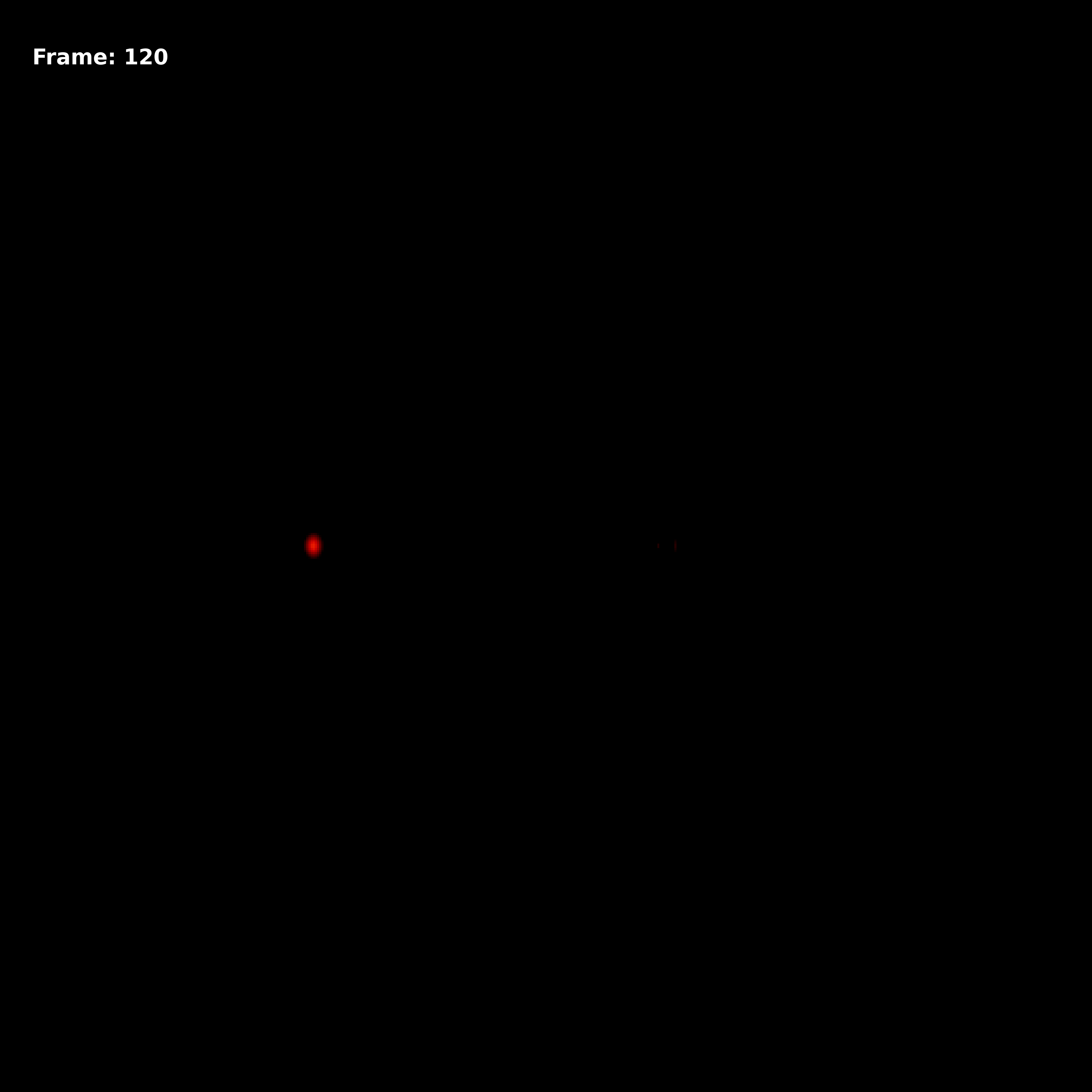}
\includegraphics[width=2.8cm]{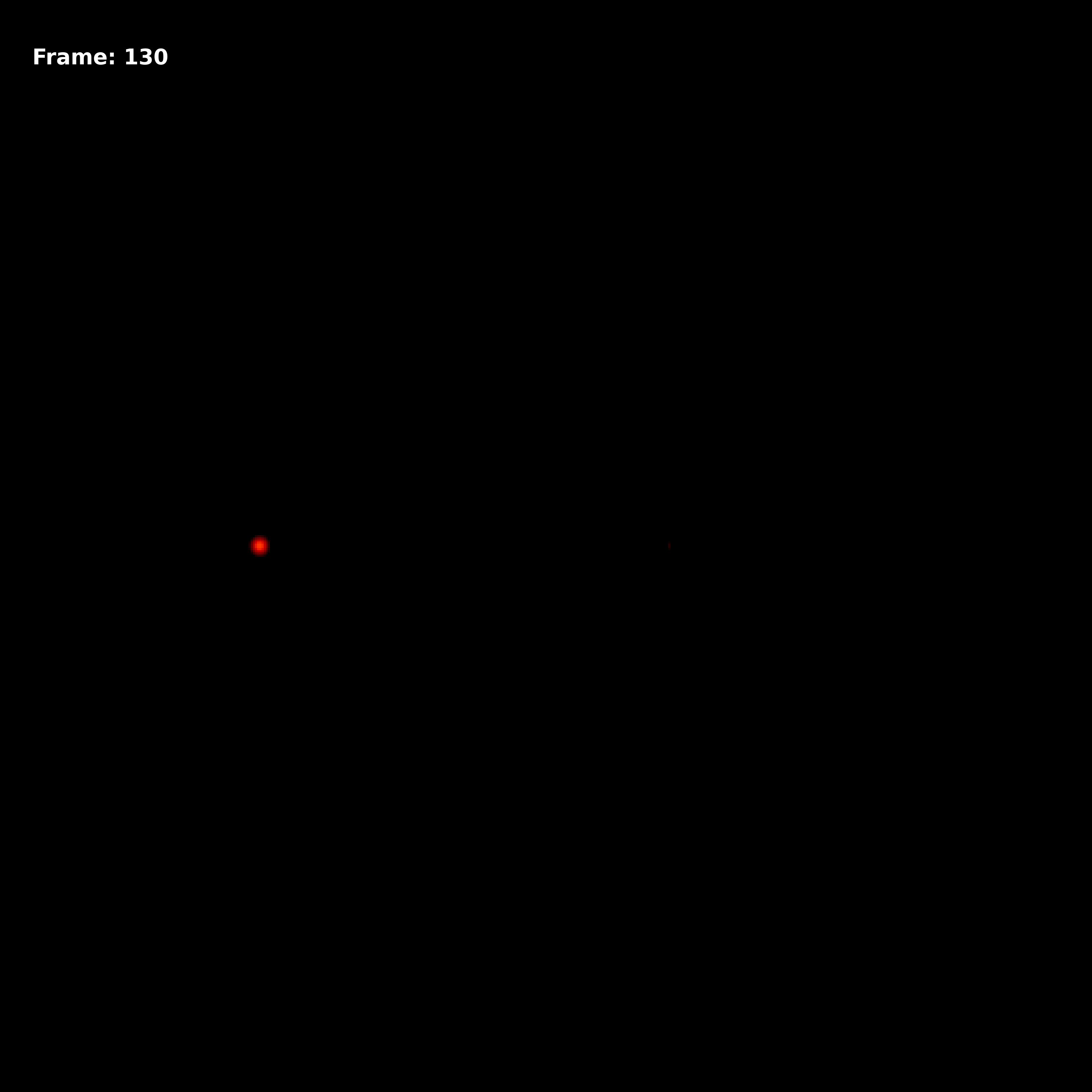}
\includegraphics[width=2.8cm]{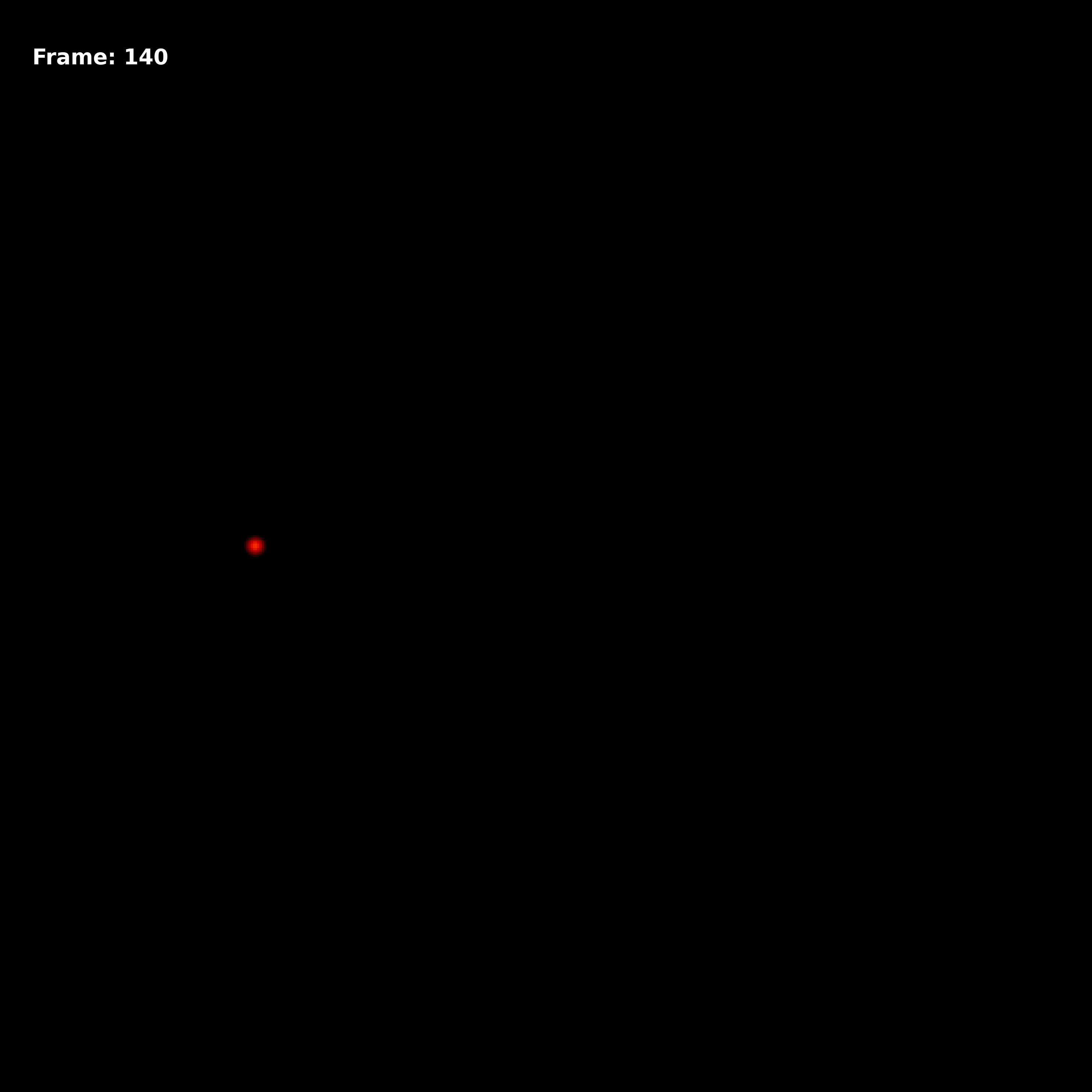}
\includegraphics[width=2.8cm]{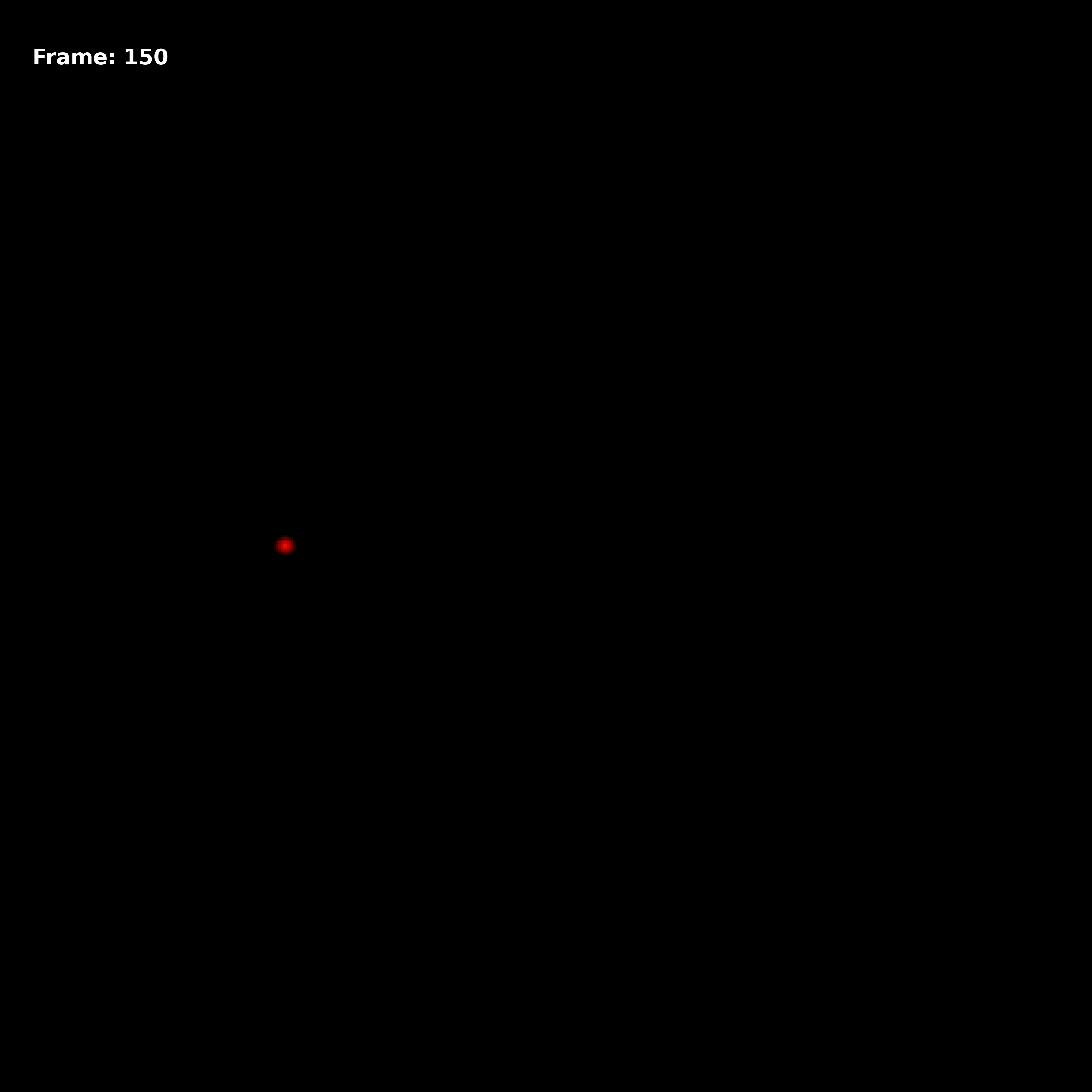}
\includegraphics[width=2.8cm]{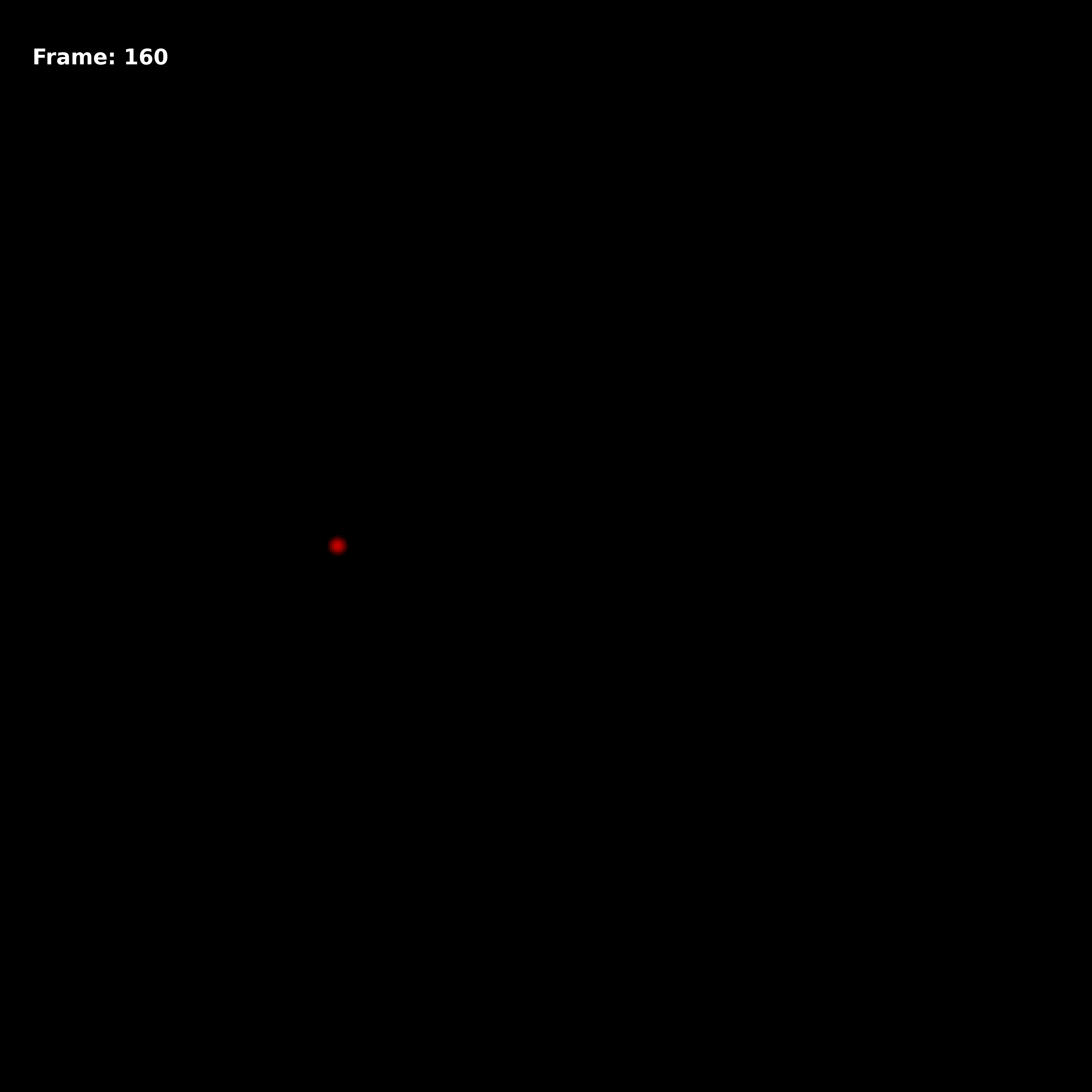}
\includegraphics[width=2.8cm]{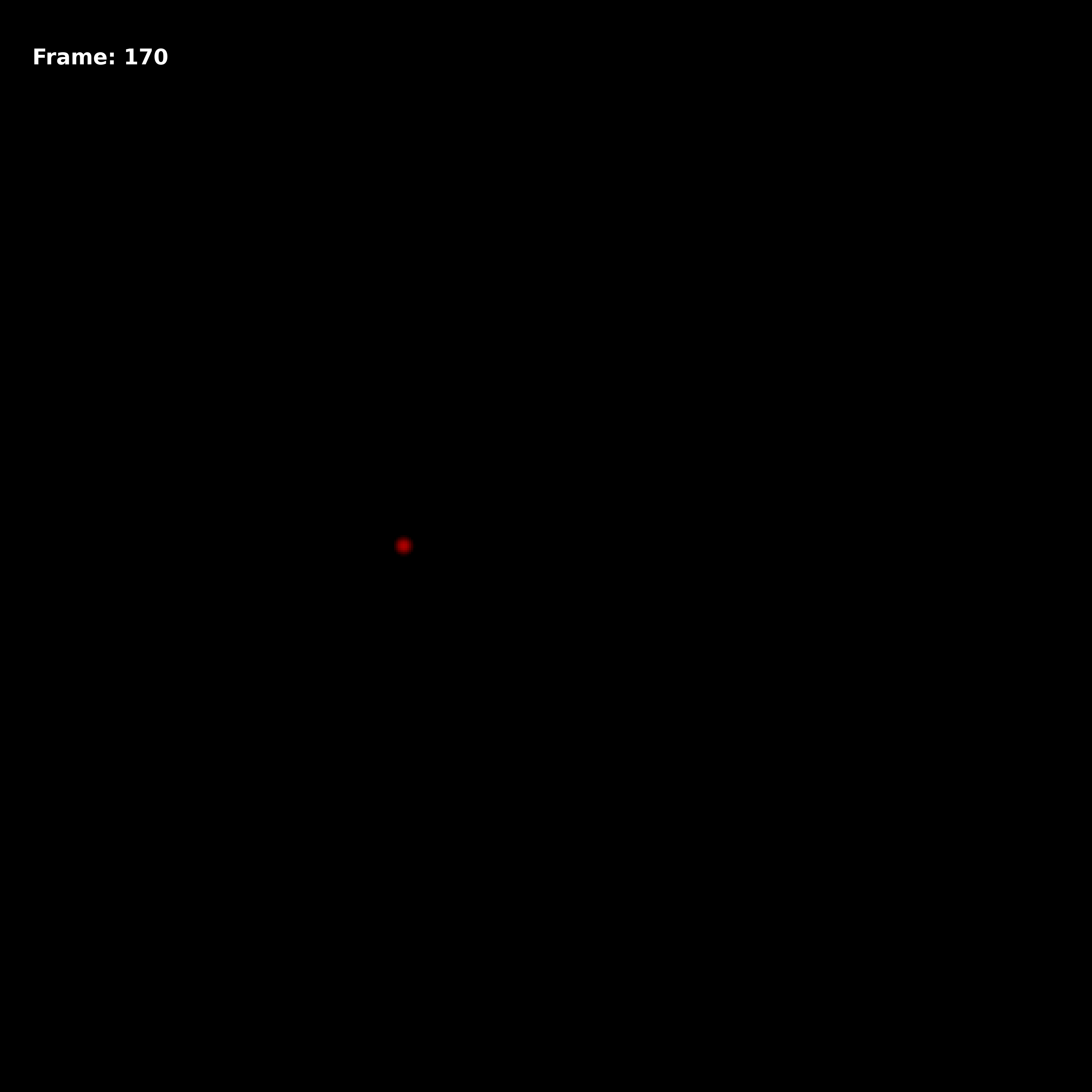}
\includegraphics[width=2.8cm]{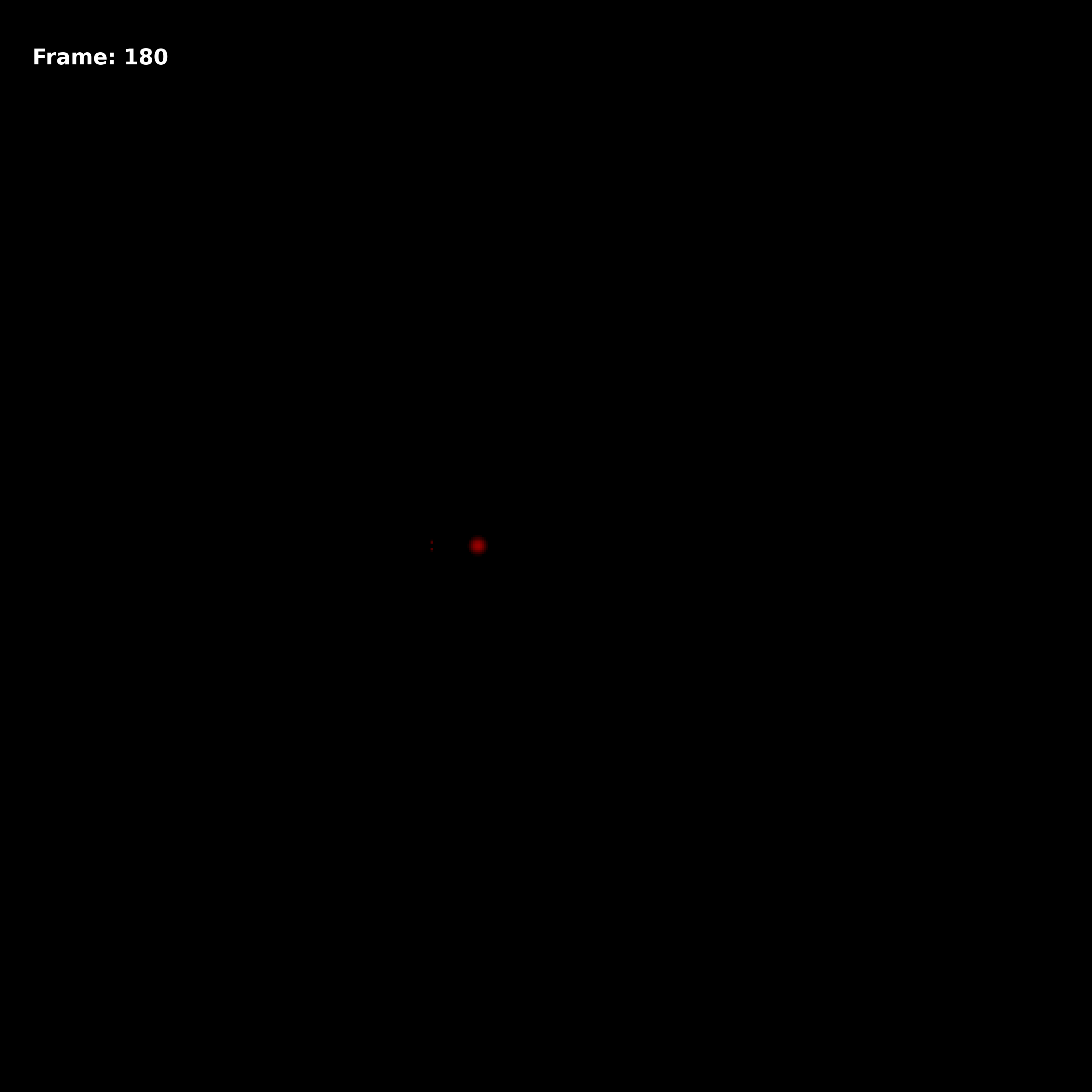}
\includegraphics[width=2.8cm]{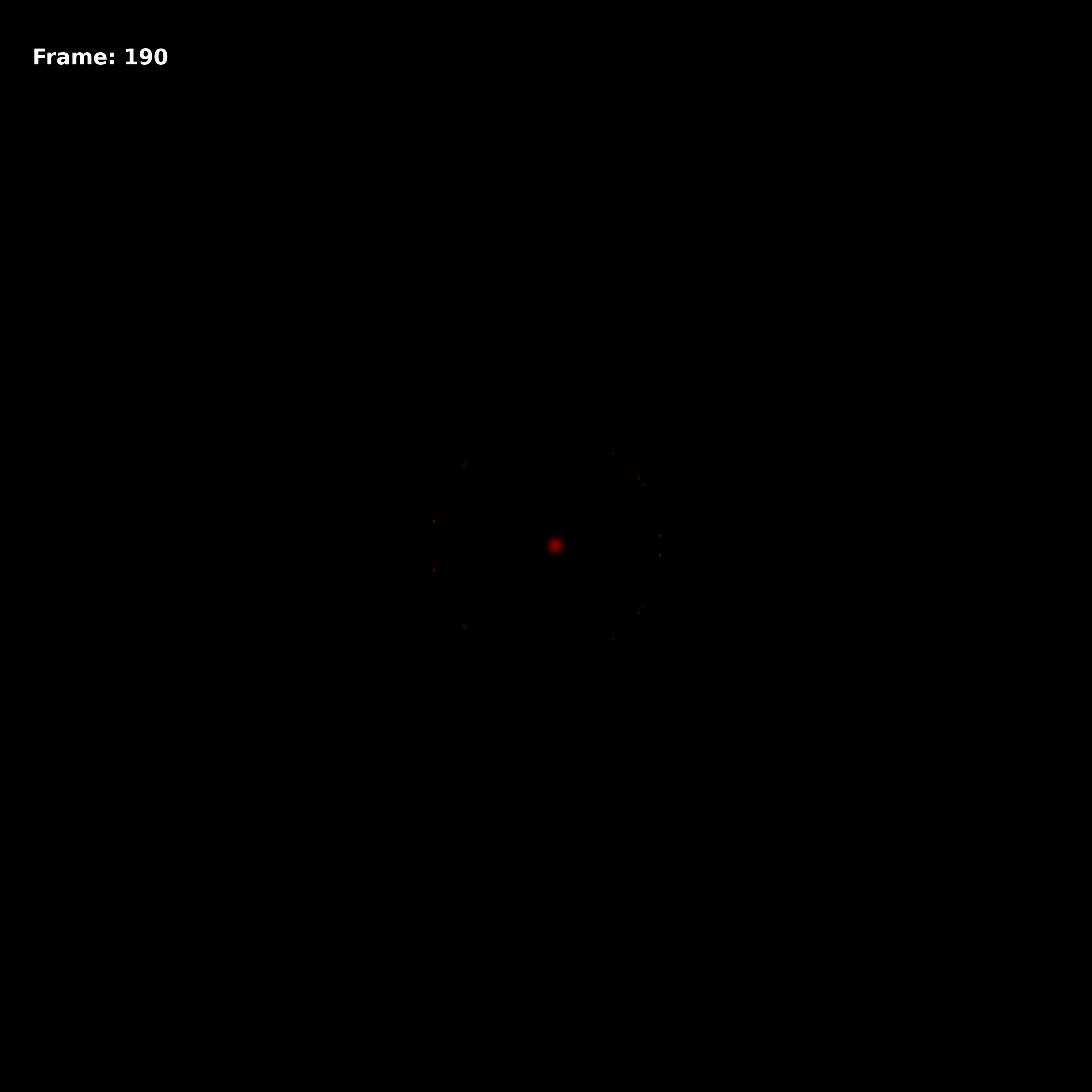}
\includegraphics[width=2.8cm]{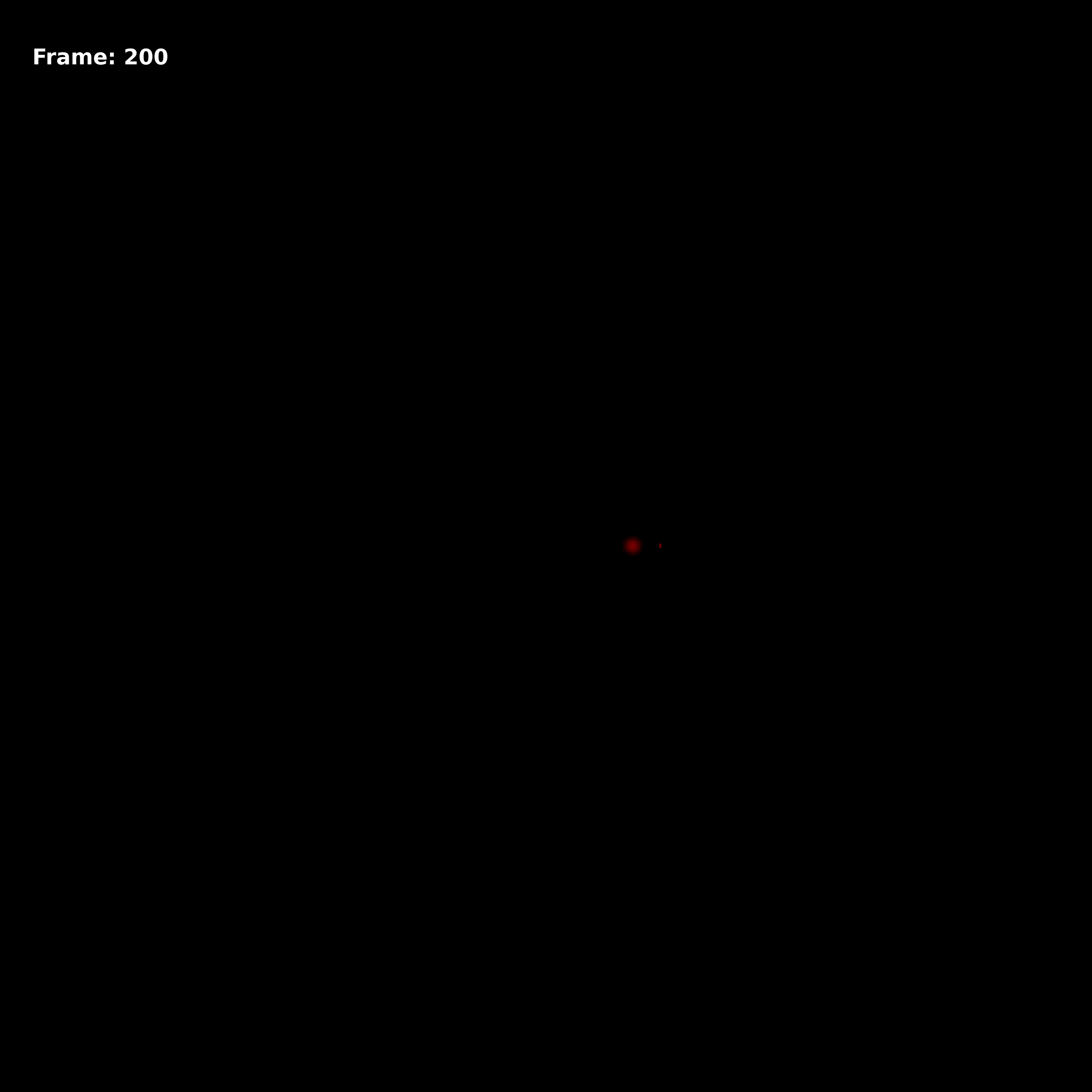}
\caption{Snapshot sequence from an animation of a hot-spot moving along Orbit 5. Each frame has a resolution of $500 \times 500$ pixels, with the observation inclination fixed at $90^{\circ}$, and the hot-spot evolution time set to $2000$ M. This sequence clearly captures two distinct ring formation events. Moreover, generating such animations provides an effective approach for studying flaring phenomena around compact objects.}}\label{fig28}
\end{figure*}

Finally, we use Orbit 5 as an example to demonstrate the algorithm's ability to generate dynamic hot-spot images. Figure 23 presents snapshots from the hot-spot animation at different time steps. It can be observed that the hot-spot image initially moves toward the right side of the field of view, gradually dimming as the source recedes from the observer. By frame 70 (second row, third column from the left), the brightness becomes nearly undetectable. Subsequently, as the hot-spot moves behind the black hole, a bright crescent-shaped image emerges slightly left of the field center. This structure eventually evolves into an asymmetric ring, exhibiting a left-right brightness contrast (frame 82, third row, first column). The hot-spot then follows an inner orbital segment back behind the black hole, producing a second bright ring (frame 110, fourth row, third column). Finally, the hot-spot returns to the outer orbital section, completing one cycle of its motion. This sequence of snapshots can be fully interpreted with the aid of figure 24, where bubbles of increasing radius indicate the position of the hot-spot in the black hole's local coordinate system for each frame shown in figure 23. The hot-spot moves counterclockwise along the gray trajectory. When it passes behind the black hole, its image forms a ring; when it travels along inner orbital segments, the image appears crescent-shaped. Additionally, the brightness variations caused by the hot-spot's recession from or approach toward the observer are straightforward to interpret.
\begin{figure*}%[tbph]
\center{
\includegraphics[width=6cm]{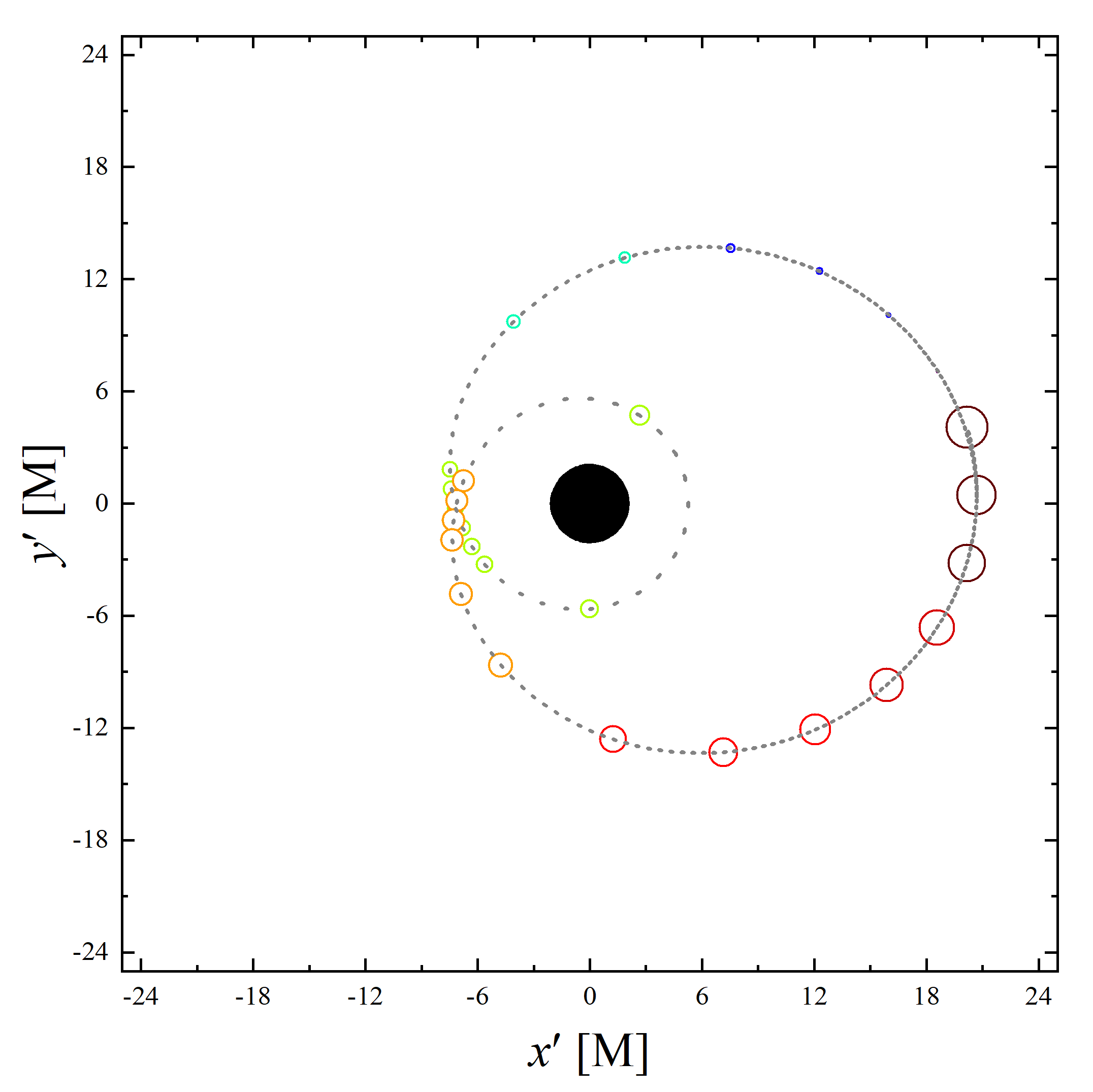}
\caption{Circles of increasing size indicate the hot-spot positions corresponding to the frame sequence in figure 23. The black dot marks the black hole, and the observer's line of sight is along the --$x^{\prime}$-axis. The hot-spot moves counterclockwise along the gray dashed trajectory (Orbit 5). When passing behind the black hole, its image appears as a ring or crescent.}}\label{fig24}
\end{figure*}
\subsection{CPU costs}
As a ray-tracing algorithm, computational efficiency is crucial, as it directly impacts both code execution and dissemination. We recorded the CPU time required by our algorithm to simulate black hole images (i.e., figures 11--13) at different resolutions, as summarized in table 3. The results show that OCTOPUS processes $10^{4}$ light rays in just 6 seconds. For most research purposes---such as accurately determining the critical curve position and inner shadow morphology---a resolution of $500 \times 500$ pixels is sufficient. Thus, our algorithm can simulate a black hole image in under 3 minutes, demonstrating commendable research efficiency.
\begin{table*}
\begin{center}
\small \caption{Computation time for simulating black hole images at different resolutions. Here, the dark matter halo parameters are fixed at $r_{\textrm{s}}=\rho_{\textrm{s}}=0.5$, with adaptive step-size parameters set to $n=1.8$, and $h_{0}=0.0002$. The maximum number of disk crossings is set to $N_{\textrm{max}}=4$, and the observation inclination is $50^{\circ}$. The computations were performed on a system with two AMD EPYC 7763 processors (128 logical cores in total), running at approximately 2.45 GHz, equipped with 64 GB of memory, under Windows 10, and without hyper-threading.} \label{t3}
\begin{tabular}{cccccccc}\hline
Resolution & $100^{2}$ & $200^{2}$ & $500^{2}$ & $1000^{2}$ & $1500^{2}$ & $2000^{2}$ & $4000^{2}$ \\
\hline
CPU costs [s] & $6$ & $24$ & $153$ & $589$ & $1328$ & $2463$ & $9452$ \\
\hline
\end{tabular}
\end{center}
\end{table*}

The resource consumption presented in table 3 represents a conservative estimate, as our algorithm retains the flexibility to further enhance computational efficiency by adjusting the adaptive step-size integration strategy---specifically, the step-size scaling exponent $n$ and the initial step-size $h_{0}$ fed into the RKF56 integrator. For instance, with $n=2$ and $h_{0}=0.001$, the convergence of RKF56 is accelerated while maintaining sufficient accuracy, leading to a significant improvement in computational performance. Table 4 lists the simulation times corresponding to different combinations of $n$ and $h_{0}$. It is noteworthy that for $n=2$ and $h_{0}=0.001$, generating a $1000 \times 1000$-resolution image requires less than 200 seconds. Moreover, we have verified that the black hole images produced with each parameter set in table 4 are consistent with those obtained at the same resolution in table 3, confirming that our chosen adaptive step-size strategy meets the precision requirements for geodesic integration.
\begin{table}[ht]
\centering
\caption{Computation time for simulating black hole images under different adaptive step-size strategies. Here, the resolution is fixed at $1000 \times 1000$ pixels, the field of view is set to $x \in [-15,15]$ M and $y \in [-15,15]$ M, the dark matter halo parameters are fixed at $ r_{\textrm{s}}=\rho_{\textrm{s}}=0.5$, and the observation inclination is set to $50^{\circ}$. We note that increasing either $n$ or $h_{0}$ effectively improves computational efficiency. Specifically, with $n=2$ and $h_{0}=0.001$, the simulation completes in under $200$ seconds. Importantly, all configurations listed in the table produce consistent and correct results, highlighting the code's significant potential for further optimization.}
\begin{tabular}{|c|c|c|}
\hline
\multicolumn{2}{|c|}{Parameters} & CPU costs [s] \\ \hline
\multirow{2}{*}{$n = 1.6$} 
& $h_{0} = 0.0001$ & 1760 \\  
& $h_{0} = 0.0005$ & 370 \\ 
& $h_{0} = 0.001$ & 223 \\ \hline
\multirow{2}{*}{$n = 1.8$} 
& $h_{0} = 0.0001$ & 1158 \\ 
& $h_{0} = 0.0005$ & 274 \\ 
& $h_{0} = 0.001$ & 195 \\ \hline
\multirow{2}{*}{$n = 2.0$} 
& $h_{0} = 0.0001$ & 816 \\  
& $h_{0} = 0.0005$ & 235 \\ 
& $h_{0} = 0.001$ & 191 \\ \hline
\end{tabular}
\end{table}
\section{Conclusions and Discussions}
This paper introduces OCTOPUS, a relativistic ray-tracing algorithm developed in Fortran with OpenMP parallelization, designed for static, spherically symmetric black hole spacetimes and publicly available on GitHub\footnote{https://github.com/Shiyang-Hu/Ray-Tracing-Code-OCTOPUS}. Using its built-in Newton iteration method, the algorithm computes key spacetime features, including the event horizon, critical photon orbit, critical impact parameter, and the ISCO radius. Furthermore, OCTOPUS employs high-precision RKF56 and RK6 integrators to efficiently simulate both null and timelike geodesics. This enables studies of accretion disk images, toroidal images, redshift factor distributions, gravitational lensing by static point sources, and hot-spot light curves. Notably, in support of multi-messenger studies correlating electromagnetic and gravitational radiation, OCTOPUS incorporates gravitational wave computations within an analytic kludge framework. All these functionalities require only the metric potential $f(r)$ and its first-, second-, and third-order radial derivatives as input. This design ensures minimal user barriers when switching between models. Additionally, the metric can be specified either analytically or numerically; in the latter case, the potential and its derivatives are obtained via interpolation functions. A separate work focusing on the numerical metric implementation in OCTOPUS is currently in preparation.

Using a Schwarzschild spacetime enveloped by a Dehnen-type dark matter halo, we comprehensively evaluated OCTOPUS in terms of accuracy, functionality, and computational efficiency. The results show that during the integration of null geodesics, OCTOPUS maintains a Hamiltonian error around $10^{-13}$ in most cases. Even in extreme regions prone to numerical singularities---such as near the black hole's polar regions or the event horizon---the Hamiltonian error remains below $10^{-6}$. Additionally, the relative error between the critical impact parameter obtained via ray-tracing and that derived analytically lies within the range of $10^{-4}$ to $10^{-3}$. These findings confirm that OCTOPUS achieves sufficient precision for scientific research. 

In terms of computational performance, OCTOPUS generates a $1000 \times 1000$-pixel black hole image in under 200 seconds, which is approximately equivalent to 244 seconds per $10^{4}$ rays per core. It should be noted that the algorithm's runtime depends on several factors, including the number of logical cores, processor frequency, memory capacity, operating system, and adaptive step-size parameters.

Using OCTOPUS, we uncover several intrinsic properties and potential observational signatures of the target black hole. We find that the event horizon radius, critical photon orbit, critical impact parameter, and ISCO radius all increase as the dark matter halo parameters grow. Notably, the scale parameter $r_{\textrm{s}}$ influences spacetime in an approximately exponential manner, while the density parameter $\rho_{\textrm{s}}$ exhibits a nearly linear effect. In terms of black hole images, increasing the dark matter halo parameters enlarges features such as the bright ring radius, inner shadow, and bright spot sizes. However, these parameters do not alter the shapes of these features---only the observation inclination can modify their morphology. We also simulate gravitational lensing by static point sources. In most configurations, the images appear crescent-shaped. When the source, observer, and black hole are perfectly aligned, an Einstein ring forms, with its radius depending on both the source position and spacetime parameters. Remarkably, for sufficiently large dark matter halo parameters, the ring can appear within the black hole shadow. These phenomena arise fundamentally because the dark matter halo acts effectively as a mass component; increasing its parameters strengthens the overall gravitational field.

Simulating hot-spot light curves and gravitational wave emissions is another key objective of OCTOPUS. We tested this capability using eight distinct timelike hot-spot orbits. The results indicate that peak characteristics in the light curves exhibit potential correlations with dark matter halo parameters and observation inclination. Additionally, the narrow dips observed in the light curves are primarily attributed to gravitational lensing effects. Regarding gravitational waveforms, signals from circular orbits are clearly distinguishable from those of quasi-periodic orbits. However, differences among various quasi-periodic orbits are subtle and mainly manifest in regions of abrupt signal variation. Extreme mass-ratio inspiral systems---composed of a hot-spot orbiting a central compact object---and their associated multi-messenger emissions represent a major topic in contemporary astrophysics. OCTOPUS provides a convenient, pipeline-ready solution for such studies, with the potential to guide future multi-messenger astronomy. Furthermore, the algorithm can generate dynamic visualizations of hot-spot motion on the observer's screen, offering a valuable tool for investigating flaring phenomena near compact objects.

In the future, we plan to extend OCTOPUS to handle polarization and enable its operation in magnetized spacetimes. Additionally, a version of OCTOPUS adapted for axisymmetric spacetimes is currently under active development and will be released in the near future.

\acknowledgments
This research has been supported by the National Natural Science Foundation of China [Grant No. 12403081]. For data visualization, we utilize the following Python libraries: NumPy, Matplotlib, Astropy, and SciPy.


\begin{thebibliography}{99}
\bibitem{Akiyama et al. (2019)}K. Akiyama et al. (Event Horizon Telescope Collaboration), \emph{First M87 Event Horizon Telescope Results. I. The shadow of the supermassive black hole}, \emph{Astrophys. J. Lett.} {\bf 875} (2019) L1.
\bibitem{Akiyama et al. (2022)}K. Akiyama et al. (Event Horizon Telescope Collaboration), \emph{First Sagittarius A$^{*}$ Event Horizon Telescope Results. I. The Shadow of the Supermassive Black Hole in the Center of the Milky Way}, \emph{Astrophys. J. Lett.} {\bf 930} (2022) L12.
\bibitem{Gralla et al. (2019)}S. E. Gralla, D. E. Holz, and R. M. Wald, \emph{Black hole shadows, photon rings, and lensing rings}, \emph{Phys. Rev. D} {\bf 100} (2019) 024018.
\bibitem{Narayan et al. (2019)}R. Narayan, M. D. Johnson, and C. F. Gammie, \emph{The Shadow of a Spherically Accreting Black Hole}, \emph{Astrophys. J. Lett.} {\bf 885} (2019) L33.
\bibitem{Tian and Zhu (2019)}S. X. Tian and Z. H. Zhu, \emph{Testing the Schwarzschild metric in a strong field region with the Event Horizon Telescope}, \emph{Phys. Rev. D} {\bf 100} (2019) 064011.
\bibitem{Zeng et al. (2020)}X. X. Zeng, H. Q. Zhang, and H. B. Zhang, \emph{Shadows and photon spheres with spherical accretions in the four-dimensional Gauss-Bonnet black hole}, \emph{Eur. Phys. J. C} {\bf 80} (2020) 872.
\bibitem{Li et al. (2020)}P. C. Li, M. Y. Guo, and B. Chen, \emph{Shadow of a spinning black hole in an expanding universe}, \emph{Phys. Rev. D} {\bf 101} (2020) 084041.
\bibitem{Zhang et al. (2020)}M. Zhang and M. Y. Guo, \emph{Can shadows reflect phase structures of black holes?}, \emph{Eur. Phys. J. C} {\bf 80} (2020) 790.
\bibitem{Hou et al. (2021a)}Y. H. Hou, M. Y. Guo, and B. Chen, \emph{Revisiting the shadow of braneworld black holes}, \emph{Phys. Rev. D} {\bf 104} (2021) 024001.
\bibitem{Peng et al. (2021)}J. Peng, M. Y. Guo, and X. H. Feng, \emph{Influence of quantum correction on black hole shadows, photon rings, and lensing rings}, \emph{Chin. Phys. C} {\bf 45} (2021) 085103.
\bibitem{Bronzwaer and Falcke (2021)}T. Bronzwaer and H. Falcke, \emph{The Nature of Black Hole Shadows}, \emph{Astrophys. J}. \textbf{920} (2021) 155.
\bibitem{Li and He (2021)}G. P. Li and K. J. He, \emph{Shadows and rings of the Kehagias-Sfetsos black hole surrounded by thin disk accretion}, \emph{J. Cosmol. Astropart. P.} {\bf 06} (2021) 037.
\bibitem{Vincent et al. (2022)}F. H. Vincent et al., \emph{Images and photon ring signatures of thick disks around black holes}, \emph{Astron. Astrophys}. \textbf{667} (2022) A170.
\bibitem{Afrin and Ghosh (2022)}M. Afrin and S. G. Ghosh, \emph{Testing Horndeski gravity from EHT observational results for rotating black holes}, \emph{Astrophys. J.} {\bf 932} (2022) 51.
\bibitem{Zeng et al. (2022)}X. X. Zeng, K. J. He, and G. P. Li, \emph{Effects of dark matter on shadows and rings of Brane-World black holes illuminated by various accretions}, \emph{Sci. China Phys. Mech} {\bf 65} (2022) 290411.
\bibitem{Hu et al. (2022a)}S. Y. Hu, C. Deng, D. Li, X. Wu, and E. W. Liang, \emph{Observational signatures of Schwarzschild-MOG black holes in scalar--tensor--vector gravity: shadows and rings with different accretions}, \emph{Eur. Phys. J. C} {\bf 82} (2022) 885.
\bibitem{Hou et al. (2022a)}Y. H. Hou, Z. Y. Zhang, H. P. Yan, M. Y. Guo, and B. Chen, \emph{Image of a Kerr-Melvin black hole with a thin accretion disk}, \emph{Phys. Rev. D} {\bf 106} (2022) 064058.
\bibitem{Hou et al. (2022b)}Y. H. Hou, P. Liu, M. Y. Guo, H. P. Yan, and B. Chen, \emph{Multi-level images around Kerr-Newman black holes}, \emph{Class. Quantum Grav.} {\bf 39} (2022) 194001.
\bibitem{Vagnozzi et al. (2023)}S. Vagnozzi, R. Roy, Y. D. Tsai, L. Visinelli, M. Afrin, A. Allahyari, P. Bambhaniya, D. Dey, S. G. Ghosh, P. S. Joshi, K. Jusufi, M. Khodadi, R. K. Walia, A. \"{O}vg\"{u}n, and C. Bambi, \emph{Horizon-scale tests of gravity theories and fundamental physics from the Event Horizon Telescope image of Sagittarius A$^{*}$}, \emph{Class. Quantum Grav.} {\bf 40} (2023) 165007.
\bibitem{Hu et al. (2023)}S. Y. Hu et al., \emph{Observational signatures of Schwarzschild-MOG black holes in scalar-tensor-vector gravity: images of the accretion disk}, \emph{Eur. Phys. J. C} \textbf{83} (2023) 264.
\bibitem{Heydari-Fard et al. (2023a)}M. Heydari-Fard, M. Heydari-Fard, and N. Riazi, \emph{Shadows and photon rings of a spherically accreting Kehagias-Sfetsos black hole}, \emph{Int. J. Mod. Phys. D} {\bf 32} (2023) 2350088.
\bibitem{Meng et al. (2023)}Y. Meng, X. M. Kuang, X. J. Wang, B. Wang, and J. P. Wu, \emph{Images from disk and spherical accretions of hairy Schwarzschild black holes}, \emph{Phys. Rev. D} {\bf 108} (2023) 064013.
\bibitem{Wang et al. (2023)}X. J. Wang, X. M. Kuang, Y. Meng, B. Wang, and J. P. Wu, \emph{Rings and images of Horndeski hairy black hole illuminated by various thin accretions}, \emph{Phys. Rev. D} {\bf 107} (2023) 124052.
\bibitem{Yang et al. (2023)}J. S. Yang, C. Zhang, and Y. G. Ma, \emph{Shadow and stability of quantum-corrected black holes}, \emph{Eur. Phys. J. C} {\bf 83} (2023) 619.
\bibitem{Gao et al. (2023)}X. J. Gao, T. T. Sui, X. X. Zeng, Y. S. An, and Y. P. Hu, \emph{Investigating shadow images and rings of the charged Horndeski black hole illuminated by various thin accretions}, \emph{Eur. Phys. J. C} {\bf 83} (2023) 1052.
\bibitem{Cao et al. (2023)}W. F. Cao, W. F. Liu, and X. Wu, \emph{Parameter constraints from shadows of Kerr-Newman-dS black holes with cloud strings and quintessence}, \emph{Gen. Relat. Gravit.} {\bf 55} (2023) 120.
\bibitem{Heydari-Fard et al. (2023b)}M. Heydari-Fard, S. G. Honarvar, and M. Heydari-Fard, \emph{Thin accretion disc luminosity and its image around rotating black holes in perfect fluid dark matter}, \emph{Mon. Not. R. Astron. Soc}. \textbf{521} (2023) 708.
\bibitem{Zhang et al. (2023)}Z. L. Zhang, S. B. Chen, and J. L. Jing, \emph{Images of Kerr-MOG black holes surrounded by geometrically thick magnetized equilibrium tori}, \emph{J. Cosmol. Astropart. P}. \textbf{09} (2024) 027.
\bibitem{Li et al. (2024a)}X. Q. Li, H. P. Yan, X. J. Yue, S. W. Zhou, and Q. Xu, \emph{Geodesic structure, shadow and optical appearance ofblack hole immersed in Chaplygin-like dark fluid}, \emph{J. Cosmol. Astropart. P.} {\bf 05} (2024) 048.
\bibitem{Li et al. (2024b)}D. Li et al., \emph{Observational features of deformed Schwarzschild black holes illuminated by an anisotropic accretion disk}, arXiv: 2409.01778.
\bibitem{Hu et al. (2024)}S. Y. Hu, D. Li, C. Deng, X. Wu, and E. W. Liang, \emph{Influences of tilted thin accretion disks on the observational appearance of hairy black holes in Horndeski gravity}, \emph{J. Cosmol. Astropart. P.} {\bf 04} (2024) 089.
\bibitem{Zhang et al. (2024a)}Z. Y. Zhang, H. P. Yan, M. Y. Guo, and B. Chen, \emph{Shadows of Kerr black holes with a Gaussian-distributed plasma in the polar direction}, \emph{Phys. Rev. D} {\bf 107} (2023) 024027.
\bibitem{Zhang et al. (2024b)}Z. Y. Zhang, Y. H. Hou, M. Y. Guo, and B. Chen, \emph{Observational signatures of rotating black holes in the semiclassical gravity with trace anomaly}, \emph{Chin. Phys. C} {\bf 48} (2024) 085106.
\bibitem{Chen et al. (2025a)}Y. F. Chen et al., \emph{Illuminating Black Hole Shadow with Dark Matter Annihilation}, \emph{Phys. Rev. Lett} {\bf 135} (2025) 121001.
\bibitem{Hu et al. (2025a)}S. Y. Hu, D. Li, and C. Deng, \emph{Novel inner shadows of the Kerr black hole with a tilted thin accretion disk}, \emph{J. Cosmol. Astropart. P}. \textbf{06} (2025) 036.
\bibitem{Li et al. (2025)}D. Li, Y. X. Zuo, S. Y. Hu, C. Deng, Y. Wang, and W. F. Cao, \emph{Shadows of three black holes in static equilibrium configuration}, \emph{Eur. Phys. J. C} \textbf{85} (2025) 905.
\bibitem{Yang et al. (2025)}C. Y. Yang, M. Israr Aslam, X. X. Zeng, and R. Saleem, \emph{Shadow images of Ghosh-Kumar rotating black hole illuminated by spherical light sources and thin accretion disks}, \emph{J. High Energy Astrop.} {\bf 46} (2025) 100345.
\bibitem{Guo et al. (2025)}S. Guo, E. W. Liang, Y. X. Huang, Y. Liang, and Q. Q. Jiang, \emph{Image of a time-dependent rotating regular black hole}, \emph{Sci. China Phys. Mech. Astron.} {\bf 68} (2025) 109512.
\bibitem{Liang et al. (2025)}Y. Liang, S. Guo, K. Lin, Y. H. Cui, and Y. X. Huang, \emph{Image of the time-dependent black hole}, \emph{Phys. Lett. B} {\bf 868} (2025) 139745.
\bibitem{Wang et al. (2025a)}P. Wang, S. Guo, L. F. Li, Z. F. Mai, and B. F. Wu, \emph{Optical images of the Kerr–Sen black hole and thin accretion disk}, {\bf 7} (2025) 747.
\bibitem{Chen et al. (2025b)}G. Chen, S. Guo, J. S. Li, Y. X. Huang, and L. F. Li, \emph{Influences of accretion flow and dilaton charge on the images of Einstein-Maxwell-dilation black holes}, \emph{Sci. China Phys. Mech. Astron.} {\bf 68} (2025) 260413.
\bibitem{Xiong et al. (2025)}Y. Xiong, J. Pu, G. P. Li, and G. M. Deng, \emph{Investigating the shadows of new regular black holes with a Minkowski core: effects of spherical accretion and core type differences}, \emph{Chin. Phys. C} {\bf 49} (2025) 095101.
\bibitem{He et al. (2025a)}K. J. He, G. P. Li, C. Y. Yang, and X. X. Zeng, \emph{Observational features of the rotating Bardeen black hole surrounded by perfect fluid dark matter}, \emph{Eur. Phys. J. C} \textbf{85} (2025) 662.
\bibitem{He et al. (2025b)}K. J. He, G. P. Li, C. Y. Yang, and X. X. Zeng, \emph{The observation image of a soliton boson star illuminated by various accretions}, \emph{J. Cosmol. Astropart. P}. \textbf{10} (2025) 003.
\bibitem{Zeng et al. (2025a)}X. X. Zeng, C. Y. Yang, M. Israr Aslam, R. Saleem, and S. Aslam, \emph{Kerr-like black hole surrounded by cold dark matter halo: the shadow images and EHT constraints}, \emph{J. Cosmol. Astropart. P}. \textbf{08} (2025) 066.
\bibitem{Chen et al. (2020)}Y. F. Chen et al., \emph{Probing Axions with Event Horizon Telescope Polarimetric Measurements}, \emph{Phys. Rev. Lett.} \textbf{124} (2020) 061102.
\bibitem{Zhang et al. (2021)}Z. L. Zhang, et al., \emph{Polarized image of a Schwarzschild black hole with a thin accretion disk as photon couples to Weyl tensor}, \emph{Eur. Phys. J. C} \textbf{81} (2021) 991.
\bibitem{Hu et al. (2022b)}Z. Z. Hu et al., \emph{Polarized images of synchrotron radiations in curved spacetime}, \emph{Eur. Phys. J. C} \textbf{82} (2022) 1166.
\bibitem{Qin et al. (2022)}X. Qin, S. B. Chen, and J. L. Jing, \emph{Polarized image of an equatorial emitting ring around a 4D Gauss--Bonnet black hole}, \emph{Eur. Phys. J. C} \textbf{82} (2022) 784.
\bibitem{Chen et al. (2022)}Y. F. Chen et al., \emph{Stringent axion constraints with Event Horizon Telescope polarimetric measurements of M87$^{*}$}, \emph{Nature Astron}. \textbf{6} (2022) 592.
\bibitem{Zhang et al. (2024c)}Z. Y. Zhang, Y. H. Hou, Z. Z. Hu, M. Y. Guo, and B. Chen, \emph{Polarized images of charged particles in vortical motions around a magnetized Kerr black hole}, \emph{J. Cosmol. Astropart. P.} {\bf 03} (2024) 013.
\bibitem{Shi et al. (2024)}H. Y. Shi and T. Zhu, \emph{Polarized image of a synchrotron emitting ring around a static hairy black hole in Horndeski theory}, \emph{Eur. Phys. J. C} \textbf{84} (2024) 814.
\bibitem{Chen et al. (2024a)}B. Chen, Y. H. Hou, Y. Song, and Z. Y Zhang, \emph{Polarization Patterns of the Hotspots Plunging into a Kerr Black Hole}, arXiv: 2407.14897.
\bibitem{Hou et al. (2025)}Y. H. Hou, J. W. Huang, M. Y. Guo, Y. Mizuno, and B. Chen, \emph{Near-horizon Polarization as a Diagnostic of Black Hole Spacetime}, \emph{Astrophys. J. Lett.} {\bf 988} (2025) L51.
\bibitem{Angelov et al. (2025)}T. Angelov, R. Bekir, G. Gyulchev, P. Nedkova, and S. Yazadjiev, \emph{Polarized equatorial emission and hot spots around black holes with a dark matter halo}, \emph{Eur. Phys. J. C} \textbf{85} (2025) 1075.
\bibitem{Wang et al. (2025b)}X. Y. Wang, S. B. Chen, M. Y. Guo, and B. Chen, \emph{Semi-analytical Study on the Polarized Images of Black Hole due to Frame Dragging}, arXiv: 2508.15178.
\bibitem{He et al. (2025c)}K. J. He, Y. W. Han, and G. P. Li, \emph{Viewing the holographic image of the Bardeen-AdS black hole}, \emph{Nucl. Phys. B} {\bf 1010} (2025) 116768.
\bibitem{Zeng et al. (2025b)}X. X. Zeng, M. Israr Aslam, R. Saleem, and X. Y. Hu, \emph{Schwarzschild-like AdS black holes: holographic imaging and LQG influences on Einstein rings}, \emph{Eur. Phys. J. C} \textbf{85} (2025) 663.
\bibitem{Chen et al. (2025c)}G. Chen, J. Y. Gui, X. X. Zeng, L. F. Li, and P. Li, \emph{Holographic Einstein ring of charged black holes in Kalb-Ramond field}, \emph{Sci. China Phys. Mech. Astron.} {\bf 55} (2025) 100411.
\bibitem{Chen et al. (2025d)}G. Chen, K. J. He, X. X. Zeng, M. J. Liang, L. F. Li, P. Li, and P. Xu, \emph{Holographic Einstein ring of charged phantom AdS black hole}, \emph{Front. Phys. (Beijing)} {\bf 20} (2025) 35203.
\bibitem{Zeng et al. (2025c)}X. X. Zeng, L. F. Li, P. Li, B. Liang, and P. Xu, \emph{Holographic images of a charged black hole in Lorentz symmetry breaking massive gravity}, \emph{Sci. China Phys. Mech. Astron.} {\bf 68} (2025) 220412.
\bibitem{Huang et al. (2024a)}J. W. Huang, Z. Y. Zhang, Y. M. Guo, and B. Chen, \emph{Images and flares of geodesic hot spots around a Kerr black hole}, \emph{Phys. Rev. D} {\bf 109} (2024) 124062.
\bibitem{Chen et al. (2024b)}Y. Q. Chen, P. Wang, and H. T. Yang, \emph{Observations of Orbiting Hot Spots around Scalarized Reissner-Nordstr\"{o}om Black Holes}, arXiv: 2401.10905.
\bibitem{Wu and Chen (2024)}T. S. Wu and Y. Q. Chen, \emph{Distinguishing the observational signatures of hot spots orbiting Reissner-Nordstr\"{o}m spacetime}, \emph{Chin. Phys. C} {\bf 48} (2024) 075103.
\bibitem{Zhu (2025a)}Q. H. Zhu, \emph{Observational signatures from higher-order images of moving hotspots in accretion disks}, \emph{Phys. Rev. D} {\bf 111} (2025) 044010.
\bibitem{Zhu (2025b)}Q. H. Zhu, \emph{Auto- and cross-correlations for multiple images of corotating hotspots in accretion disks}, \emph{Phys. Rev. D} {\bf 112} (2025) 064021.
\bibitem{Zhang et al. (2025)}Z. Y. Zhang, Y. H. Hou, M. Y. Guo, Y. Mizuno, and B. Chen, \emph{Autocorrelation signatures in time-resolved black hole flare images: secondary peaks and convergence structure}, arXiv: 2503.17200.
\bibitem{Gralla and Lupsasca (2021)}S. E. Gralla and A. Lupsasca, \emph{Lensing by Kerr black holes}, \emph{Phys. Rev. D} {\bf 101} (2020) 044031.
\bibitem{Gao and Xie (2021)}Y. X. Gao and Y. Xie, \emph{Gravitational lensing by hairy black holes in Einstein-scalar-Gauss-Bonnet theories}, \emph{Phys. Rev. D} {\bf 103} (2021) 043008.
\bibitem{Wu et al. (2021)}W. H. Wu, C. Y. Zhang, C. G. Shao, and W. L. Qiang, \emph{Onset of chaotic gravitational lensing in non-Kerr rotating black holes with quadrupole mass moment}, \emph{Chin. Phys. C} {\bf 47} (2023) 085102.
\bibitem{He et al. (2024)}G. S. He, Y. Xie, C. H. Jiang, and W. B. Lin, \emph{Gravitational lensing of massive particles by a black-bounce-Schwarzschild black hole}, \emph{Phys. Rev. D} {\bf 110} (2024) 064008.
\bibitem{He et al. (2025d)}G. S. He, W. B. Lin, and Y. Xie, \emph{Leading-order gravitational time delay of massive particles by a moving Schwarzschild lens}, \emph{Phys. Rev. D} {\bf 111} (2025) 124055.
\bibitem{Wang et al. (2025c)}X. Wang, W. B. Lin, G. Mustafa, and G. S. He, \emph{Leading-order deflection of particles by a moving Schwarzschild lens with a two-dimensional velocity}, \emph{Phys. Rev. D} {\bf 111} (2025) 084009.
\bibitem{Wang et al. (2025d)}X. Wang, W. B. Lin, B. Yang, and G. S. He, \emph{Gravitational deflection of light in the polar-axis plane of a moving Kerr-Newman black hole}, \emph{Commun. Theor. Phys.} {\bf 77} (2025) 075403.
\bibitem{Huang et al. (2025)}Y. Huang, D. J. Liu, and H. S. Zhang, \emph{Lensing and light rings of parity-odd rotating boson stars}, \emph{Sci. China Phys. Mech. Astron.} {\bf 68} (2025) 280411.
\bibitem{Porth et al. (2019)}O. Porth, K. Chatterjee, R. Narayan, et al., \emph{The Event Horizon General Relativistic Magnetohydrodynamic Code Comparison Project}, \emph{Astrophys. J. Suppl. S.} {\bf 243} (2019) 26.
\bibitem{Wu et al. (2006)}K. Wu, S. V. Fuerst, K. Lee, and G. Branduardi-Raymont, \emph{General Relativistic Radiative Transfer: Emission from Accreting Black Holes in AGN}, \emph{Chin. J. Astron. Astrophys.} {\bf 6} (2006) 205-220.
\bibitem{Younsi et al. (2012)}Z. Younsi, K. Wu, and S. V. Fuerst, \emph{General relativistic radiative transfer: formulation and emission from structured tori around black holes}, \emph{Astron. Astrophys}. \textbf{545} (2012) A13.
\bibitem{Gammie and Leung (2012)}C. F. Gammie and P. K. Leung, \emph{A Formalism for Covariant Polarized Radiative Transport by Ray Tracing}, \emph{Astrophys. J.} {\bf 752} (2012) 123.
\bibitem{Dexter (2016)}J. Dexter, \emph{A public code for general relativistic, polarised radiative transfer around spinning black holes}, \emph{Mon. Not. R. Astron. Soc}. \textbf{462} (2016) 115-136.
\bibitem{Takahashi and Umemura (2017)}R. Takahashi and M. Umemura, \emph{General relativistic radiative transfer code in rotating black hole space-time: ARTIST}, \emph{Mon. Not. R. Astron. Soc}. \textbf{464} (2017) 4567-4585.
\bibitem{Pelle et al. (2022)}J. Pelle, O. Reula, F. Carrasco, and C. Bederian, \emph{Skylight: a new code for general-relativistic ray-tracing and radiative transfer in arbitrary space-times}, \emph{Mon. Not. R. Astron. Soc}. \textbf{515} (2022) 1316-1327.
\bibitem{Kawashima et al. (2023)}T. Kawashima, K. Ohsuga, and H. R. Takahashi, \emph{RAIKOU: A General Relativistic, Multiwavelength Radiative Transfer Code}, \emph{Astrophys. J.} {\bf 949} (2023) 101.
\bibitem{Sharma et al. (2023)}A. Sharma, L. Medeiros, C. Chan, G. Halevi, P. D. Mullen, J. M. Stone, G. N. Wong, \emph{Mahakala: a Python-based Modular Ray-tracing and Radiative Transfer Algorithm for Curved Space-times}, arXiv: 2304.03804.
\bibitem{Huang et al. (2024b)}J. W. Huang, L. H. Zheng, M. Y. Guo, and B. Chen, \emph{Coport: a new public code for polarized radiative transfer in a covariant framework}, \emph{J. Cosmol. Astropart. P.} {\bf 11} (2024) 054.
\bibitem{Gralla et al. (2020)}S. E. Gralla, A. Lupsasca, and D. P. Marrone, \emph{The shape of the black hole photon ring: A precise test of strong-field general relativity}, \emph{Phys. Rev. D} {\bf 102} (2020) 124004.
\bibitem{Chael et al. (2021)}A. Chael, M. D. Johnson, and A. Lupsasca, \emph{Observing the Inner Shadow of a Black Hole: A Direct View of the Event Horizon}, \emph{Astrophys. J.} {\bf 918} (2021) 6.
\bibitem{Hou et al. (2021b)}Y. H. Hou, Z. Y. Zhang, M. Y. Guo, and B. Chen, \emph{A new analytical model of magnetofluids surrounding rotating black holes}, \emph{J. Cosmol. Astropart. P.} {\bf 02} (2024) 030. 
\bibitem{Zhou et al. (2025)}L. H. Zhou et al., \emph{Forward Ray Tracing and Hot Spots in Kerr Spacetime}, \emph{Phys. Rev. D} {\bf 111} (2025) 064075.
\bibitem{Karas et al. (1992)}V. Karas, D. Vokrouhlick\'{y}, and G. Polnarev, \emph{In the vicinity of a rotating black hole a fast numerical code for computing observational effects}, \emph{Mon. Not. R. Astron. Soc}. \textbf{259} (1992) 569-575.
\bibitem{Vincent et al. (2011)}F. H. Vincent, T. Paumard, E. Gourgoulhon, and G. Perrin, \emph{GYOTO: a new general relativistic ray-tracing code}, \emph{Class. Quantum Grav.} {\bf 28} (2011) 225011.
\bibitem{Johannsen and Psaltis (2010a)}T. Johannsen and D. Psaltis, \emph{Testing the No-hair Theorem with Observations in the Electromagnetic Spectrum. I. Properties of a Quasi-Kerr Spacetime}, \emph{Astrophys. J.} {\bf 716} (2010) 187.
\bibitem{Johannsen and Psaltis (2010b)}T. Johannsen and D. Psaltis, \emph{Testing the No-hair Theorem with Observations in the Electromagnetic Spectrum. II. Black Hole Images}, \emph{Astrophys. J.} {\bf 718} (2010) 446.
\bibitem{Psaltis and Johannsen (2012)}D. Psaltis and T. Johannsen, \emph{A Ray-tracing Algorithm for Spinning Compact Object Spacetimes with Arbitrary Quadrupole Moments. I. Quasi-Kerr Black Holes}, \emph{Astrophys. J.} {\bf 745} (2012) 1.
\bibitem{Baubock et al. (2012)}M. Baub\"{o}ck, D. Psaltis, F. \"{O}zel, and T. Johannsen, \emph{A Ray-tracing Algorithm for Spinning Compact Object Spacetimes with Arbitrary Quadrupole Moments. II. Neutron Stars}, \emph{Astrophys. J.} {\bf 753} (2012) 175.
\bibitem{Chen et al. (2013)}C. K. Chen, D. Psaltis, and F. \"{O}zel, \emph{GRay: A Massively Parallel GPU-based Code for Ray Tracing in Relativistic Spacetimes}, \emph{Astrophys. J.} {\bf 777} (2013) 13.
\bibitem{Chen et al. (2018)}C. K. Chen, L. Medeiros, F. \"{O}zel, and D. Psaltis, \emph{GRay2: A General Purpose Geodesic Integrator for Kerr Spacetimes}, \emph{Astrophys. J.} {\bf 867} (2018) 59.
\bibitem{Bambi (2012)}C. Bambi, \emph{A code to compute the emission of thin accretion disks in non-Kerr space-times and test the nature of black hole candidates}, \emph{Astrophys. J.} {\bf 761} (2012) 174.
\bibitem{Chen et al. (2015)}B. Chen, R. Kantowski, X. Y. Dai, E. Baron, and P. Maddumage, \emph{Algorithms and Programs for Strong Gravitational Lensing In Kerr Space-time Including Polarization}, \emph{Astrophys. J. Suppl. S.} {\bf 218} (2015) 4.
\bibitem{Yang and Wang (2013)}X. L. Yang and J. C. Wang, \emph{YNOGK: A new public code for calculating null geodesics in the Kerr spacetime}, \emph{Astrophys. J. Suppl. S.} {\bf 207} (2013) 6.
\bibitem{Luminet (1979)}J. -P. Luminet, \emph{Image of a spherical black hole with thin accretion disk}, \emph{Astron. Astrophys}. \textbf{75} (1979) 228.
\bibitem{Alejandro et al. (2023)}A. C\'{a}rdenas-Avenda\~{n}o, A. Lupsasca, and H. R. Zhu, \emph{Adaptive analytical ray tracing of black hole photon rings}, \emph{Phys. Rev. D} {\bf 107} (2023) 043030.
\bibitem{Wang et al. (2024)}K. Wang, C. J. Feng, and T. Wang, \emph{Image of Kerr-de Sitter black holes illuminated by equatorial thin accretion disks}, \emph{Eur. Phys. J. C} \textbf{84} (2024) 457.
\bibitem{Guo et al. (2024)}S. Guo, Y. X. Huang, E. W. Liang, Y. Liang, and Q. Q. Jiang, \emph{Image of the Kerr–Newman Black Hole Surrounded by a Thin Accretion Disk}, \emph{Astrophys. J.} {\bf 975} (2024) 237.
\bibitem{Younsi et al. (2016)}Z. Younsi, A. Zhidenko, L. Rezzolla, R. Konoplya, and Y. Mizuno, \emph{New method for shadow calculations: application to parametrized axisymmetric black holes}, \emph{Phys. Rev. D} {\bf 94} (2016) 084025.
\bibitem{Pu et al. (2016)}H. Y. Pu, K. Yun, Z. Younsi, and S. J. Yoon, \emph{Odyssey: a public GPU-based code for general-relativistic radiative transfer in Kerr spacetime}, \emph{Astrophys. J.} {\bf 820} (2016) 105.
\bibitem{Lin et al. (2022)}F. L. Lin, A. Patel, and H. Y. Pu, \emph{Black hole shadow with soft hairs}, \emph{J. High Energy Phys.} {\bf 09} (2022) 117.
\bibitem{Kimpson et al. (2019)}T. Kimpson, K. Wu, and S. Zane, \emph{Spatial dispersion of light rays propagating through a plasma in Kerr space-time}, \emph{Mon. Not. R. Astron. Soc}. \textbf{484} (2019) 2411-2419.
\bibitem{Cunha et al. (2015)}P. V. P. Cunha, C. A. R. Herdeiro, E. Radu, and H. F. R\'{u}narsson, \emph{Shadows of Kerr Black Holes with Scalar Hair}, \emph{Phys. Rev. Lett} {\bf 115} (2015) 211102.
\bibitem{Cunha et al. (2016)}P. V. P. Cunha, C. A. R. Herdeiro, E. Radu, and H. F. R\'{u}narsson, \emph{Shadows of Kerr black holes with and without scalar hair}, \emph{Int. J. Mod. Phys. D} \textbf{25} (2016) 1641021.
\bibitem{Hu et al. (2021)}Z. Z. Hu, Z. Zhong, P. C. Li, M. Y. Guo, and B. Chen, \emph{QED effect on a black hole shadow}, \emph{Phys. Rev. D} {\bf 103} (2021) 044057.
\bibitem{Zhong et al. (2021)}Z. Zhong, Z. Z. Hu, H. P. Yan, M. Y. Guo, and B. Chen, \emph{QED effects on Kerr black hole shadows immersed in uniform magnetic fields}, \emph{Phys. Rev. D} {\bf 104} (2021) 104028.
\bibitem{Dexter and Agol (2009)}J. Dexter and E. Agol, \emph{A Fast New Public Code for Computing Photon Orbits in a Kerr Spacetime}, \emph{Astrophys. J.} {\bf 696} (2009) 1616.
\bibitem{Muller and Frauendiener (2012)}T. M\"{u}ller and J. Frauendiener, \emph{Interactive visualization of a thin disc around a Schwarzschild black hole}, \emph{Eur. J. Phys.} \textbf{33} (2012) 955.
\bibitem{Muller (2014)}T. M\"{u}ller, \emph{GeoViS---Relativistic ray tracing in four-dimensional spacetimes}, \emph{Comput. Phys. Commun.} {\bf 185} (2014) 2301.
\bibitem{James et al. (2015)}O. James, E. von Tunzelmann, P. Franklin, and K. S. Thorne, \emph{Gravitational lensing by spinning black holes in astrophysics, and in the movie Interstellar}, \emph{Class. Quantum Grav.} {\bf 32} (2015) 065001.
\bibitem{Ayzenberg and Yunes (2018)}D. Ayzenberg and N. Yunes, \emph{Black hole shadow as a test of general relativity: quadratic gravity}, \emph{Class. Quantum Grav.} {\bf 35} (2018) 235002.
\bibitem{Huang et al. (2018)}Y. Huang, Y. P. Dong, and D. J. Liu, \emph{Revisiting the shadow of a black hole in the presence of a plasma}, \emph{Int. J. Mod. Phys. D} \textbf{27} (2018) 1850114.
\bibitem{Cadavid et al. (2022)}J. M. Vel\'{a}squez-Cadavid et al., \emph{OSIRIS: a new code for ray tracing around compact objects}, \emph{Eur. Phys. J. C} \textbf{82} (2022) 103.
\bibitem{Davelaar and Haiman (2022a)}J. Davelaar and Z. Haiman, \emph{Self-lensing flares from black hole binaries: General-relativistic ray tracing of black hole binaries}, \emph{Phys. Rev. D} {\bf 105} (2022) 103010.
\bibitem{Garnier (2023)}A. Garnier, \emph{Motion equations in a Kerr–Newman–de Sitter spacetime: some methods of integration and application to black holes shadowing in Scilab}, \emph{Class. Quantum Grav.} {\bf 40} (2023) 135011.
\bibitem{Gou et al. (2010)}L. Gou, et al., \emph{The spin of the black hole in the soft X-ray transient A0620-00}, \emph{Astrophys. J. Lett.} {\bf 718}, (2010) L122.
\bibitem{Steiner et al. (2012)}J. F. Steiner, J. E. McClintock, and M. J. Reid, \emph{The Distance, Inclination, and Spin of the Black Hole Microquasar H1743-322}, \emph{Astrophys. J. Lett.} {\bf 745}, (2012) L7.
\bibitem{Steiner et al. (2014)}J. F. Steiner, et al., \emph{The low-spin black hole in LMC-X3}, \emph{Astrophys. J. Lett.} {\bf 793}, (2014) L29.
\bibitem{Bohn et al. (2015)}A. Bohn, W. Throwe, F. H\'{e}bert, K. Henriksson, D. Bunandar, M. A. Scheel, and N. W. Taylor, \emph{What does a binary black hole merger look like?} \emph{Class. Quantum Grav.} {\bf 32} (2015) 065002.
\bibitem{Genzel et al. (2003)}R. Genzel, R. Sch\"{o}del, T. Ott, et al., \emph{Near-infrared flares from accreting gas around the supermassive black hole at the Galactic Centre}, \emph{Nature} {\bf 425} (2003) 934.
\bibitem{Li et al. (2014)}Z. L. Li, L. Y. Kong, and C. Bambi, \emph{Testing the Nature of the Supermassive Black Hole Candidate in SgrA$^{*}$ with Light Curves and Images of Hot Spots}, \emph{Astrophys. J.} {\bf 787} (2014) 152.
\bibitem{Baubock et al. (2020)}M. Baub\"{o}ck, J. Dexter, R. Abuter, et al., \emph{Modeling the orbital motion of Sgr A$^{*}$'s near-infrared flares}, \emph{Astron. Astrophys.} {\bf 635} (2020) A143.
\bibitem{Matsumoto et al. (2020)}T. Matsumoto, C. H. Chan, and T. Piran, \emph{The origin of hotspots around Sgr A$^{*}$: orbital or pattern motion?} \emph{Mon. Not. R. Astron. Soc.} {\bf 497} (2020) 2385.
\bibitem{Ball et al. (2021)}D. Ball, F. \"{O}zel, P. Christian, et al., \emph{A plasmoid model for the Sgr A$^{*}$ flares observed with Gravity and CHANDRA}, \emph{Astrophys. J.} {\bf 917} (2021) 8.
\bibitem{Davelaar and Haiman (2022b)}J. Davelaar and Z. Haiman, \emph{Self-Lensing Flares from Black Hole Binaries: Observing Black Hole Shadows via Light Curve Tomography}, \emph{Phys. Rev. Lett.} {\bf 128} (2022) 191101.
\bibitem{Yfantis et al. (2024)}I. Yfantis, M. A. Mo\'{s}cibrodzka, M. Wielgus, et al., \emph{Fitting the light curves of Sagittarius A$^{*}$ with a hot-spot model Bayesian modeling of QU loops in the millimeter band}, \emph{Astron. Astrophys.} {\bf 685} (2024) A142.
\bibitem{Hu et al. (2025b)}S. Y. Hu, D. Li, and C. Deng, \emph{Light Curves of Chaotic Charged Hot-Spots in Curved Spacetime: Opening an Observational Window to Chaos }, arXiv: 2508.17384.
\bibitem{Levin and Perez-Giz (2008)}J. Levin and G. Perez-Giz, \emph{A periodic table for black hole orbits}, \emph{Phys. Rev. D} {\bf 77} (2008) 103005.
\bibitem{Liu et al. (2019)}C. Q. Liu, C. K. Ding, and J. L. Jing, \emph{Periodic Orbits Around Kerr Sen Black Holes}, \emph{Commun. Theor. Phys.} {\bf 71} (2019) 1461.
\bibitem{Deng (2020a)}X. M. Deng, \emph{Geodesics and periodic orbits around quantum-corrected black holes}, \emph{Phys. Dark Universe} {\bf 30} (2020) 100629.
\bibitem{Deng (2020b)}X. M. Deng, \emph{Periodic orbits around brane-world black holes}, \emph{Eur. Phys. J. C} {\bf 80} (2020) 489.
\bibitem{Wang et al. (2022)}R. F. Wang, F. B. Gao, and H. X. Chen, \emph{Periodic orbits around a static spherically symmetric black hole surrounded by quintessence}, \emph{Ann. Phys-New York} {\bf 447} (2022) 169167.
\bibitem{Lin and Deng (2023)}H. Y. Lin and X. M. Deng, \emph{Precessing and periodic orbits around hairy black holes in Horndeski's Theory}, \emph{Eur. Phys. J. C} {\bf 83} (2023) 311.
\bibitem{Tu et al. (2023)}Z. Y. Tu, T. Zhu, and A. Z. Wang, \emph{Periodic orbits and their gravitational wave radiations in a polymer black hole in loop quantum gravity}, \emph{Phys. Rev. D} {\bf 108} (2023) 024035.
\bibitem{Huang and Deng (2024)}L. Huang and X. M. Deng, \emph{Can a particle's motion distinguish scale-dependent Planck stars from renormalization group improved Schwarzschild black holes?} \emph{Phys. Rev. D} {\bf 109} (2024) 124005.
\bibitem{Babak et al. (2007)}S. Babak, H. Fang, J. R. Gair, K. Glampedakis, and S. A. Hughes, \emph{``Kludge'' gravitational waveforms for a test-body orbiting a Kerr black hole}, \emph{Phys. Rev. D} {\bf 75} (2007) 024005.
\bibitem{Bertone et al. (2005)}G. Bertone, D. Hooper, and J. Silk, \emph{Particle dark matter: evidence, candidates and constraints}, \emph{Phys. Rep.} {\bf 405} (2005) 279-390.
\bibitem{Clifton et al. (2012)}T. Clifton, P. G. Ferreira, A. Padilla, and C. Skordis, \emph{Modified gravity and cosmology}, \emph{Phys. Rep.} {\bf 513} (2012) 1.
\bibitem{Planck (2020)}Planck Collaboration, \emph{Planck 2018 results. VI. Cosmological parameters}, \emph{Astron. Astrophys.} {\bf 641} (2020) A6.
\bibitem{Chakrabarti et al. (2025)}S. Chakrabarti, P. Chang, S. Profumo, and P. Craig, \emph{Constraints on a dark matter sub-halo near the Sun from pulsar timing}, arXiv: 2507.16932.
\bibitem{Gu et al. (2025)}G. Gu, X. M. Wang, Y. T. Wang, et al., \emph{Dynamical dark energy in light of the DESI DR2 baryonic acoustic oscillations measurements}, \emph{Nat. Astron.} (2025) https://doi.org/10.1038/s41550-025-02669-6
\bibitem{Xu et al. (2018)}Z. Y. Xu, X. Hou, X. B. Gong, and J. C. Wang, \emph{Black hole space-time in dark matter halo}, \emph{J. Cosmol. Astropart. P.} {\bf 09} (2018) 038.
\bibitem{Xu et al. (2020)}Z. Y. Xu, X. B. Gong, and S. N. Zhang, \emph{Black hole immersed dark matter halo}, \emph{Phys. Rev. D} {\bf 101} (2020) 024029.
\bibitem{Gohain et al. (2024)}M. M. Gohain, P. Phukon, and K. Bhuyan, \emph{Thermodynamics and null geodesics of a Schwarzschild black hole surrounded by a Dehnen type dark matter halo}, \emph{Phys. Dark Universe} {\bf 46} (2024) 101683. 
\bibitem{Jha (2024)}S. K. Jha, \emph{Shadow, ISCO, Quasinormal modes, Hawking spectrum, Weak Gravitational lensing, and parameter estimation of a Schwarzschild Black Hole Surrounded by a Dehnen Type Dark Matter Halo}, arXiv: 2407.18509.
\bibitem{Al-Badawi and Shaymatov (2024)}A. Al-Badawi and S. Shaymatov, \emph{Quasinormal modes and shadow of Schwarzschild black holes embedded in a Dehnen type dark matter halo exhibiting string cloud}, arXiv: 2412.20037.
\end{thebibliography}
\end{document}